\documentclass{aa}  

\usepackage{graphicx}
\usepackage{txfonts}

\usepackage{amsmath}
\usepackage{amssymb}
\usepackage{sistyle}
\usepackage{enumitem}
\usepackage{multirow}
\usepackage{xcolor}
\usepackage{natbib}

\newcommand{\cmc}{\mbox{$\mbox{cm}^{-3}$}}

\newcommand{\msun}{\mbox{$M_\odot$}}

\newcommand{\micron}{\mbox{$\mu$m}}
\newcommand{\sapt}{\itshape Saptarsy\upshape}
\newcommand{\dvodpk}{{\itshape DVODPK}}
\newcommand{\radmc}{{\itshape RADMC-3D}}
\newcommand{\pando}{{\itshape Pandora}}
\newcommand{\Td}{$T_\textnormal{\tiny{d}}$}

\newcommand{\Go}{$G$}
\newcommand{\Jnu}{$J_\nu$}
\newcommand{\hii}{H{\sc ii}}
\newcommand{\nh}{$n_\textnormal{H}$}

\newcommand{\teff}{\mbox{$T_{\mbox{\tiny eff}}$}}
\newcommand{\Ltot}{\mbox{$L_{\mbox{\tiny tot}}$}}
\newcommand{\av}{A$_\mathrm{V}$}

\newcommand{\hcn}{HC$^{15}$N}
\newcommand{\hnc}{HN$^{13}$C}
\newcommand{\hcop}{HCO$^+$}
\newcommand{\form}{H$_2$CO}
\newcommand{\meth}{CH$_3$OH}
\newcommand{\ntohp}{N$_2$H$^+$}

\begin{document} 

   \title{Chemical modeling of internal photon-dominated regions surrounding deeply embedded HC/UC\hii\ regions.}

   \author{G.~St\'ephan\inst{1,2}
          \and
          P.~Schilke\inst{1}
          \and
          J.~Le Bourlot\inst{2,3} 
          \and
          A.~Schmiedeke\inst{1,4}
          \and
          R.~Choudhury\inst{1,4}
          \and
          B.~Godard\inst{2}
          \and
          \'A. S\'anchez-Monge\inst{1}
          }

   \institute{I. Physikalisches Institut, Universit\"at zu K\"oln, Z\"ulpicherstrasse 77, 50937 K\"oln, Germany \\
              \email{stephan@ph1.uni-koeln.de \thanks{Current affiliation: Department of Chemistry, University of Virginia, McCormick Road, Charlottesville, VA 22904, USA -- \email{gs6dz@virginia.edu}}} 
         \and
             LERMA, Observatoire de Paris, 5 place Jules Janssen, 92190, Meudon, France
         \and
             Universit\'e Paris-Diderot, Sorbonne Paris-Cit\'e, 75013 Paris, France
         \and
             Max-Planck-Institut f\"ur extraterrestrische Physik, Giessenbachstrasse 1, 85748, Garching, Germany
             }

   \date{Received February 16, 2017; accepted February 09, 2018}

 
  \abstract
   {}
   {We aim to investigate the chemistry of internal photon-dominated regions (PDRs) surrounding deeply embedded hypercompact (HC) and ultracompact (UC) \hii\ regions. We search for specific tracers of this evolutionary stage of massive star formation that can be detected with current astronomical facilities.}
   {We modeled hot cores with embedded HC/UC\hii\ regions (called \hii\ region models in the article despite the fact that we do not model the \hii\ region itself), by coupling the astrochemical code \sapt\ to a radiative transfer framework obtaining the spatio-temporal evolution of abundances as well as time-dependent synthetic spectra. In these models where we focused on the internal PDR surrounding the \hii\ region, the gas temperature is set to the dust temperature and we do not include dynamics thus the density structure is fixed. We compared this to hot molecular core (HMC) models and studied the effect on the chemistry of the radiation field which is included in the \hii\ region models only during the computation of abundances. In addition, we investigated the chemical evolution of the gas surrounding \hii\ regions with models of different densities at the ionization front, different sizes of the ionized cavity and different initial abundances.}
   {We obtain the time evolution of synthetic spectra for a dozen of selected species as well as ratios of their integrated intensities. We find that some molecules such as C, N$_2$H$^+$, CN, and HCO do not trace the inner core and so are not good tracers to distinguish the \hii/PDR regions to the HMCs phase. On the contrary, C$^+$ and O trace the internal PDRs, in the two models starting with different initial abundances, but are unfortunately currently unobservable with the current achievable spatial resolution because of the very thin internal PDR ($\Delta$ r$_{\rm PDR}$ < 100~AU). The emission of these two tracers is very dependent on the size of the \hii\ region and on the density in the PDR.
   In addition, we find that the abundance profiles are highly affected by the choice of the initial abundances, hence the importance to properly define them.}
   {}


   \maketitle
%
\section{Introduction}

The chemical evolution of hot molecular cores (HMCs) has been the focus of many studies \citep{Garrod&Herbst2006, Vasyunin&Herbst2013, Garrod2013a} as these objects often present a very rich and complex chemistry. HMCs are compact ($\leq$ 0.10~pc), very dense (\nh\ $\geq$ 10$^6$~\cmc), hot (T $\geq$ 100~K) and are a characteristic early stage of massive star formation \citep{Kurtz2000}. They are transient objects with a lifetime, although not well constrained, of about 10$^4$--10$^5$~years \citep{Herbst&VanDishoeck2009}. It is still unclear if they are formed before the hyper compact \hii\ (HC\hii) region stage or coexist with it \citep{Gaume1995, Kurtz2005, Beuther2007}. HC\hii\ regions are compact (<~0.01~pc) and very dense (>~10$^6$~\cmc) ionized gas regions (with T$_{\rm e}$ $\sim$ 10$^4$~K) surrounding recently born high-mass stars still deeply embedded in their parental clouds (e.g., \citealt{Kurtz2000, Hoare2007, SanchezMonge2013}). The small sizes are likely due to the high accretion rates and high densities of the first evolutionary stages. HC\hii\ regions are expected to evolve into ultracompact \hii\ (UC\hii) regions as the embedded proto-star(s) become more luminous and gradually ionize their surroundings, and the accretion rates and surrounding density decrease. This evolution is not necessarily homogeneous as \hii\ regions may flicker as proposed theoretically by \cite{Peters2010a, Peters2010b} and found observationally by \cite{DePree2014}. UC\hii\ regions are slightly larger (0.1~pc) with densities of about 10$^4$~\cmc\ (e.g., \citealt{Wood&Churchwell1989, Kurtz2005}).

The small size of HC and UC\hii\ regions results in very strong ultraviolet (UV) radiation fields at the ionization front, that is, at the interface between the ionized gas and the molecular gas. This results in photon-dominated regions (PDRs) formed at the border between the \hii\ region and the molecular dense cloud where they are embedded. These so-called internal PDRs are very dense (\nh\ $\geq$ 10$^6$~\cmc) and very thin because of the fast decrease of the radiation field ($\Delta$ r$_{\mathrm{PDR}}$ $\approx$ 200~AU, corresponding to \av\ = 1.6~mag obtained with the standard Milky-Way extinction curve). Most chemical models do not extend their studies to the evolutionary stages associated with HC\hii\ and UC\hii\ regions. So far, only a few studies have investigated internal PDRs (e.g., \citealt{Bruderer2009c, Bruderer2009b, Lee2015}) in the case of an outflow cavity around low-mass stars with weak UV fields.

Observationally, HMCs and deeply embedded HC and UC\hii\ regions have comparable spectral energy distributions in the infrared (IR) and sub-mm regime (e.g., \citealt{Churchwell2002, Cesaroni2015}). However during the HMC phase the proto-star is not sufficiently evolved and does not radiate large enough quantities of UV photons to have a bright free-free thermal emission from ionized gas. 
Thus, HMCs are weak or even unobservable in the radio-continuum regime \citep{Hoare2007, Beuther2007}. This is the main observational difference between these two evolutionary stages of massive star formation. Do other observational methods perform better in differentiating these two phases? Can these phases be characterized by some specific chemical tracers? Can the presence of the internal PDRs be detected and how?

In this work we study the chemistry of the gas associated with HC/UC\hii\ regions and their internal PDRs, without looking at the emission from the gas in the ionized cavity but only the gas in the PDR, as well as in HMCs before they are largely affected by UV radiation. We study the spatio-temporal evolution of the radiation field in \hii\ region models as well as the evolution of the dust temperature, and how these physical parameters affect the chemistry and emissivity of different atomic and molecular tracers. Density is a parameter fixed in time in these models but we investigate the variation in the chemistry due to the density at the ionization front (i.e., density in the neutral gas just beyond the front), the different sizes of the ionized cavity, the density structure and the initial elementary abundances.

The paper is structured as follows: In Sect.~\ref{sec:method} we describe the HMC and \hii\ regions models, the physical parameters as well as the chemical code and the different modifications implemented since the first paper: \cite{Choudhury2015}. 
In Sect.~\ref{sec:results} we present first the different models as well as the temperature and radiation field intensity variations. Then we present the spatio-temporal evolution of abundances for selected species, the dissociation front properties derived from H$_2$ abundances and the line emission of the selected species. The results are finally discussed in Sect.~\ref{sec:discussion}.
Finally we summarize our findings in Sect.~\ref{sec:conclusion} and give some future directions for this work.

\section{Method}
\label{sec:method}

\begin{figure}[t]
        \centering
        \includegraphics[width=0.49\textwidth]{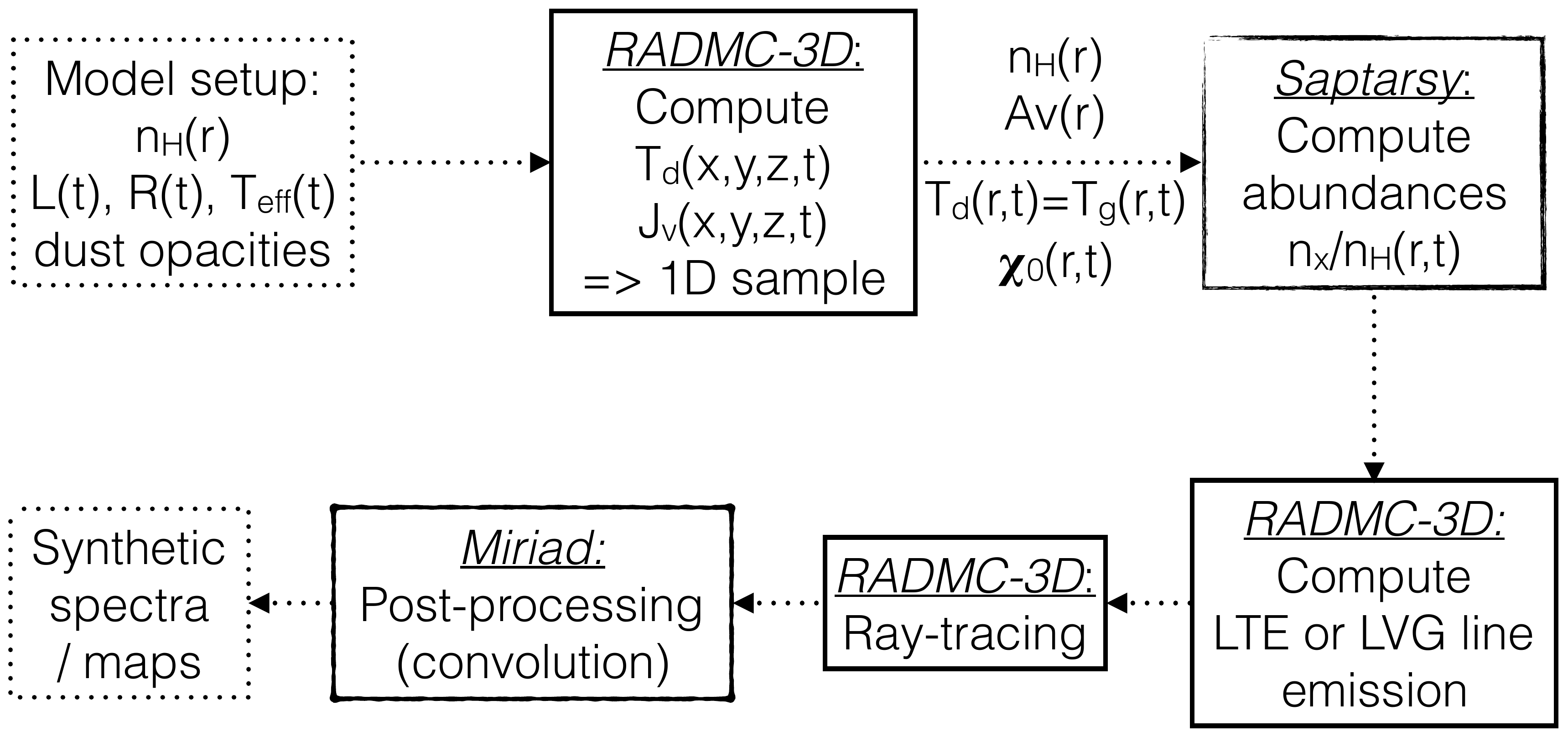}
        \caption{Flowchart showing  the setup of the variant of the modeling framework \pando\ \citep{Schmiedeke2016} employed here.}
        \label{fig:flowchart}
\end{figure}

The chemical modeling was performed with \sapt\ (see Sect.~\ref{sec:chemical code} and \cite{Choudhury2015}). 
\sapt\ works alongside \radmc\footnote{\url{http://www.ita.uni-heidelberg.de/~dullemond/software/radmc-3d/}} \citep{Dullemond2012} through a \textit{Python} based wrapper script \citep{Choudhury2015}, chemical mode of a former version of the framework \pando\footnote{For consistency reasons we employ the same predecessor version of \pando\ that Choudhury used in their benchmark study. A full integration into Pandora is currently underway.} \citep{Schmiedeke2016}. This was done to obtain synthetic spectra using the detailed modeling of the spatio-temporal evolution of the chemical abundances.  
The modeling framework is presented in the flowchart in Fig.~\ref{fig:flowchart} and is summarized as follows: 
first, based on a density profile and the evolution of stellar parameters, \radmc\ calculates both the dust temperature and the mean intensity of the radiation field self-consistently. The gas temperature is assumed to be equal to the dust temperature in all models. This approximation will be lifted in a future work since it is known that the gas temperature may be much higher than the dust temperature at the inner edge of the PDR as well as in low density environments such as the envelope in our models \citep{Lesaffre2005}. 
In a second step, the spatial evolution of the density and the spatio-temporal evolution of the dust temperature and radiation field intensity are used by \sapt\ to compute relative abundances. Finally, we used the abundances as inputs for \radmc\ to produce time-dependent synthetic spectra using the LTE approximation. The spectra can then be post-processed (convolved) to be compared to observations. The structure of the models and the chemical code are described in more detail below.

\subsection{HMC, Hollow HMC and \hii\ region models}
\label{sec:new HMC model}

HMCs are modeled as spherical symmetric cores containing a single central protostar. The star is surrounded by gas and dust\footnote{The gas to dust mass ratio is assumed to be 100 in the models.} with a density profile defined as follows: 
\begin{equation}
        \label{eq:plummer}
        n(r) = n_c \left(1 + \left(\frac{r}{r_p}\right)^2\right)^{-\gamma},
\end{equation}
where $n_c$ is the central density and $r_p$ is the size of the inner flat portion of the density profile. The exponent $\gamma$ defines the steepness of the Plummer-like function. 

\begin{figure}[t]
        \centering
        \includegraphics[width=0.495\textwidth]{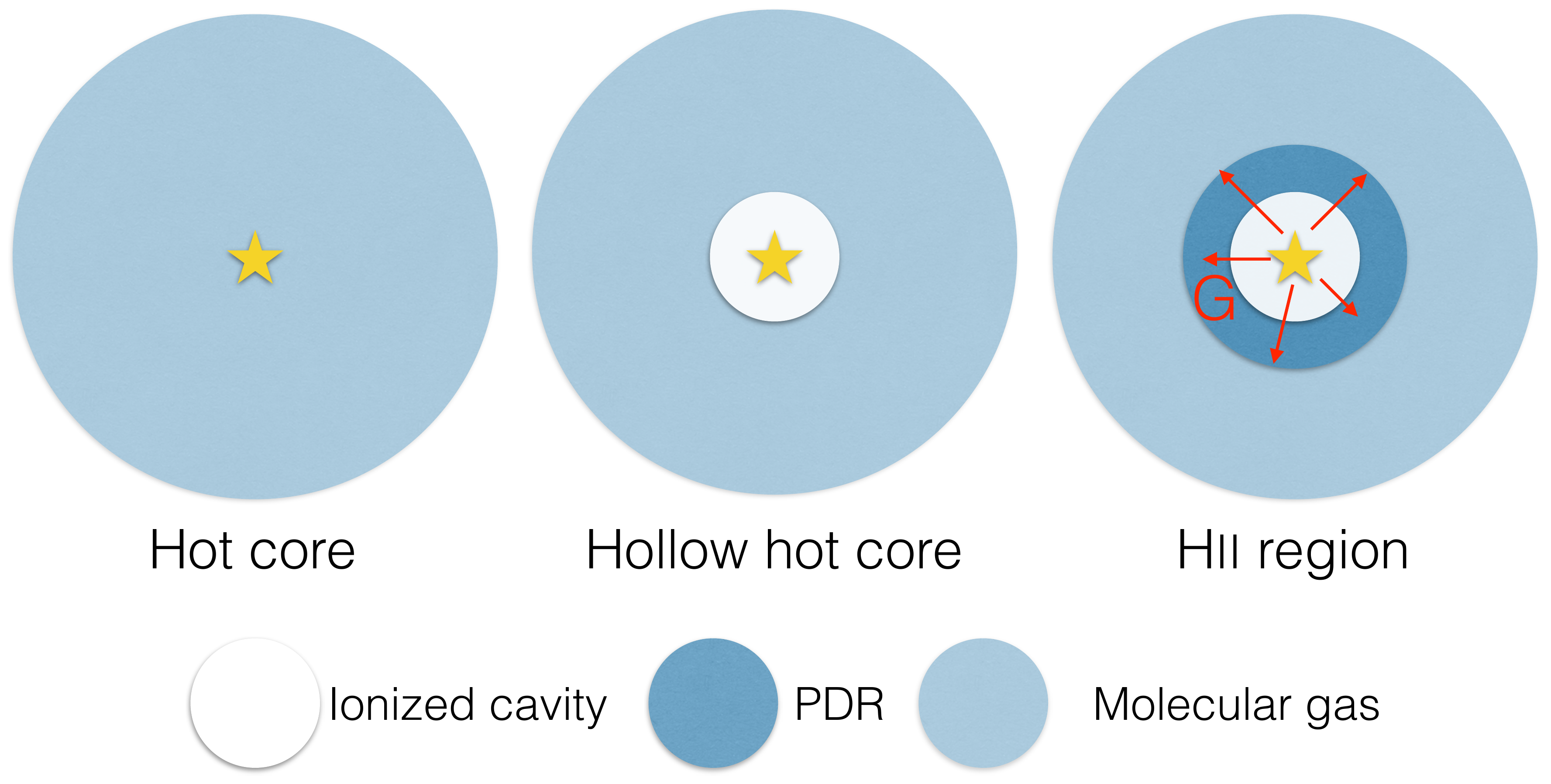}
        \caption{Scheme of the HMC (left), HHMC (middle), and \hii\ region (right) models. The spatio-temporal evolution of the radiation field is included in the computation of the relative abundances only for the \hii\ region models.}
        \label{fig:scheme}
\end{figure}

\begin{figure}[t]
        \centering
        \includegraphics[width=0.27\textwidth, trim={25 20 25 20}, clip]{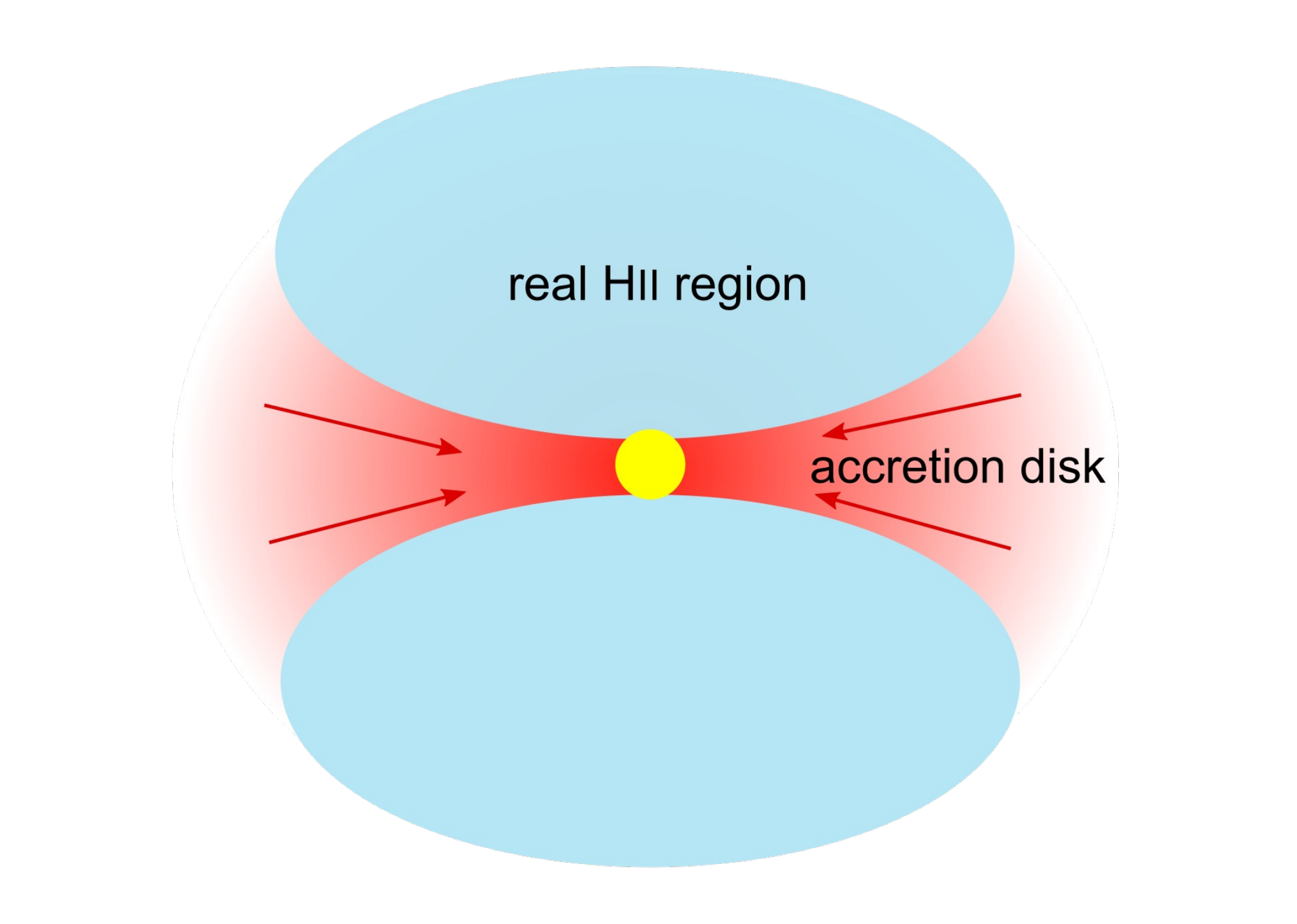}        
        \includegraphics[width=0.19\textwidth, trim={0 0 0 0}, clip]{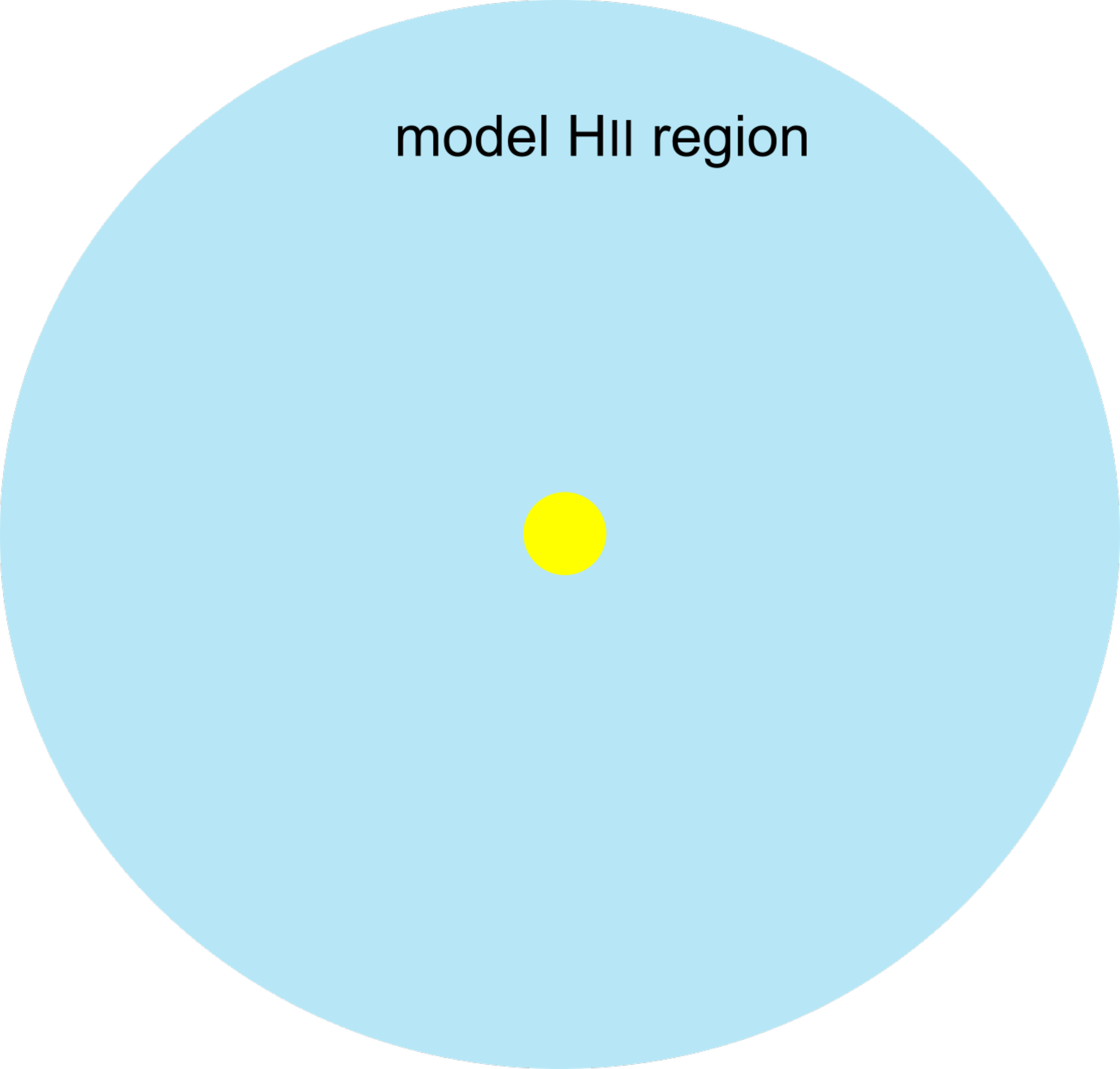}    
        \caption{On the left, sketch of the real geometry of the \hii\ region in the presence of accretion (based on models by \citealt{Peters2010a, Peters2010b}), and on the right our approximation by a spherical \hii\ region filled with electrons and ions, but no dust. The PDR and the molecular gas surrounding the \hii\ region are not represented in the sketches. Since the accretion disk does not cover a large fraction of the volume, and does not shield a large part of the spherical surface, we deem the approximation by a spherical region a reasonable approach, which greatly facilitates the calculation, without deviating too much from the real situation. While \citealt{Peters2010a, Peters2010b} show that the situation is dynamic and the shape varies on short timescales, following the true dynamics is beyond the current computational capabilities. We believe that, while the current static model is not strictly correct, it still provides valuable insights into the process, even though it eventually will have to be superseded with a model incorporating the dynamics.}
        \label{fig:HIIsketch}
\end{figure}

The \hii\ region and its surroundings\footnote{Hereafter we will use \hii\ region model to refer to this model although the ionized gas is only used as a boundary condition.} are modeled as if they were an HMC (see above) with a spherical cavity around the proto-star filled only with ionized hydrogen and electrons uniformly distributed, in other words, a Str\"omgren sphere (see also \citealp{Schmiedeke2016}). In Fig.~\ref{fig:HIIsketch} we represent an HII region with its real geometry (light blue) and its geometry as modeled in this work (dark blue). The cavity is assumed to be dust free, the spatial distribution of the dust remains unknown for HC/UCHII regions and models of more evolved HII regions suggest that big grains such as the ones used in this work are swept out of the cavity \citep{Akimkin2015}. But we note that if radiation pressure pushes the grains out
of the \hii\ region, then the PDR will have enhanced dust opacity.

In order to compare the chemical evolution of HMCs and \hii\ regions we created a third set of models called "Hollow HMC" (hereafter HHMC). The HHMC model has the same density structure as the HMC model, but it contains a cavity in the center matching the size of the \hii\ region cavity. With this artifice we ensure that the density profile as well as the total column density are the same as in the \hii\ region models. Similarly, the temperature profile in the HHMC and \hii\ region models are the same. Thus, the line emission in these two models is directly comparable. The difference between the \hii\ region model and the HHMC model is the emission of UV photons from the proto-star, which is modeled in \sapt\ with the implementation of the spatio-temporal evolution of the radiation field in the \hii\ region model. Therefore, we can focus on the effect of the radiation field on the chemistry and line emission of several selected species. In Fig.~\ref{fig:scheme} we show a scheme for the three models and their characteristics.

\subsection{Physical parameters}
\label{sec:physical parameters}

The ionized cavity in the HHMC and \hii\ region models is assumed to be filled only with ionized hydrogen and electrons. The temperature is set to the typical \hii\ region temperature of 10$^4$~K and the electron density is equal to 10$^5$~\cmc. This density is typical of ultracompact \hii\ regions (e.g., \citealt{Wood&Churchwell1989}). The cavity is present from the beginning of the time evolution and is static in time as we consider a \hii\ region whose expansion is limited by the high accretion flow \citep{Peters2010a}. In order to compute the dust temperature and the radiation field intensity using \radmc\ we defined the stellar luminosity, the density, the grain properties and the modeling grid.

\begin{figure}[t]
        \centering
        \includegraphics[width=0.48\textwidth]{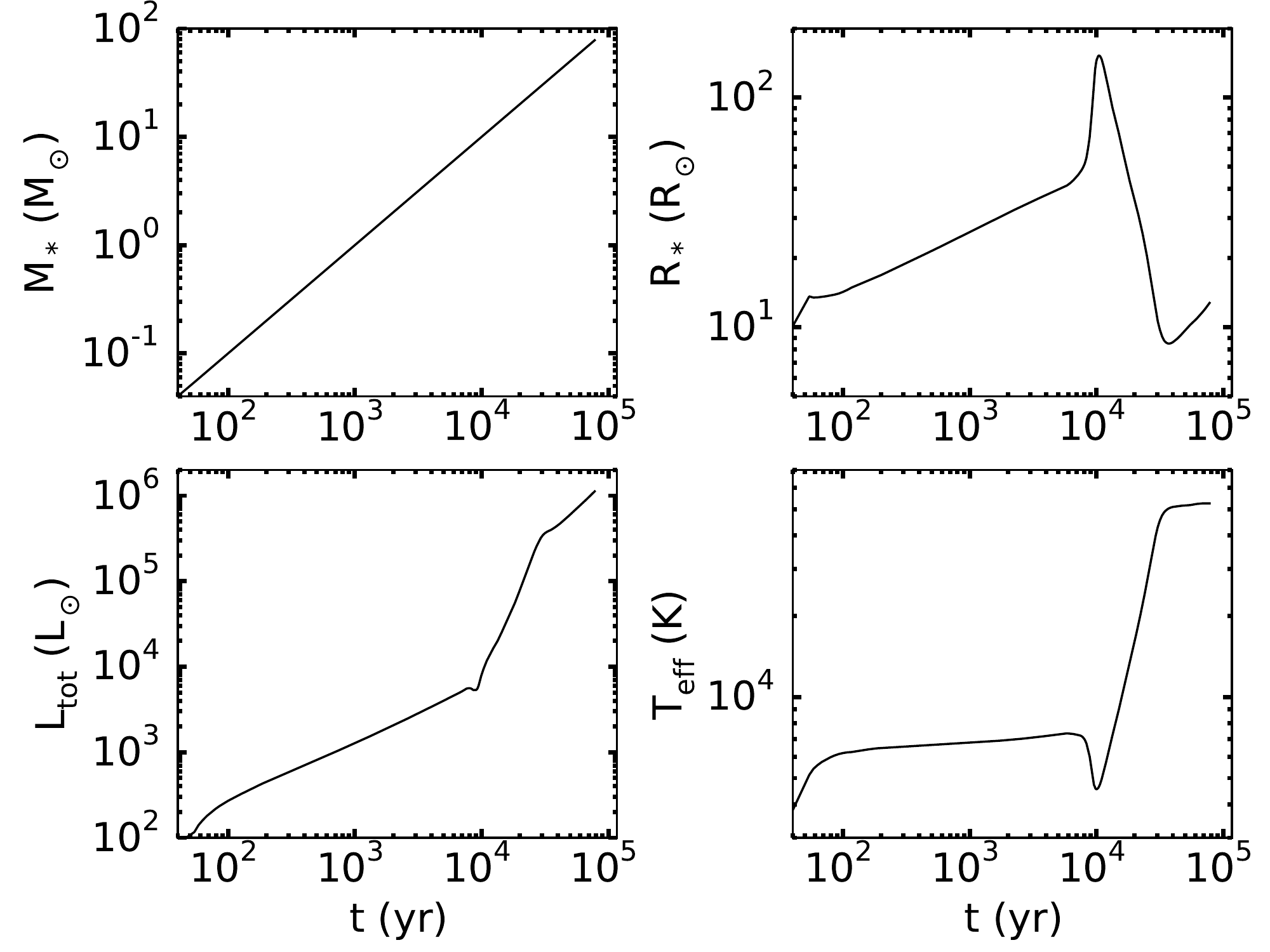}
        \caption{Time evolution of the parameters for an accreting massive proto-star with an accretion rate of $10^{-3}$~\msun\,s$^{-1}$ \citep{Hosokawa&Omukai2009}. The different panels show the mass of the star $M_*$ (top left), the star radius $R_*$ (top right), the total luminosity \Ltot\ (bottom left) and the effective temperature \teff\ (bottom right).}
        \label{Hosokawa_params}
\end{figure}

\textbf{Stellar luminosity:} We considered the model of \cite{Hosokawa&Omukai2009} which predicts the time evolution of a spherically accreting massive proto-star with a constant mass accretion rate of $\dot M$ = 10$^{-3}$~\msun\,yr$^{-1}$. The proto-star is the only heating source of the core, because no external radiation field is included, and its radiation is isotropic. 
Different to the model presented in \cite{Choudhury2015}, the effective temperature \teff\ of the proto-star is no longer a constant parameter but changes with time as follows:
\begin{equation}
        \teff(t) = \left(\displaystyle{\frac{L_{\textnormal{tot}}(t)}{4\,\pi\,R_*^4(t)\,\sigma}}\right)^{1/4},
\end{equation}
where $\sigma$ is the Stefan-Boltzmann constant, $R_*$ is the star radius and \Ltot\ is the total luminosity. The latter parameters are both provided by T. Hosokawa (priv. comm.; see Fig.~\ref{Hosokawa_params}). The time evolution of the stellar parameters covers the range from 54 to 7.8$\times$10$^4$~years. A linear interpolation was used from 0 to 54~years, assuming that at the beginning all the parameters are set to 0 except for the dust temperature which is set to a constant value of 10~K. 
Beyond 10$^5$~years, the stellar parameters are constant in time resulting in a constant dust temperature and radiation field intensity.

\begin{figure}[t]
        \centering
        \includegraphics[width=0.242\textwidth]{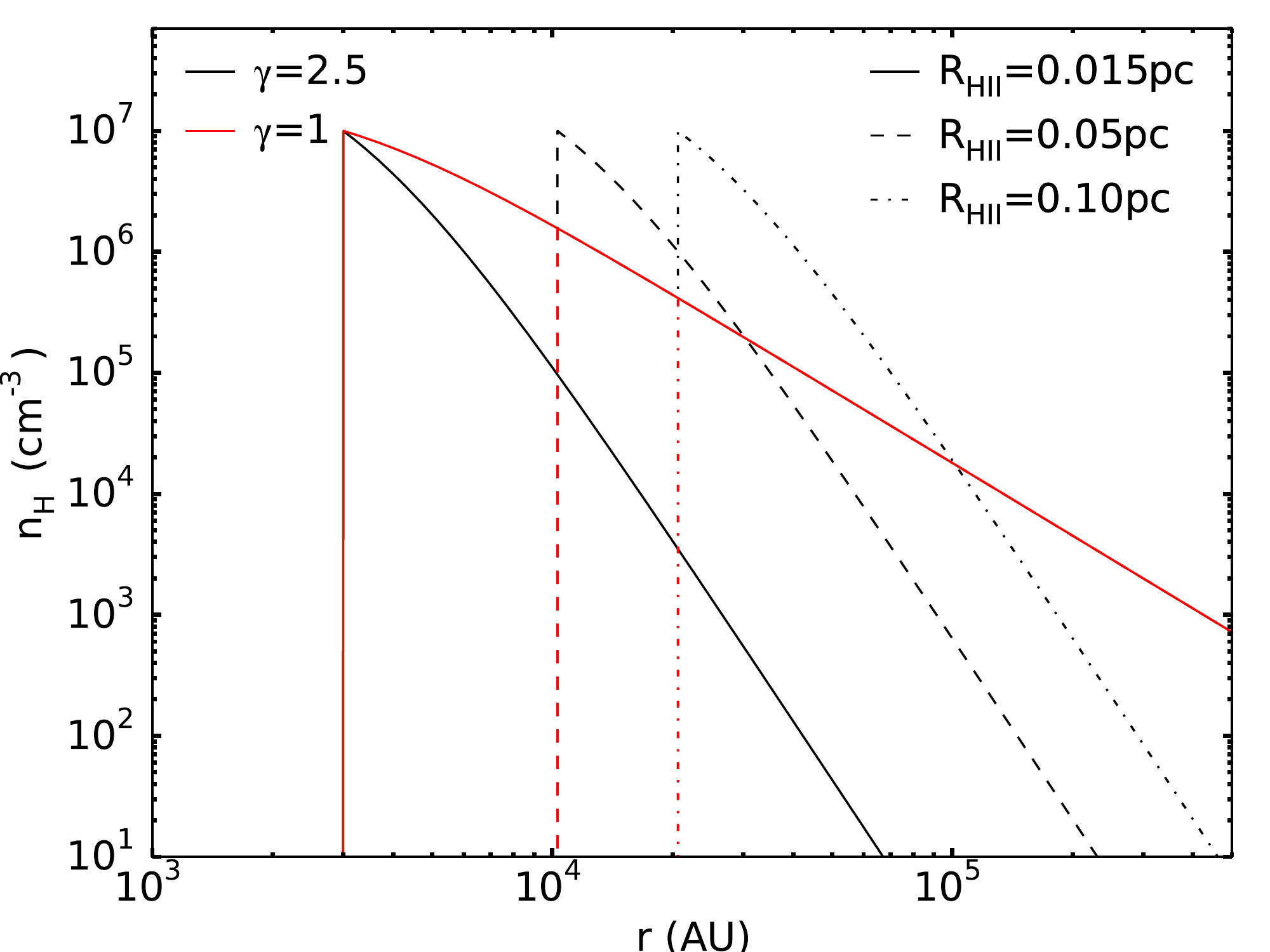}
        \includegraphics[width=0.242\textwidth]{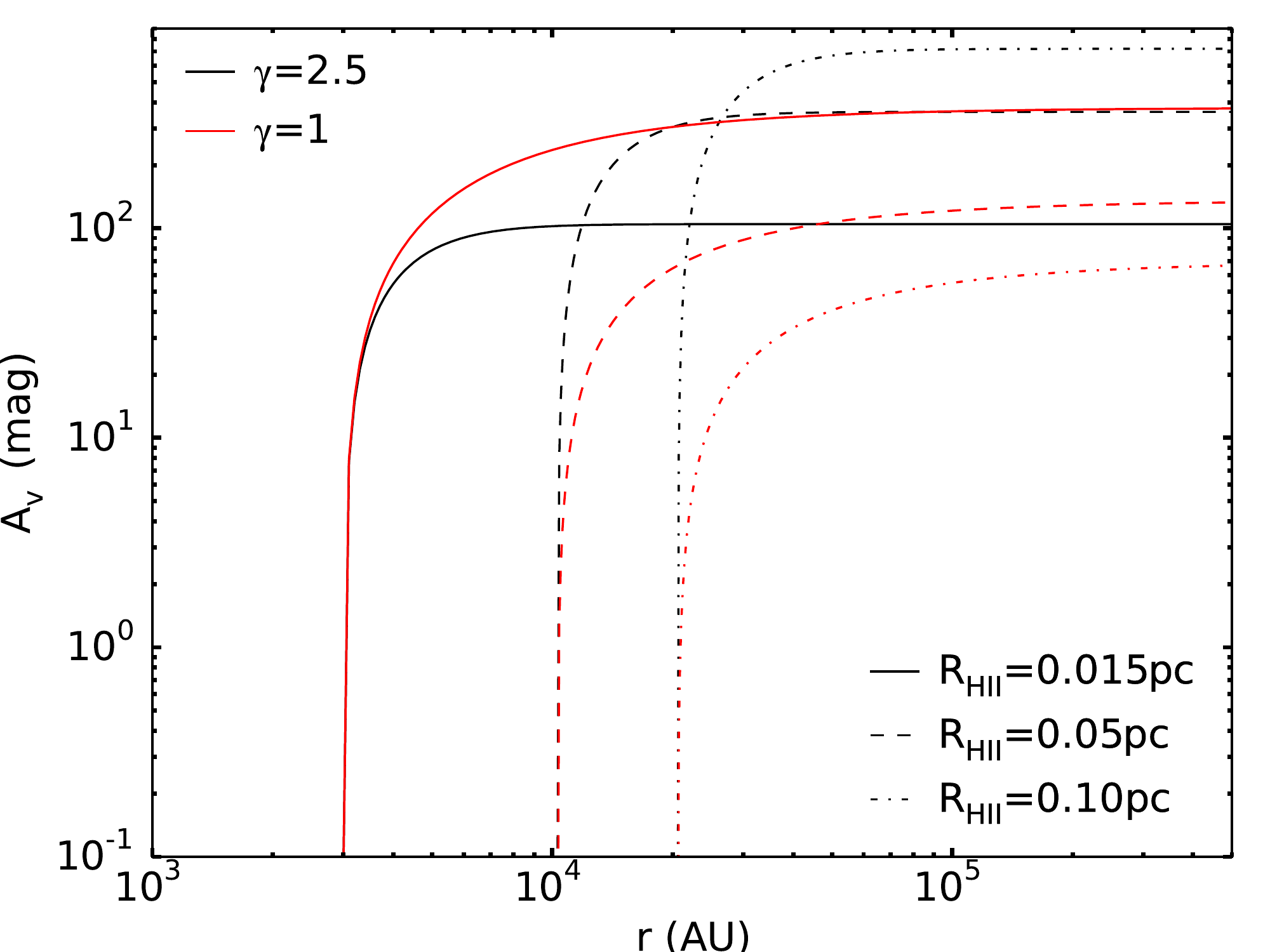}
        \caption{Left panel: Density distributions as a function of radius for a Plummer function with exponent $\gamma$ = 2.5 (black) and $\gamma$ = 1 (red) for the three sizes of \hii\ regions: 0.015 (solid), 0.05 (dashed), 0.10~pc (dash-dotted). Right panel: Visual extinction as a function of radius.}
        \label{fig:densDistrib}
\end{figure}

\textbf{Density distribution:}
We considered a Plummer-like density distribution as defined in Eq.~\ref{eq:plummer}, with two possible exponents $\gamma$ (see Fig.~\ref{fig:densDistrib}). In the first case (black lines in the figure), $\gamma$ is equal to 2.5 like the hot core models in \cite{Choudhury2015}. This factor results in a asymptotical radial dependence of the density of r$^{-5}$, consistent with what is found toward massive dense cores, for example, the dense core Sgr~B2(N)-SMA1 that is located close to the UC\hii\ region K2 \citep{Qin2011}. 
The value $r_p$ corresponds to the size of a \hii\ region, while $n_c$ is defined so that the density at the ionization front (density of the neutral gas just after the ionized cavity) is 10$^7$~\cmc\ (or 10$^6$~\cmc\ for the model with a lower density, see Table~\ref{tab:models} for a summary of the different models). In a second case (red lines), $\gamma$ is equal to 1 which results in an asymptotically radial dependence of r$^{-2}$. This is similar to the results of \cite{Didelon2015} for the dust density distribution profile found toward the Mon~R2 star forming cloud, which contains an embedded UC\hii\ region surrounded by PDRs (e.g., \citealt{Trevino-Morales2014, Trevino-Morales2016}). 
In this case, $r_p$ is fixed and equal to 3$\times$10$^3$~AU whatever the size of the \hii\ region, and $n_c$ is equal to 10$^7$~\cmc. Thus, when the size of the \hii\ region increases the density at the ionization front decreases. For a \hii\ region of 0.05~pc the density at the ionization front is then 1.56$\times$10$^6$~cm$^{-3}$ and for a \hii\ region of 0.10~pc the density is 4.15$\times$10$^5$~cm$^{-3}$. For simplicity we still annotate the models \textit{rXn7p1} (see the abbreviations of the different models in Table~\ref{tab:models}) despite the fact that the density at the ionization front is 10$^7$~cm$^{-3}$ only for model \textit{r0.015n7p1}.

\textbf{Grain properties:} We adopted the grains absorption and scattering coefficients from \cite{Laor&Draine1993}. The grains are chosen to be 100\% carbonaceous, spherical symmetric and with the classical grain size of 0.1~\micron\ \citep{Hasegawa1992}. 
Despite the fact that \radmc\ can consider a size distribution and different types of dust grains (e.g., carbonaceous and silicate), the current version of the chemical code \sapt\ considers only one type of grain. This is, however enough to calculate the chemistry because we only need the total grain surface as there is no real radiative transfer treatment.

\textbf{Modeling grid:} \radmc\ computes the dust temperature and the radiation field intensity in the entire modeling grid and for each time step given by the time evolution of the stellar luminosity. We used a 3D spherical grid. The edges of the modeling grid are reached when the density has decreased to 10~\cmc. The size of the grid is a parameter that can vary, for example, to filter out the extended emission by decreasing the size of the grid. 

In summary, the \hii\ region and the HHMC models have the same structure: same temperature, same opacity and a cavity of ionized hydrogen with no dust. The difference between these two models is the UV radiation field which is accounted for in the case of the \hii\ region models, but is not included in the HHMC model, during the computation of abundances performed by \sapt. For the HMC model, the density near the star is high enough to completely attenuate the UV radiation. 

\begin{table}[t]
        \caption{Model abbreviations. The reference model values are shown in bold. The results for the last two items are shown in the Appendix. For the models \textit{p1} the density at the ionization front is 10$^7$~cm$^{-3}$ for the \hii\ region with a radius of 0.015~pc. The density at the ionization front is lower for the other cases (1.56$\times$10$^6$~cm$^{-3}$ for 0.05~pc, and 4.15$\times$10$^5$~cm$^{-3}$ for 0.1~pc)}
        \centering
        \begin{tabular}{c|l|r|r}
                \hline
                Items & Parameter & Value & Abbreviation \\
                \hline
                \hline
                1& HC/UC\hii\ region model & & {\bf \textit{mHII}} \\
                 & HHMC model & & \textit{mHHMC} \\
                 & HMC model & & \textit{mHMC} \\
                 & & &  \\
                2 & Size of the \hii\ region & 0.015 & \textit{r0.015} \\
                 & (pc) & 0.05 & {\bf \textit{r0.05}} \\
                 & & 0.10 & \textit{r0.10} \\
                 & & & \\
                3 & Density at the & 10$^6$ & \textit{n6} \\
                 & ionization front (\cmc) & 10$^7$& \textit{n7} \\             
                 & & & \\
                4 & Initial conditions & & {\bf \textit{ini1}} \\
                 &  & & \textit{ini2} \\
                 & & & \\
                5 & Cut-off density (\cmc) & 10 & {\bf \textit{c1}} \\
                 &  &  10$^6$ & \textit{c6} \\
                 & & & \\
                App.~\ref{apsec:size-network} & Number of species & 183 & {\bf \textit{s183}} \\
                 & in the network & 334 & \textit{s334} \\
                 & & & \\
                App.~\ref{apsec:density-profile} & Plummer exponent $\gamma$ & 1 & \textit{p1} \\
                 & & 2.5 & {\bf \textit{p2.5}} \\
                 \hline
        \end{tabular}
        \label{tab:models}
\end{table}

\subsection{Chemical code}
\label{sec:chemical code}

\sapt\ \citep{Choudhury2015} is a gas-grain code based on the rate equation approach solving both the spatial and temporal evolution of chemical abundances. It uses the Netlib library solver \dvodpk\footnote{Differential Variable-coefficient Ordinary Differential equation solver with the Preconditioned Krylov method GMRES for the solution of linear systems; http://www.netlib.org/ode} to solve the ordinary differential equations and the MA28 solver\footnote{http://www.hsl.rl.ac.uk/} for sparse systems of linear equations from the Harwell Mathematical Software Library. 

We extended the use of the code presented in \cite{Choudhury2015} to study HC/UC\hii\ regions and their associated internal PDRs. To do this, we included the spatio-temporal evolution of the local radiation field intensity \Go(r,t) in Draine unit \citep{Draine1978}, which is obtained using the mean intensity of the radiation field $J_{\lambda}$(r) computed by \radmc\ for several time steps given by the time evolution of the stellar parameters. The computation of $J_{\lambda}$(r) accounts for the attenuation by dust once the UV radiation penetrates the PDR.
\begin{equation}
        G(r) = \frac{1}{1.08\times10^{-13}} \int_{912\angstrom}^{2400\angstrom} \frac{4 \pi}{c}J_{\lambda}(r)\,d\lambda \;\; \textnormal{(Draine unit)}.
\end{equation}

\begin{table}[t]
    \caption{Desorption energies: the second column shows the energies used in the current models; the third column shows the energies used in the previous work \citep{Choudhury2015}.}
    \centering
    \begin{tabular}{l c r}
            \hline
            Species & $E_\textnormal{D}$ (K) & Old $E_\textnormal{D}$ (K) \\
            \hline
            \hline
            H$_2$ & 430\tablefootmark{e} & 23 \\
            O & 1660\tablefootmark{a} & 800 \\
            O$_2$ & 930\tablefootmark{b} & 1000 \\
            O$_3$ & 1833\tablefootmark{b} & 1800 \\
            SO & 1745\tablefootmark{c} & 2600 \\
            NO & 2460\tablefootmark{c} & 1600 \\
            HNO & 2910\tablefootmark{c} & 2050 \\
            H$_2$CO & 3300\tablefootmark{d} & 2050 \\
            \hline
        \end{tabular}
        \tablefoot{
            \tablefoottext{a}{ \cite{He2015} }
            \tablefoottext{b}{ \cite{Jing2012} }
            \tablefoottext{c}{ Derived from $E_\textnormal{D}$(O) }
            \tablefoottext{d}{ \cite{Noble2012b} }
            \tablefoottext{e}{ \cite{Garrod2013a} }
                      }
    \label{table_ED}
\end{table}

In this work we have used a chemical network, containing 183 species, based on a reduced version of the OSU network \citep{Garrod2008}, including new reactions for HCN and HNC \citep{Graninger2014, Loison2014}. 
The reduced version of the OSU network speeds up the computational time, and excludes species not relevant for the current work, such as those species bearing Cl, P, Na, and Mg, as well as all anions and long carbon chains with more than four carbon atoms. For Si and Fe, only the neutral, ionized, and grain surface atomic form were included in the network. In addition, the desorption energies $E_\textnormal{D}$ of different species have been updated, and the values used can be found in Table~\ref{table_ED}. Since measurements of diffusion barriers $E_\mathrm{b}$ only exist for a few species, we considered them to be related to the desorption energy as $E_\mathrm{b} / E_\mathrm{D} = 0.50$ \citep{Vasyunin&Herbst2013}.

\begin{table}[t]
    \caption{Initial abundances of the principal species used in the models are shown in the second and third columns. The fifth column shows the elemental abundances, of the gas and ice phases and not of the refractory dust, taken from \citet{Wakelam&Herbst2008} that are used to obtain the initial abundances shown in the first columns. The values a(b) refers to a$\times$10$^\textnormal{b}$ through the paper. The grain surface species are indicated by the 's-' before their name.}
    \centering
    \begin{tabular}{l l l   ||  l l }
            \hline
            Species & $n_\textnormal{X}$/\nh & $n_\textnormal{X}$/\nh & Elemental & $\delta_{X}$ \\
             &  \textit{ini1} & \textit{ini2} & species &  \\
            \hline
            \hline
             H$_2$ & 4.99(-1) & 4.99(-1) & H & 1.00 \\
             He & 9.00(-2) & 9.00(-2) & He & 9.00(-2)\\ 
             s-H$_2$ & 1.24(-4) & 1.24(-4) & O & 2.56(-4)\\
             s-H$_2$O & 9.87(-5) & 2.46(-4) & N & 7.60(-5) \\
             s-CO & 9.19(-5) & 9.58(-6) & C$^+$ & 1.20(-4) \\ 
             s-OH & 1.91(-5) & 7.24(-14) & S$^+$ & 1.50(-5) \\
             s-HNO & 2.89(-5) & 3.12(-7) & Si$^+$ & 1.70(-6) \\
             s-N$_2$ & 1.33(-5) & 6.14(-12) & Fe$^+$ & 2.00(-7) \\  
             s-O & 1.16(-5) & 2.74(-16) &&\\
             s-HCN & 6.23(-6) & 4.17(-7) &&\\
             s-CH$_4$ & 6.20(-6) & 1.10(-4) &&\\
             s-NH$_3$ & 8.03(-6) & 7.53(-5) &&\\
             s-H$_2$CO & 3.20(-7) & 7.12(-9) &&\\
             s-CH$_3$OH & 1.29(-7) & 2.67(-17) &&\\
             s-CO$_2$ & 9.83(-8) & 5.91(-10) &&\\
             s-H$_2$S & 3.09(-8) & 7.93(-8) &&\\
             s-Si & 8.00(-9) & 8.00(-9) &&\\
             s-Fe & 3.00(-9) & 3.00(-9) &&\\
             \hline
    \end{tabular}
    \label{tab:IniAbun}
\end{table}

Photon processes and secondary photon processes were not used in the previous version of \sapt\ \citep{Choudhury2015}. For the present work, we have added photo-dissociation reactions in the gas phase and on the grain surface as described in \cite{Semenov2010}. 
Photo-desorption is also included for 11 species\footnote{The prefix s- indicates grain surface species, we will use this notation throughout the paper} (s-H, s-H$_2$, s-O, s-O$_2$, s-OH, s-H$_2$O, s-CO, s-HCO, s-H$_2$CO, s-N$_2$, and s-CO$_2$) as defined in \cite{LeBourlot2012} in the case of H$_2$. In the associated reaction rate the photon flux also takes into account the secondary photon flux. The direct dissociation by cosmic-rays and the secondary photon processes were treated the same way in the first version of \sapt. Secondary photon processes are now treated as expressed in \cite{LePetit2006}. Desorption via exothermic surface reactions, also called reactive desorption, with an efficiency of 1\% has been added following \cite{Garrod2007}. Furthermore, self-shielding is an important process in internal PDRs. We included a simple treatment of the self-shielding for H$_2$ and CO following \cite{Lee1996} (see also \citealt{Draine&Bertoldi1996} and \citealt{Sternberg2014} for H$_2$ and \citealt{Panoglou2012} for CO), which is used in the \hii\ region models. Appendix~\ref{app:further modifications} describes other modifications done in the chemical code \sapt.

The initial abundances, noted \textit{ini1} (see Table~\ref{tab:models} for the abbreviations), are obtained by simulating the collapse of a prestellar core in two steps. We started with the elemental abundances from \cite{Wakelam&Herbst2008}, a core with a temperature of 10~K, an extinction of 10~mag and no radiation field. In the first step, the core has a density of 10$^4$~\cmc\ and we let the chemistry evolve up to 10$^5$~years. We then increased the density to 10$^7$~\cmc\ to simulate the density increase during the process of collapse, and repeated the process using the results of the first step. This means that for all the models (HMC, HHMC, and \hii\ region), we started with a medium where molecules are depleted onto grains except for hydrogen which is found in its molecular form H$_2$. The carbon is mostly found in s-CO, the oxygen is in s-H$_2$O, and the nitrogen is mainly in s-N$_2$ as seen in the second column of Table~\ref{tab:IniAbun}. We note that carbon is also found in s-CH$_4$ and s-HCN, oxygen in s-O and s-OH and nitrogen in s-HNO and s-NH$_3$. 
In order to simplify the comparison between models, we used the same initial abundances, except for models that have the second set of initial abundances \textit{ini2} (see column 3 of Table~\ref{tab:IniAbun}). This second set of initial abundances \textit{ini2} is obtained for the steady state of a prestellar core model at a density of 10$^7$~\cmc. In these conditions the species are also frozen on grains, except hydrogen which is in its molecular form in the gas phase. Carbon is found mostly in s-CH$_4$, oxygen in s-H$_2$O, and nitrogen in s-NH$_3$.

\section{Results and analysis}
\label{sec:results}

\subsection{Models}
\label{sec:models}

In the following we present the parameters used for the different models considered. Our reference model is an \hii\ region with a size of 0.05~pc (\textit{r0.05}), a density of 10$^7$~\cmc\ at the ionization front (\textit{n7}) and a Plummer exponent of 2.5 (\textit{p2.5}). The model uses the first set of initial abundances (\textit{ini1}), a chemical network with 183 species (\textit{s183}) and a cut-off density of 10$^1$~\cmc\ (\textit{c1}).
The size and density of the \hii\ region is consistent with the values commonly found for HC and UC\hii\ regions (see \citealt{Kurtz2005}). 
In addition to this model, we have changed different parameters to see their effects on the chemistry. The different models are: 
\begin{enumerate}
\item We consider first the reference model of \hii\ region and the associated HHMC and HMC models.
\item Models with different sizes of \hii\ regions are produced. In addition to R$_{{\rm HII}}$ = 0.05~pc (10300~AU) we used R$_{{\rm HII}}$ = 0.015~pc and 0.10~pc (or 3000 and 20600~AU). The model where R$_{{\rm HII}}$ = 0.015~pc corresponds to a HC\hii\ region, while the one of 0.10~pc is at the transition from a UC to a compact \hii\ region \citep{Kurtz2005, SanchezMonge2013b}. In Appendix~\ref{apsec:density-profile} we present the results obtained for the same models but with a different density profile where the exponent of the Plummer-like function (Eq.~\ref{eq:plummer}) is set to 1. 
\item We also considered models with a lower density at the ionization front: 10$^6$~cm$^{-3}$. This value is typically found in the dense gas surrounding UC\hii\ regions. 
\item In addition to the initial abundances \textit{ini1}, we also considered a second set of initial conditions -- \textit{ini2}. These initial conditions are shown in the second column of Table~\ref{tab:IniAbun}.  
\item Finally, we considered a model for which the modeling grid is stopped at a higher cut-off density of 10$^6$~\cmc (instead of 10~\cmc). This results in a smaller grid that contains only the emission from the inner core, and therefore excludes the extended and more diffuse envelope. 
\item Additional models are presented in Appendix~\ref{apsec:density-profile} where we set the exponent of the Plummer-like function to 1 (model \textit{p1}) and \ref{apsec:size-network} where we used a chemical network with 334 species (model \textit{s334}).
\end{enumerate}
 
To distinguish between the different models we use the abbreviations presented in Table~\ref{tab:models} and the different comparison points are referred as ``Item i'' from the list of Sect.~\ref{sec:models}. We annotate the reference model as \textit{mHII:r0.05n7ini1c1p2.5s183} and unless indicated otherwise, all parameters are considered to be set to their reference values in the different models. In the following, we discuss how the temperature, density structure, abundances and line intensities change from model to model.

\subsection{Temperature and radiation field variations}
\label{sec:Td_and_G0}

\begin{figure}[t]
        \centering
        \includegraphics[width=0.242\textwidth]{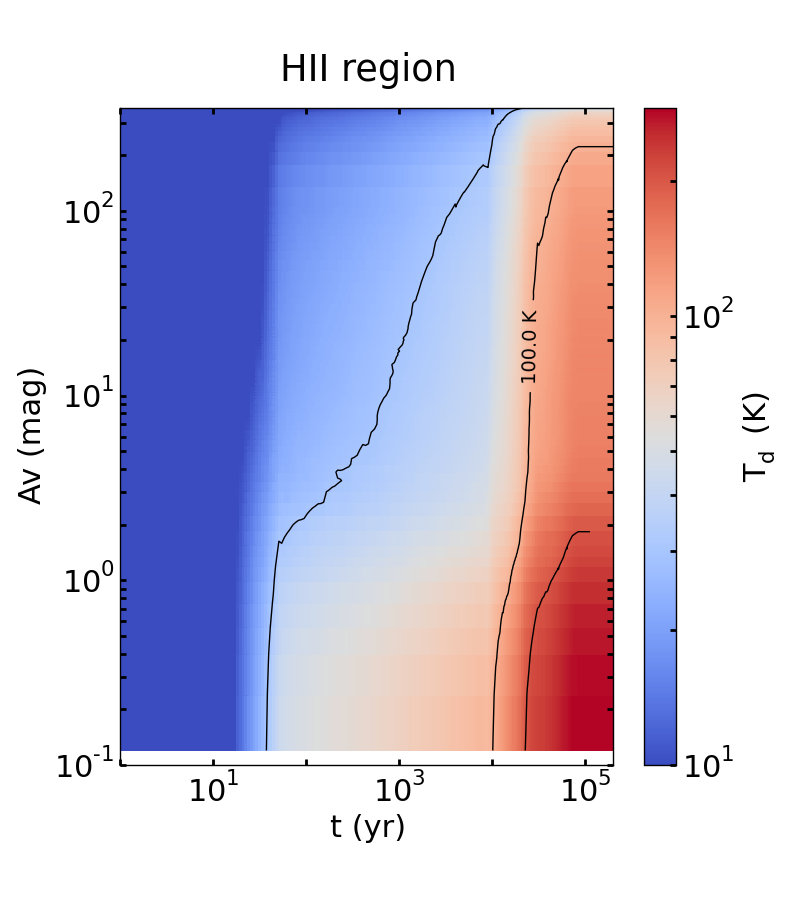}
        \includegraphics[width=0.242\textwidth]{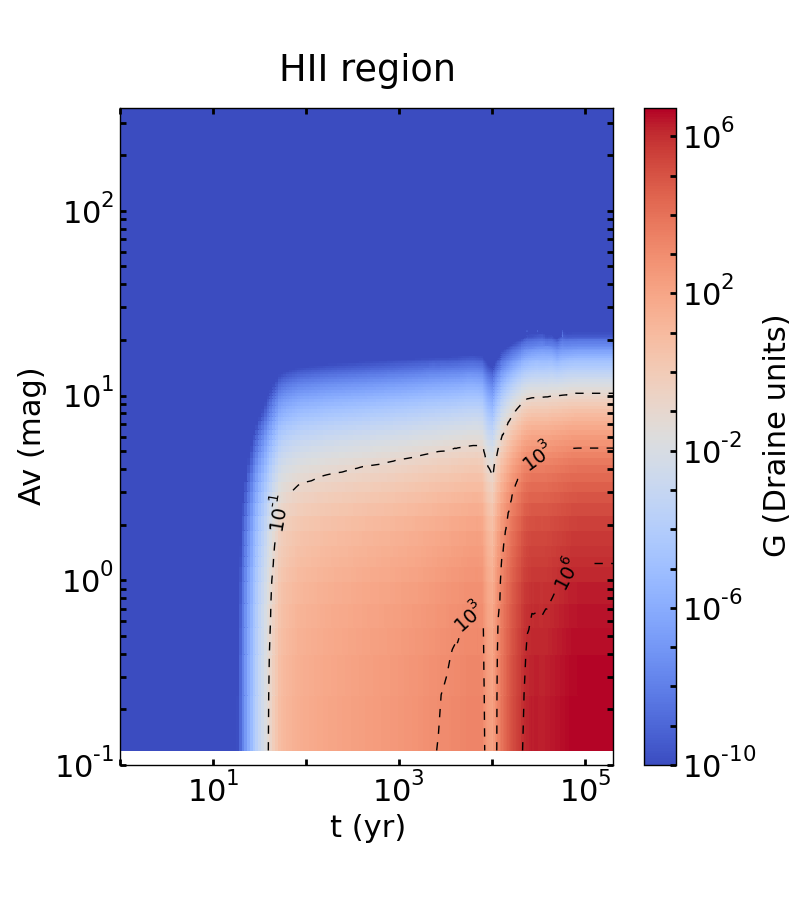}
        \caption{Spatio-temporal evolution of the dust temperature (left) and radiation field intensity (right) in the case of the reference model \textit{mHII:r0.05n7ini1c1p2.5s183}. Contours are plotted: solid line for \Td\ (30, 100 and 200~K) and dashed lines for \Go\ (10$^{-1}$, 10$^3$ and 10$^6$~Draine unit).}
        \label{fig:3DTd-G0}
\end{figure}

The spatio-temporal evolution of the dust temperature and the radiation field intensity for the reference model \textit{mHII:r0.05n7ini1c1p2.5s183} are presented in Fig.~\ref{fig:3DTd-G0}. 
The variation of these two parameters can be used to determine different chemical transitions during the core evolution.  
We started with the initial conditions (\Td\ = 10~K and \Go\ = 10$^{-20}$~Draine unit) for the first 50~years, which correspond to the grain surface region where all the molecules are frozen onto the grain surface. 
After this time, \Go\ is equal to 10$^{-1}$ and \Td\ is higher than 50~K. We enter into a molecular region as the species are released into the gas phase. 
Deeper in the cloud the transition from grain surface to gas phase might be delayed because the temperature is still too low. From approximately 10$^4$~years, the radiation field strength increases but only for extinctions up to 10 -- 20~mag. The temperature is also the highest in this region. It is an atomic-ionized region, i.e., a PDR, although the spatial resolution and the approximate treatment of ionization and thermal balance induce some inaccuracies that disappear further in. 
For the latest phase of the evolution (after 10$^4$~years) and deep in the cloud (\av\ > 20~mag), the temperature is the only parameter influencing the chemistry due to absence of the radiation field. The temperature is high enough, $\sim$ 100~K, to ensure that all molecules are in the gas phase. 

\subsection{Abundances}
\label{sec:abundances}

In this section we present the spatio-temporal profiles of the abundances, relative to $n$(H), obtained for the different models detailed in Sect.~\ref{sec:models}. We obtained these abundances with \sapt\ for all species present in the chemical network. We focus principally on a few species such as HCN, HNC, water and some others that are more relevant in HMCs including methanol and in HC/UC\hii\ regions like ionized and atomic species.

\subsubsection{Item 1 --  HMC/HHMC/\hii\ region:}
\label{subsubsec:abunItem1}

\begin{figure}[t]
        \centering
        \includegraphics[width=0.49\textwidth, trim={0 0 0 20mm},clip]{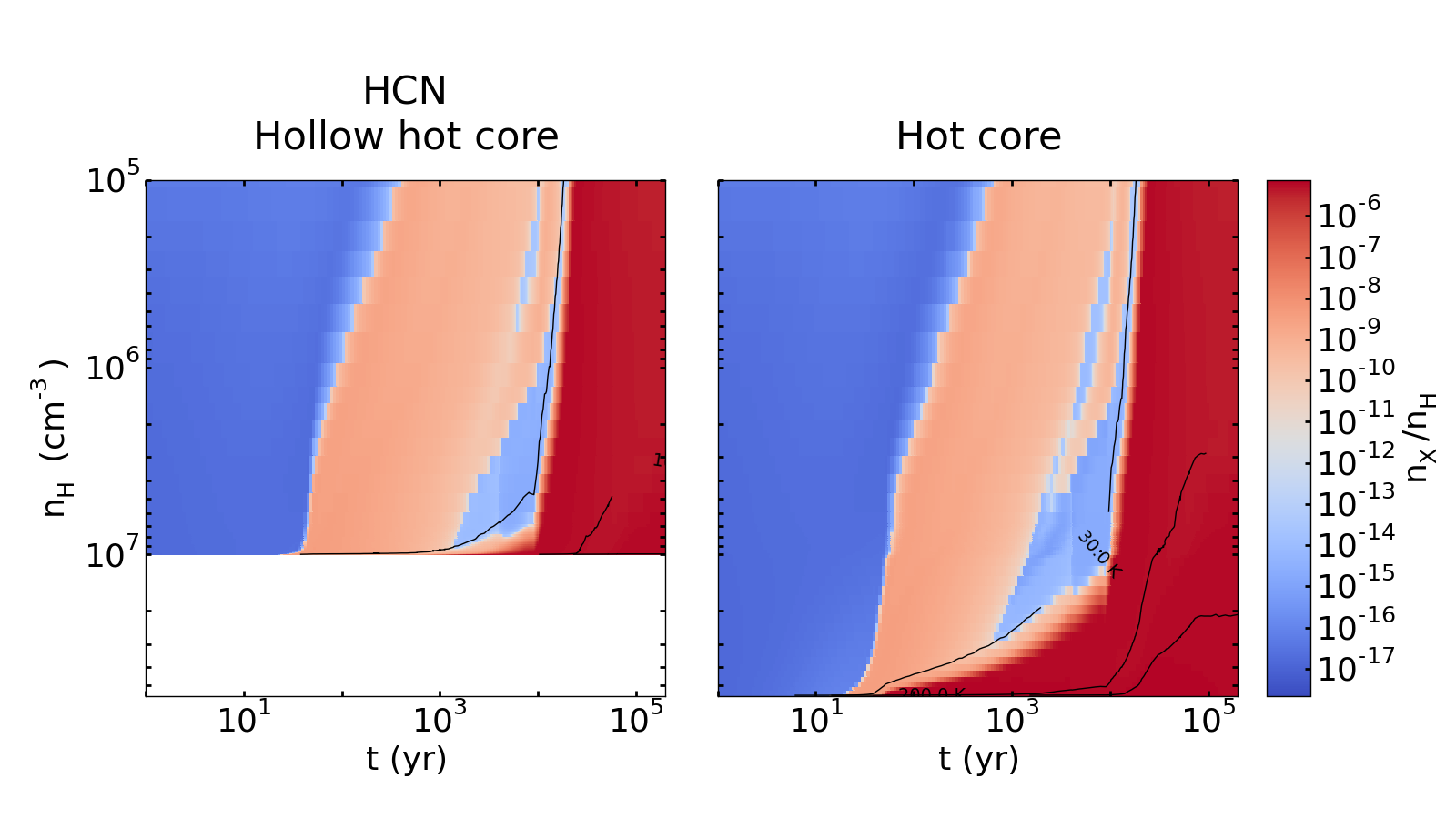}
        \caption{Spatio-temporal evolution of the abundance of HCN for the model \textit{mHHMC:r0.05n7ini1c1p2.5s183} (left panels) and the corresponding model \textit{mHMC} (right). Contours are plotted in solid line for \Td\ (30, 100 and 200~K).}
        \label{fig:3Dabun_compHHMC-HMC}
\end{figure}

In Fig.~\ref{fig:3Dabun_compHHMC-HMC} we compare the abundance of HCN for the HHMC and HMC models with the following parameters \textit{r0.05n7ini1c1p2.5s183}. To help for the comparison and because the extinctions are different for both models the abundances are shown as a function of the density in y-axis, from 10$^7$ to 10$^4$~\cmc\ for the HHMC model and from 5.6$\times$10$^7$~\cmc\ for the HMC model which does not have the ionized cavity around the proto-star. We notice that the 100~K temperature at late times is reached at a density higher than 3$\times$10$^6$~\cmc\ for the HHMC model whereas it occurs for a lower density in the HMC model. The 150~K temperature is reached for slightly higher densities in the HMC model. However, we see that from 10$^7$ to 10$^4$~\cmc\ the abundances and their spatio-temporal evolution are consistent in both models. The same is found for the other species of the chemical network. 

\begin{figure}[t]
        \centering
        \includegraphics[width=0.49\textwidth, trim={0 0 0 20mm},clip]{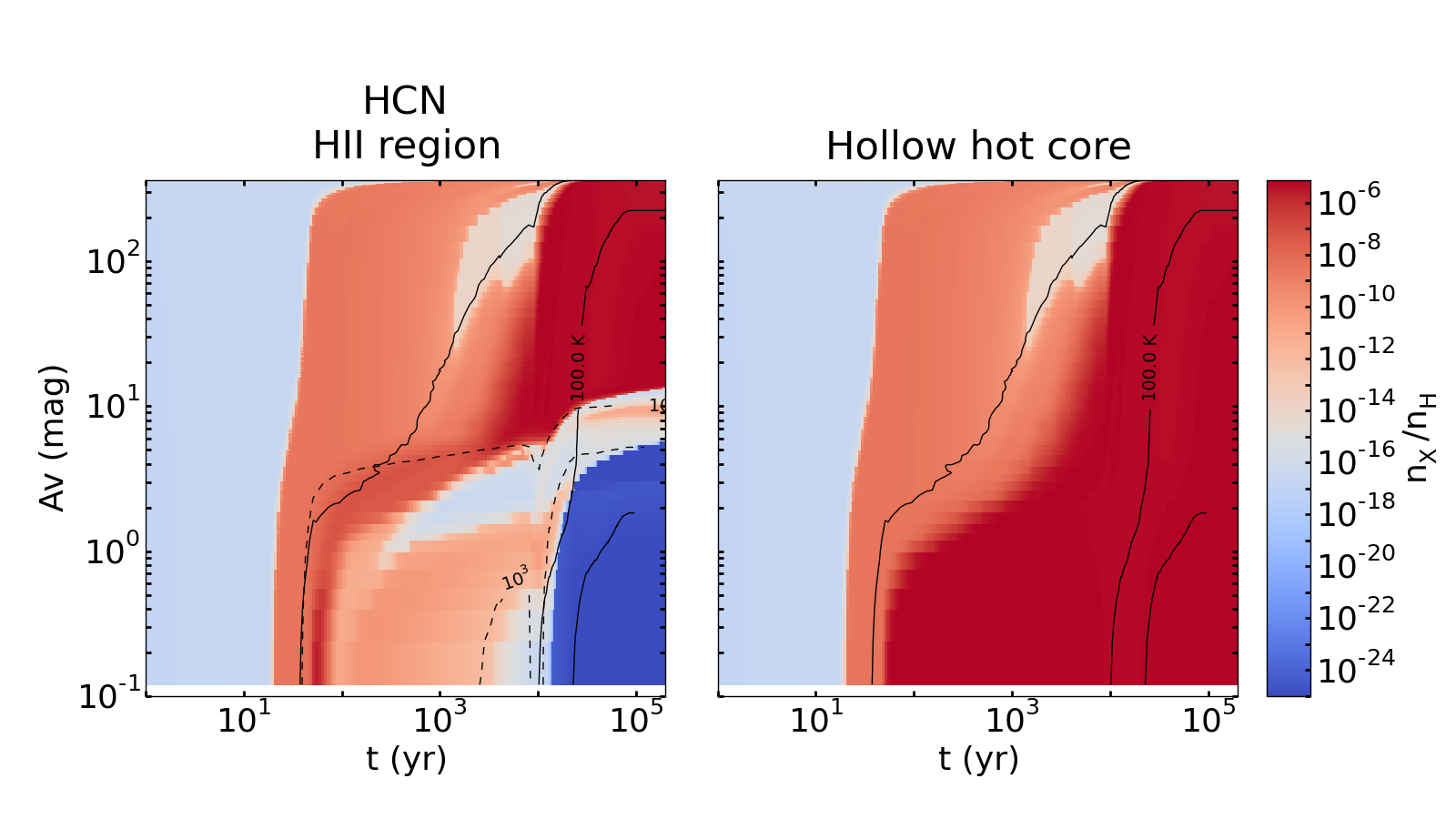}
        \includegraphics[width=0.49\textwidth, trim={0 0 0 20mm},clip]{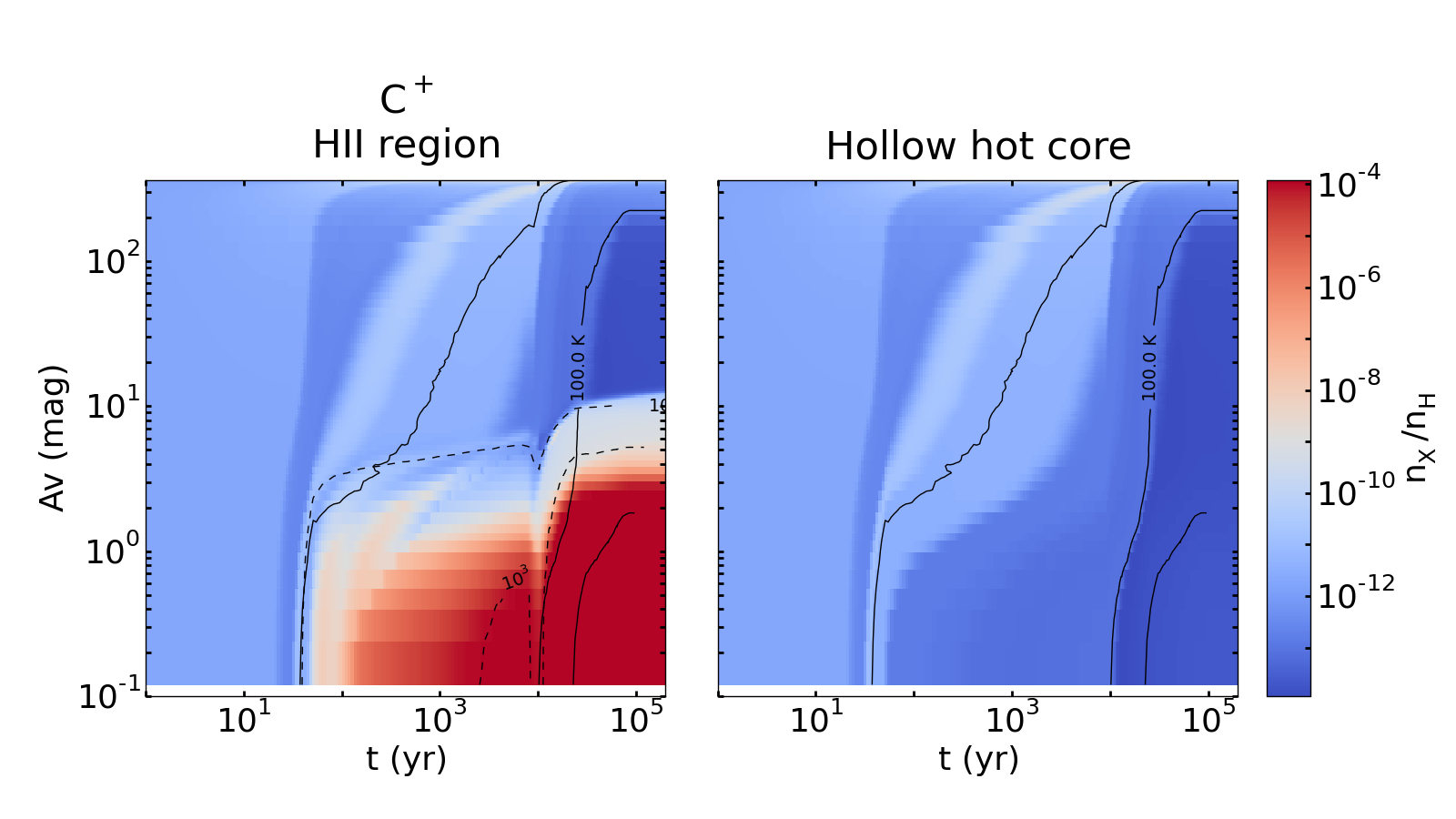}
        \caption{Spatio-temporal evolution of the abundance of HCN (top panels) and C$^+$ (bottom) for the reference model \textit{mHII} (left panels) and model \textit{mHHMC} (right). Contours are plotted: solid line for \Td\ (30, 100 and 200~K) and dashed lines for \Go\ (10$^{-1}$ and 10$^3$~Draine unit).}
        \label{fig:3Dabun_compHII-HotC}
\end{figure}

We also compare the distribution for the \hii\ region model and the HHMC model to investigate the differences in the abundances when a UV radiation field intensity is added to the model. Figure~\ref{fig:3Dabun_compHII-HotC} shows the abundance profile of HCN and C$^+$ for the reference \hii\ region model \textit{mHII:r0.05n7ini1c1p2.5s183} (left panels) and for the one of the HHMC model \textit{mHHMC:r0.05n7ini1c1p2.5s183} (right). We choose these two species because they present the two different behaviors happening in presence of a radiation field. The well known PDR tracer C$^+$ is formed by photo-dissociation and photo-ionization in regions where the radiation field is strong. On the contrary, HCN is a more fragile molecule against UV radiation and therefore more easily destroyed \citep{Fuente1993}. However, HCN can also be found in these high temperature and dense environments \citep{Paglione1995, Paglione1997}, particularly in highly excited energy states \citep{Ziurys&Turner1986}.

\begin{figure}[t]
        \centering
        \includegraphics[width=0.242\textwidth]{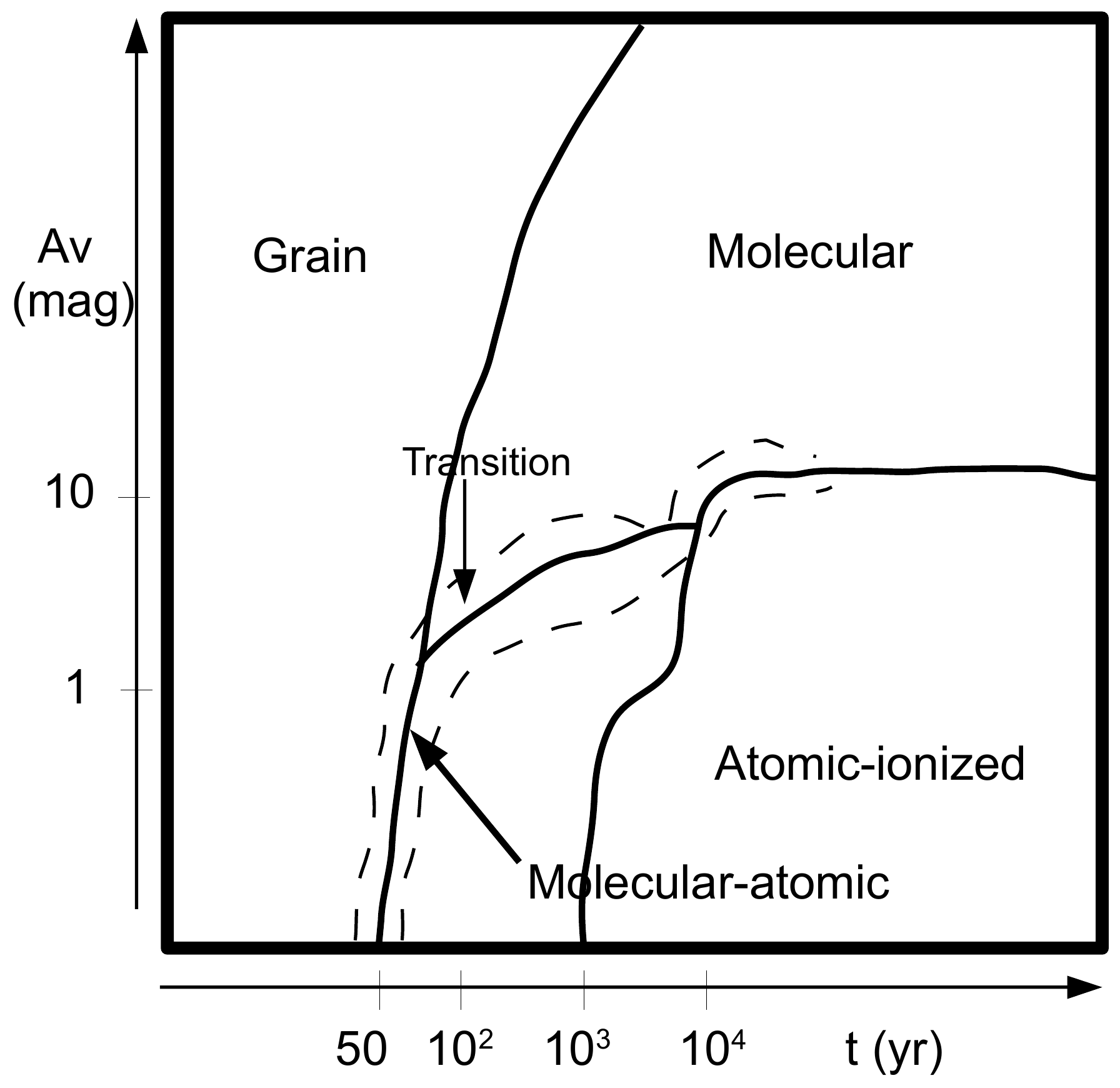}
        \includegraphics[width=0.242\textwidth]{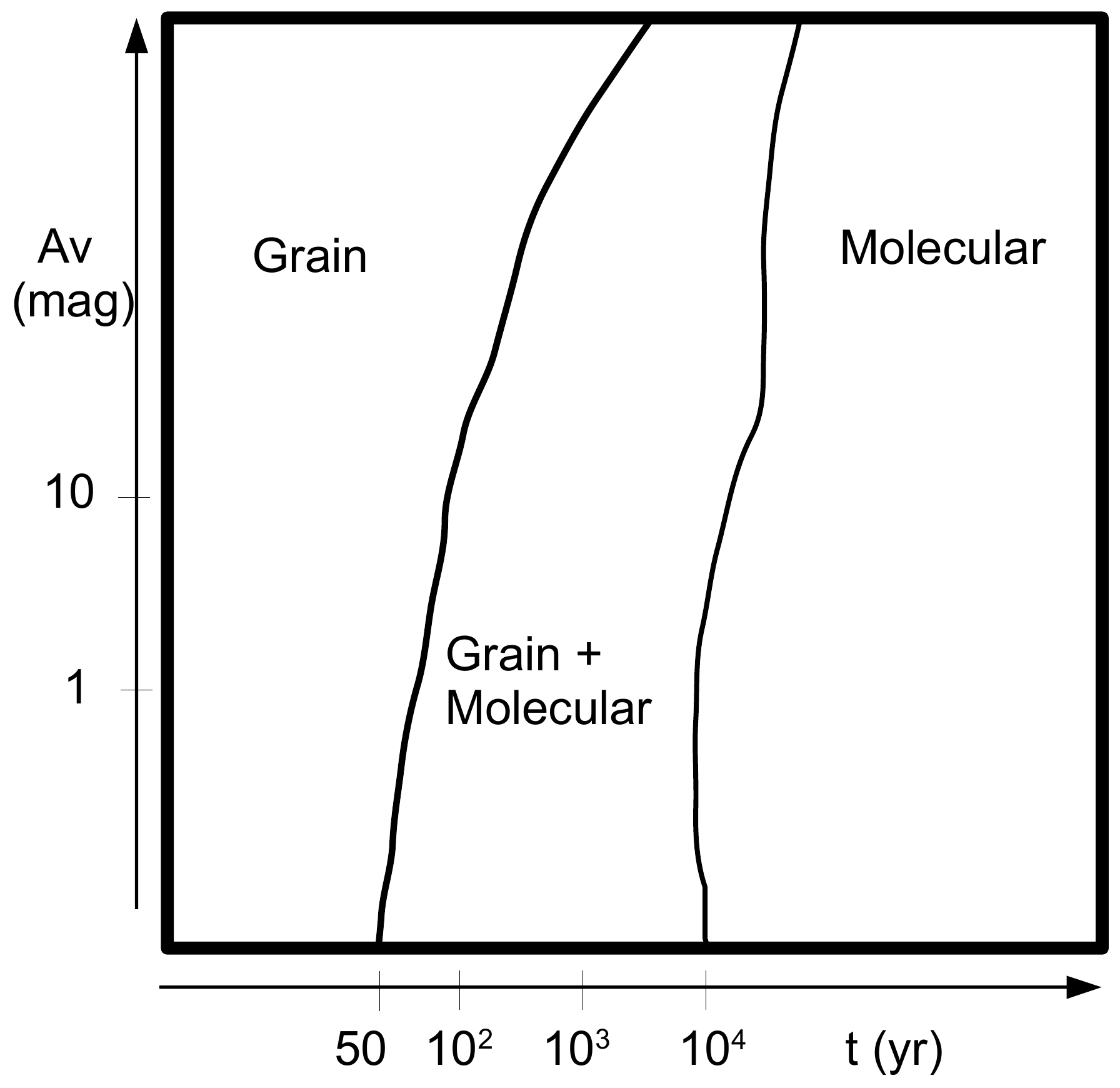}
        \caption{Schematics showing the different regions in the \hii\ region models (left) and in the HMC/HHMC models (right). 
        }
        \label{fig:regions}
\end{figure}

The ion C$^+$ is formed in the \hii\ region model, at $\sim$ 100~years when the UV radiation becomes strong (see dashed contours), while HCN disappears. From the comparison of these abundance profiles, and considering other species, we identified the grain-surface chemistry region, as well as the molecular gas and the atomic-ionized gas regions indicated previously in Sect~\ref{sec:Td_and_G0}. In addition, a region partly molecular partly atomic can be seen as well as a transition region between this molecular-atomic region and the molecular region. Schemes of the different regions seen in the \hii\ region and the HMC/HHMC models are shown in Fig.~\ref{fig:regions}. The molecular-atomic and atomic-ionized regions constitute the internal PDR in the \hii\ region models. It is created by the UV radiation field. 
In the HMC/HHMC models, the grain surface chemistry region can be larger close to the edge of the cavity depending on the considered species (see the right panel of Fig.~\ref{fig:regions}). For instance, in the case of methanol, thermal desorption occurs before 100~years in model \textit{mHII} whereas it starts at around 10$^4$~years in models \textit{mHMC} and \textit{mHHMC}. This is not the case for species with low desorption energy such as CO. For those molecules desorption happens at the same time as in the \hii\ region models, but with a smoother transition. 

\subsubsection{Item 2 -- Size of the ionized cavity:} 
\label{subsubsec:abunItem2}

The size of the ionized cavity affects the duration of the grain surface chemistry. Models with larger \hii\ region cavities have a grain surface chemistry region (as shown in Fig.~\ref{fig:regions}) that extends deeper into the core, 
stopping at 100~years for the \textit{mHII:r0.015} and lasting until 10$^4$~years for \textit{mHII:r0.10} (see the high extinctions of the abundance profile of HCN in top panels of Fig.~\ref{apfig:3Dabun_compHIIsize}). s-H$_2$O seems to stay on the grain surface even after 10$^5$~years when the size of the cavity increases. This is because the temperature increases more slowly in the core due to the distance from the star. 

In the molecular-atomic region, the abundance of most gas phase species becomes larger as the core evolves for models with larger ionized cavities models because of the smaller radiation field reaching the gas surface. This is also the case for grain surface species with high desorption energies such as methanol. The desorption of more complex molecules like methanol is delayed when we increase the size of the cavity. In addition, the size of the atomic-ionized region is independent of the temperature and only depends on \Go. 

Furthermore, when \av\ > 10~mag and between 10$^3$ and 10$^4$~years, we see a decrease in the abundance of HCN, as well as HNC and CH$_3$OH, before a sudden increase of the abundances due to thermal desorption (Fig.~\ref{fig:3Dabun_compHII-HotC} and top panels of Fig.~\ref{apfig:3Dabun_compHIIsize}). This decrease, mainly due to a significant accretion of the species on the grain, appears closer to the ionization front when the size of the \hii\ region increases.

\subsubsection{Item 3 -- Density at the ionization front:} 
\label{subsubsec:abunItem3}

The density at the ionization front is another parameter that affects the chemistry. The abundance of a few ions such as N$^+$, NH$^+$ and O$^+$ is higher for models with lower density during the entire evolution, especially in the atomic-ionized region. 
For a lower density and at the beginning of the evolution, the abundances of He$^+$ is larger and therefore it reacts to form more of these ions. The higher abundances of these ions are maintained throughout the evolution. 
From 100 to 10$^4$~years and for extinctions up to 2 -- 3~mag, the abundance of C$^+$ and O are also higher for the model with a lower density, hence the abundance of gas phase molecules, such as HCN, NO, or NH, as well as grain species is lower (see Fig.~\ref{apfig:3Dabun_compHIIdens}).

\begin{figure}[t]
        \centering
        \includegraphics[width=0.5\textwidth, trim={0 33 0 35}, clip]{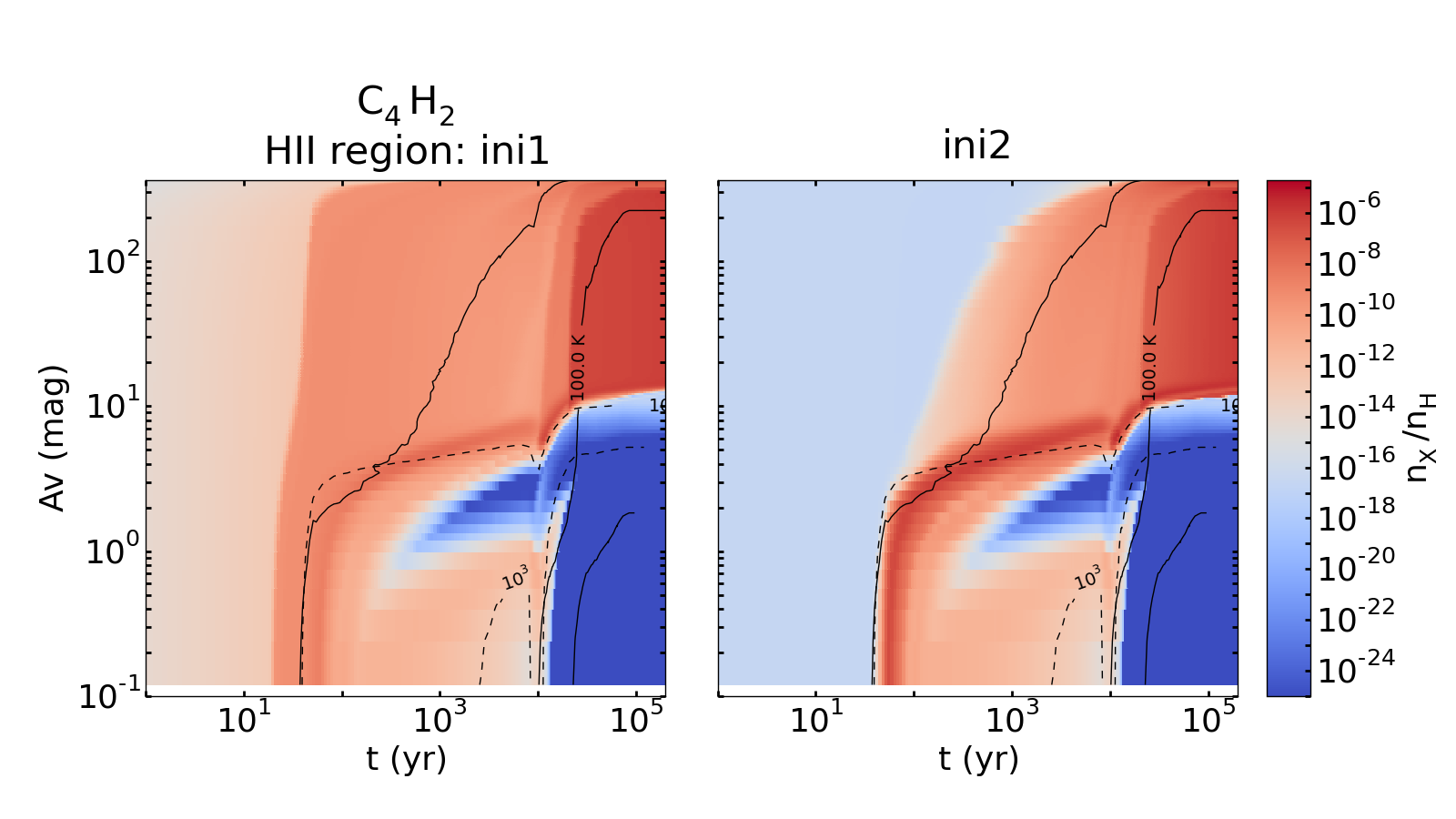}
        \includegraphics[width=0.5\textwidth, trim={0 33 0 35}, clip]{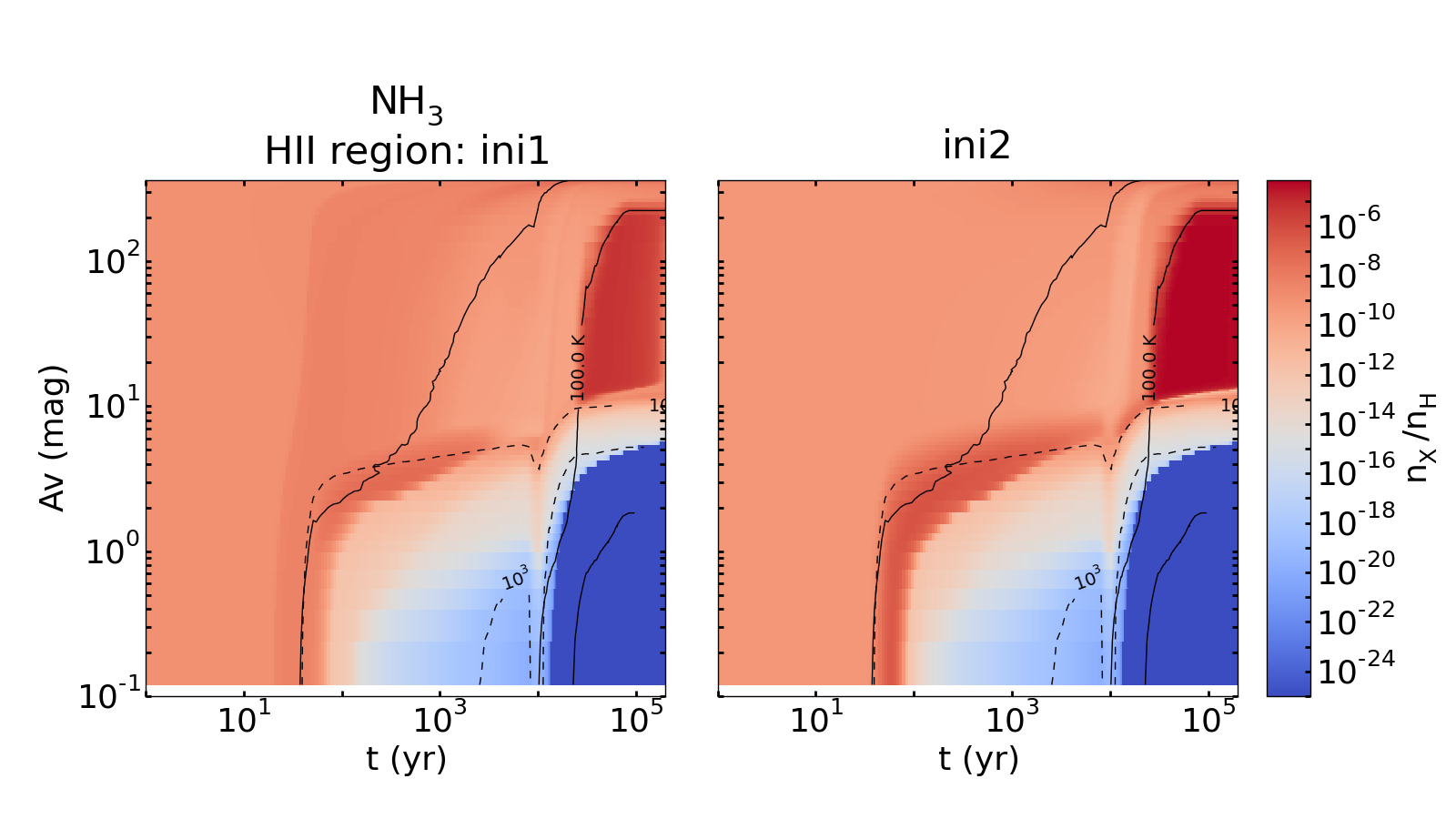}                            
        \caption{Spatio-temporal evolution of the abundance of C$_4$H$_2$ and NH$_3$ for the reference model \textit{mHII:ini1} (left panel) and model \textit{mHII:ini2} (right panel). Contours are plotted: solid line for \Td\ (20, 30, 100 and 150~K) and dashed lines for \Go\ (10$^{-1}$ and 10$^3$~Draine unit).}
        \label{fig:3Dini1-ini2}
\end{figure}

\subsubsection{Item 4 -- Initial abundances:} 
\label{subsubsec:abunItem4}

Using different initial abundances leaves the dust temperature and radiation field intensity profiles unchanged. In Fig.~\ref{fig:3Dini1-ini2} we show the abundances of C$_4$H$_2$ and NH$_3$ for model \textit{ini1} (left panel) and \textit{ini2} (right). In model \textit{ini2} we start with about 15 times more s-CH$_4$ and about ten times less s-CO as shown in Table~\ref{tab:IniAbun}. The evolution starts with less complex molecules, their precursors and carbon chains in the gas phase and on grain surfaces in model \textit{ini2}. Above a dust temperature of 30~K carbon chains form on the grain surface until they desorb at a temperature of about 100~K. In the gas phase, they begin to form only above 20~K instead of 15~K for model \textit{ini1} (see the case of C$_4$H$_2$ in the top panels of Fig.~\ref{fig:3Dini1-ini2}). Therefore, in model \textit{ini2}, up to three orders of magnitude more carbon chains are produced, for a longer period that lasts about 2 -- 3$\times$10$^6$~years longer (see also the warm carbon-chains chemistry (WCCC) model in \cite{Sakai2009} for further discussions). 

In the molecular region (cf.\ Fig.~\ref{fig:regions}), from 10$^4$ to 10$^6$~years, the abundance of CH$_4$, H$_2$O and NH$_3$ is higher in model \textit{ini2}, 15, 5, and 10 times higher respectively (see the case of NH$_3$ in the bottom panels of Fig.~\ref{fig:3Dini1-ini2}). The abundance of radicals leading to these species is also higher but in a lesser extent. For instance the abundances of CH and OH are 10 and 3 times, respectively, higher for model \textit{ini2}. On the contrary, CO, O$_2$ and N$_2$ are less abundant, approximatively 10, 7, and 25 times less, up to 10$^6$~years when they become the dominant species. Additional results are presented in Appendix~\ref{apsec:ini2}.

\subsubsection{Item 5 -- Cut-off density:} 
\label{subsubsec:abunItem5}

When changing the cut-off density in the model from 10~\cmc\ (\textit{c1}) to 10$^6$~\cmc\ (\textit{c6}), the abundance profiles remain the same. The only difference is that they do not extend as deep into the envelope for \textit{c6} because the grid is cut out at a higher density and so at a much closer radius from the star.

\subsection{Dissociation front}
\label{subs:dissociation-front}

\begin{table}[t]
    \centering
    \caption{H$_2$ dissociation front position from the ionization front at 10$^5$~years for the different models. The dissociation front position is defined as $\frac{n(\mathrm{H_2})}{n_\mathrm{H}}$ = 0.1. The asterisk (*) marks the reference model. A$_{\rm V, df}$ and $n$$_{\rm H, df}$ are the visual extinction and density at the dissociation front (annotated ``df'') and $\Delta$ r$_{\rm PDR}$ is the size of the PDR from the ionization front to the dissociation front.}
    \begin{tabular}{lcrc}
            \hline
            Models & \av$_{\rm , df}$ & $\Delta$ r$_{\rm PDR}$ & \nh$_{\rm , df}$ \\
            & (mag) & (AU) & (\cmc) \\
            \hline
            \hline
            \textit{r0.015n7ini1p2.5s183} & 6.8 & 88.7 & 9.26$\times$10$^6$ \\
            \textit{r0.05n7ini1p2.5s183} (*) & 5.2 & 64.2 & 9.82$\times$10$^6$ \\
            \textit{r0.10n7ini1p2.5s183} & 4.8 & 59.2 & 9.90$\times$10$^6$ \\
            \textit{r0.05n6ini1p2.5s183} & 5.5 & 758.3 & 8.30$\times$10$^5$ \\
            \textit{r0.015n7ini1p1s183} & 6.8 & 86.4 & 9.72$\times$10$^6$ \\
            \textit{r0.05n7ini1p1s183} & 5.5 & 455.3 & 1.44$\times$10$^6$ \\
            \textit{r0.10n7ini1p1s183} & 4.7 & 1508.2 & 3.62$\times$10$^5$ \\
            \hline
    \end{tabular}
    \label{tab:dissociation-front-values}
\end{table}

We used the abundance profiles to study the dissociation front and the size of the PDR. In Table~\ref{tab:dissociation-front-values} we present the position of the dissociation front located where $\frac{n(\mathrm{H_2})}{n_\mathrm{H}}$ = 0.1. 
In all the cases the width of the PDR ($\Delta$ r$_{\rm PDR}$) is extremely small, about tow orders of magnitude, compared to the size of the \hii\ region. This results in the PDR being a very thin shell surrounding the \hii\ region. 
For models \textit{mHII:p2.5} the size of the PDR decreases when the size of the \hii\ region increases, and the density at the dissociation front is higher, because the radiation field is less strong at the ionization front. On the contrary and as we could expect, the size of the PDR increases when the density at the ionization front decreases because the UV radiation can penetrate deeper into the core. 
For the models \textit{mHII:p1} the distance between the ionization front and the dissociation front increases with the size of the \hii\ region despite the lower incident radiation field due
to a decrease in density at the ionization front. 

We notice here that the H/H$_2$ transition occurs for higher visual extinction (> 5~mag) compared to typical PDRs where it happens at \av\ < 2~mag. This is due to the extremely strong radiation field in the models and also to the limited spatial resolution that could underestimate the H$_2$ self-shielding. For example, it is about two orders of magnitude higher than what can found in the Orion bar. For models with lower incident radiation field the H/H$_2$ happens for lower extinction, for example, if we divide the radiation field intensity by ten in the reference model the dissociation front is located at $\approx$ 3.9~mag. If divided by 100 it is at 2.6~mag.

\subsection{Line emission}
\label{sec:synthetic spectra}

\begin{table}[t]
    \caption{List of selected species, one of their transition and the corresponding frequency in GHz.}
    \vspace{0.25cm}
    \centering
    \begin{tabular}{c|c|c}
            \hline
            Molecules & Transition & Frequency (GHz) \\
            \hline
            \hline
             NH$_3$ & 5(5,1) -- 5(5,0) & 24.53 \\
             HC$^{15}$N & 1 -- 0 & 86.06 \\
             HCO & 1(0,1) -- 0(0,0) & 86.71 \\
             HN$^{13}$C & 1 -- 0 & 87.09 \\ 
             HCO$^+$ & 1 -- 0 & 89.19 \\
             N$_2$H$^+$ & 1 -- 0 & 93.17 \\
             H$_2$CO & 2(0,2) -- 1(0,1) & 101.33 \\ 
             CN & 1 -- 0 & 113.12 \\             
             CH$_3$OH & 2(2,1) -- 3(1,2) & 335.13 \\
             H$_2$$^{18}$O & 4(1,4) -- 3(2,1) & 390.61 \\
             C & $^3$P$_{1}$ -- $^3$P$_{0}$ & 492.16 \\ 
             HCN & 7 -- 6  v$_2$ = 1 & 623.36 \\ 
             HNC & 7 -- 6 v$_2$ = 1 & 638.91 \\
             C$^+$ & $^2$P$_{3/2}$ -- $^2$P$_{1/2}$ & 1900.54 \\
             O & $^3$P$_{1}$ -- $^3$P$_{2}$ & 4744.78 \\
             \hline
            \end{tabular}
    \label{tab:selected-molecules}
\end{table}

We studied the line emission for the molecular and atomic transitions presented in Table~\ref{tab:selected-molecules}. We have selected species such as C$^+$ and O that can be produced in the internal PDR due to the UV radiation field, and other species such as HCN and CH$_3$OH that can be either destroyed or not formed. 
We also study vibrationally excited HCN and HNC as they trace mainly hot gas found close to the \hii\ regions \citep{Rolffs2010}. Most of the transitions are chosen because the lines are optically thin and unblended by other lines from the same species. Thus, we used \hcn\ and \hnc\ instead of HCN and HNC\footnote{\hcn\ is optically thick from 10$^4$ to 3$\times$10$^4$~years.}. 
No isotopologues are included in the chemical network, so to obtain their spectra we used the local ISM isotopologue ratios from \cite{Wilson&Rood1994}: $^{12}$C/$^{13}$C = 70, $^{14}$N/$^{15}$N = 450 and $^{16}$O/$^{18}$O = 500.  

We used \radmc\ (see Sect.~\ref{sec:method}) to produce synthetic cubes from which we can extract synthetic maps and spectra for the different models listed in Table~\ref{tab:models}. In the following, we describe the results obtained for each of them.

\begin{figure}[t]
        \centering
        \includegraphics[width=0.45\textwidth]{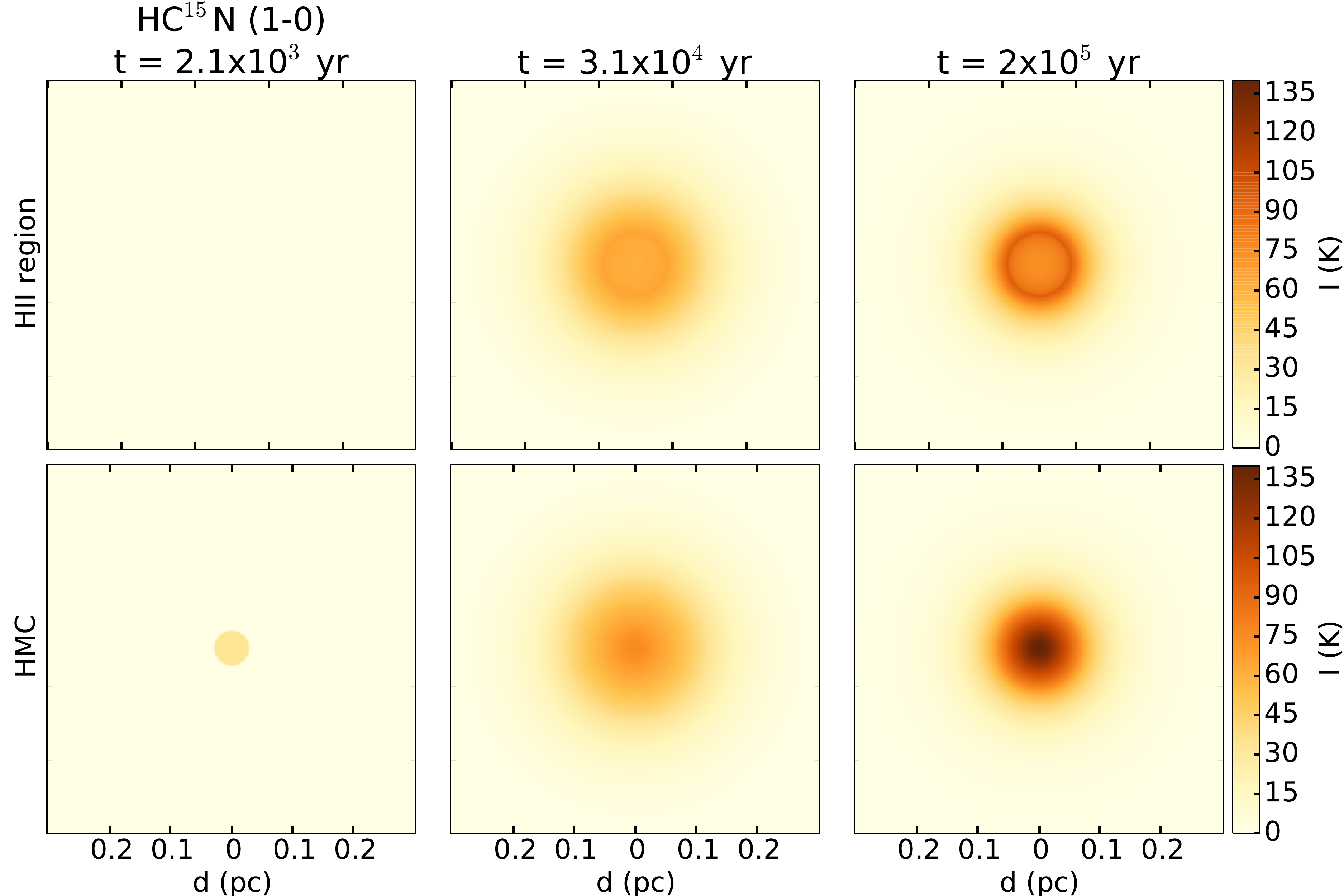}
        \includegraphics[width=0.45\textwidth, trim={0 0 0 21mm},clip]{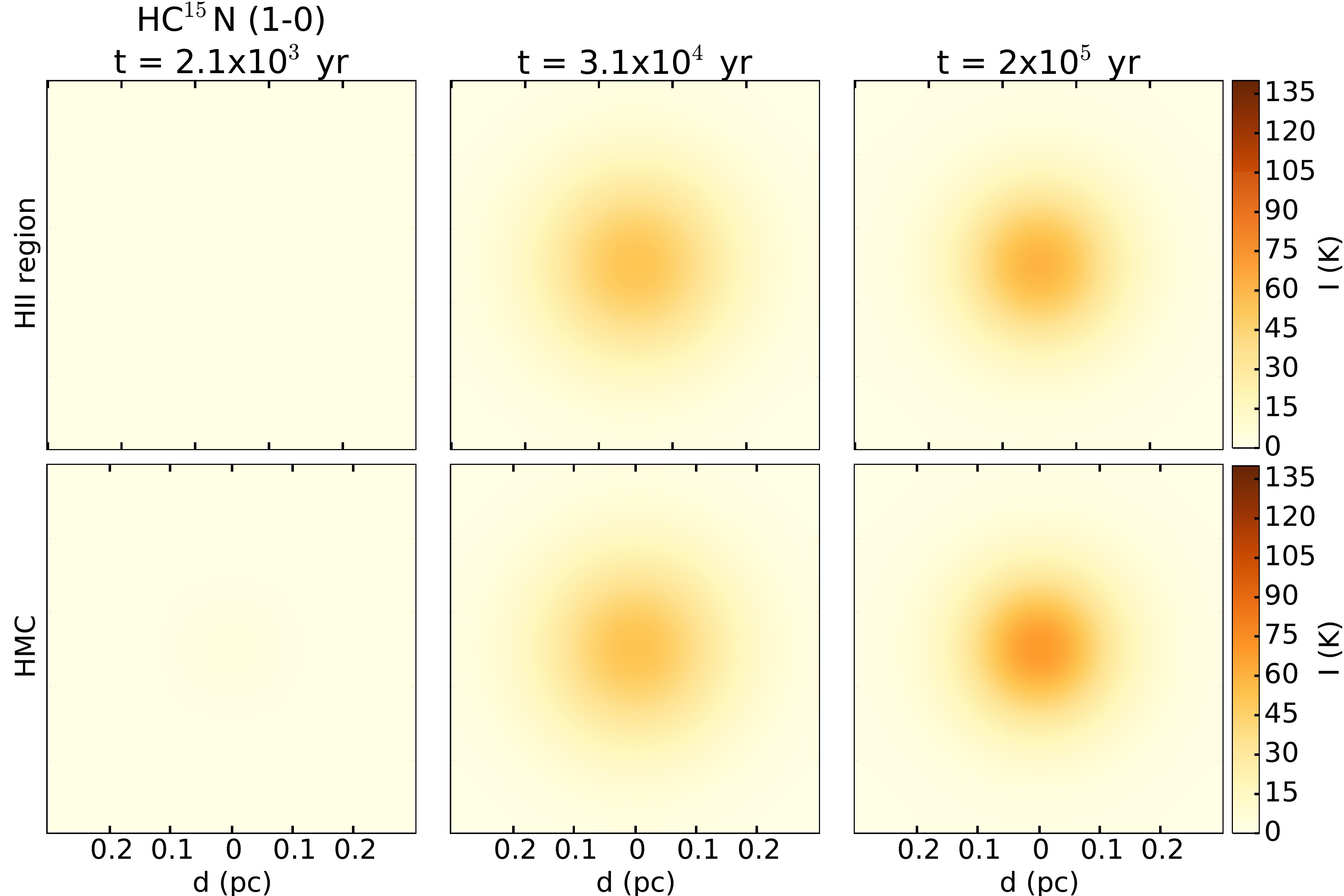}
        \caption{Time evolution of the \hcn\,(1--0) maps for the reference model \textit{mHII} (top) and the associated model \textit{mHMC} (bottom). The y-axis is the same as the x-axis. The bottom maps represent the same spectra convolved to the beam of the IRAM 30~m telescope. The convolved spectra have a weaker intensity.}
        \label{fig:hc15n_maps}
\end{figure}

\begin{figure}[t]
        \centering
        \includegraphics[width=0.45\textwidth]{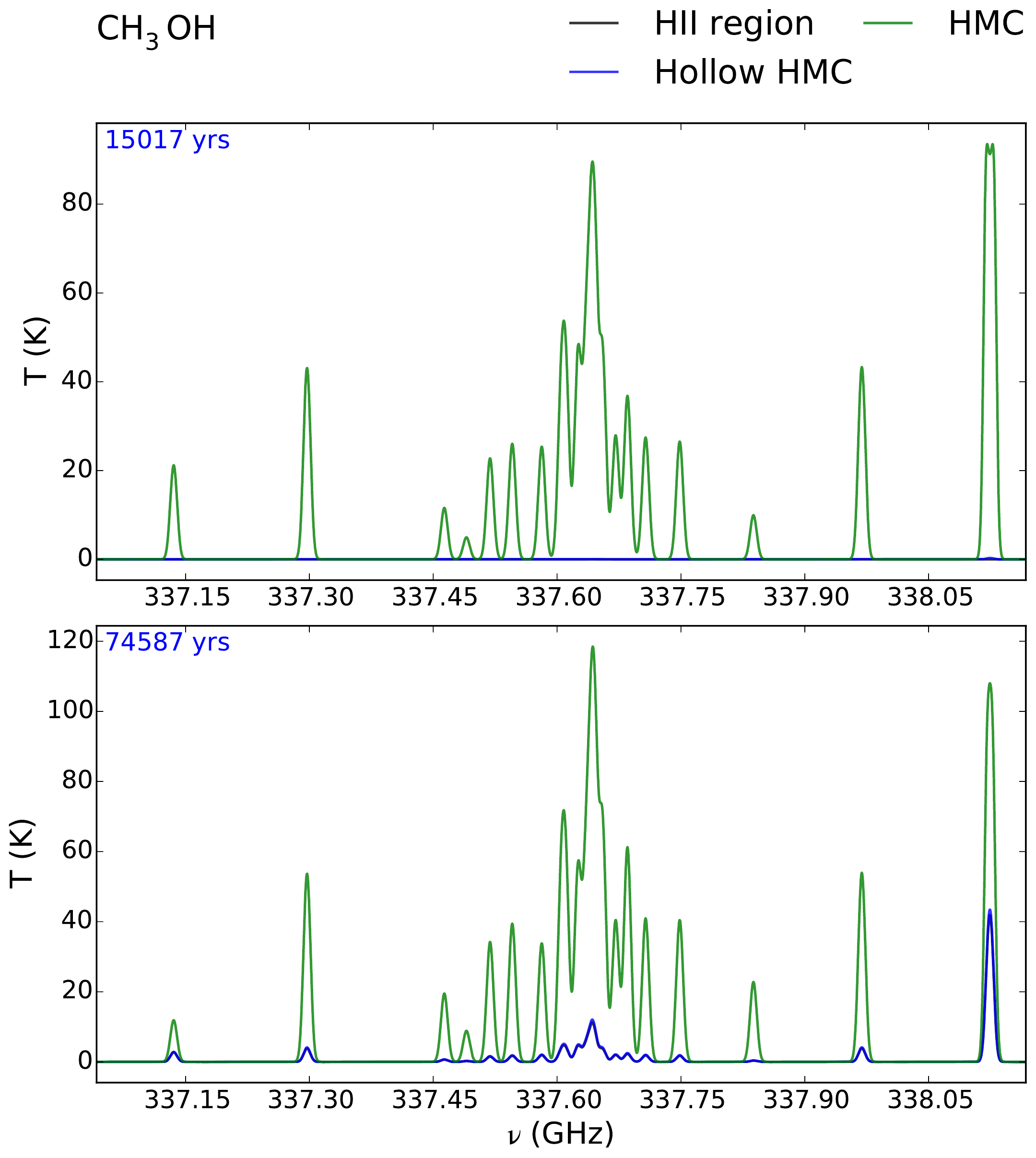}
        \caption{Example of the time evolution of CH$_3$OH spectra around 337.6~GHz for the reference model \textit{mHII} (black) and the associated model \textit{mHHMC} (blue) and \textit{mHMC} (green). Top: t = 1.5$\times$10$^4$~years. Bottom: t = 7.46$\times$10$^4$~years.}
        \label{fig:ch3oh_spec}
\end{figure}

\subsubsection{Item 1 -- HMC/HHMC/\hii\ region:}  
\label{subsubsec:specItem1}

In Fig.~\ref{fig:hc15n_maps} we show the maps of the peak intensity of \hcn\,(1--0) for the reference model \textit{mHII:r0.05n7ini1c1p2.5s183} (top panels) and model \textit{mHMC:r0.05n7ini1c1p2.5s183} (bottom). 
The maps are shown for different times between 10$^3$ and 10$^5$~years and the x-axis represents the distance in parsec from the proto-star. The top six maps have the original resolution of the model of 0.5~$''$ for an object located at 1~kpc (corresponds to a pixel size of 500~AU), while the six bottom panels show the spectra convolved to the beam of the IRAM 30~m telescope at the frequency of HC$^{15}$N\,(1--0), which corresponds to 29.22$''$. The maps for the HHMC model are not shown in Fig.~\ref{fig:hc15n_maps} because the results are very similar to the \hii\ region maps for these three particular time steps. We see that the emission of \hcn\ increases with time and is weaker for the \hii\ region compared to the HMC model. It also presents a ring structure around the central proto-star. This ring structure is not seen in the convolved maps\footnote{The free-free continuum was not included, which would result in the lines being seen in absorption in the \hii\ region model.}.

Synthetic spectra obtained for CH$_3$OH, centered at 337.6~GHz, are shown at 1.5$\times$10$^4$ and 7.5$\times$10$^4$~years in Fig.~\ref{fig:ch3oh_spec} for model \textit{mHII} (black), \textit{mHHMC} (blue) and \textit{mHMC} (green). Contrary to the maps presented in Fig.~\ref{fig:hc15n_maps}, these spectra (as well as all the others used in this work) have a pixel size of 100~AU (0.1$''$ for an object at 1~kpc) and are not convolved to the beam of any observations. For the HHMC and \hii\ region models there is no methanol emission at about 10$^4$~years. Later, the emission of these two models increases but still remains weaker by almost one order of magnitude compared to the HMC model. After that time, both the HHMC and \hii\ region models have similar integrated intensities for these species. This high difference between the HMC model and the two others is due to the much higher column density in the HMC model as there is no cavity.

\begin{figure*}[htbp]
        \includegraphics[width=0.32\textwidth]{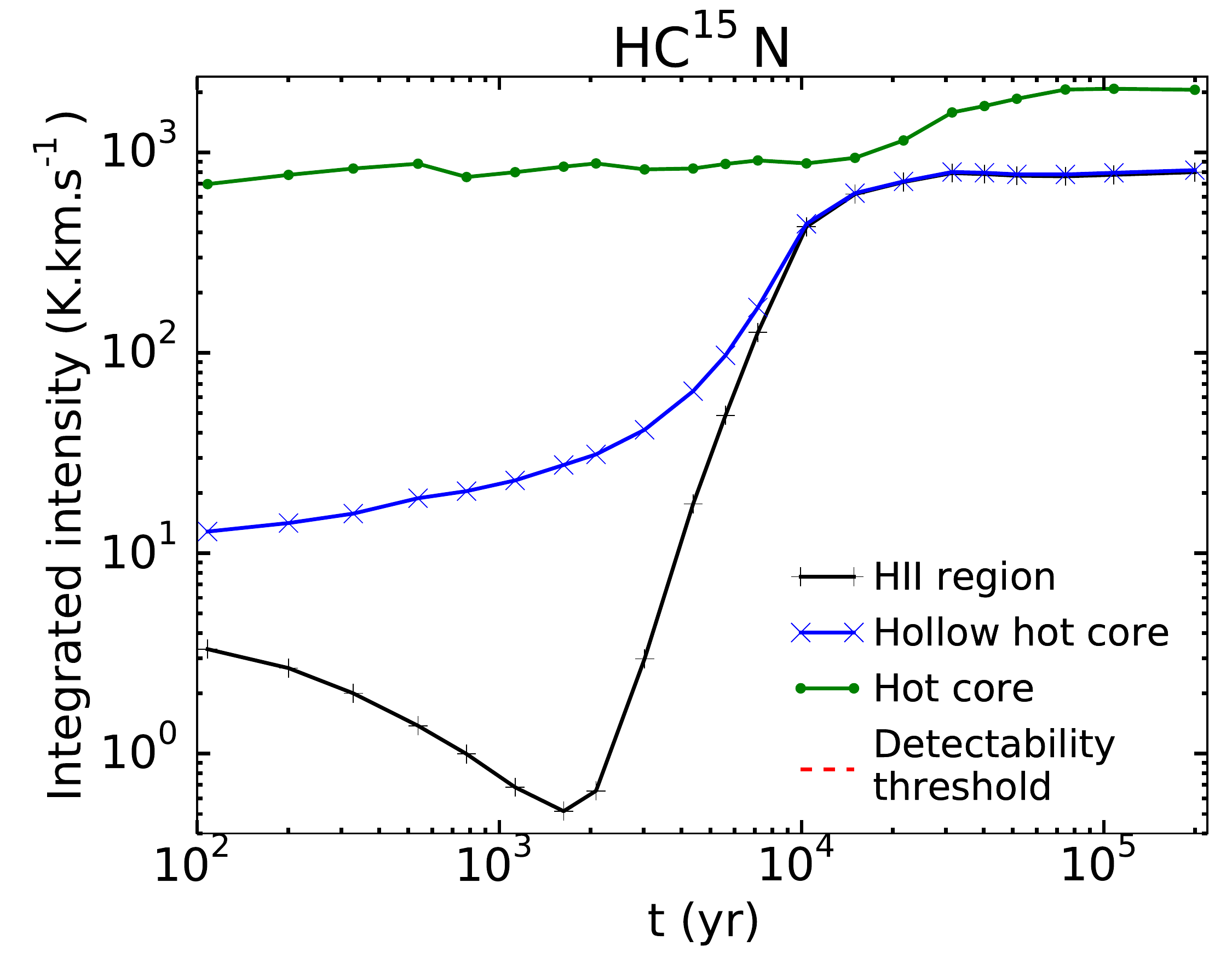}
        \includegraphics[width=0.32\textwidth]{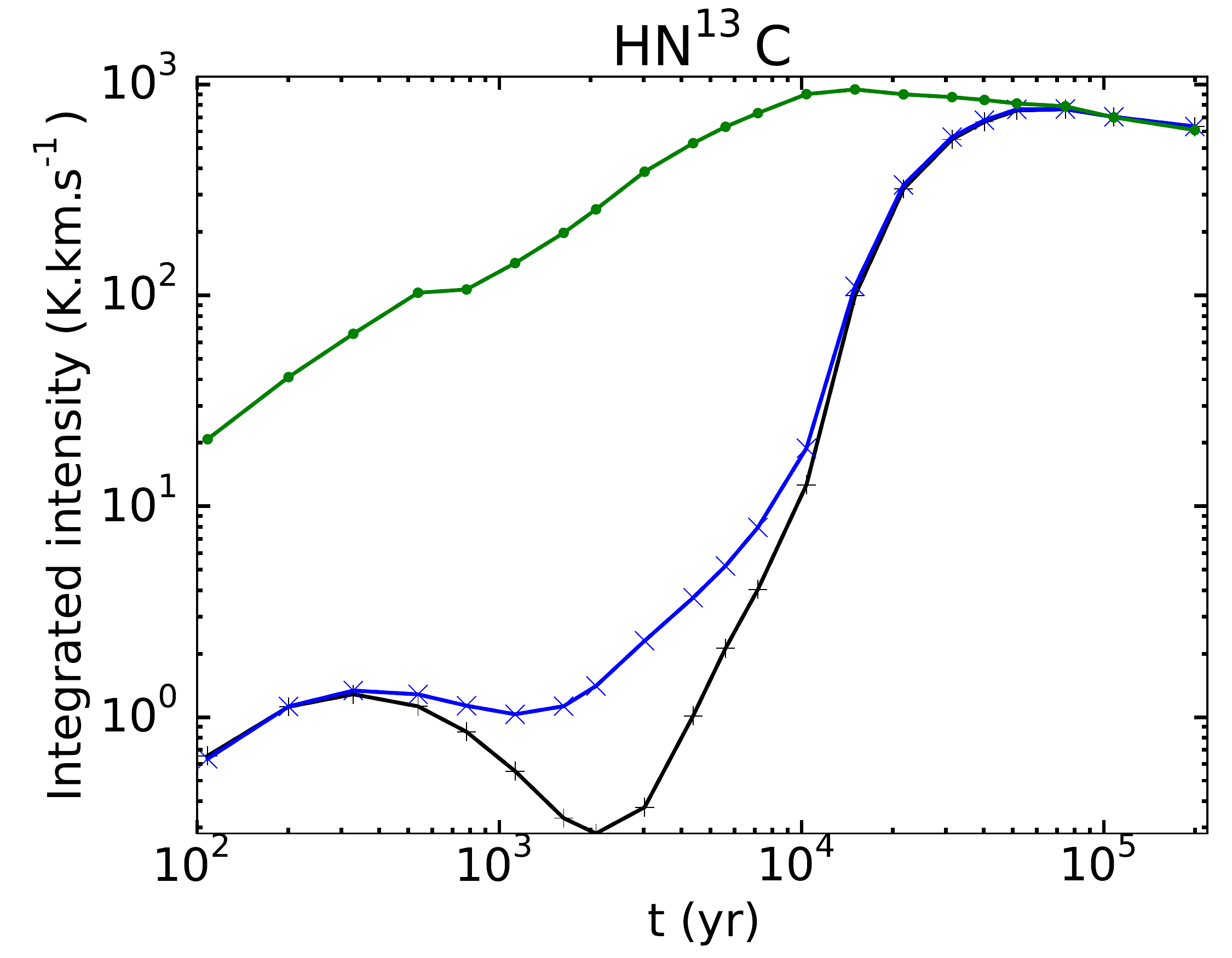}  
        \includegraphics[width=0.32\textwidth]{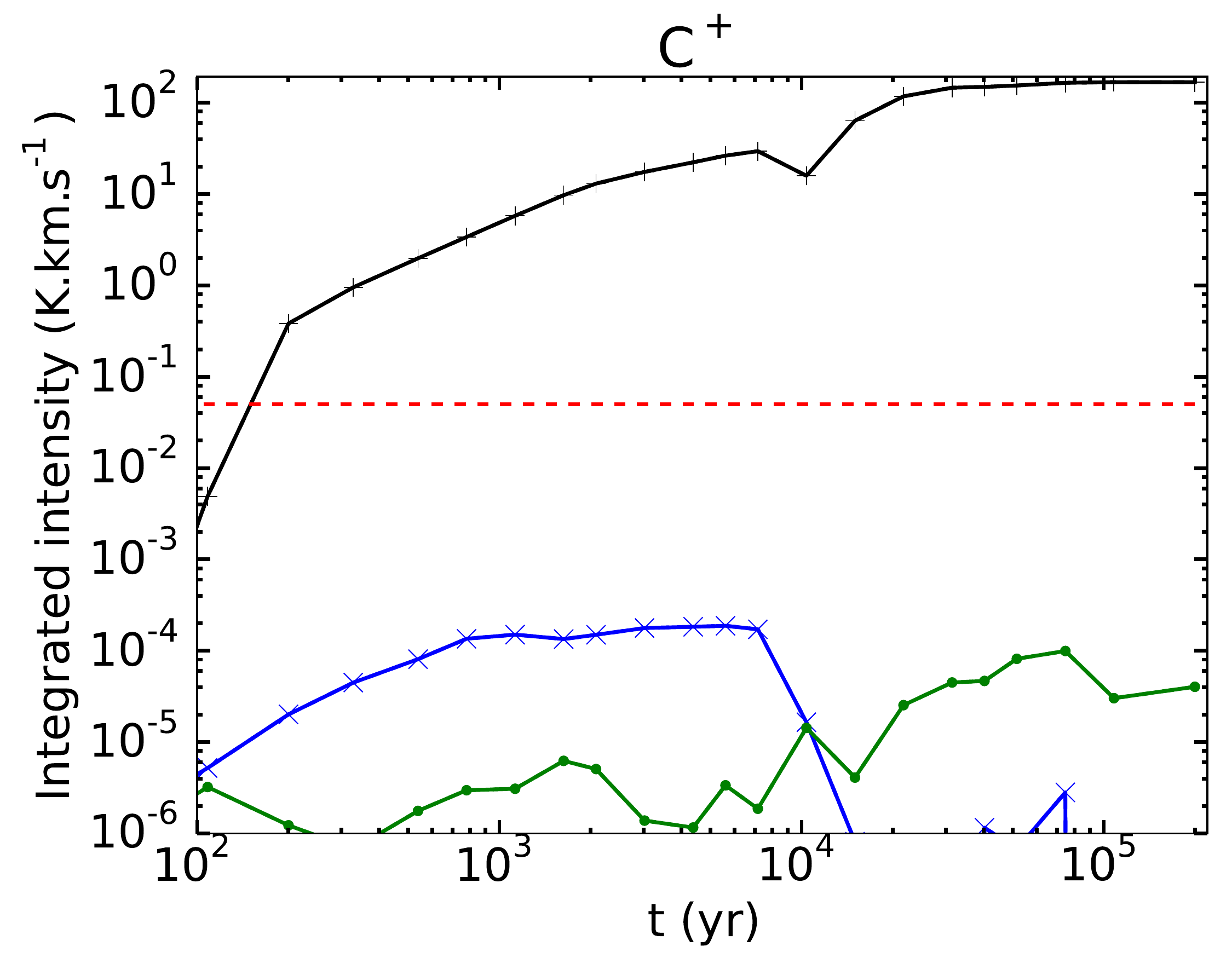} 

        \includegraphics[width=0.32\textwidth]{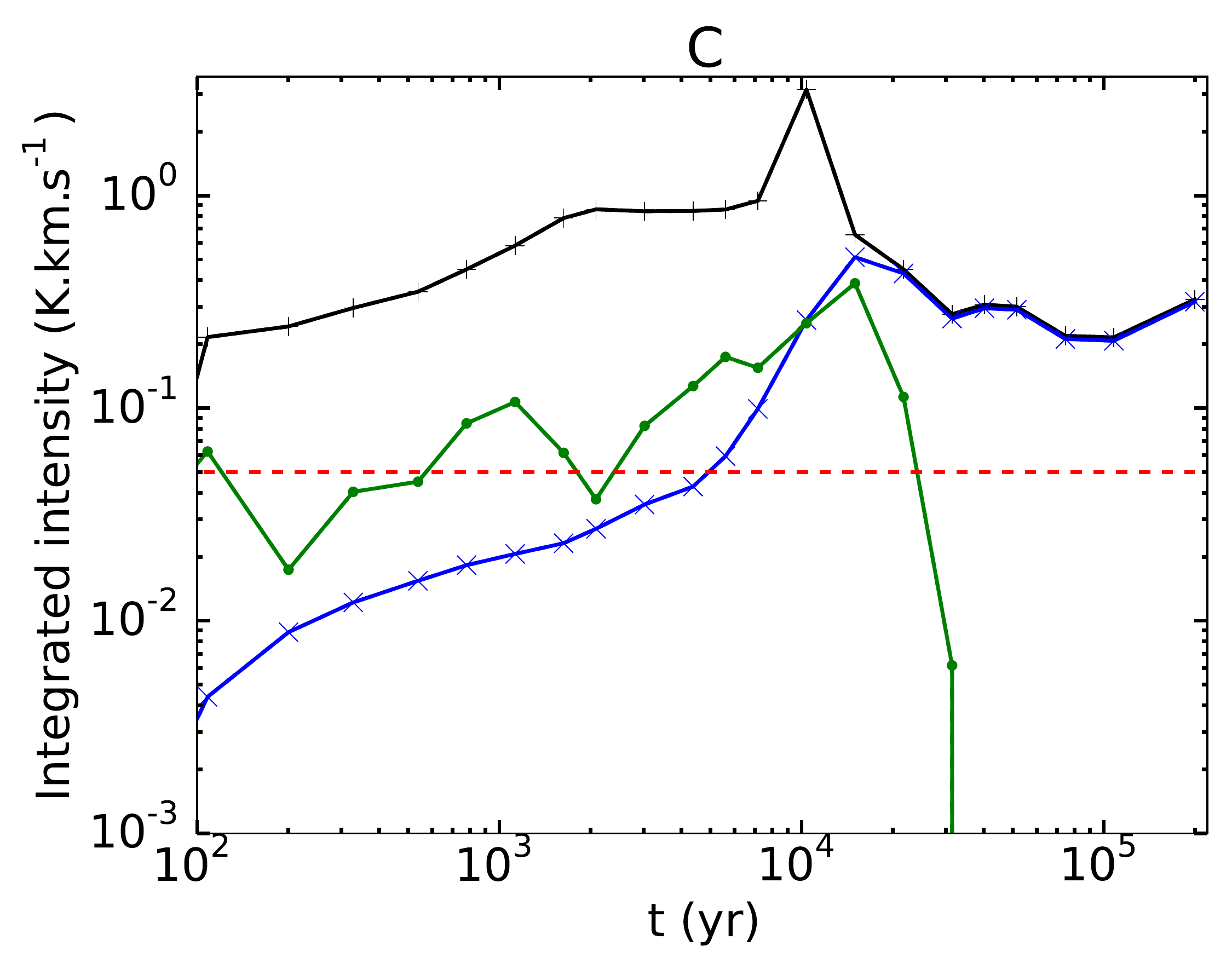}
        \includegraphics[width=0.32\textwidth]{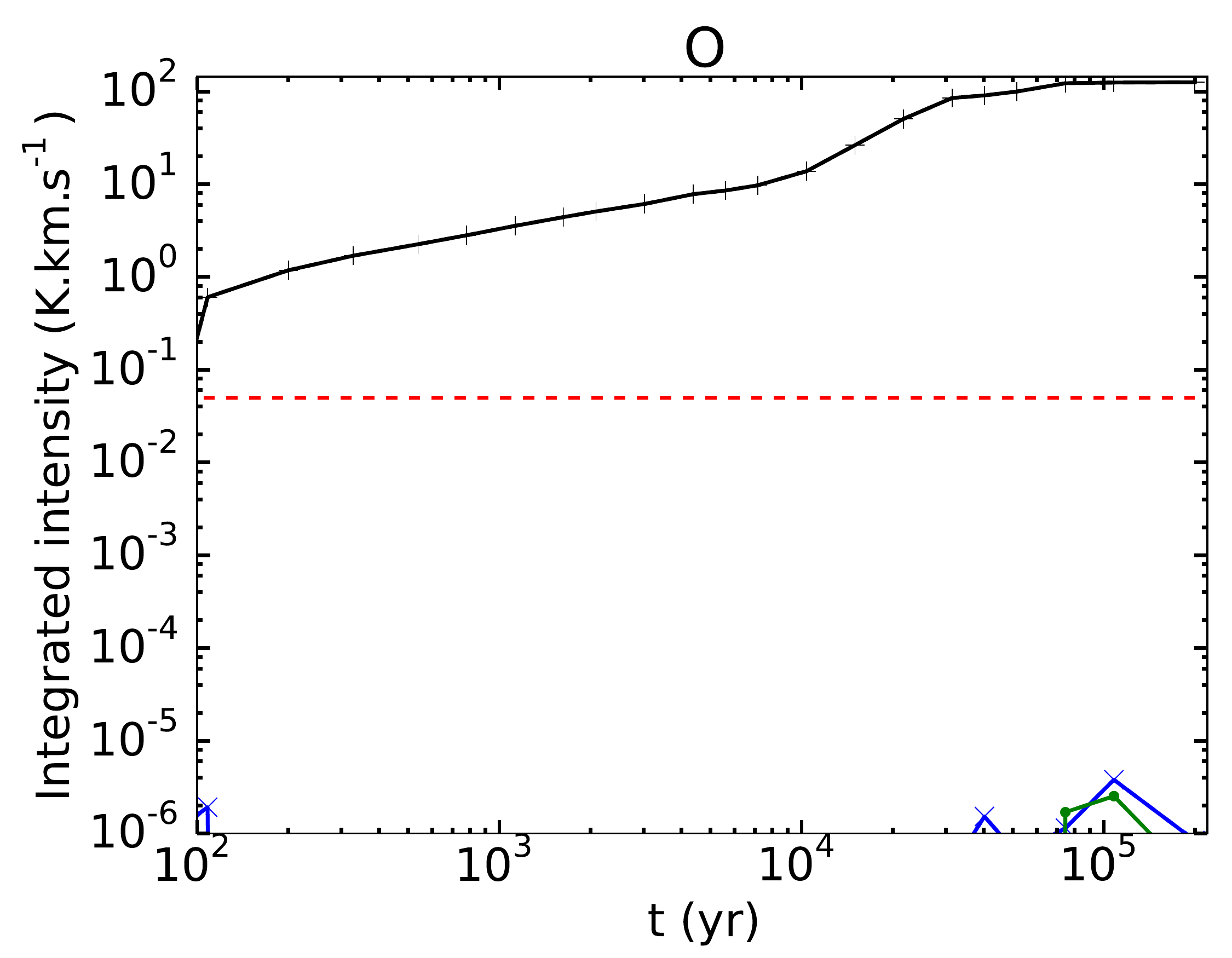}
        \includegraphics[width=0.32\textwidth]{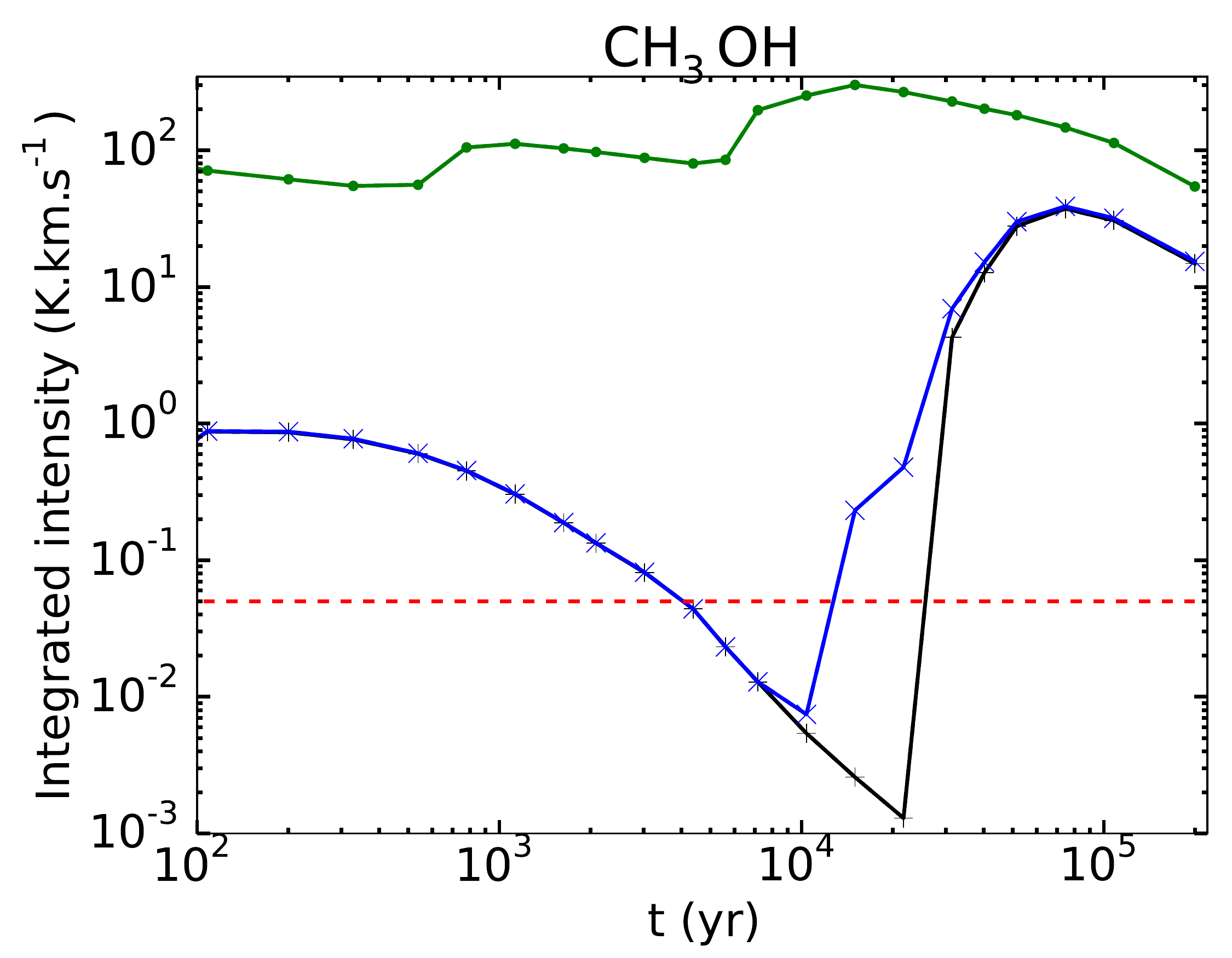} 

        \includegraphics[width=0.32\textwidth]{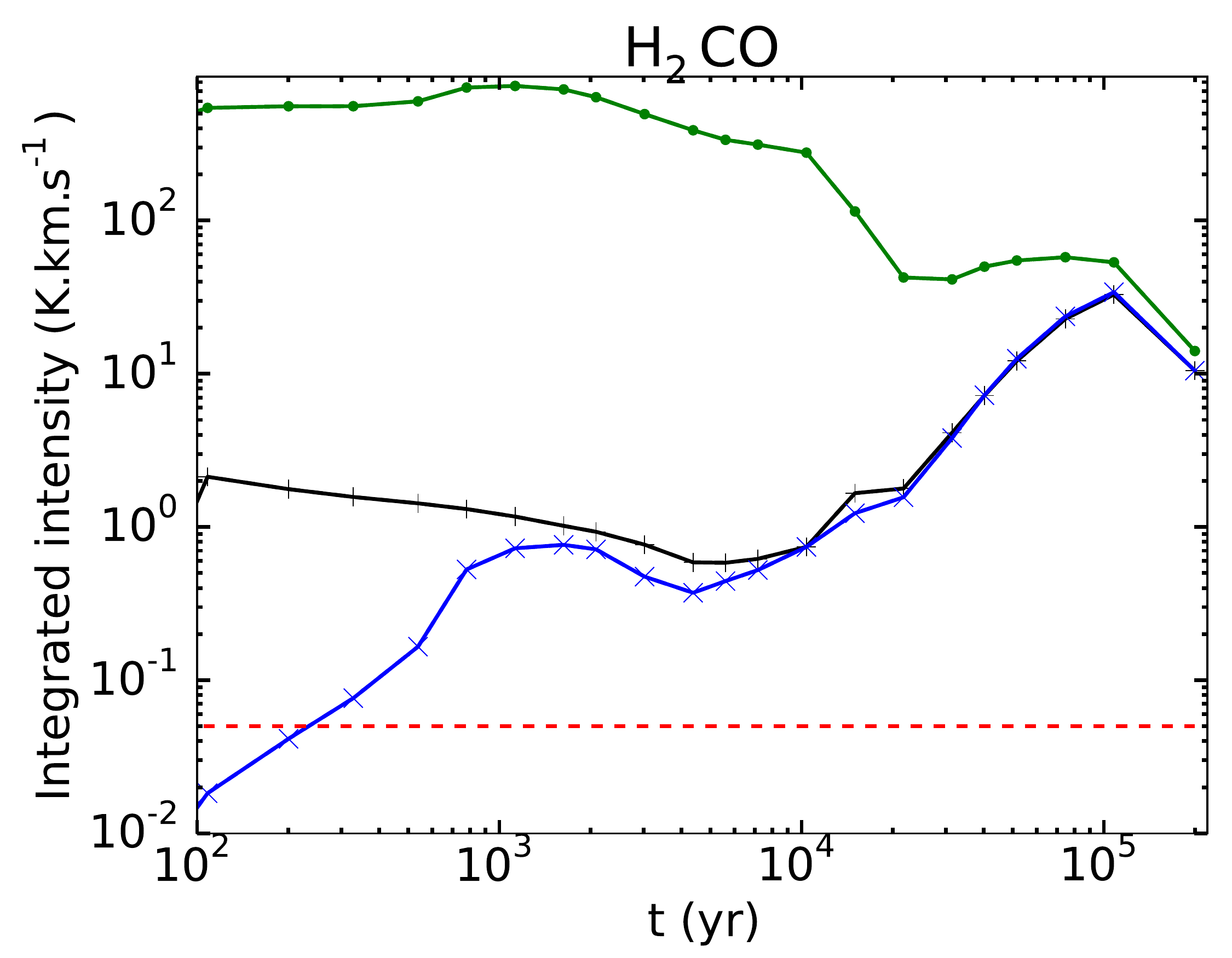}
        \includegraphics[width=0.32\textwidth]{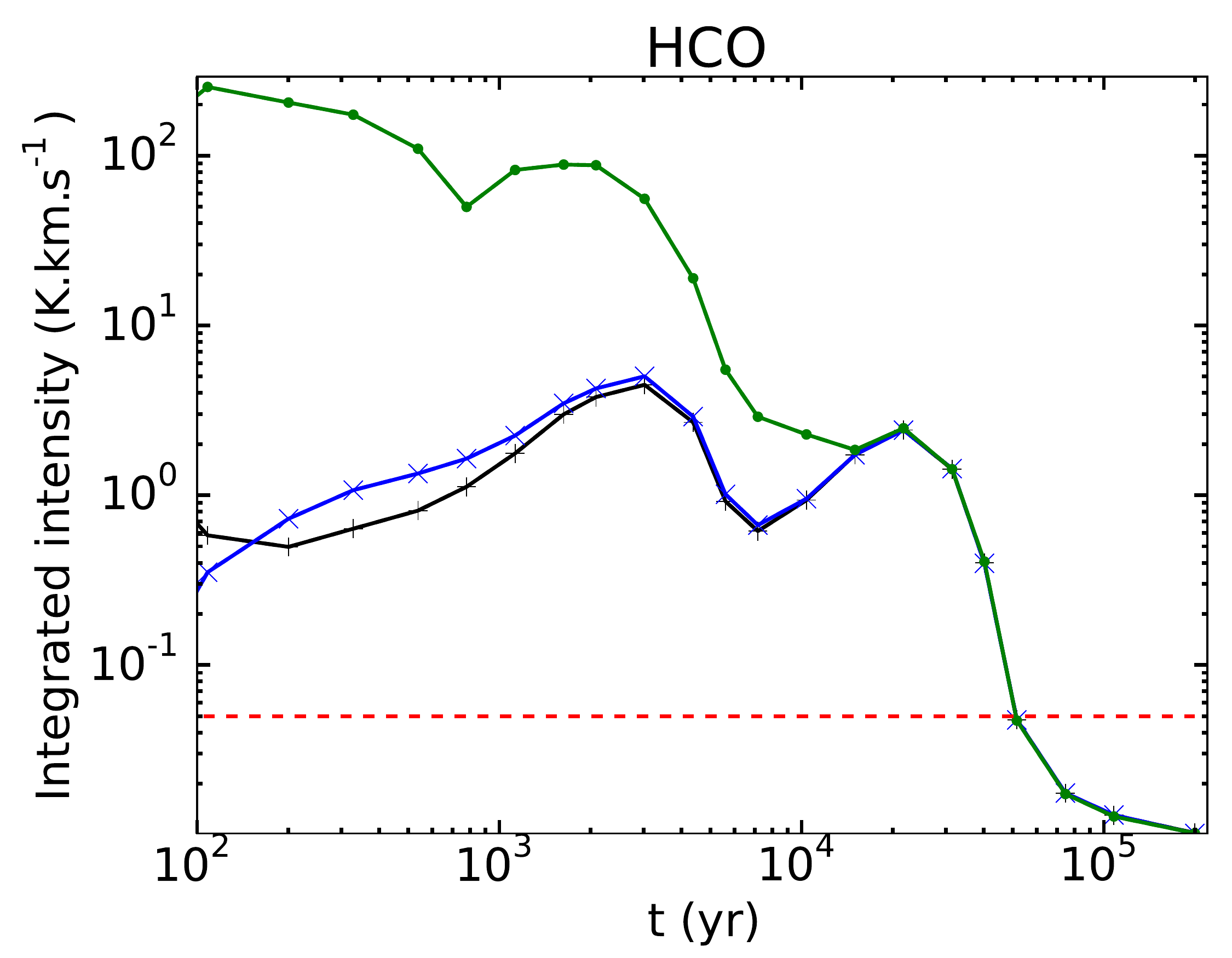}
        \includegraphics[width=0.32\textwidth]{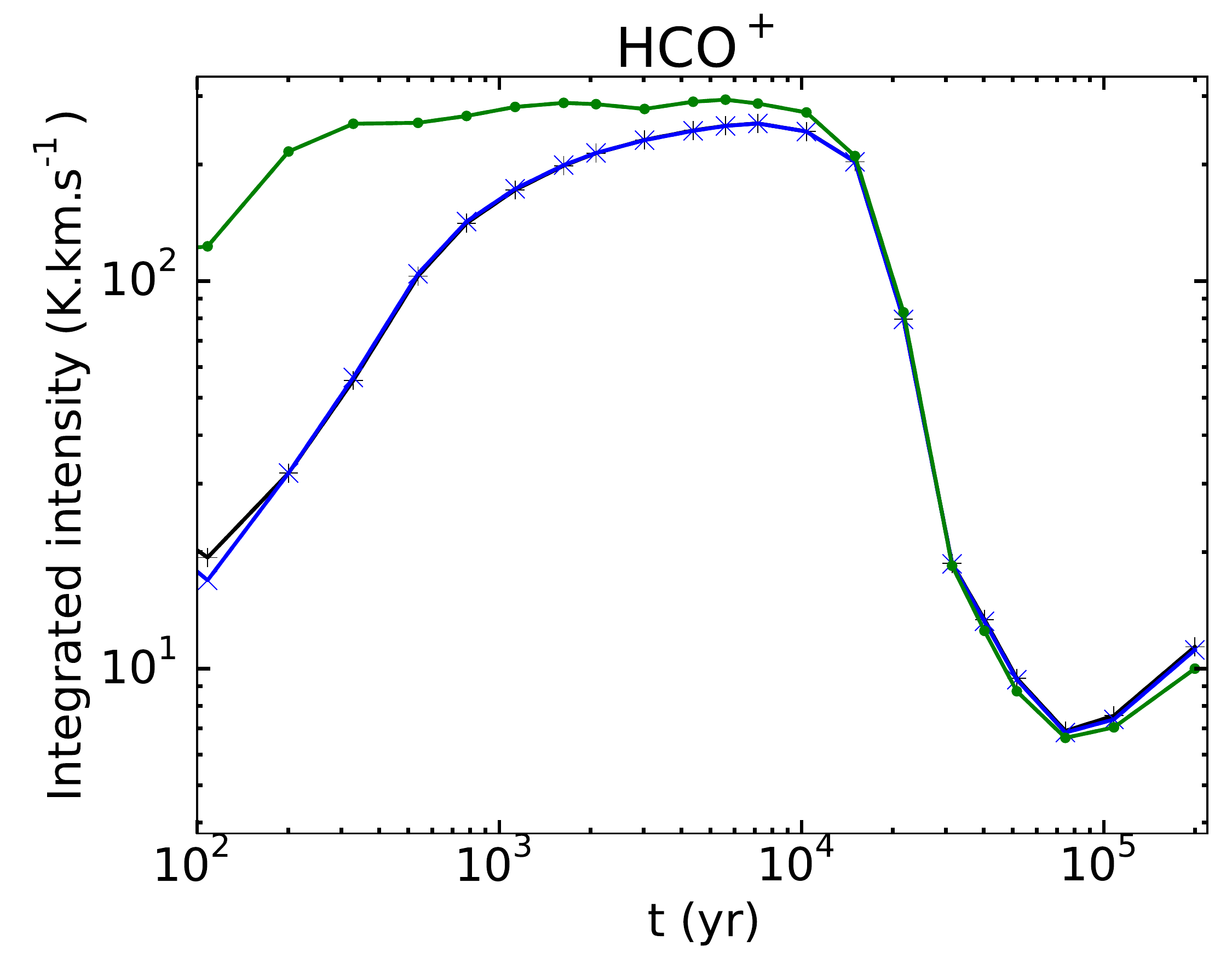} 

        \includegraphics[width=0.32\textwidth]{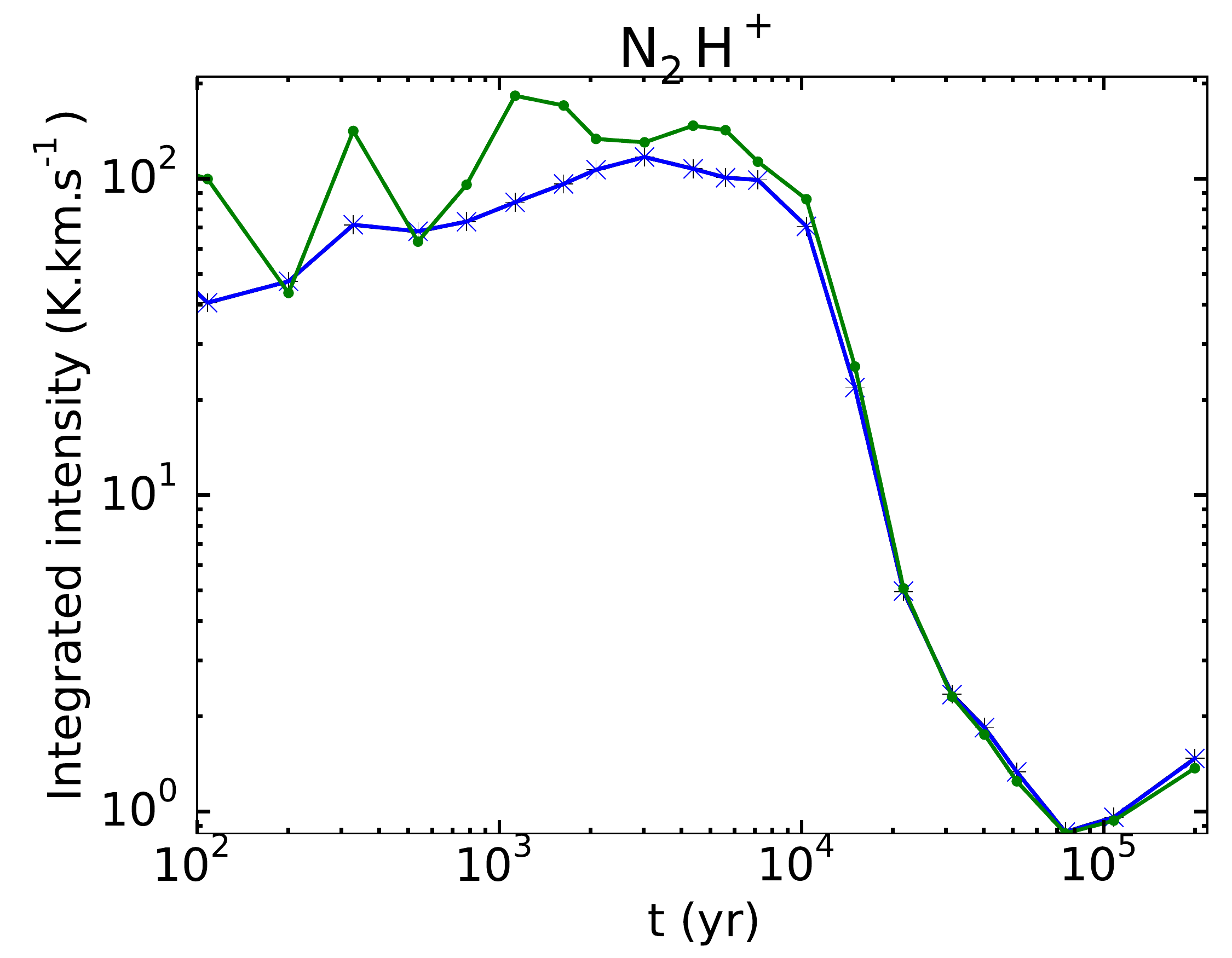}
        \includegraphics[width=0.32\textwidth]{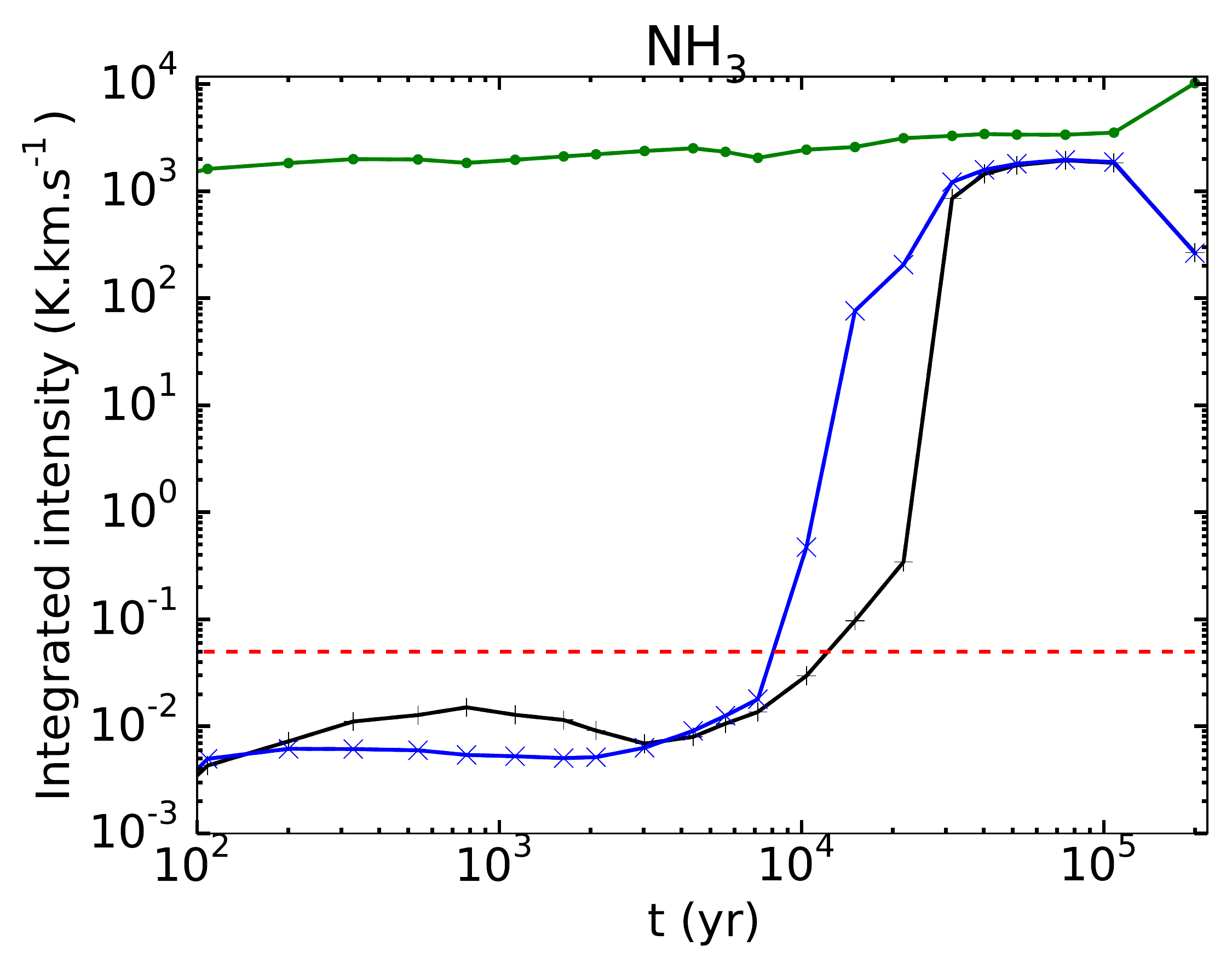}
        \includegraphics[width=0.32\textwidth]{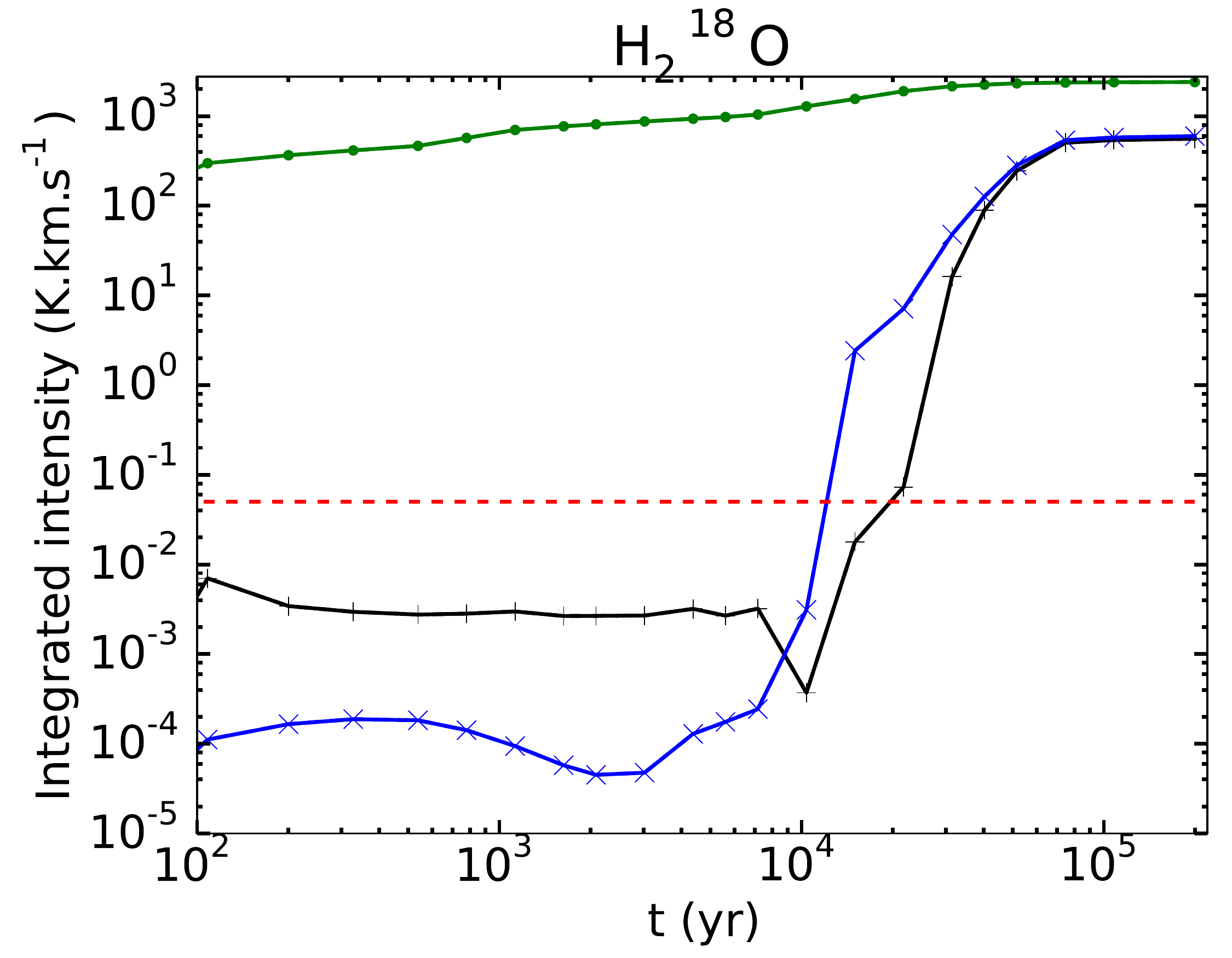} 

        \includegraphics[width=0.32\textwidth]{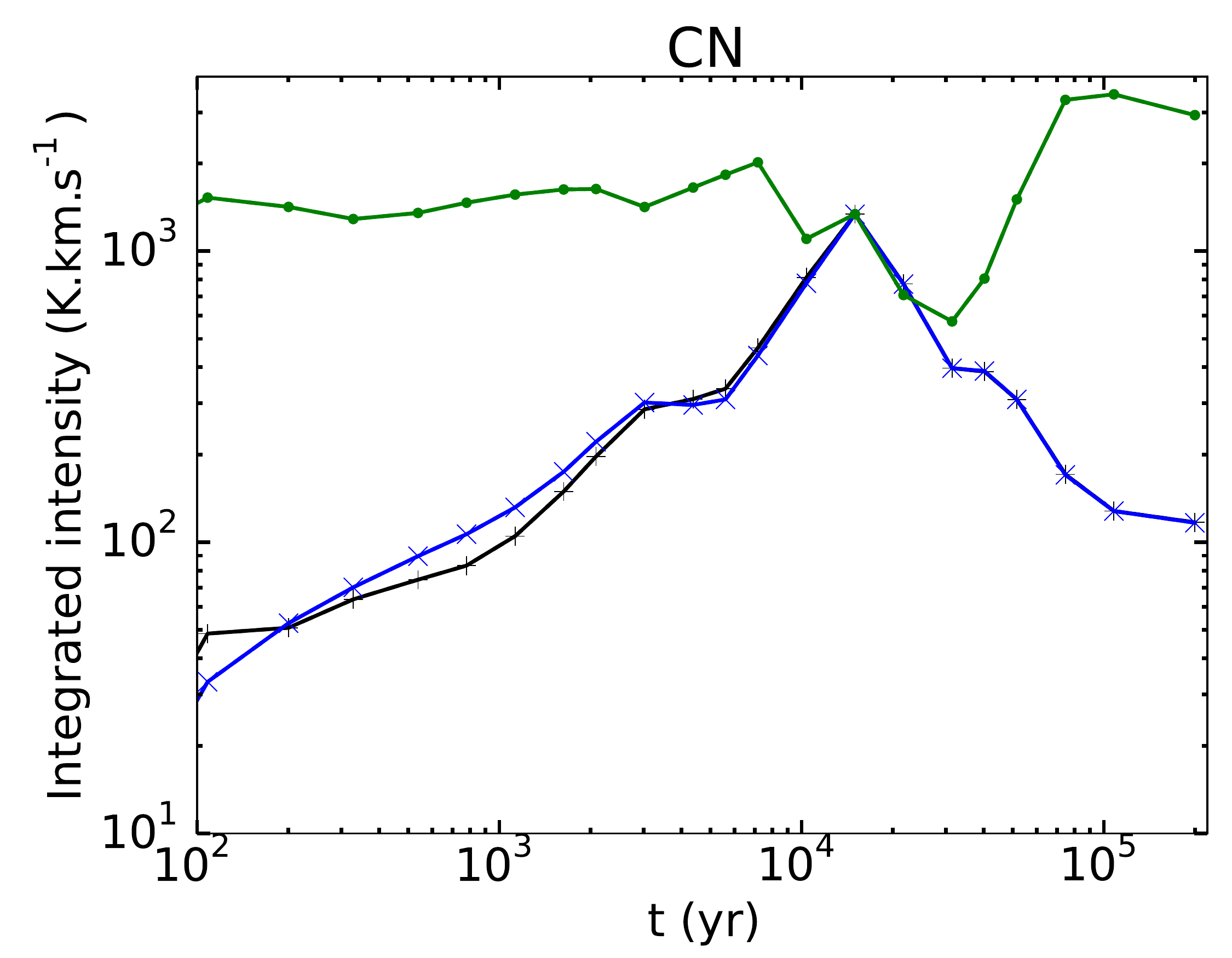}
        \includegraphics[width=0.32\textwidth]{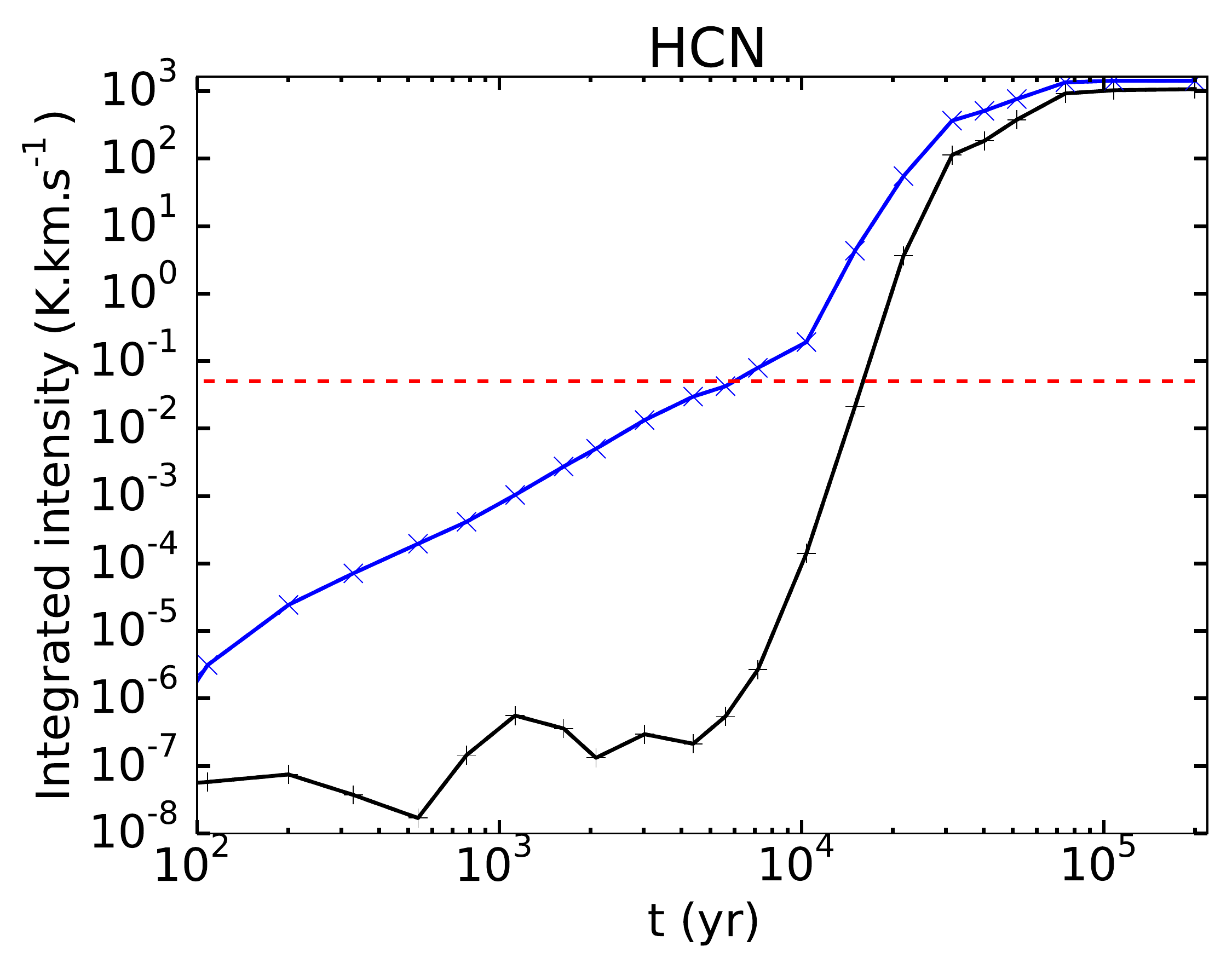}
        \includegraphics[width=0.32\textwidth]{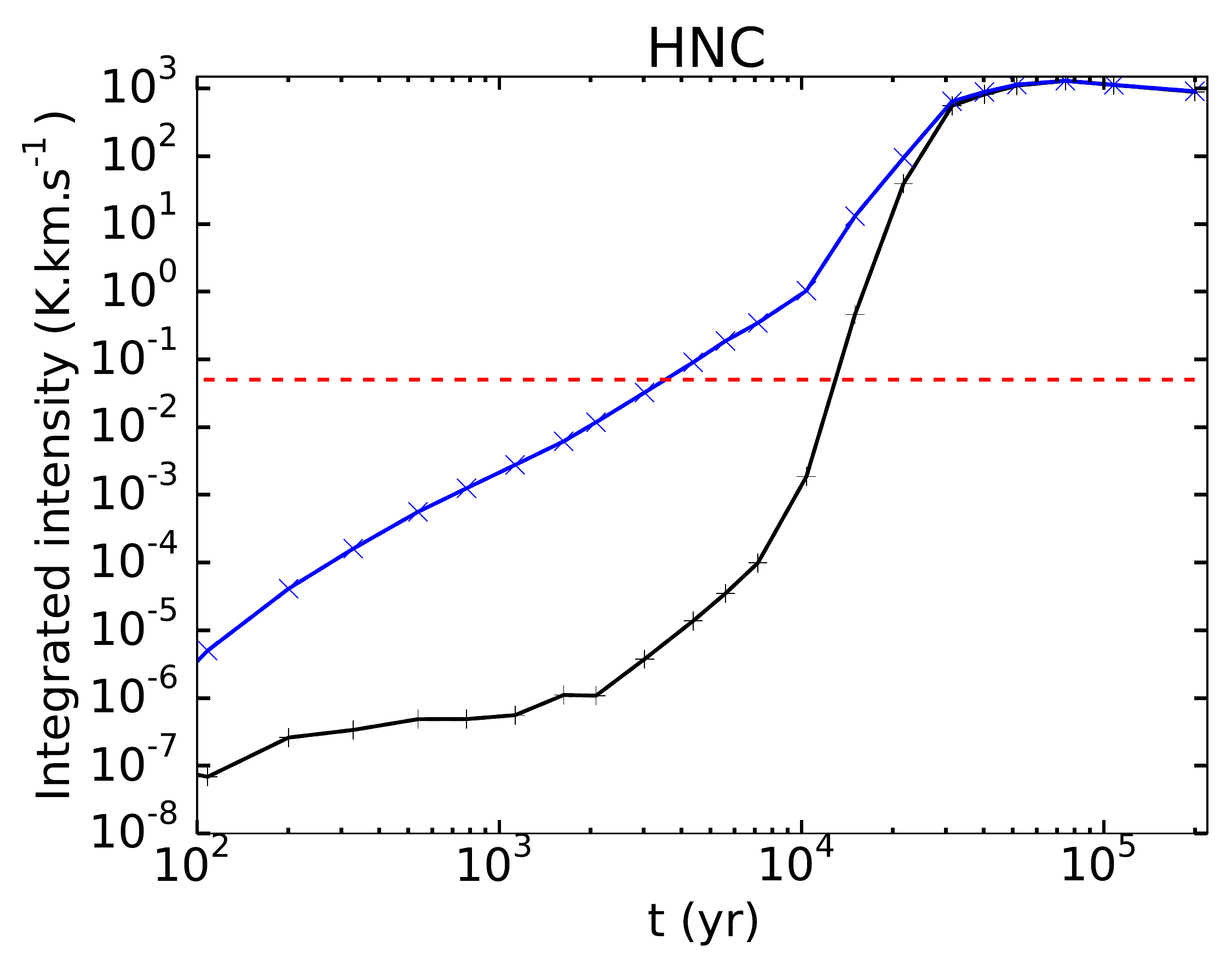} 
        \caption{Time evolution of the integrated intensities of all the species presented in Table~\ref{tab:selected-molecules}. Model \textit{mHII} is represented in black, model \textit{mHHMC} in blue and model \textit{mHMC} in green. The red dashed line represents the detectability threshold of 5$\times$10$^{-2}$~K\,km\,s$^{-1}$ obtained with an assumed rms noise of 3~mK.}
        \label{fig:intInt_comp_HII-HHMC-HMC}
\end{figure*}

The temporal evolution of integrated intensities for all the species and their transitions listed in Table~\ref{tab:selected-molecules} is shown in Fig.~\ref{fig:intInt_comp_HII-HHMC-HMC}. The emission of \hnc\ is similar to the ones of \hcn. Until 10$^4$~years, the emission of the HHMC model is also stronger than the emission in the \hii\ region model by approximately one order of magnitude for \hcn\ and a bit less for \hnc. After that time, both models have similar integrated intensities for most of the species. The emission increases in time for all models as the species are released from the dust grains and then formed in the gas phase. The presence of the cavity has a strong effect on the intensities of the HHMC model which are in general lower than the intensities of the HMC model due to a smaller column density. On the contrary, the emission for C$^+$ is stronger in the \hii\ region model and appears to be undetectable for the HMC and HHMC models with an integrated intensity lower than 4$\times$10$^{-4}$~K\,km\,s$^{-1}$. 

In order to determine if the lines are detectable with current astronomical facilities, we assumed an arbitrary detectability threshold for all lines derived from observations with the IRAM 30~m telescope\footnote{We refer several times in this paper to observations with the IRAM 30~m telescope because of the spectral survey of the UC\hii\ region Monoceros R2 made by \cite{Trevino-Morales2014, Trevino-Morales2016} with this telescope. An outlook of this work will be to produce a model of Monoceros R2 and compare the results with the observations.}. 
Considering a minimum value of the rms noise of about 3~mK \citep{Trevino-Morales2014, Trevino-Morales2016} in the 3~mm wavelength range where most of the selected lines are located, all lines with a peak intensity lower than 15~mK or with an integrated intensity below 5$\times$10$^{-2}$~K\,km\,s$^{-1}$ (with an assumed line width of 5~km\,s$^{-1}$) can be considered undetectable.

There is no general trend in the evolution of the integrated intensities from one species to another. The emission of oxygen in the different models behaves like the emission of C$^+$. For C, the \hii\ region emission is stronger than the two other models but it becomes equal to the HHMC emission at 10$^4$~years and  the HMC emission is stronger than the HHMC emission up to 10$^4$~years before it decreases to become undetectable. For HCO$^+$ and N$_2$H$^+,$ the HHMC and \hii\ region emissions are equal. For H$_2$CO, H$_2$$^{18}$O, and NH$_3$, the \hii\ region emission is stronger than the HHMC emission up to 10$^4$~years. And then it is slightly weaker until 4$\times$10$^4$~years for H$_2$$^{18}$O and NH$_3$ before they become equal. For CH$_3$OH, HCO, and CN the behavior is the same as \hnc\ but the HMC emission becomes stronger again after 2$\times$10$^4$~years in the case of CN. In the case of the vibrationally excited state of the transition 7--6 of HCN and HCN, the HHMC emission is stronger than the \hii\ region but only until 3$\times$10$^4$~years for HNC. 

We also note that the emission of some species is considered undetectable because the integrated line intensity is below the detectability threshold of 5$\times$10$^{-2}$~K\,km\,s$^{-1}$. This is the case for H$_2$$^{18}$O and NH$_3$ in the HHMC and \hii\ region models before 10$^4$~years. The vibrationally excited lines of HCN and HNC are undetectable before about 3$\times$10$^3$~years for the HHMC model and until 10$^4$~years for the \hii\ region model. CN can be considered undetectable before 10$^3$~years for the HMC and HHMC models. The HMC emission for C is lower than the threshold after 2$\times$10$^4$~years. In the case of CH$_3$OH, the line is undetectable between 10$^4$ and 2$\times$10$^4$~years approximately for the \hii\ region model.

\subsubsection{Item 2 -- Size of the ionized cavity:} 
\label{subsubsec:specItem2}

\begin{figure*}[htbp]
        \includegraphics[width=1\textwidth]{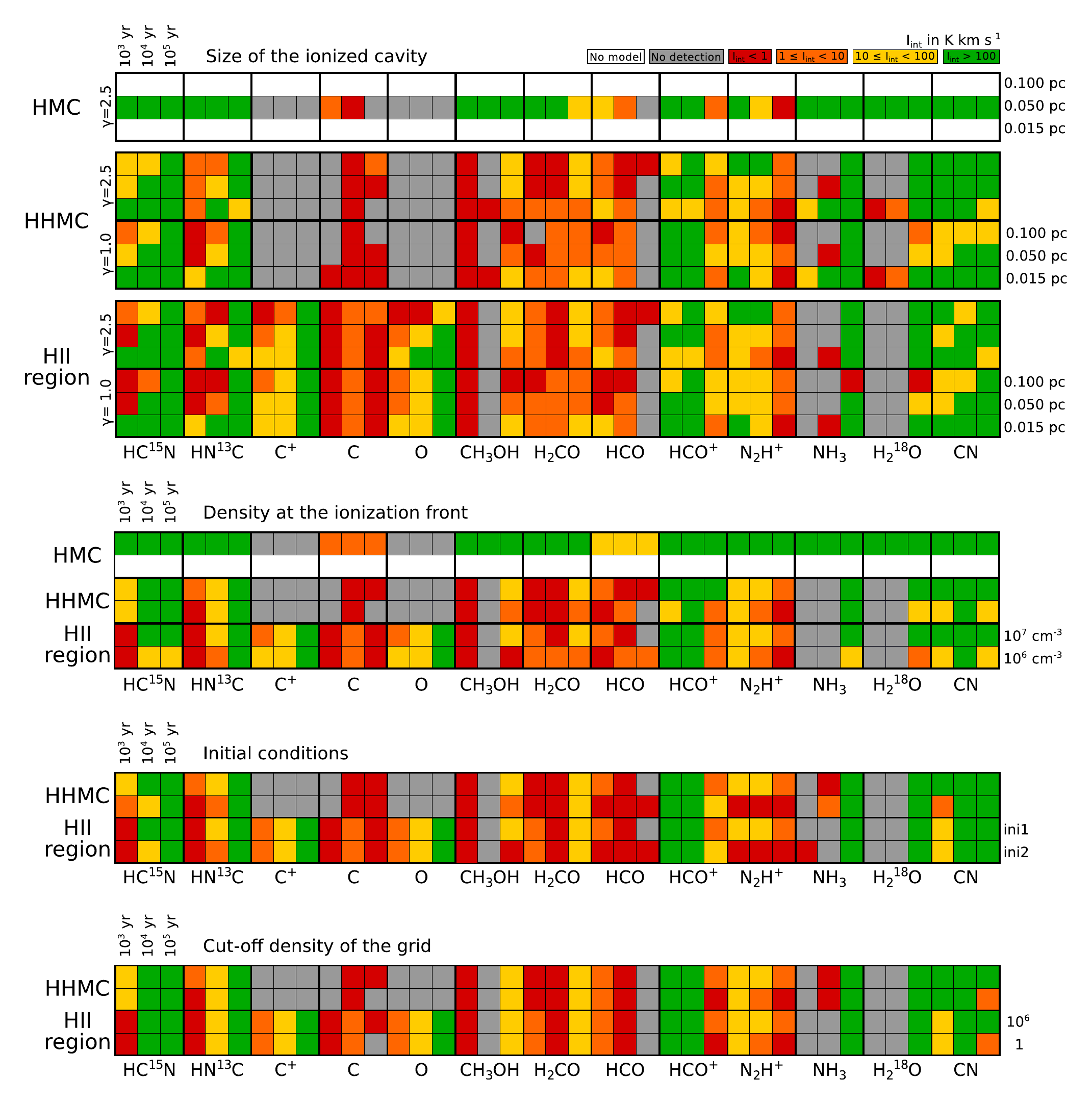}     
        \caption{Schematic summarizing the results of the time evolution of the integrated intensities for the different models: size of the ionized cavity and density profile (top panel), density at the ionization front (second panel), initial abidances (third panel) and cut-off density (bottom panel).
Lines with the integrated intensity $I_{int} > 100$ are represented in green, $10 < I_{int} < 100$ in yellow, $1 < I_{int} < 10$ in orange and $I_{int} < 1~{\rm K}\,{\rm km}\,{\rm s}^{-1}$ in red. Undetected lines, with integrated intensities lower than the detectability threshold of 5$\times$10$^{-2}$~K\,km\,s$^{-1}$, are represented in gray.}
        \label{fig:intInt_summaryPlot}
\end{figure*}

We also investigated the variation of the integrated intensities when changing the size of the ionized cavity. In the top panel of Fig.~\ref{fig:intInt_summaryPlot} we summarize the main behavior of the integrated intensities for the selected species. The detailed time evolution of the integrated intensities are shown in Fig.~\ref{apfig:intInt_comp_HIIsize}. 

The species C$^+$ and O are brighter for smaller \hii\ regions. The opposite is true for N$_2$H$^+$. For the other species the integrated intensity is larger for the smallest \hii\ region at the beginning of the time evolution and for the biggest \hii\ region at the end of the evolution. Species desorb earlier in models with a smaller ionized cavity. Later they are abundant in the three models, but the column density is higher in models with a bigger cavity, thus the intensity is stronger. 

In the cases of \hcn, \hnc, \form,\ and \meth\ the integrated intensity slightly decreases due to decrease of the abundances mentioned in Sect.~\ref{subsubsec:abunItem2}. Then it increases with a delay when the size of the \hii\ region increases. This is due to the delay of the desorption of grain species.
The results found for a density profile where $\gamma$ = 1 (model \textit{p1}) are presented in Appendix.~\ref{apsec:density-profile} but are also summarized in the top panel of Fig.~\ref{fig:intInt_summaryPlot}.

\subsubsection{Item 3 -- Density at the ionization front:} 
\label{subsubsec:specItem3}

The variation of the density also influences the integrated intensities (see second panel of Fig.~\ref{fig:intInt_summaryPlot} and Fig.~\ref{apfig:intInt_comp_HIIdens}). Therefore, the emission of \textit{mHHMC} at this density is always stronger except before 500~years. For the other density models \textit{n7} and \textit{n6}, the emission of all species is stronger for the model \textit{n7} except for C$^+$ and O as well as for H$_2$CO until 10$^4$~years. 

\subsubsection{Item 4 -- Initial abundances:} 
\label{subsubsec:specItem4}

When we compare models with different initial abundances (see third panel of Fig.~\ref{fig:intInt_summaryPlot} and Fig.~\ref{apfig:intInt_comp_HIIini}), we obtain different results except for C$^+$ and O. The integrated intensities are also very similar for molecules like H$_2$$^{18}$O and HCO$^+$ and C. The largest difference occurs for N$_2$H$^+$ for which model \textit{ini2} gives integrated intensities that are more than two orders of magnitude lower. This is due to the fact that there is a difference of about five orders of magnitude in the initial abundance of N$_2$H$^+$. This difference is one of the largest and is due to the destruction of N$_2$H$^+$ and other N-bearing species to form NH$_3$ after 10$^9$~years in the prestellar core model \textit{ini2}. We also obtain lower intensities for \hnc, \hcn\ as well as for HCO and CN up to about 10$^4$~years.

\subsubsection{Item 5 -- Cut-off density:} 
\label{subsubsec:specItem5}

When comparing the integrated intensities for models with different sizes of the modeling grid \textit{c1} and \textit{c6} (see bottom panel of Fig.~\ref{fig:intInt_summaryPlot} and Fig.~\ref{apfig:intInt_comp_HIIcut}), we observe that C, HCO, HCO$^+$, CN, and N$_2$H$^+$ integrated intensities are lower in model \textit{c6}, about ten times lower after 10$^4$~years suggesting that the emission of these species is important in the outer region of the model. The methanol and formaldehyde emission is slightly affected. It is less than a factor of two lower for model \textit{c6}, between 10$^3$ and 10$^4$~years approximately for the first molecule and between 10$^4$ and 3$\times$10$^4$~years for the second one suggesting that the emission of these species is mainly coming from the inner core.

\section{Discussion}
\label{sec:discussion}

From the abundance profiles of the \hii\ region and HMC/HHMC models we have defined different regions depicted in Fig~\ref{fig:regions}. The evolution starts with molecules frozen onto the grains surface, the grain surface chemistry region (cf. Fig.\ref{fig:regions}). It ends when the proto-star is evolved enough to emit IR as well as UV photons in the \hii\ region model. Then, as the proto-star is switched on, the temperature and radiation field increase. The IR photons penetrate deeper into the cloud and heat up the grains leading to the evaporation of some grain surface species between 15 and 30~K for high extinctions but it takes more time than close to the \hii\ region. Molecules with high desorption energy thermally desorb later. We note that the inclusion of photo-desorption reactions may accelerate the desorption of some species. 
The molecular-atomic and atomic-ionized regions - the internal PDR - is created by the radiation field which dissociates and ionizes the species. This formation of the PDR is the main difference between the \hii\ region and HMC/HHMC models. In addition, the decrease in the effective temperature of the proto-star (see bottom right panel of Fig.~\ref{Hosokawa_params}) happening at 10$^4$~years affects the temperature and radiation field intensity. They both decrease at this time and thus affect the abundance and line emission of the species.

We have seen that C$^+$ and O emit only in the \hii\ region models and their emission is not affected when we remove the envelope. This indicates that they are possible tracers of the internal PDR. Furthermore, we have seen that the internal PDR is extremely thin with a size ranging between 50 and 1500~AU depending on the model parameters. Assuming a source at a distance of 1~kpc, we would need an angular resolution of maximum 0.04 -- 0.05$''$ to be able to resolve the minimal size of the internal PDR. Thus, to trace it and avoid extended emission we need the high spatial resolution interferometric observations which are now possible with ALMA. However, there are no instruments to conduct the observations in the THz frequency regime that are necessary to resolve the C$^+$ and O emission associated with the internal PDRs. 
In addition, limitations of the models, such as the absence of an external radiation field, might cause the underestimation of the emission of these atoms coming from the envelope. Interferometric observations are thus absolutely mandatory to filter out the extended emission from the envelope. 

The emission from the envelope affects the emission of some species. C could also trace the internal PDR as the emission is stronger in the \hii\ region in the early stage of the evolution until 10$^4$~years. This is more difficult to observe due to the short time spent by the object in this phase. In addition, C emission is affected by the envelope mostly after that time as the HMC emission is in absorption and the intensity of the line in the \hii\ region model decreases when we filter out the extended emission in the late times of the evolution. Since atomic carbon can be observed with ALMA band 8, this is an interesting test to try. For the species \ntohp, HCO, \hcop,\ and CN, the emission is also strongly affected by the envelope. Due to the decrease in their emission they might not be good tracers to distinguish between \hii\ regions and HMC as they do not trace the inner core. NH$_3$ emission is also affected when removing the extended emission but in a lesser extent. 

Molecules like \meth, H$_2$$^{18}$O or NH$_3$ appear to be detectable (I$_{\mathrm{peak}}$ $\geq$ 15~mK) around 10$^4$~years for the HHMC models and only around 4$\times$10$^4$~years for the \hii\ region models. The intensities of methanol in the HHMC models yet remain quite low (I$_{\mathrm{peak}}$ $\sim$ 0.1~K) during the time when the \hii\ region emission is still too weak. These molecules are likely destroyed mainly during the desorption process when the UV radiation field increases. The few molecules that desorb are photo-dissociated in a short time ($\sim$ 50~years). They appear again after 10$^4$~years due to thermal desorption outside the PDR before being destroyed at 10$^6$~years.
However in the HMC model, i.e., the model without the ionized cavity around the proto-star, the abundance of these molecules is high at the center inducing high column densities and strong intensities from 100~years, compared to the HHMC model and thus the \hii\ region model. These molecules might help in tracing the HMC phase. Similarly, H$_2$CO is always detectable for the three models but the emission for the HMC model is much stronger despite that they tend to be equal once it thermally desorbs. In addition, a ring structure seems to appear in the \hii\ region model in the raw synthetic maps. But the convolution to the beam of the IRAM 30~m telescope for example attenuates this effect. 
The vibrationally excited levels of HCN trace the hot gas and could be used to distinguish the hot core and the \hii\ region as the emission is stronger in the HHMC model. This is not the case for vibrationally excited HNC. This difference between HCN and HNC could happen because the abundance of HCN is high near the ionized cavity ($\geq$ 10$^{-6}$) for the HHMC model whereas the HNC abundance is about four magnitudes lower. The region up to an extinction of 5~mag has then more influence on the intensity for HCN. 

Our results show that the region where UV radiation influences chemistry is spread in time but remains thin, mainly for high density ($\geq$ 10$^6$~\cmc). Hence, we do not obtain certain molecules which appear to be abundant in observations. HCO$^+$ is usually optically thick in observations of hot cores or \hii\ regions but in our models it is destroyed around 10$^4$~years. Destruction of HCO$^+$ is mainly due to dissociative recombination forming H and CO and by reactions with HCN and H$_2$O which are really abundant because of thermal desorption. In addition, CO$^+$ is an important species in PDRs and its abundance is enhanced when irradiated by strong UV fields (e.g., \citealt{Trevino-Morales2016}). However we do not produce enough CO$^+$ in the models compared to observations. CO$^+$ is rapidly destroyed to form C$^+$. The medium is too dense and the UV radiation cannot penetrate deep enough into the cloud and thus the PDR remains too thin. This issue has been discussed in \cite{Bruderer2009b}. They circumvent it by modeling a proto-star embedded in a cavity produced by a bipolar outflow. It provides a larger surface to the UV radiation.

There is also an important dependence of the results on the initial abundances. Obtaining good initial conditions is thus critical. In this work, the first set of initial abundances, from the collapsing cold core in two steps, seems to be more realistic than the second set. However, implementing a real collapse of the core by adding the temporal evolution of the density (similar to \citealt{Garrod&Herbst2006}) would be the next step. 

The spectra are obtained without including the free-free emission which might be an important physical parameter for transitions at lower frequencies in the \hii\ region models. In our models, the continuum level for the \hii\ region models' spectra is low (close to zero) for the lower frequencies, while \hii\ regions may be bright up to 100 -- 200~GHz. Another aspect that might influence the emission of the species is the position from which the synthetic spectra is obtained. In this work, all synthetic spectra used to compute the integrated intensities are chosen at the center of the synthetic maps, where the proto-star is located, in order to probe the gas in the entire region. Choosing spectra located at the position of the internal PDR could induce different variations for the \hii\ region models compared to the HHMC models.

\subsection{{\rm [HNC]/[HCN]}}

\begin{figure}[t]
        \centering
        \includegraphics[width=0.242\textwidth]{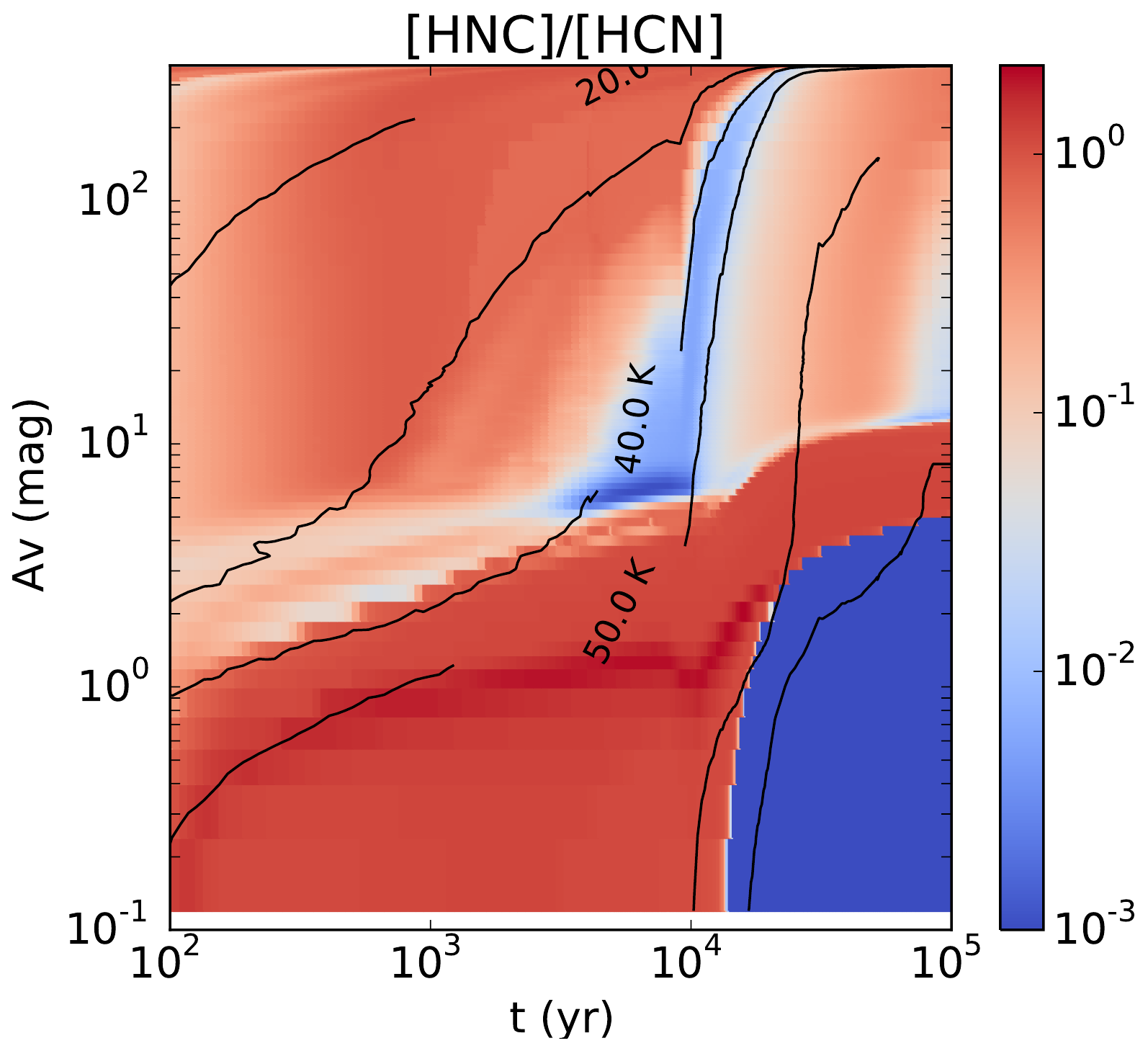}
        \includegraphics[width=0.242\textwidth]{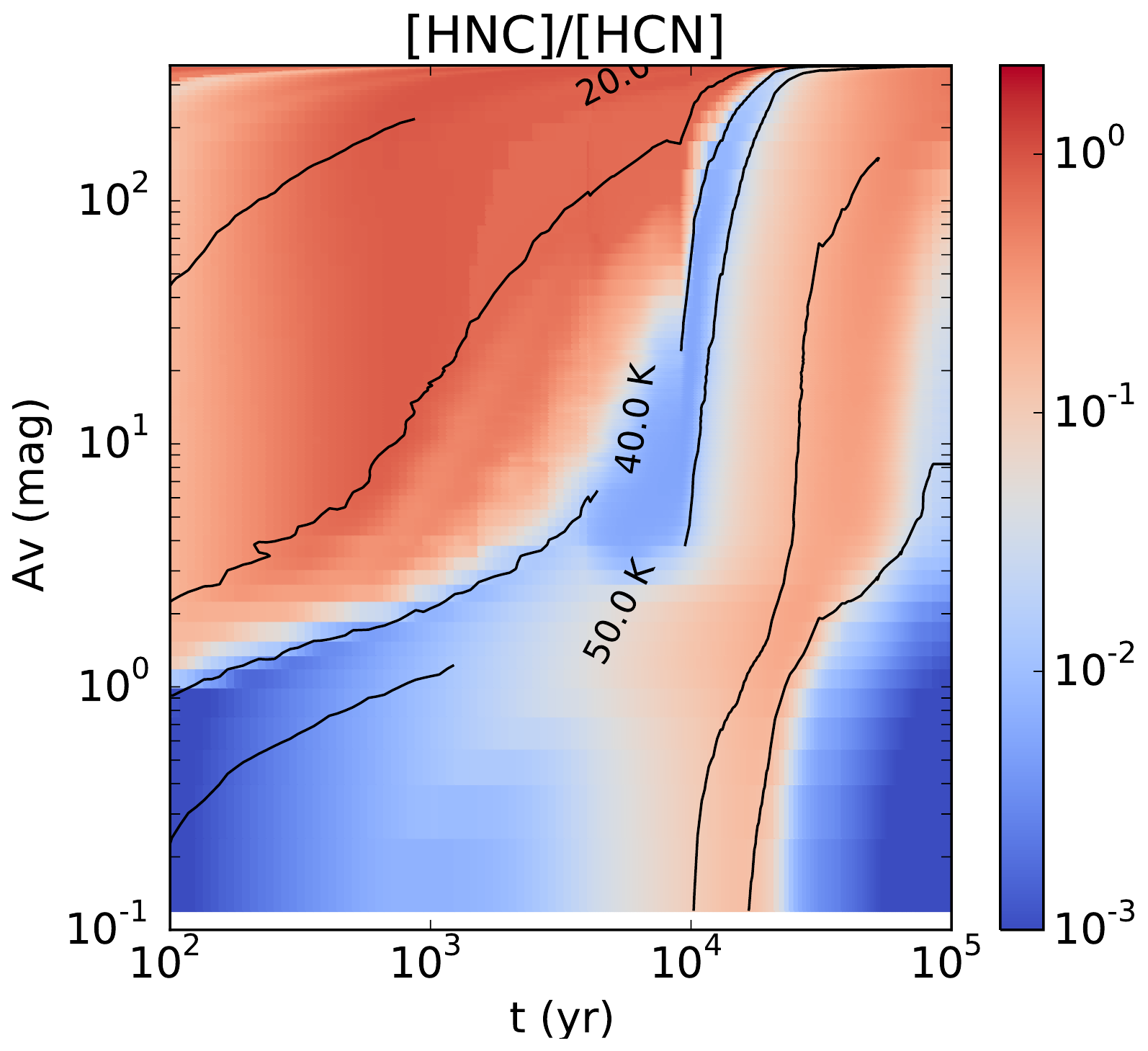}
        \includegraphics[width=0.242\textwidth]{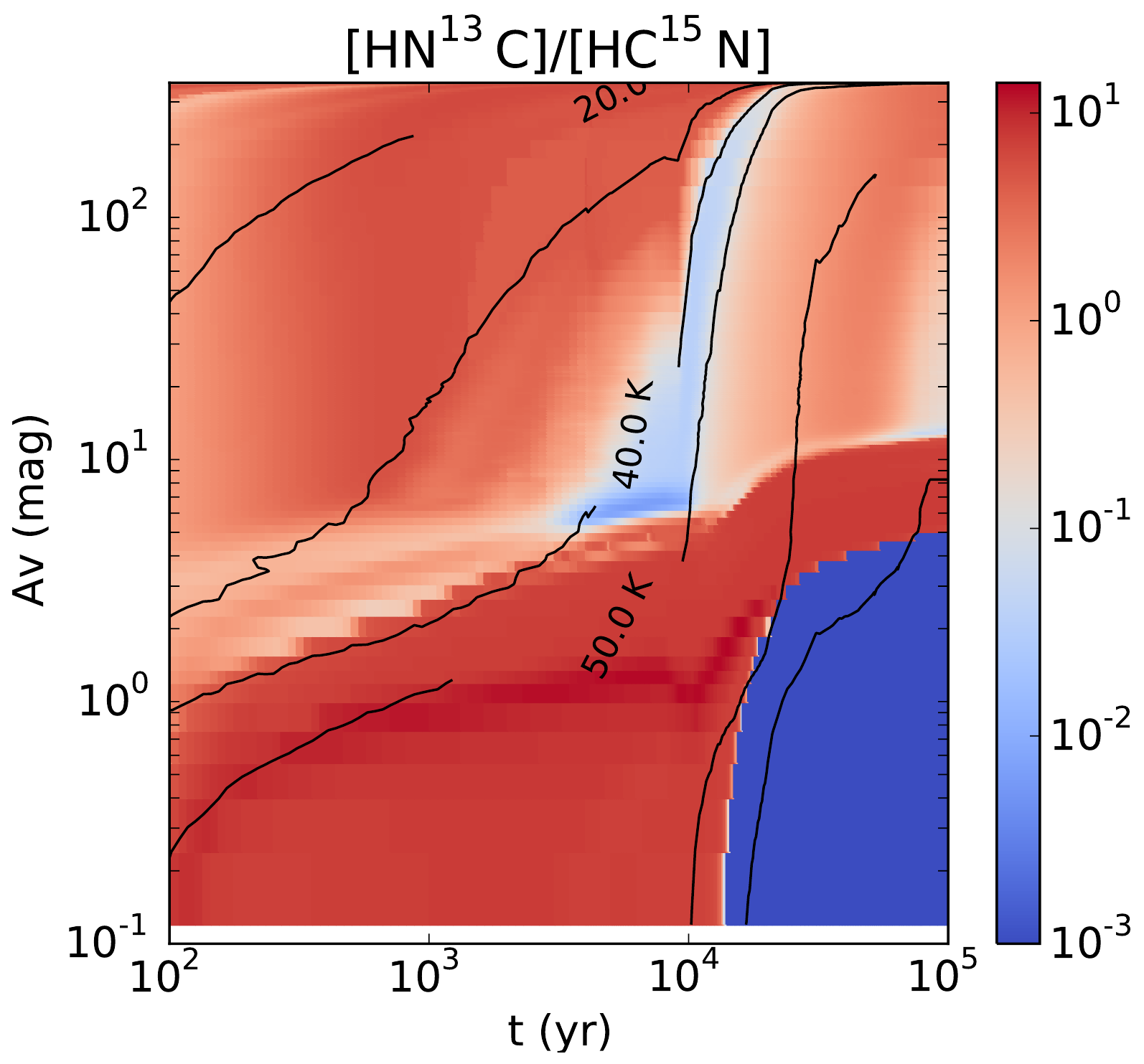}
        \includegraphics[width=0.242\textwidth]{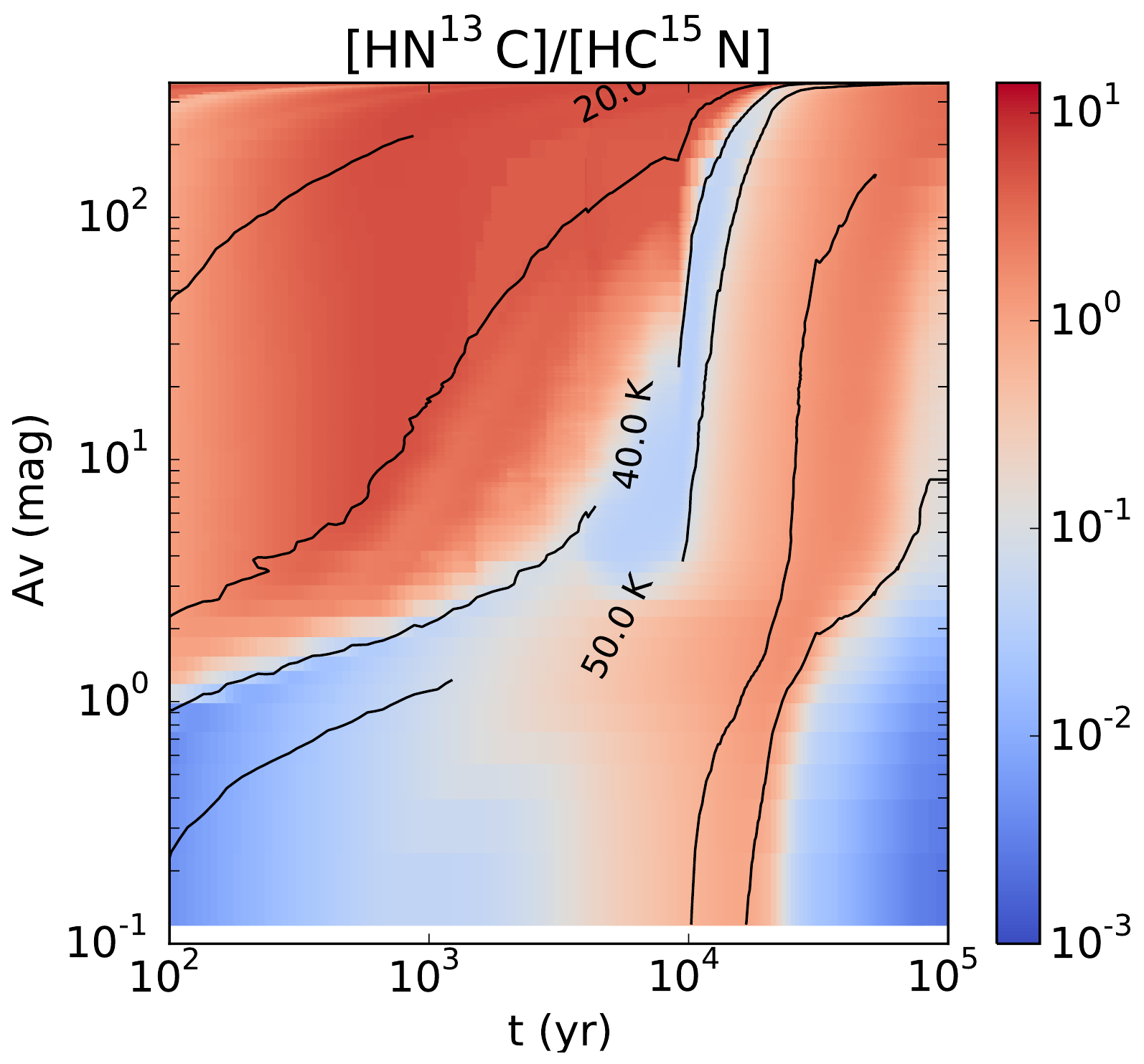}
        \caption{Spatio-temporal evolution, focused between 100 and 10$^5$~years, of the abundance ratio [HNC]/[HCN] for the reference model \textit{mHII} (left) and the corresponding \textit{mHHMC} model (right). Contours are plotted for \Td\ (20, 30, 40, 50, 100 and 150~K). The bottom panels show the same results for [\hnc]/[\hcn].}
        \label{fig:Abun_ratio}
\end{figure}

\begin{figure}[t]
        \centering
        \includegraphics[width=0.45\textwidth]{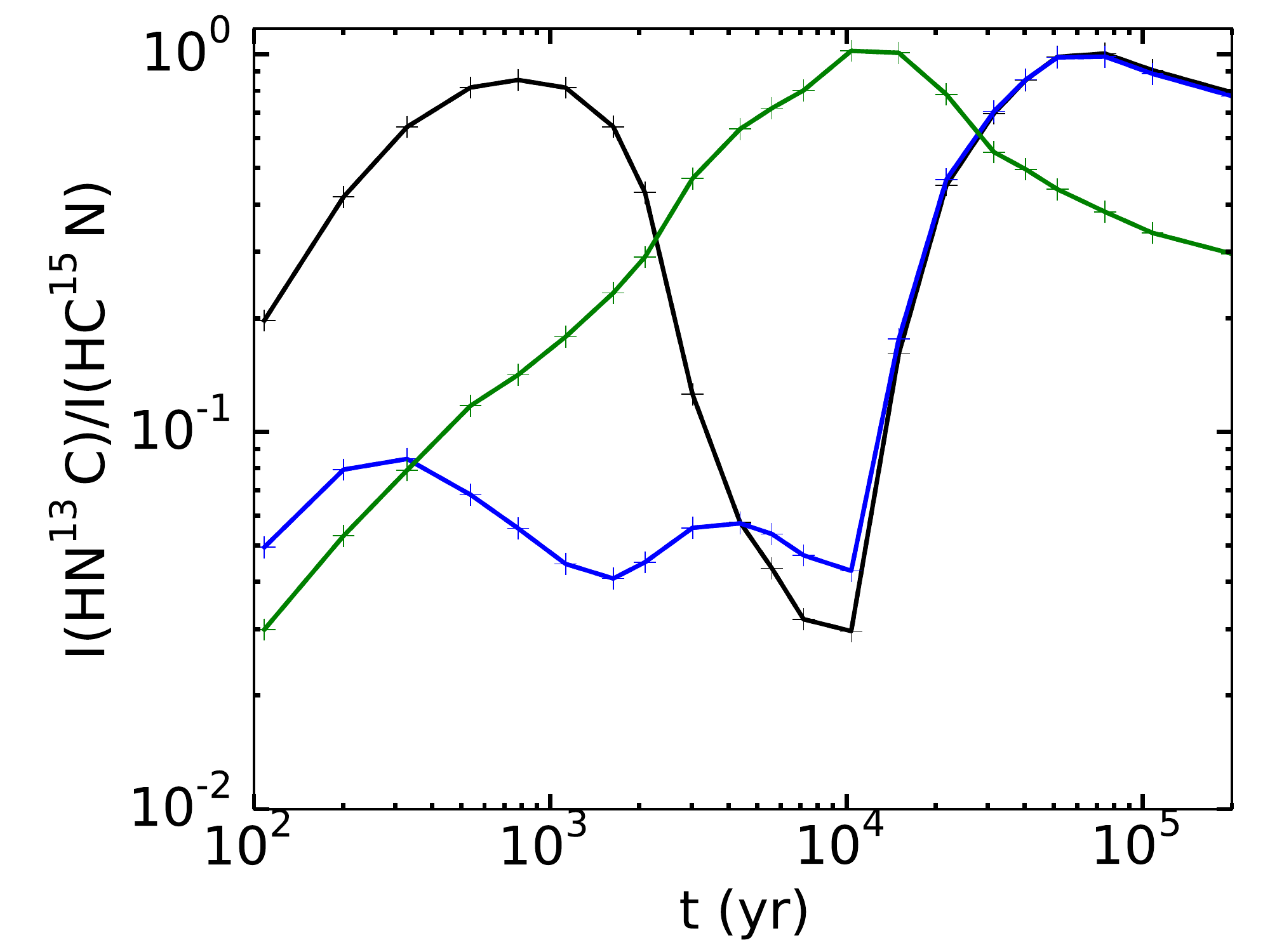}
        \caption{Time evolution for the reference model \textit{mHII} (black), \textit{mHHMC} (blue) and \textit{mHMC} (green) of integrated intensity ratio (HN$^{13}$C\,(1--0) / HC$^{15}$N\,(1--0))}
        \label{fig:ratios_HII-HC_size}
\end{figure}

In Fig.~\ref{fig:Abun_ratio} we present the spatio-temporal variations of the abundance ratio [\hnc]/[\hcn] and [HNC]/[HCN] for the \hii\ region (left) and HHMC (right) models. The colorbar varies from 10$^{-3}$ and 14 because the ratio varies from approximately 10$^{-2}$ to 14, if we do not consider the PDR region from 10$^4$~years and \av\ < 5~mag. The maximum ratio appears in the \hii\ region model. In the HHMC model the maximum is around 6.5. The variations for the ratio [HNC]/[HCN] are from 10$^{-3}$ to 1.07 for \textit{mHHMC} and to 2.16 for \textit{mHII}.
The variations in the abundance ratio result into temporal variations of the line intensity ratio seen in Fig.~\ref{fig:ratios_HII-HC_size}. The main contribution to the intensity ratio comes from the part of the core where the extinction is higher than 4~mag approximately, whatever the size of the \hii\ region. The ratio is barely affected by the evolution of the radiation field which influences the core only for extinctions lower than 5~mag. Different works on the [HNC] to [HCN] ratio show that it depends on the kinetic temperature in the cloud \citep{Goldsmith1986, Schilke1992, Wang2009, Jin2015}. In the model this ratio increases until the temperature is around 20~K, then it decreases. It goes through a minimum when the temperature of the core is between 40 and 50~K. The ratio increases again until the temperature reaches about 150~K. The behavior found for [\hnc]/[\hcn] is the same for [HNC]/[HCN]. This strong decrease around 40~K is due to the thermal desorption of s-HCN which is faster than for s-HNC. The thermal desorption of s-HNC is slowed down due to the decrease of the abundance of s-HNC around 3$\times$10$^3$~years because it reacted with s-H to form s-HCN or HCN via reactive desorption. 

In \cite{Schilke1992} the minimum  [HNC]/[HCN] for a cloud with a density of 10$^7$~\cmc\ appears at 30~K and in \cite{Graninger2014} the abundance ratio is found to go through a minimum for a temperature around 30 -- 40~K. This ratio also depends on the density of the core as the integrated intensity ratio is smaller for lower density due to a lower abundance ratio but the minimum is the same.

The intensity ratio of the vibrationally excited line of HNC and HCN decreases from 10$^4$ to 10$^5$~years. Before these times the ratio is unrealistic because the line emission is extremely low or undetectable. The decrease of the ratio indicates that the major contribution to the line emission for these transitions comes from the inner core where the temperature is high. The abundance ratio decreases close to the ionized cavity for \textit{mHII} and \textit{mHHMC} affecting in a similar manner the intensity ratio.

\subsection{Gas temperature in the PDR}

For all the models presented here we have assumed that the gas temperature is equal to the dust temperature. We know that this is not the case in the internal PDR where the gas temperature is usually approximately one order of magnitude higher than the dust temperature. We know that some species not investigated in this paper such as mid and high-$J$ transitions of CO may be strongly affected by a gas temperature of a thousands Kelvin and trace the hot gas and thus the internal PDR (e.g., \citealt{Lee2015}). But for the species and transitions investigated in this paper the assumption $T_{\rm g} = T_{\rm d}$ seems to be fine. We have produced one model of a core with an embedded \hii\ region similar to the reference model but where the gas temperature is increased for the abundances' computation with \sapt\ (model \textit{mHII-Tg}). The gas temperature is defined as follows:
\begin{equation}
        \label{eq_Tg}
        T_{\rm g} = T_{\rm d}\,(1 + \beta \exp(-\alpha \,A_V))
\end{equation}
where $\beta$ = 20 and $\alpha$ = 1.5. This equation is obtained from a fit of a model, with physical conditions close to our \hii\ region model, produced by the Meudon PDR code \citep{LePetit2006}.
 
\begin{figure}[t]
        \centering
        \includegraphics[width=0.242\textwidth]{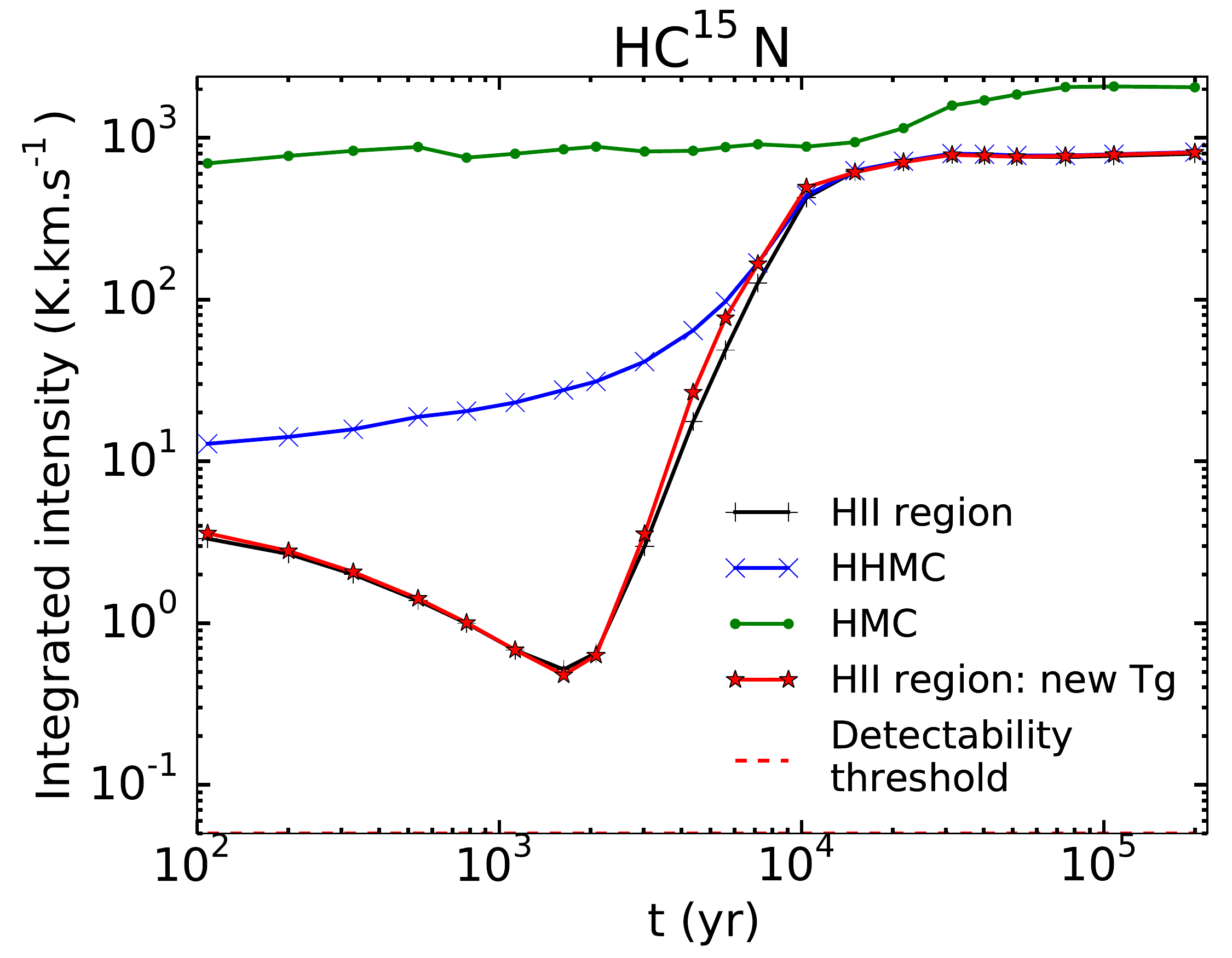}
        \includegraphics[width=0.242\textwidth]{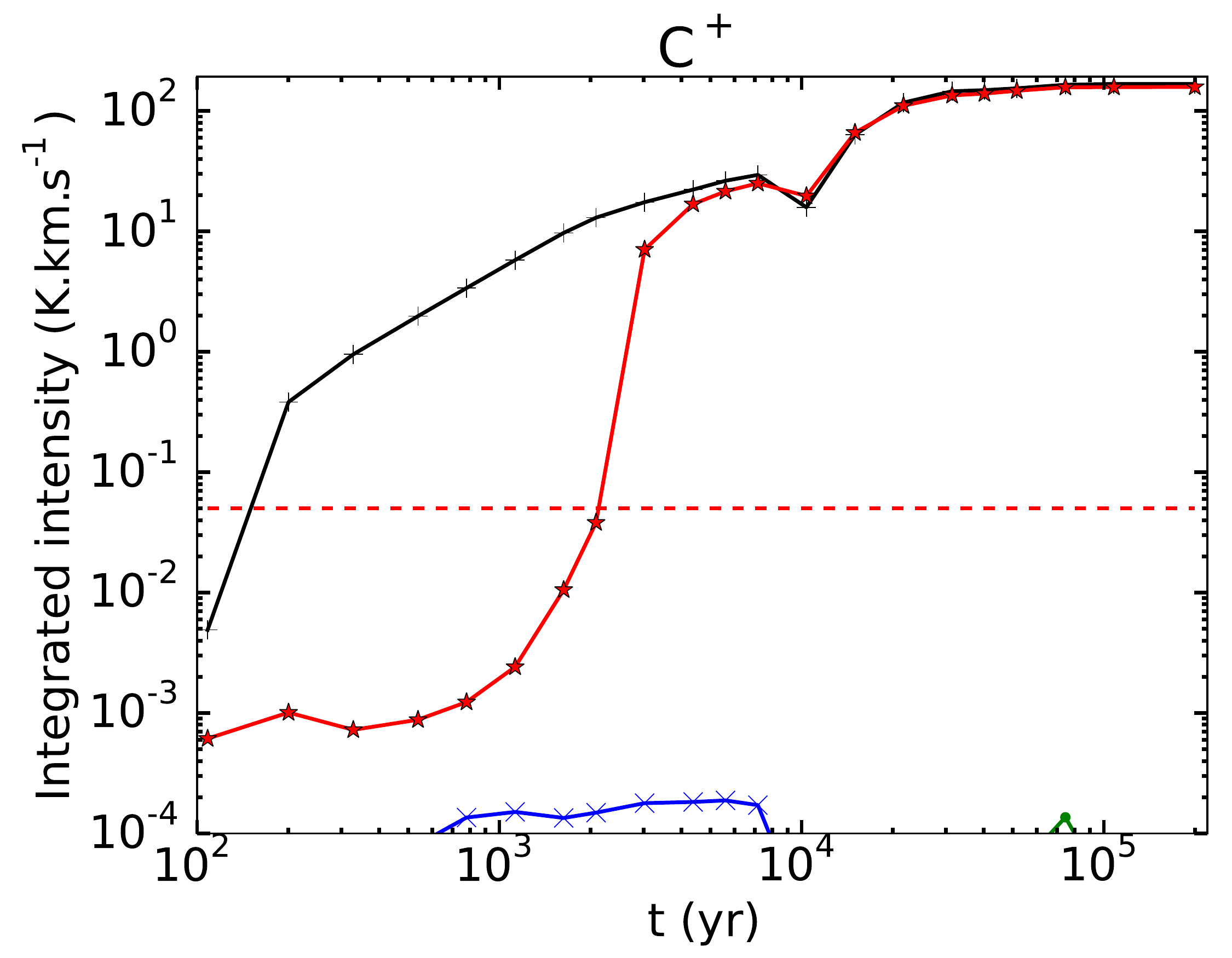}
        \includegraphics[width=0.242\textwidth]{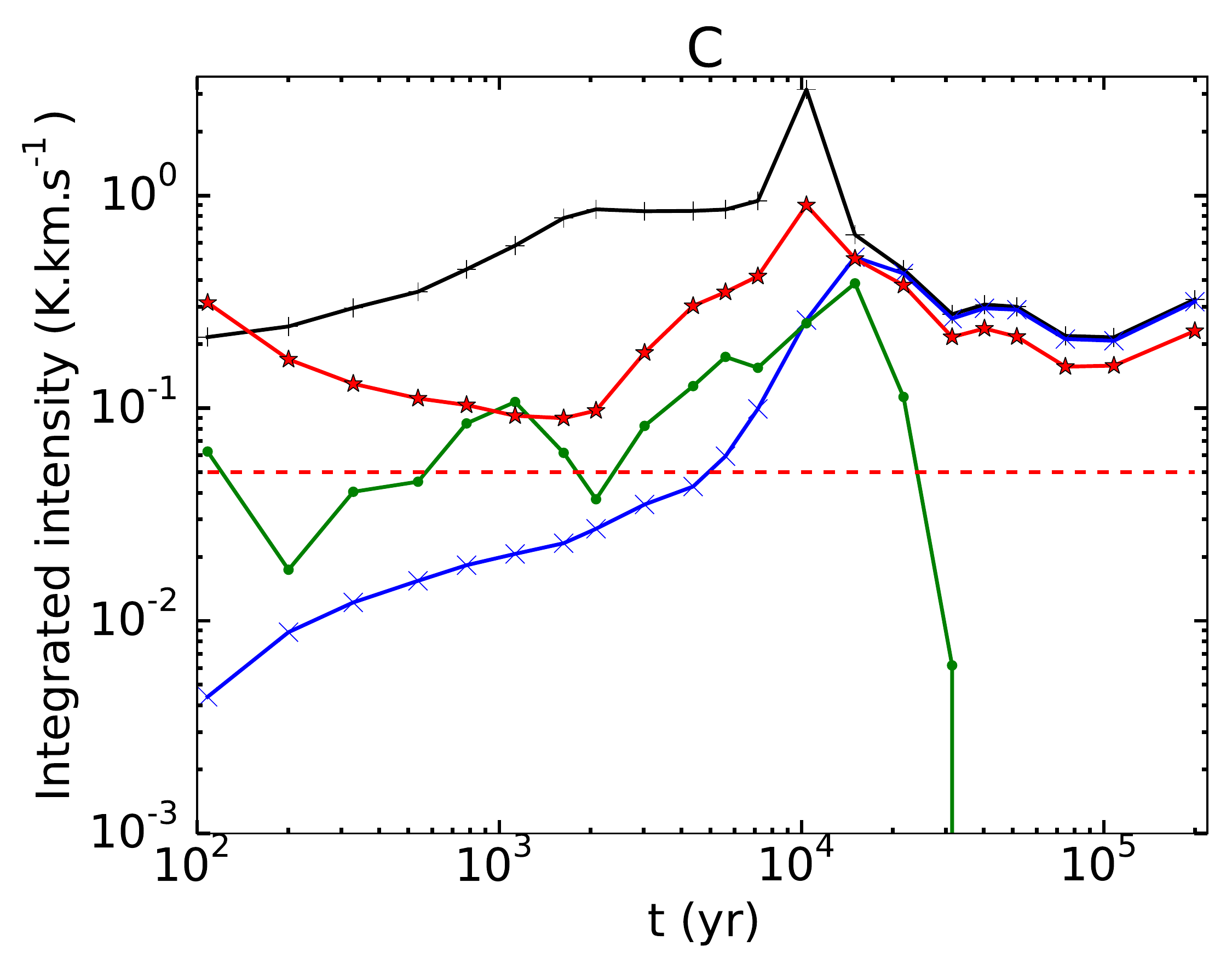}
        \includegraphics[width=0.242\textwidth]{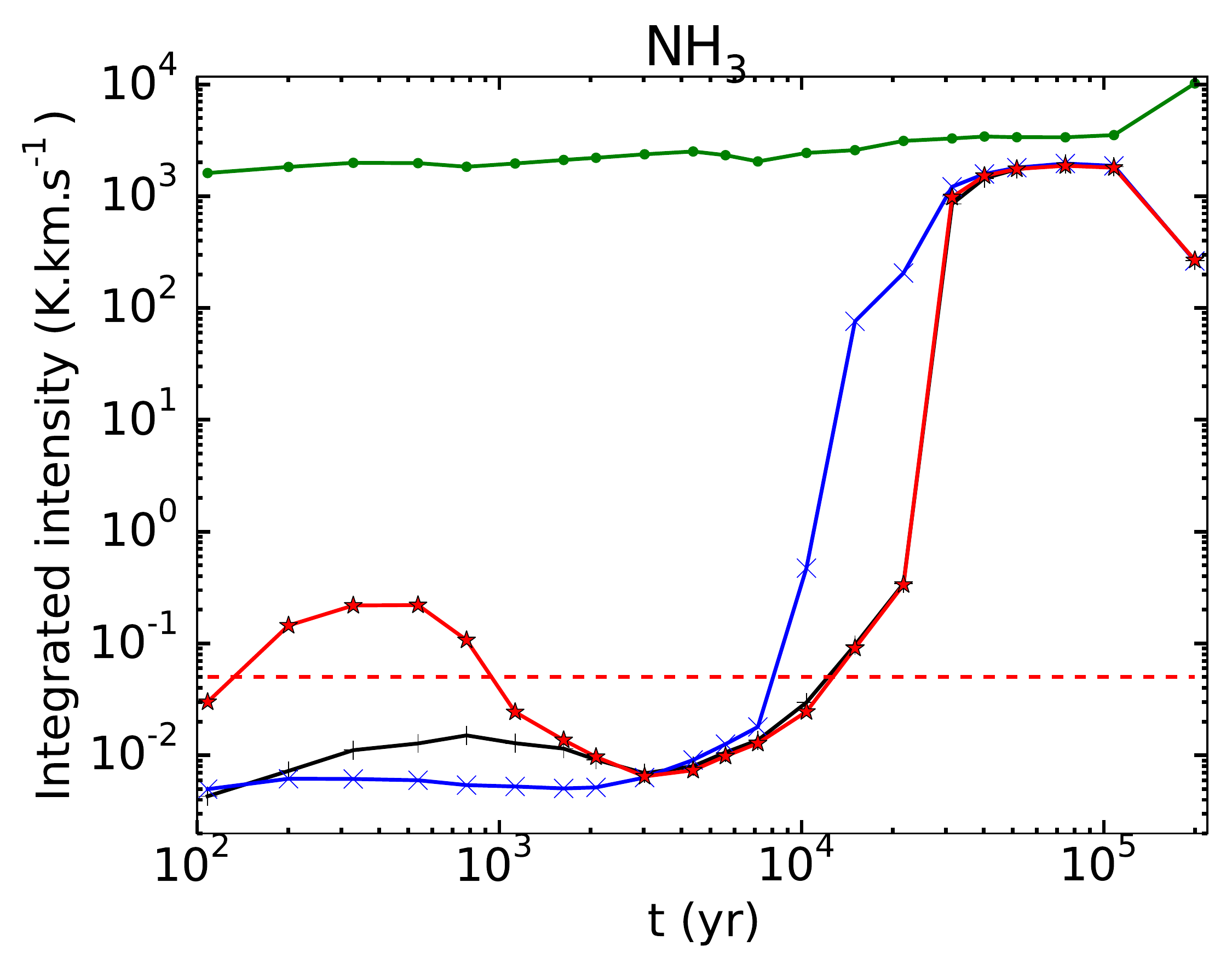}
        \caption{Time evolution of the integrated intensities of HC$^{15}N$, C$^+$, C and NH$_3$ (see the corresponding transition in Table~\ref{tab:selected-molecules}). The reference model \textit{mHII} is represented in black, model \textit{mHHMC} in blue and model \textit{mHMC} in green and the new \hii\ region model in red. The red dashed line represents the detectability threshold of 5$\times$10$^{-2}$~K\,km\,s$^{-1}$ obtained with an assumed rms noise of 3~mK.}
        \label{fig:intInt_newTg}
\end{figure}

In Figure~\ref{fig:intInt_newTg} we show the time evolution of the integrated intensities for model \textit{mHII} (black), \textit{mHHMC} (blue), \textit{mHMC} (green) and model \textit{mHII-Tg} (red), for HC$^{15}N$, C$^+$, C and NH$_3$. Apart from HC$^{15}N$ these are the species the most affected by the increase of the gas temperature. C$^+$ and NH$_3$ present different behaviors from the reference \hii\ region model only before 3000~years. Objects of this age do not yet present an internal PDR as it appears at around 10$^4$ years in the models and it seems unlikely that we would observe such young objects. Atomic carbon is then the species of most concern here because its intensity is lower than a model assuming $T_{\rm g} = T_{\rm d}$ all along the time evolution.

\section{Conclusion}
\label{sec:conclusion}

\subsection{Summary}

We investigated the spatio-temporal evolution of the chemistry in HC/UC\hii\ regions and hot molecular cores using the spatio-temporal evolution of the dust temperature and the UV radiation field. this is done in order to obtain the detailed modeling of relative abundances and the temporal evolution of line emission of selected species. 

All the abundance profiles clearly show a difference between these two evolutionary stages of massive star formation. But the typical tracers of the HC/UC\hii\ region phases, C$^+$ and O, appear to be unobservable due to the capabilities of the current instruments which do not allow to resolve the internal PDR in the THz range, although atomic carbon might be a good tracer to be observed with ALMA. We find the internal PDR to be extremely thin, with a size ranging from about 50 to 1500~AU. This barely affects the emission of the other species. The HMC emission is stronger than the \hii\ region during most of the time evolution for some species: \hcn, \hnc, H$_2$CO, H$_2$$^{18}$O, NH$_3,$ and CH$_3$OH. This is due to the increased column density. More species could exhibit this behavior and remain to be investigated.

\subsection{Outlook}

In addition, the models can be improved as indicated below:
\begin{enumerate}
\item We would also like to investigate HC/UC\hii\ regions in the context of ionized cavities \citep{Peters2010a} or PDRs on outflow cavities \citep{Visser2012, Bruderer2009b} which create large surfaces and thus more easily observable chemistry. Therefore, we would have to increase the complexity of the source structure. 
\item We are currently improving the treatment of the grain chemistry by incorporating a multilayer ice mantle.
\item Finally, we plan to implement the treatment of dynamics in the models (e.g., including infall and following the parcels of gas or by post-processing data obtained with a MHD code). It appears essential to obtain dynamical models for the prestellar phase in order to fix the initial conditions.
\end{enumerate}

We plan to compare our results with molecular lines found in the main PDR surrounding the UC\hii\ region of Monoceros R2. This UC\hii\ region is the closest with a distance of 830~pc \citep{Herbst&Racine1976} and is irradiated by the main infrared B0-type star IRS~1. In this region densities are around 10$^6$--10$^7$~\cmc\ and the UV field is superior to 10$^5$~Habing unit which fits our models parameters. In \cite{Trevino-Morales2015PhD}, integrated intensities for all lines observed with IRAM 30~m telescope are tabulated (see also \citealt{Trevino-Morales2014, Trevino-Morales2016}).

\begin{acknowledgements}
	We thank the anonymous referee for insightful comments that greatly improved this paper.
      Part of this work was supported by the Collaborative Research Centre 956, 
      sub-projects Astrochemistry [C3] and High-mass star formation [A6], 
      funded by the German Deut\-sche For\-schungs\-ge\-mein\-schaft (DFG), 
      by the French CNRS national program PCMI 
      and by COST Action "Our Astro-Chemical History" CM1401.
      We kindly thank T. Hosokawa for providing his stellar model results in digital format to us. We furthermore thank C.P. Dullemond for useful discussions. 
      This research has made use of NASA's Astrophysics Data System.
\end{acknowledgements}

%
%
\bibliographystyle{aa}
\bibliography{/Users/gstephan/mybiblio.bib}

\begin{appendix} 

\section{\sapt: further modifications}
\label{app:further modifications}

In addition to using the spatio-temporal evolution of the mean intensity in \sapt\ we made some other modifications in the code. We chose to give to the solver the natural logarithm of the chemical abundances of each species instead of the abundances in order to improve the stability of the code. The ordinary differential equations as well as the Jacobian matrix are thus computed differently. In the first version of \sapt\ \citep{Choudhury2015} we computed the chemical abundances of the different species $n_{\rm x}$. In the new version of \sapt\ we compute $\ln{(n_{\rm x})}$.

The gas and dust temperatures as well as the radiation field intensity are defined as time-dependent polynomials using the values computed by \radmc. A spline interpolation is performed to determine the function parameters and thus the time derivatives and Jacobian parameters for these variables which are then used by the solver. Thanks to the interpolation we are no longer restricted to using the same number of time steps than those used in \pando\ for the computation of \Td\ and \Jnu. Thus, running the models is less time consuming.

\section{Additional results and models}
\label{app:additional-results}

\subsection{Models with a larger chemical network}
\label{apsec:size-network}

We also produced models with an increased number of species included in the chemical network, from 183 (model \textit{s183}) to 334 species (model \textit{s334}). More complex molecules such as methyl formate (HCOOCH$_3$), ethanol (CH$_3$CH$_2$OH), and dimethyl ether (CH$_3$OCH$_3$) in the gas phase, in their neutral or ionized form and on the grain surface are in this network. The models do not use the initial abundances used with a network of 183 species but are computed again. For the species common to both networks the initial abundances are similar (see Table~\ref{aptab:IniAbun}) but the elements are spread out on a larger number of species and thus some small variations appear.

\begin{table}[t]
    \caption{Initial abundances of the principal species used in the models are shown in Column 2 for a chemical network of 183 species and in Column 3 for 334 species.}
    \centering
    \begin{tabular}{l l l}
            \hline
            Species & $n_\textnormal{X}$/\nh & $n_\textnormal{X}$/\nh \\
             & 183 species & 334 species \\
            \hline
            \hline
             H$_2$ & 4.99(-1) & 4.99(-1) \\
             He & 9.00(-2) & 9.00(-2) \\ 
             s-H$_2$ & 1.24(-4) & 1.24(-4)\\
             s-H$_2$O & 9.87(-5) & 1.00(-4) \\
             s-CO & 9.19(-5) & 9.22(-6)  \\ 
             s-OH & 1.91(-5) & 1.90(-5) \\
             s-HNO & 2.89(-5) & 2.68(-5) \\
             s-N$_2$ & 1.33(-5) & 1.38(-5) \\  
             s-O & 1.16(-5) & 1.15(-5)  \\
             s-HCN & 6.23(-6) & 4.62(-6) \\
             s-CH$_4$ & 6.20(-6) & 6.05(-6) \\
             s-NH$_3$ & 8.03(-6) & 7.29(-6) \\
             s-H$_2$CO & 3.20(-7) & 3.09(-7) \\
             s-CH$_3$OH & 1.29(-7) & 1.38(-7) \\
             s-CO$_2$ & 9.83(-8) & 9.94(-8) \\
             s-H$_2$S & 3.09(-8) & 2.78(-8) \\
             s-Si & 8.00(-9) & 7.82(-9) \\
             s-Fe & 3.00(-9) & 3.00(-9) \\
             \hline
    \end{tabular}
    \label{aptab:IniAbun}
\end{table}

The most affected molecules included in both chemical networks are CH$_2$OH, s-HCO and s-O$_2$. The abundance of CH$_2$OH is larger by up to sene or eight orders of magnitude in model \textit{s334} from 10$^4$ to 10$^6$~years for extinctions higher than 10~mag. This is mainly due to the destruction of complex organic molecules such as CH$_3$OCH$_3$ or HCOOCH$_3$. On the contrary, s-HCO has a few orders of magnitude smaller abundance in model \textit{s334} for high extinctions and from 10$^2$ to 10$^4$~years. s-O$_2$ has also a smaller abundance in model \textit{s183} up to 10$^4$~years and for all radii due to the very different initial abundances for this molecule -- 10$^4$ times higher in model \textit{s183}. However, the abundances of most species change by a factor of at most two or three during most of the spatio-temporal evolution and for most species. In general, the main differences appear first between 100 and 10$^4$~years for extinctions up to 10~mag. Several species, for example, CO$^+$, (s-)CH$_3$OH, (s-)NH$_3$ and (s-)CO$_2$, have larger abundances in model \textit{s334}. They can be larger by up to two to three orders of magnitude but only on a very short period of time such as a few hundred years for CH$_3$OH. On the contrary, molecules like O and C$^+$ are less abundant in model \textit{s334} during this time period due to the presence of more complex organic molecules where O and C are locked in. A second relevant difference appears around 10$^6$ -- 10$^7$~years for higher extinctions. For example, in model \textit{s334} the abundances are lower for species like H$_2$CO, HCNH$^+$ or C$_2$H$_4$, and higher for species like CN, HNC$^+$. In summary, these differences between the two chemical networks are not significant and the smaller network is thus the only network used to study the emission of species of particular interest for this work such as C$^+$ or HCN. 

We also have more complex molecules in the \textit{mHHMC:s334} model compared to \textit{mHII:s334}. In addition, the abundance of HCOOCH$_3$ decreases around 10$^5$~years from the edge of the ionized cavity to the core. This decrease is only seen for this complex molecule. We also observed a decrease of the abundances in models with larger \hii\ cavities. Finally, up to 3~mag and from 10$^2$ to 10$^3$~years, more complex molecules are present in models \textit{s334} with lower densities.

\subsection{Models with a different density profile: $\gamma$ = 1}
\label{apsec:density-profile}

Hot cores and \hii\ regions environments do not all have the same density structure. We also studied the impact of the exponent of the density function (see Eq.~\ref{eq:plummer}) on the chemistry. When changing the exponent of the density profile from 2.5 (model \textit{p2.5}) to 1 (model \textit{p1}) the density at the ionization front changes. Model \textit{mHII:r0.015p1} has a density at the ionization front of 10$^7$~\cmc\ and models \textit{mHII:r0.05p1} and \textit{mHIIr0.10p1} have a density of 1.56$\times$10$^6$~\cmc and 4.15$\times$10$^5$~\cmc\ respectively (instead of 10$^7$~\cmc\ in models \textit{mHII:p2.5}, whatever the size of the ionized cavity). The temperature and radiation field profiles are also slightly different in models \textit{p1} compared to \textit{p2.5}. At the same radius the temperature and radiation field are higher for model \textit{p1} and the difference increases deeper into the core. It also increases when the size of the ionized cavity increases because the density at the ionization front is lower in model \textit{p1} for models \textit{r0.05} and \textit{r0.10}. 

\begin{figure*}[htbp]
        \centering
                \includegraphics[width=1\textwidth, trim={0 55 0 40}, clip]{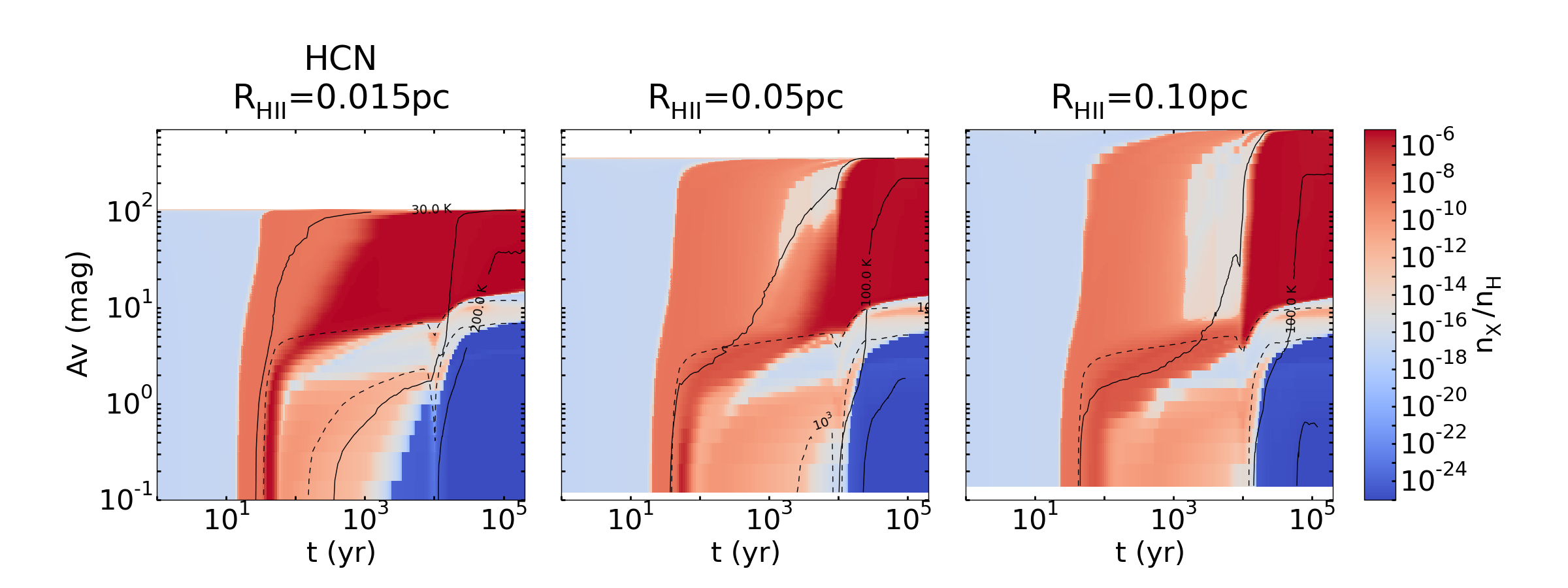}
                \includegraphics[width=1\textwidth, trim={0 15 0 140}, clip]{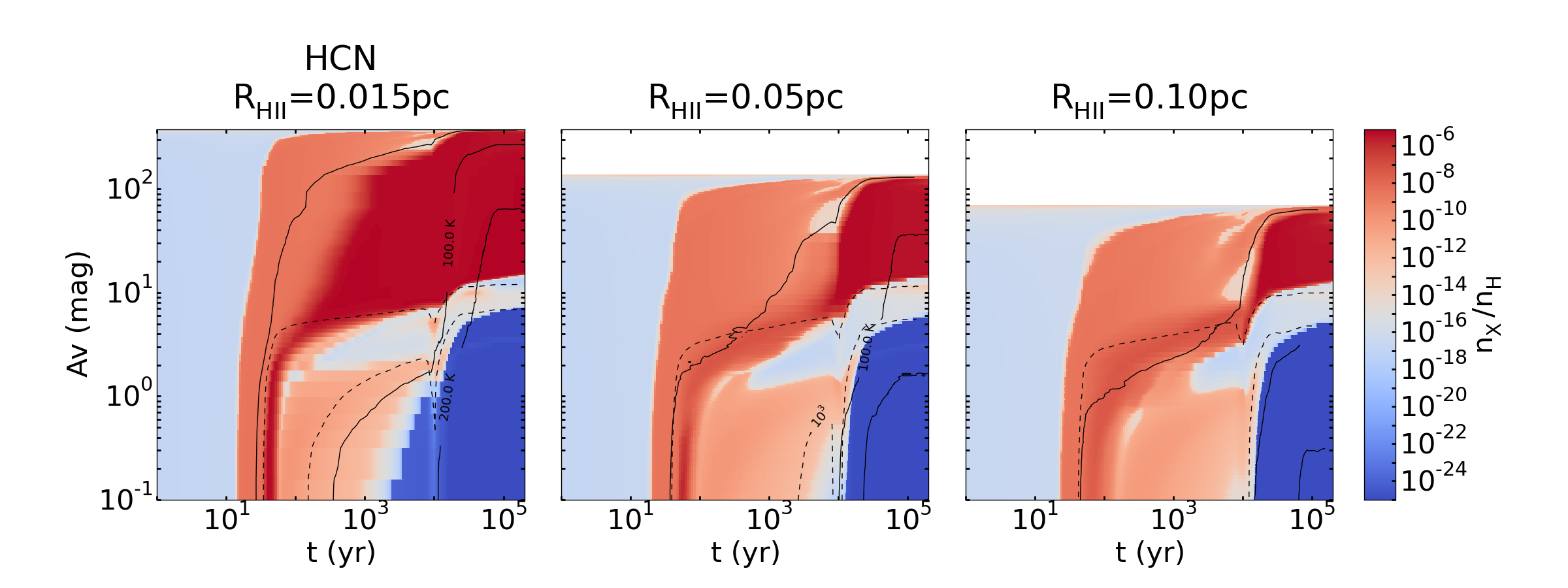}
        \caption{Spatio-temporal evolution of the abundances for HCN for models \textit{mHII:p2.5} (top panels) and \textit{mHII:p1} (bottom panels) with models \textit{r0.015} (left panels), models \textit{r0.05} (middle panels) and models \textit{r0.10} (right panels). Contours are plotted: solid line for the \Td\ (30, 100 and 150~K) and dashed lines for \Go\ (10$^{-1}$ and 10$^3$~Draine unit).}
        \label{apfig:3Dabun_compHIIsize}
\end{figure*}

The decrease in the abundance of species like HCN between 10$^3$ and 10$^4$~years as seen in Section \ref{subsubsec:abunItem2} is less extended in time and radius (see bottom panels of Fig.~\ref{apfig:3Dabun_compHIIsize}). The abundances of a few ions is also larger in the internal PDR as seen in Section \ref{subsubsec:abunItem3}.

\begin{figure*}[htbp]
        \includegraphics[width=0.32\textwidth]{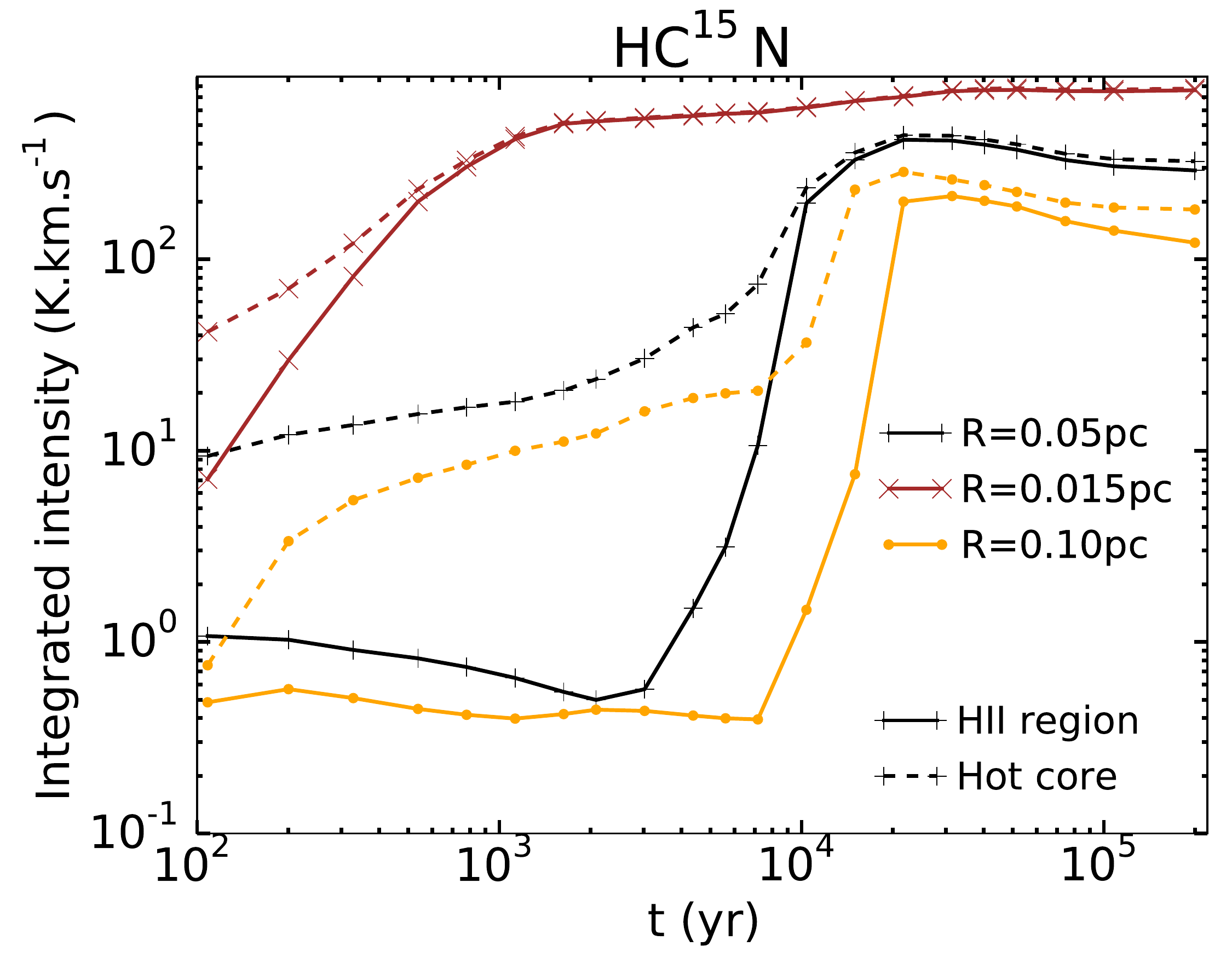}
        \includegraphics[width=0.32\textwidth]{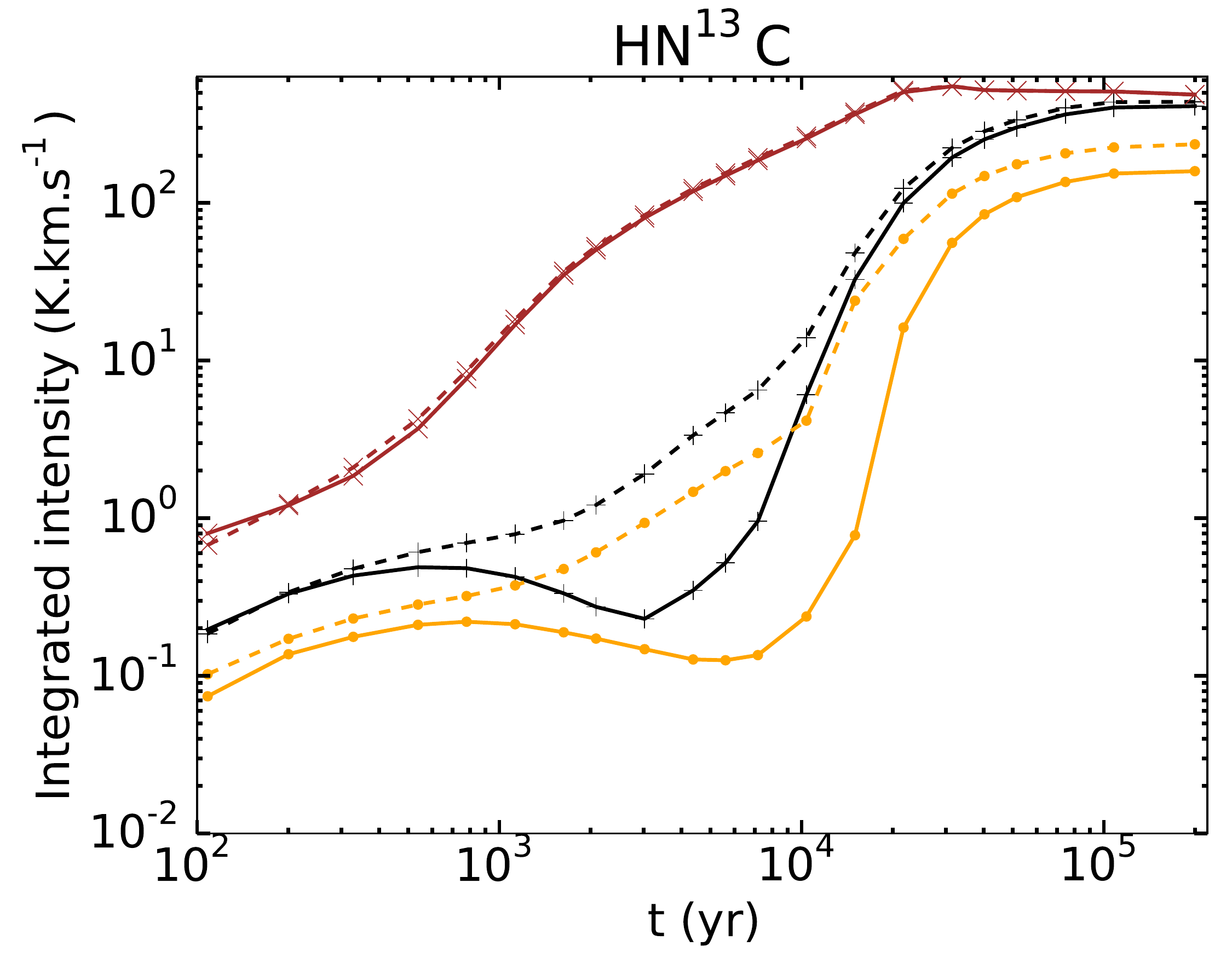}   
        \includegraphics[width=0.32\textwidth]{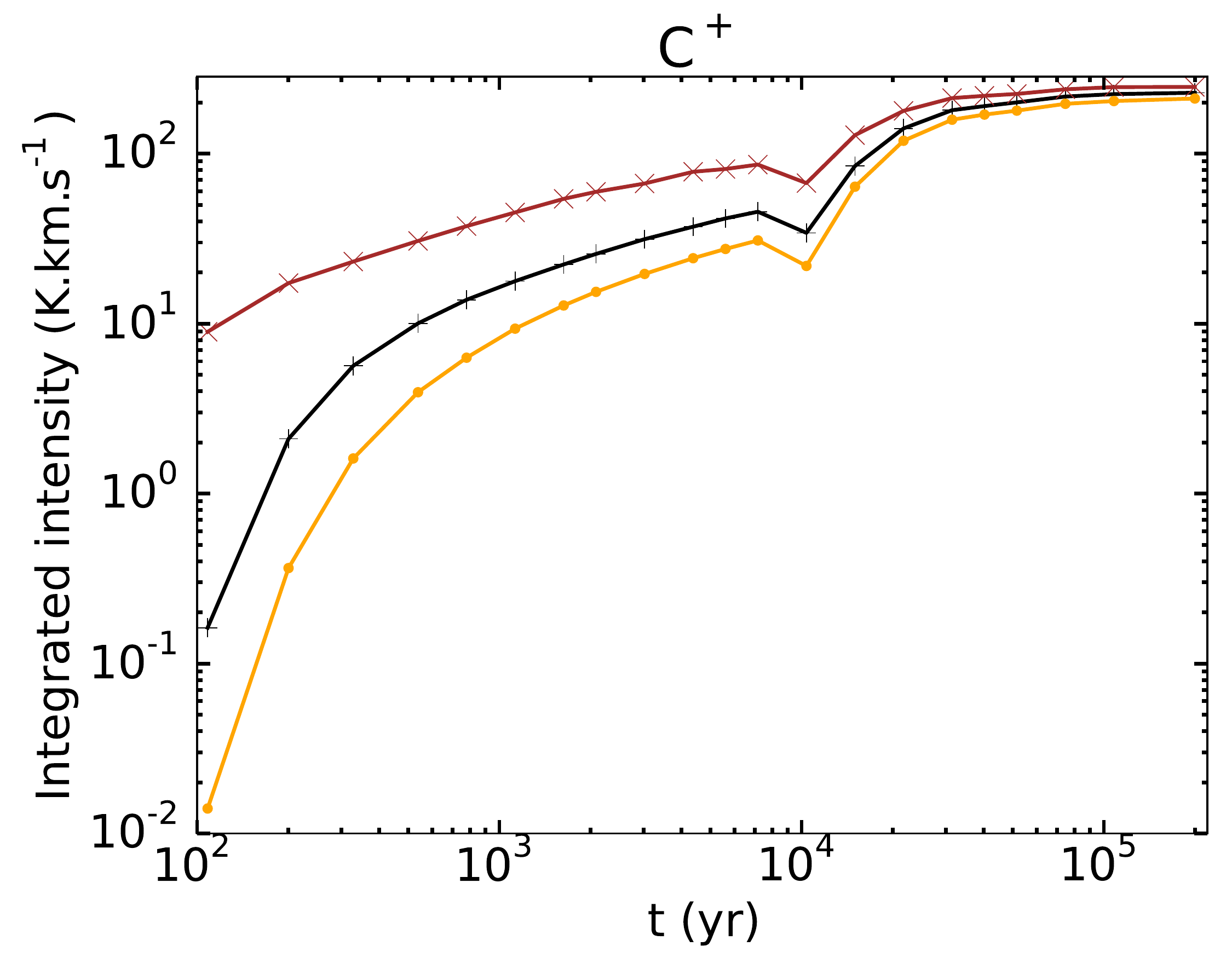} 

        \includegraphics[width=0.32\textwidth]{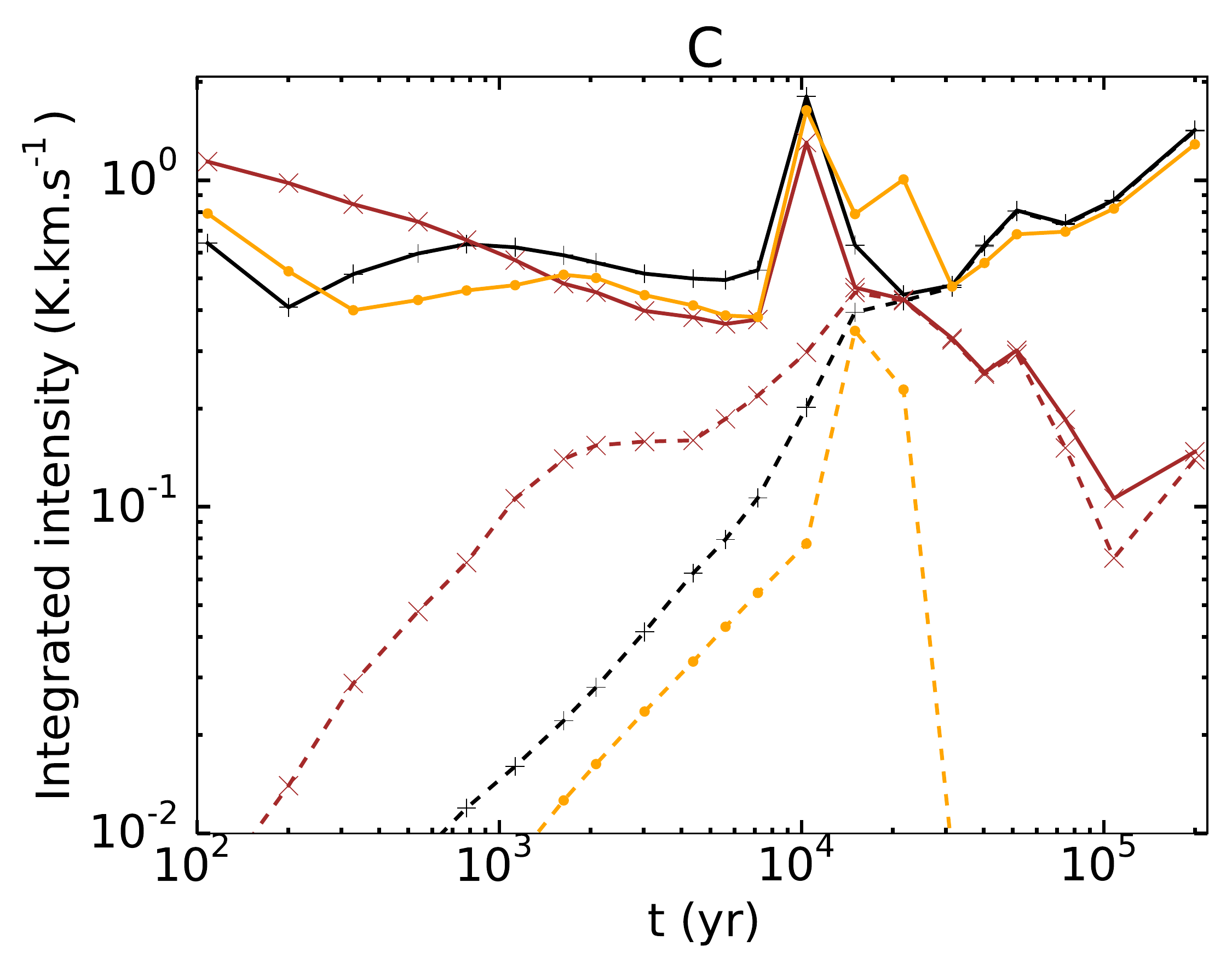}
        \includegraphics[width=0.32\textwidth]{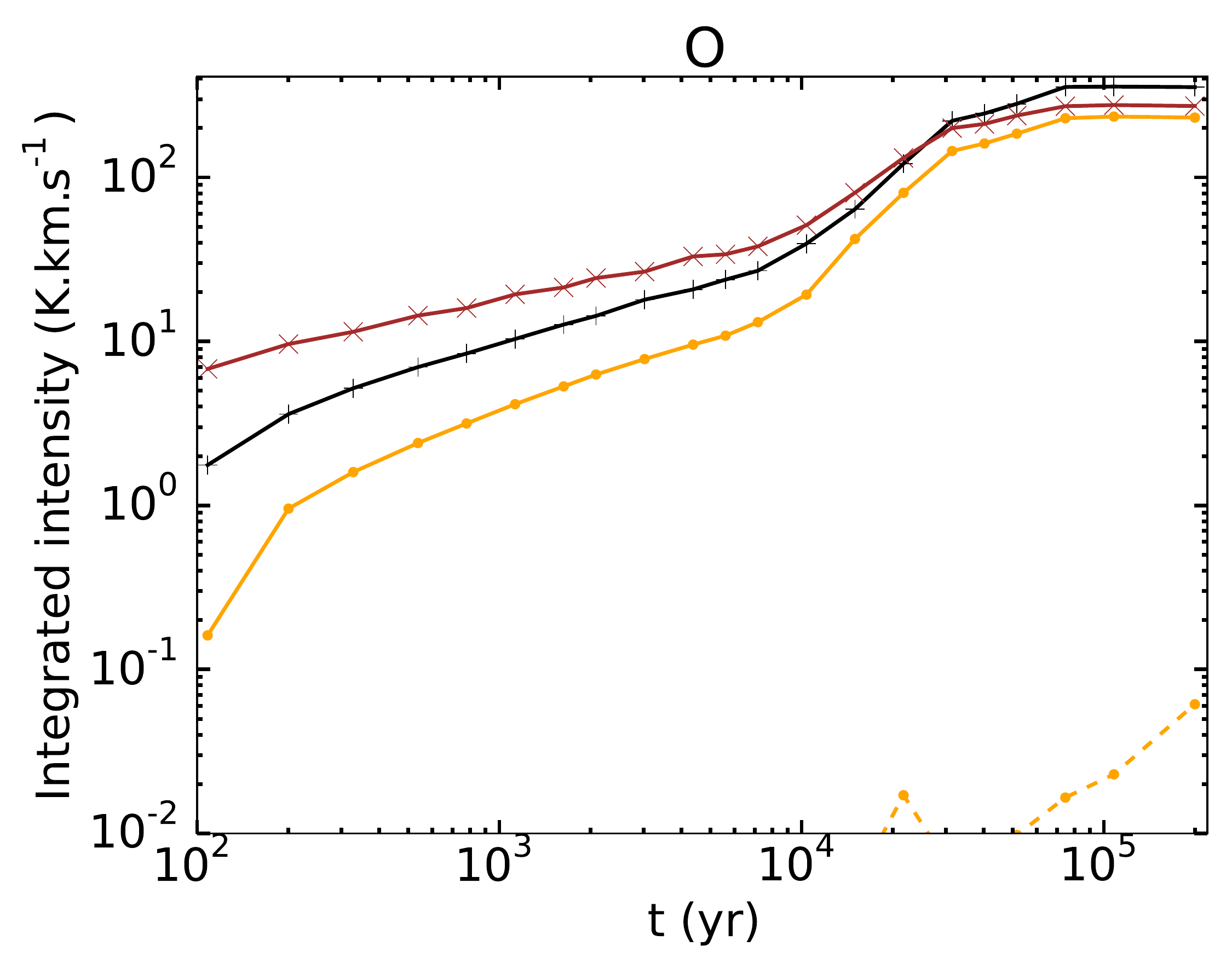}
        \includegraphics[width=0.32\textwidth]{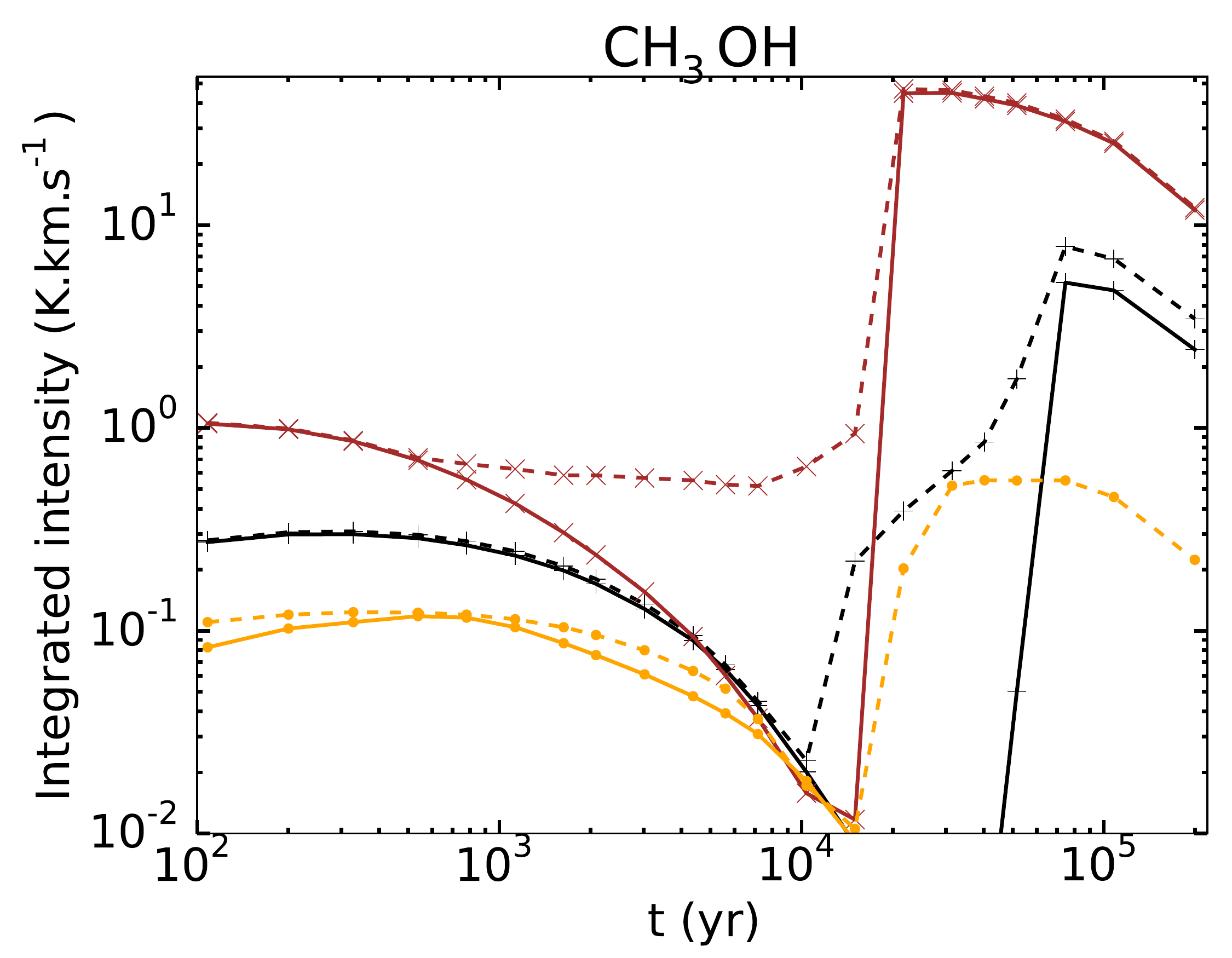} 

        \includegraphics[width=0.32\textwidth]{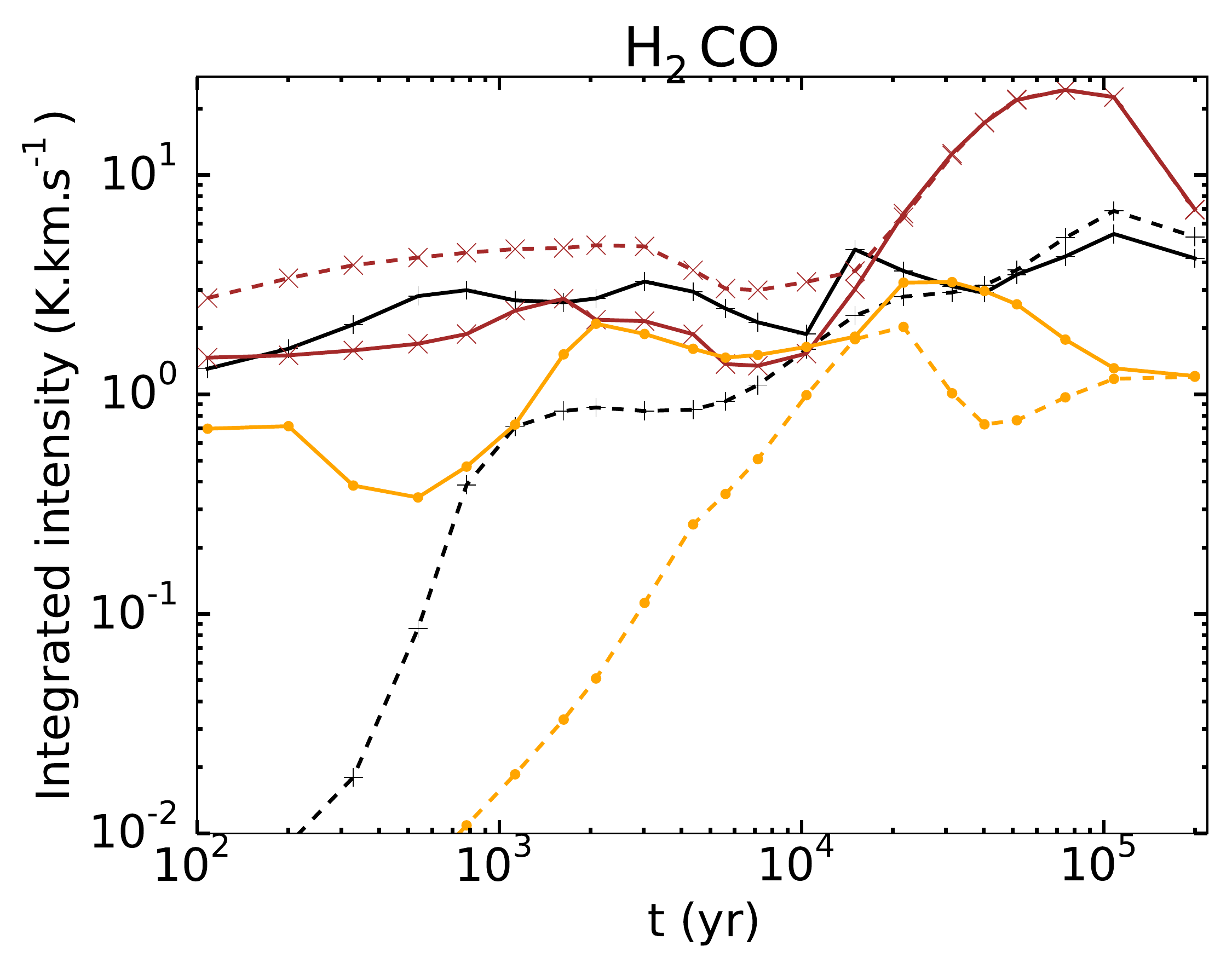}
        \includegraphics[width=0.32\textwidth]{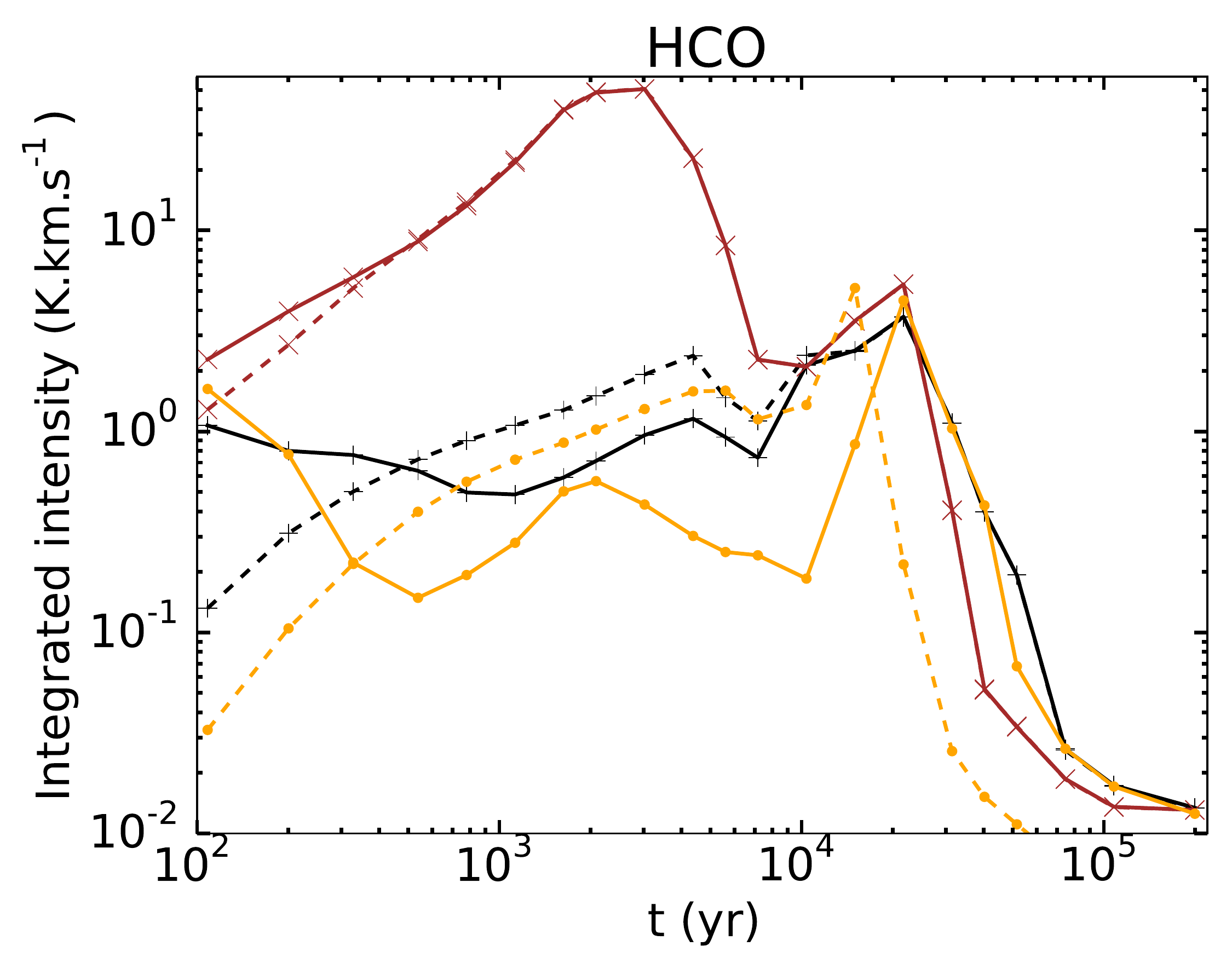}
        \includegraphics[width=0.32\textwidth]{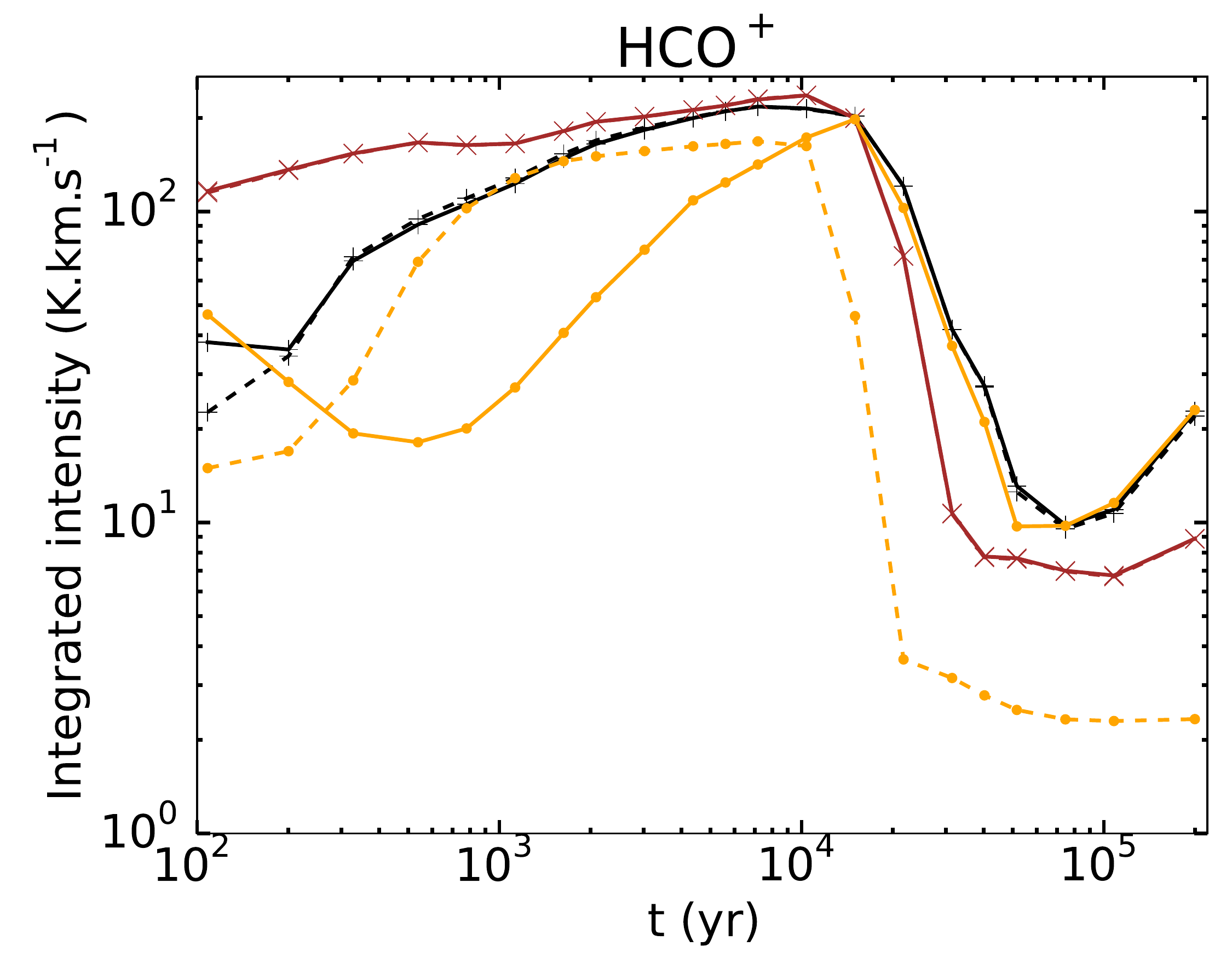} 

        \includegraphics[width=0.32\textwidth]{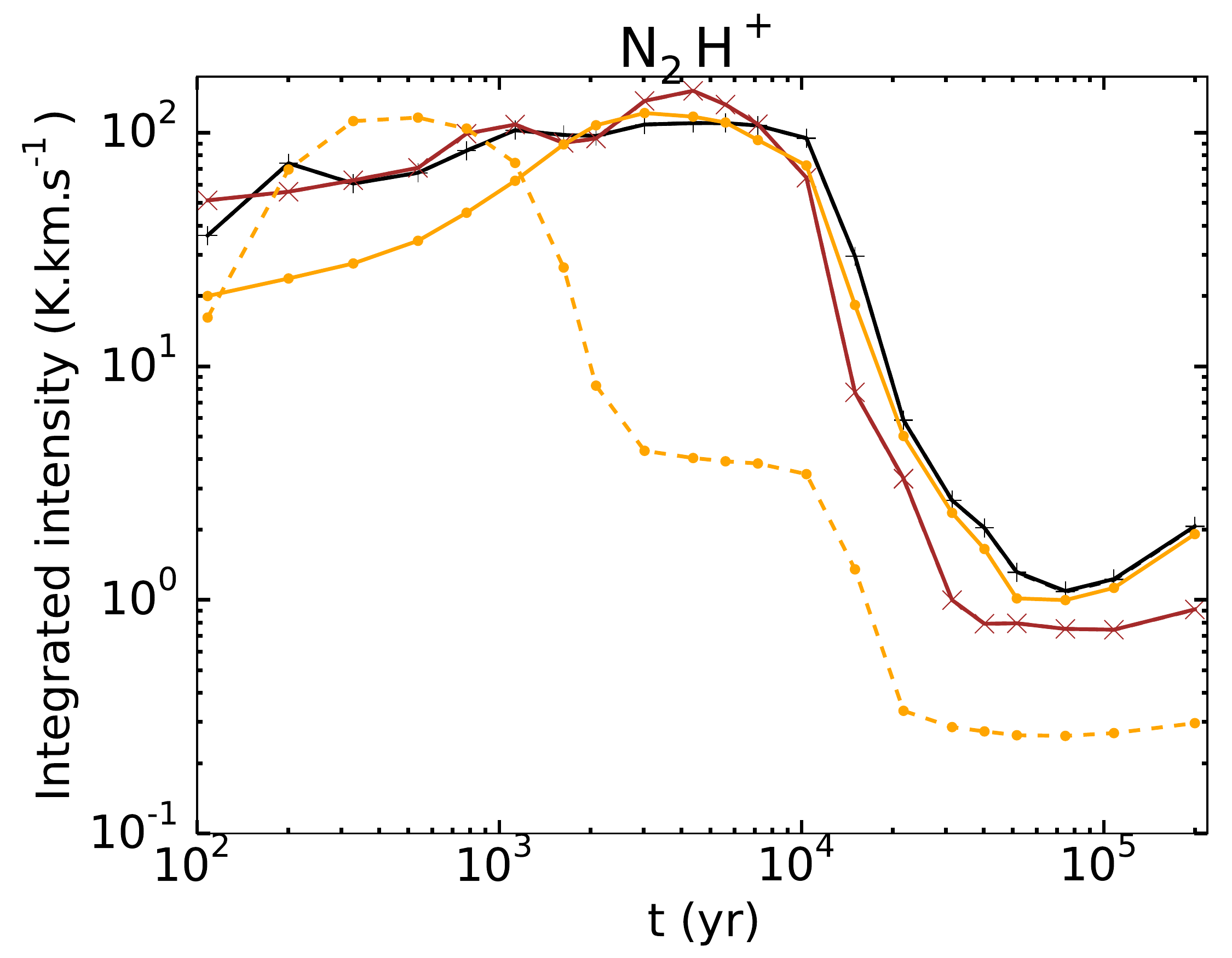}
        \includegraphics[width=0.32\textwidth]{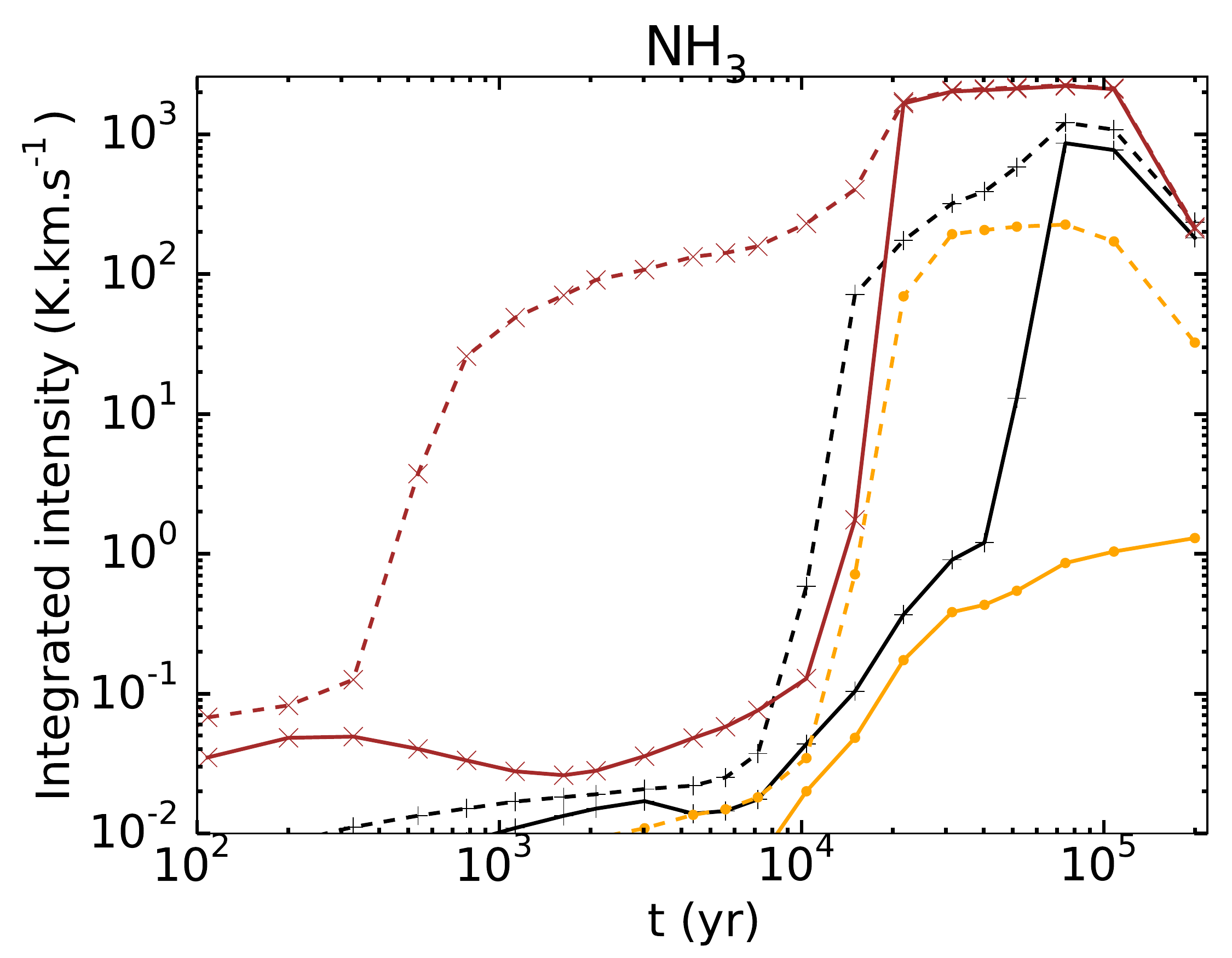}
        \includegraphics[width=0.32\textwidth]{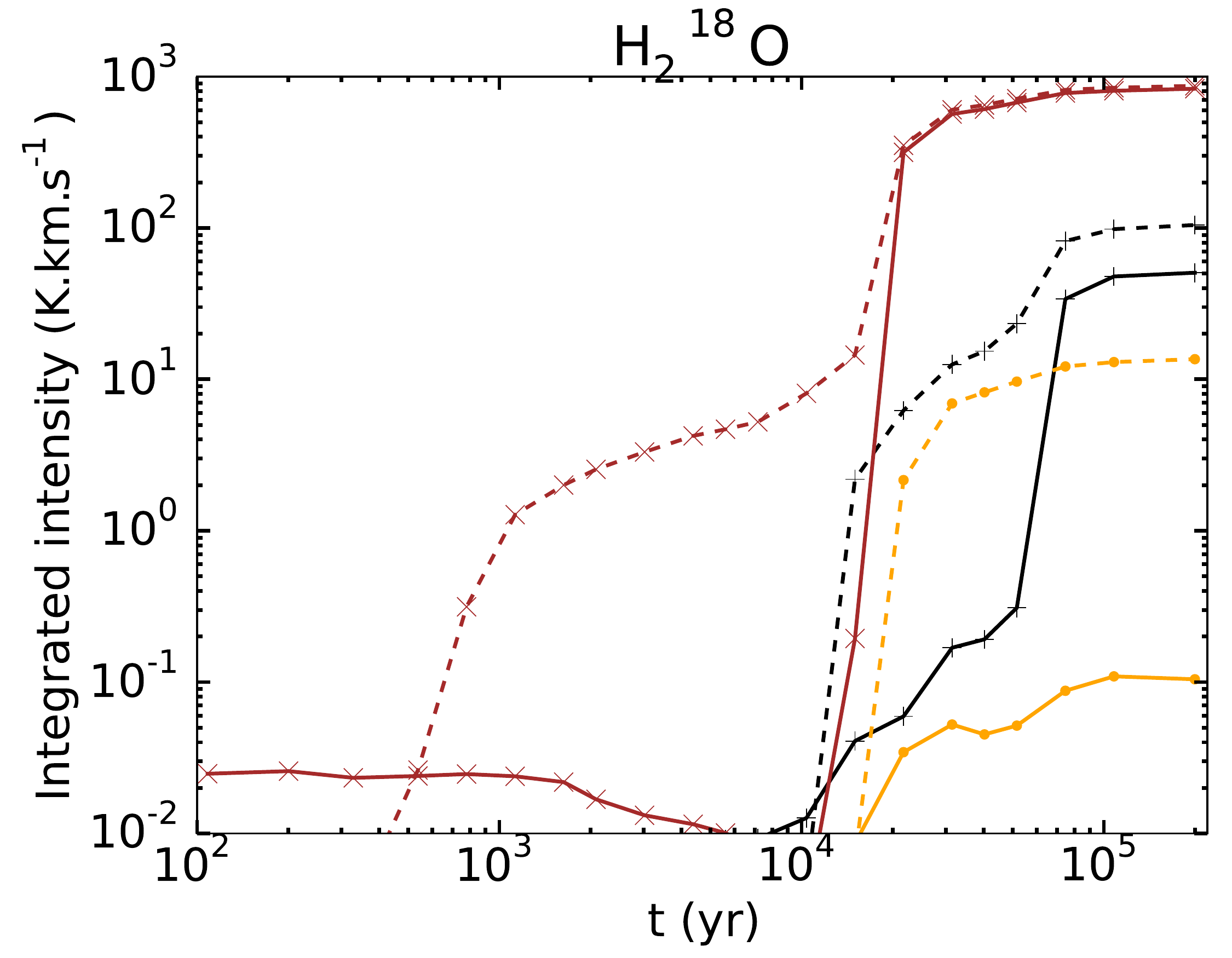} 

        \includegraphics[width=0.32\textwidth]{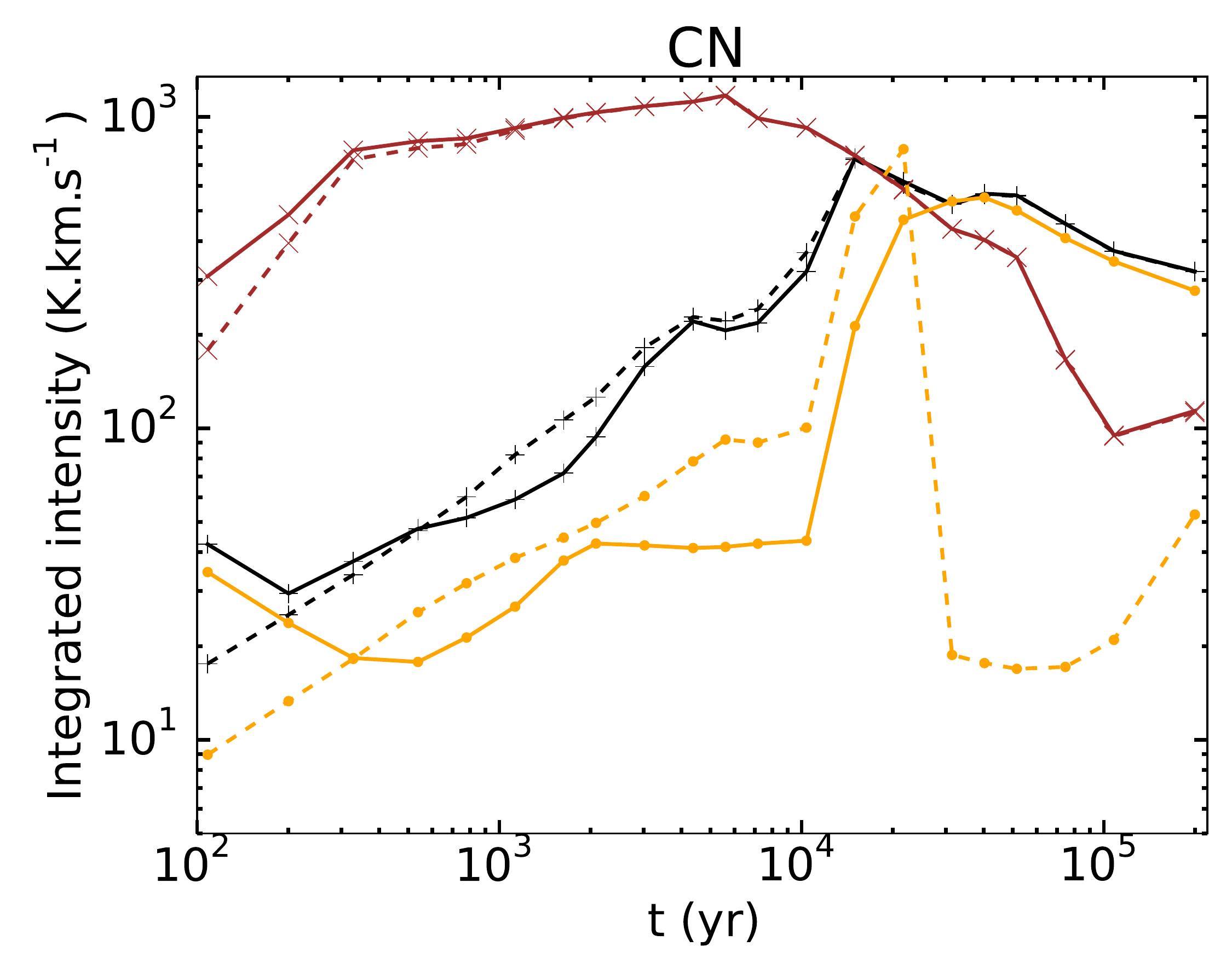} 
        \caption{{\bf Change of the density profile \textit{p1}:} Time evolution of the integrated intensities of the selected species listed in Table~\ref{tab:selected-molecules}. Model \textit{mHII:p1} is represented with solid line and model \textit{mHHMC:p1} with dashed line. Model \textit{r0.015} is represented in brown, model \textit{r0.05} in black, and model \textit{r0.10} in yellow.}
        \label{apfig:intInt_comp_HIIsize_mod2}
\end{figure*}

With the Plummer-like density profile \textit{p1} (see top panel of Fig.~\ref{fig:intInt_summaryPlot} and Fig.~\ref{apfig:intInt_comp_HIIsize_mod2}), when we increase the \hii\ region size we also decrease the density. The model with the smallest \hii\ region size \textit{r0.015} has the highest molecular emission as its density is higher. This is not the case for the late evolution of C$^+$ and HCO for which the intensities are equal for the different models \textit{mHII:p1}. For CN the intensity is three times lower for the model with the largest \hii\ region and it is two times lower for \ntohp\ and \hcop. In addition, \hcop\ and \ntohp\ emissions are quite similar for models \textit{mHHMC} and \textit{mHII} except for model \textit{r0.10} where the HHMC emission is ten times lower.

\subsection{Initial abundances: \textit{ini2}}
\label{apsec:ini2}

\begin{figure}[t]
        \centering
        \includegraphics[width=0.5\textwidth, trim={0 0 0 25mm},clip]{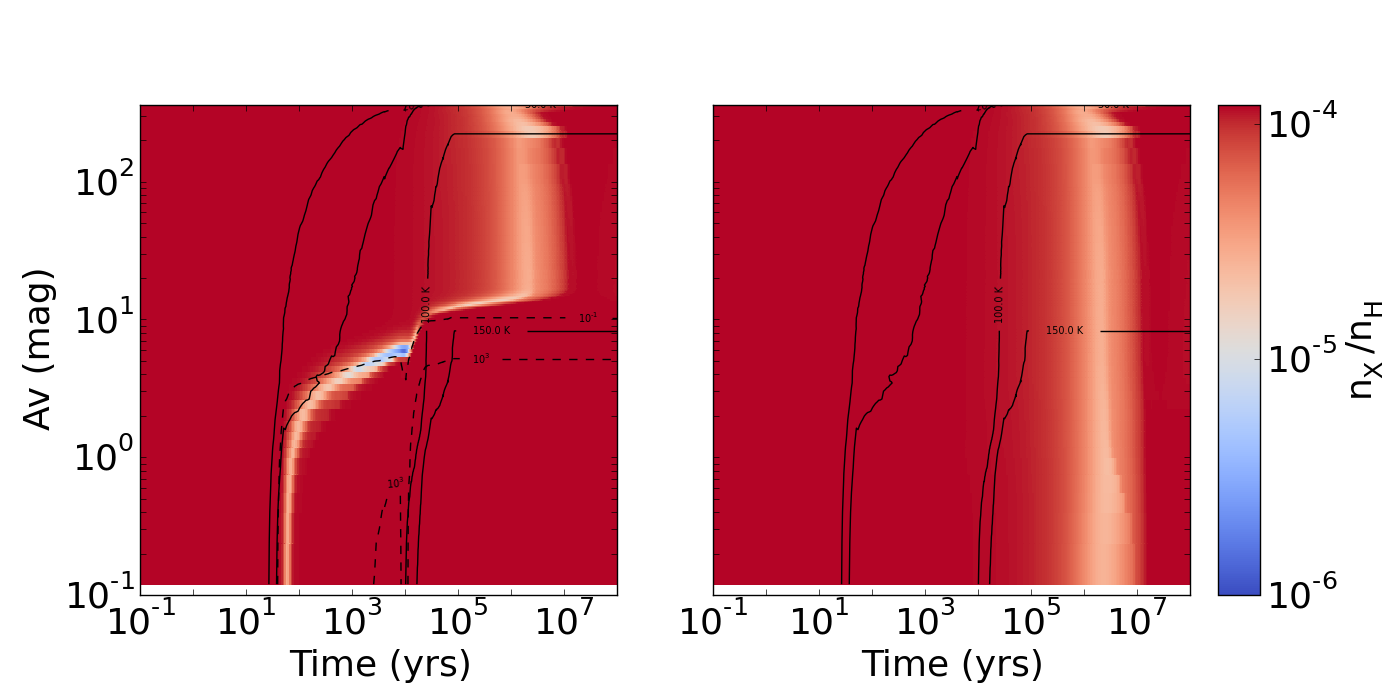}                
        \caption{Spatio-temporal evolution of the sum of the abundances of the main C-bearing species (C$^+$, C, CO, CH$_4$, s-CO, and s-CH$_4$) for model \textit{mHII:ini2} (left panel) and model \textit{mHHMC:ini2} (right panel).}
        \label{fig:3Dini1-ini2_sum}
\end{figure}

An interesting result in model \textit{ini2} concerning C-bearing species is that there is an area where the main species do not dominate in the core. This is shown in Fig.~\ref{fig:3Dini1-ini2_sum} where we sum up the relative abundances of the following species: C, C$^+$, CO, CH$_4$, s-CH$_4,$ and s-CO. Some radicals such as CH or CH$_3$ appear to have larger abundances in the transition area, up to 5~mag (see left panel of Fig.~\ref{fig:regions}). The frontier between different chemical regimes is marked by a dispersion of the elements over a large number of species. 
But given the short period of time when they pop up it is unlikely to detect them. Between 5 and 10~mag, more complex grain species like s-CO$_2$, s-C$_2$H$_2$, s-C$_3$H$_2$ or s-CH$_3$OH prevail. At higher extinctions other molecules such as HCN are the dominant species regarding the carbon content. For \av\ > 300~mag, carbon chains, mostly s-C$_4$H$_2$, are formed on the grain surface as there is no radiation field but the temperature is still higher than 30~K. 

Furthermore, when we used the chemical network with 334 species, the sum of the main C-bearing species is 10$^{-5}$ higher in abundance. In general, the abundance of the molecules at the frontier between the chemical regimes in model \textit{ini2} is lower with the larger network probably due to the presence of more complex species as well as the different initial abundances.  

\section{Additional figures}
\label{app:additional figures}

\begin{figure*}[htbp]
        \centering
        \includegraphics[width=1\textwidth, trim={0 15 0 30}, clip]{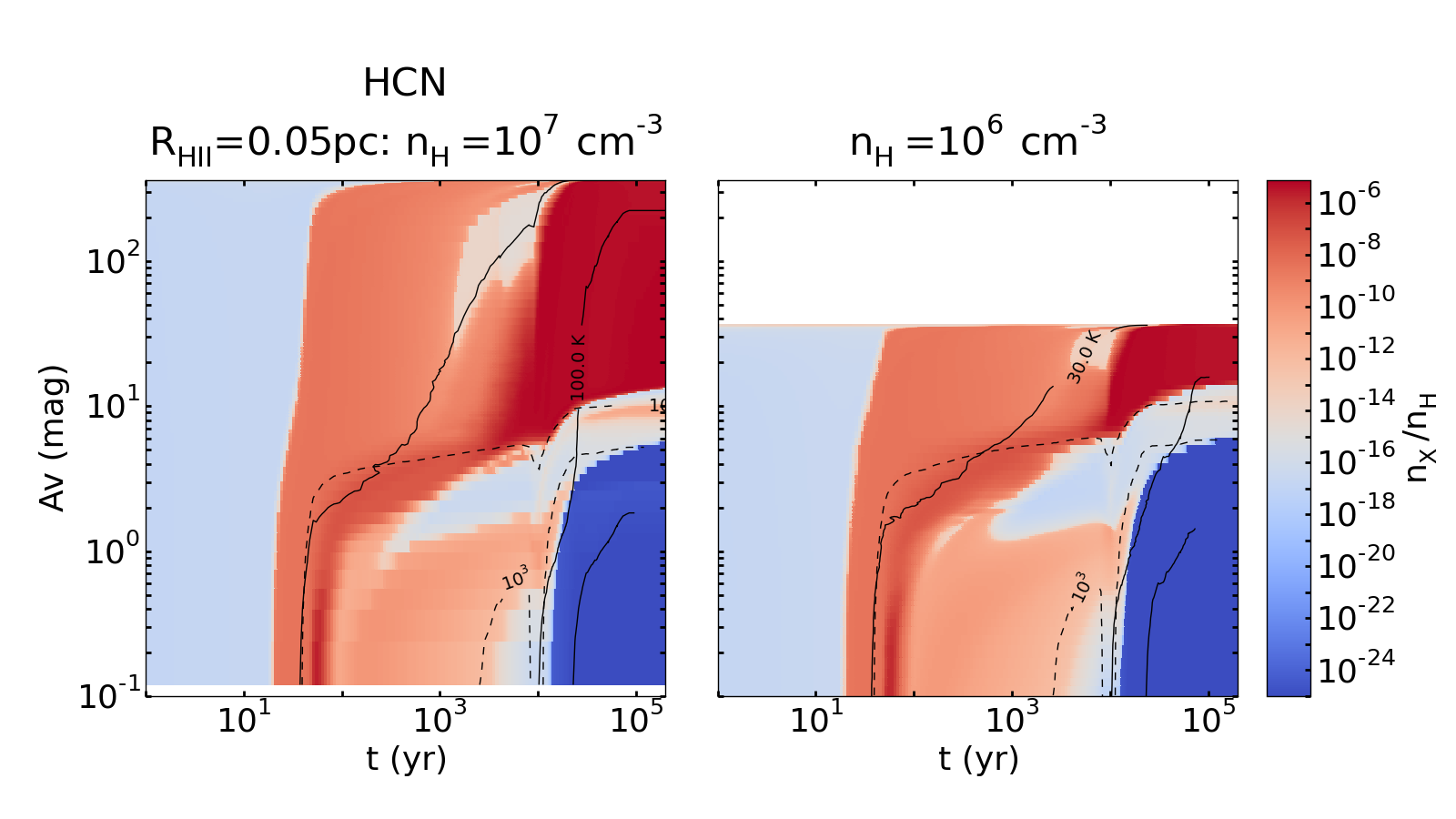}
        \caption{Spatio-temporal evolution of the abundances for HCN for the reference model \textit{mHII:n7} (left panels) and models \textit{n6} (right panels). Contours are plotted: solid line for the \Td\ (30, 100 and 150~K) and dashed lines for \Go\ (10$^{-1}$ and 10$^3$~Draine unit).}
        \label{apfig:3Dabun_compHIIdens}
\end{figure*}

\begin{figure*}[htbp]
        \includegraphics[width=0.32\textwidth]{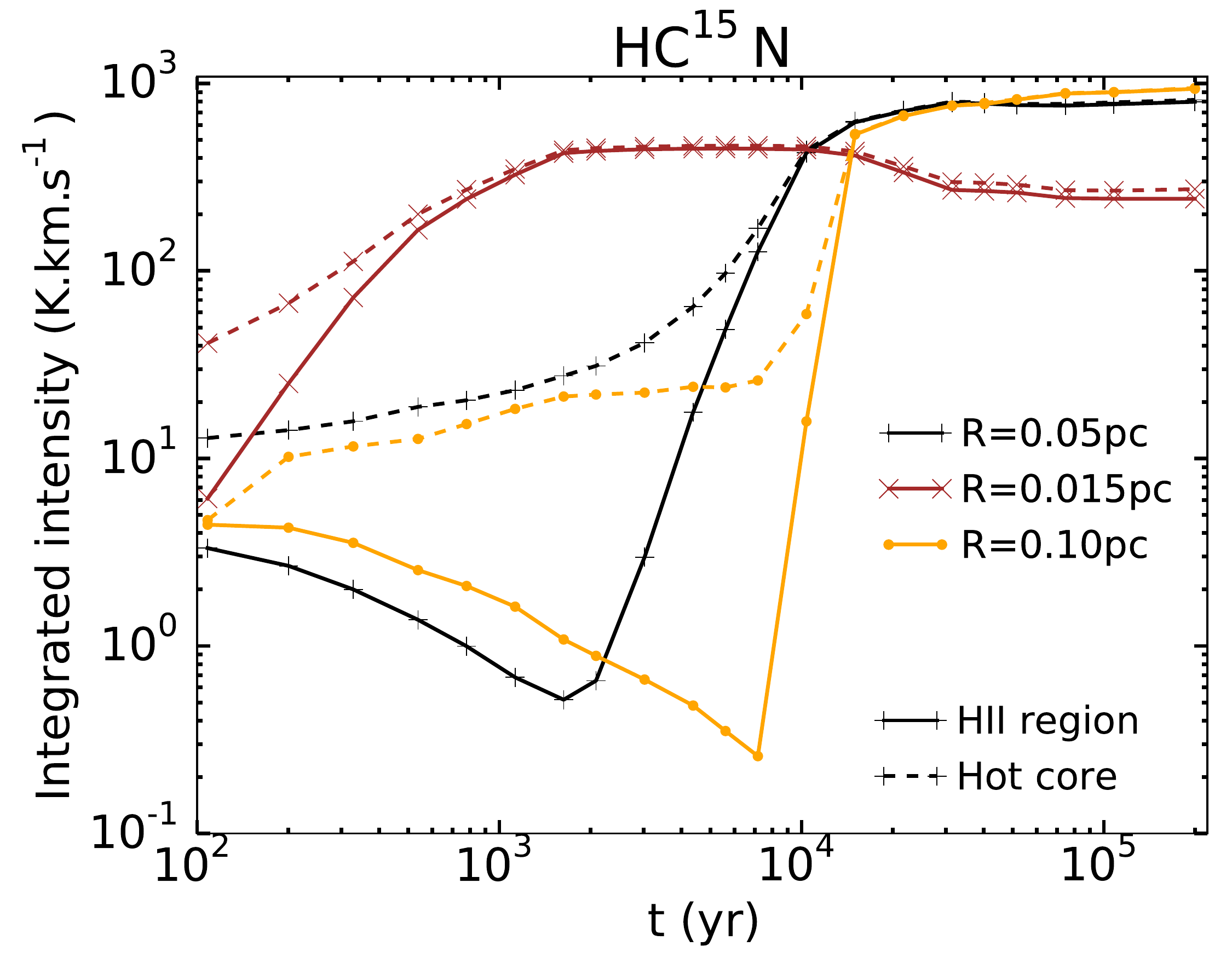}
        \includegraphics[width=0.32\textwidth]{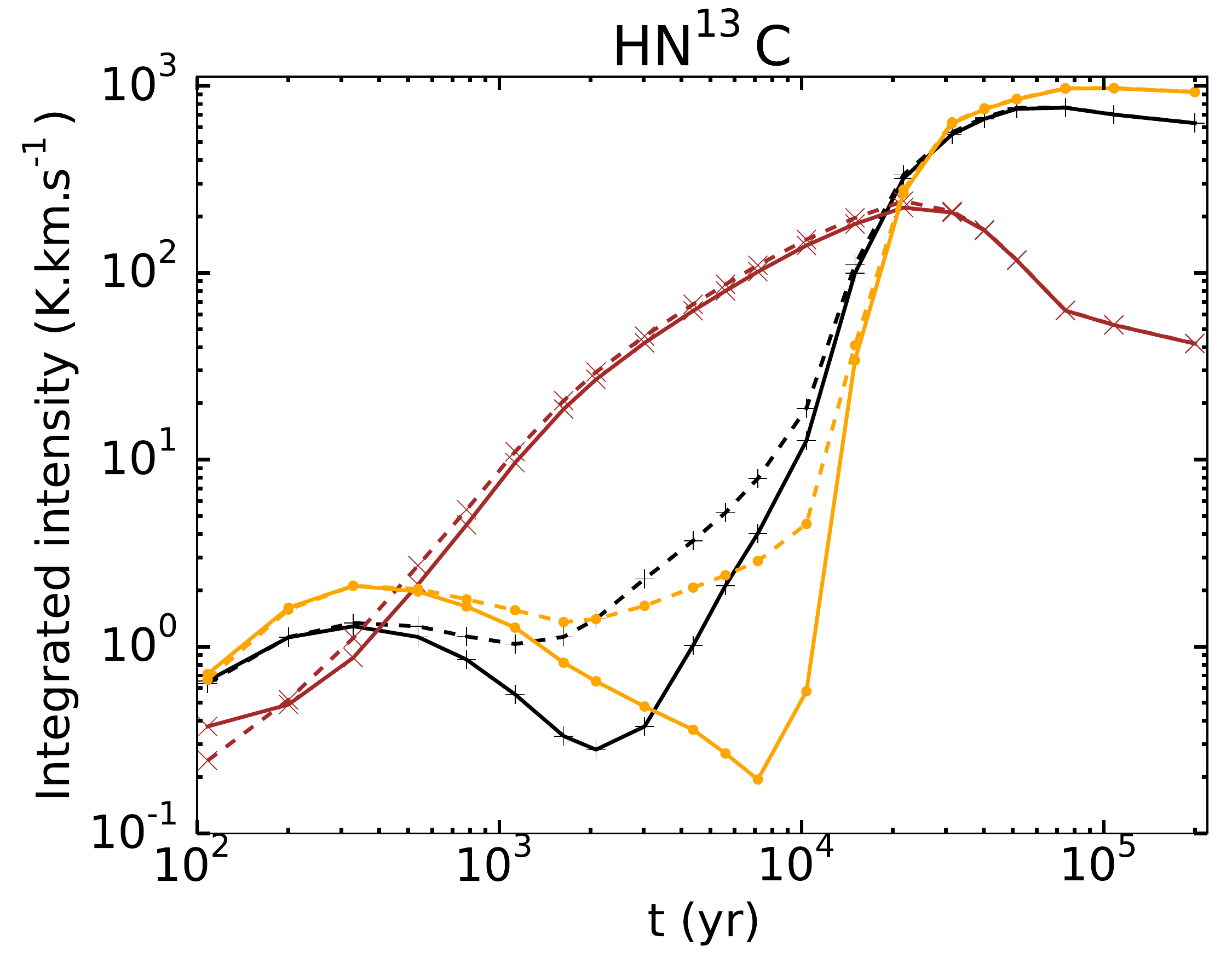}     
        \includegraphics[width=0.32\textwidth]{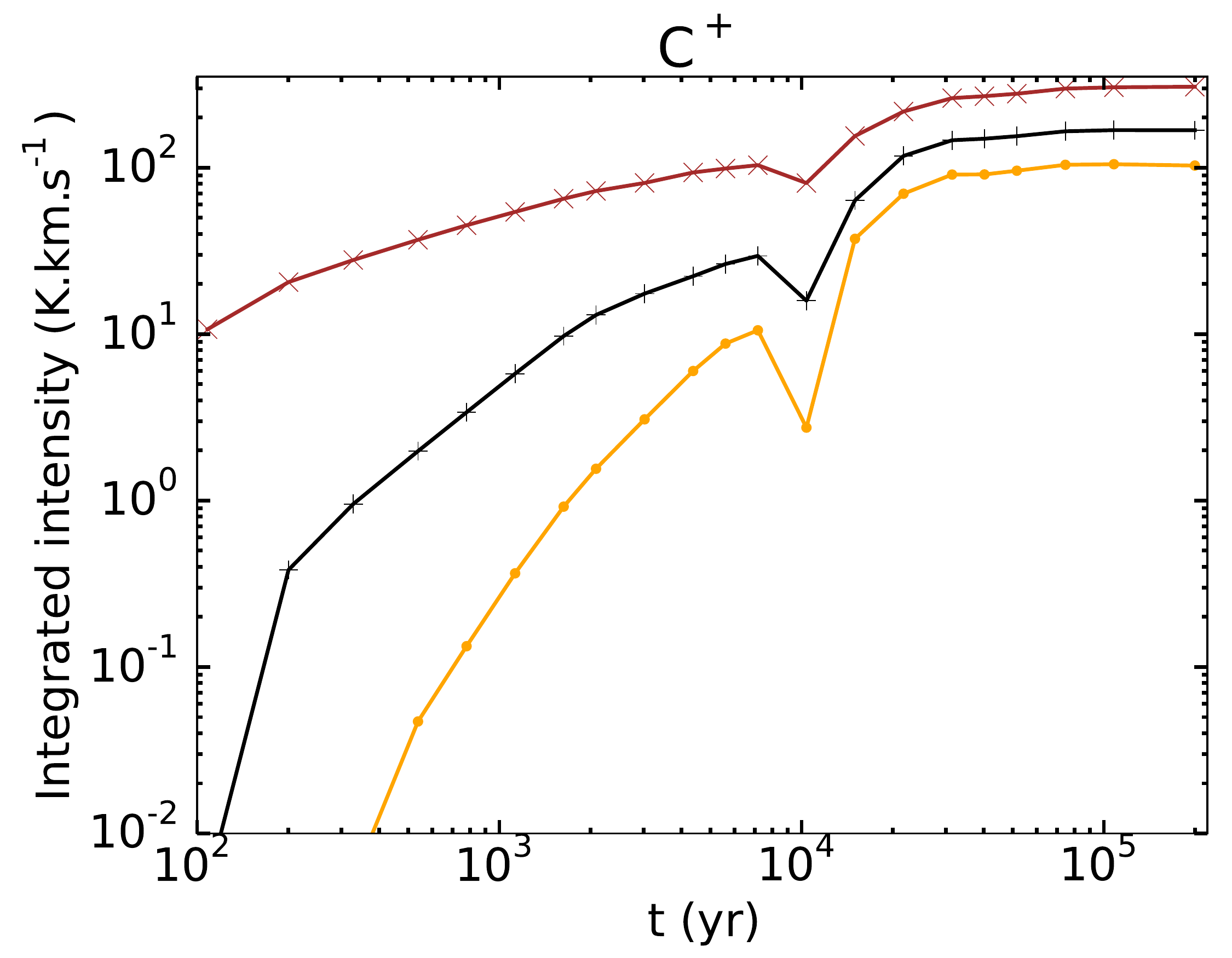} 

        \includegraphics[width=0.32\textwidth]{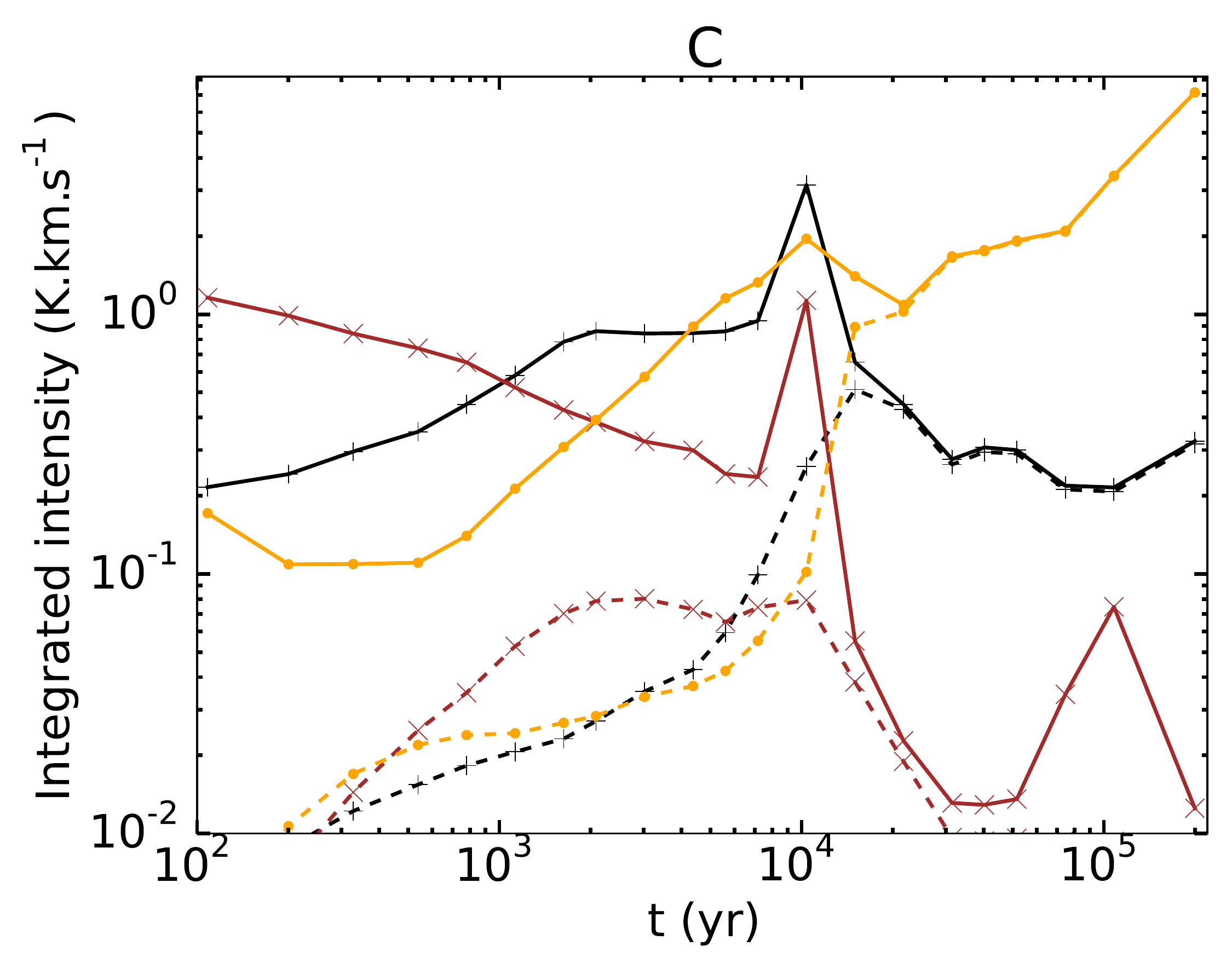}
        \includegraphics[width=0.32\textwidth]{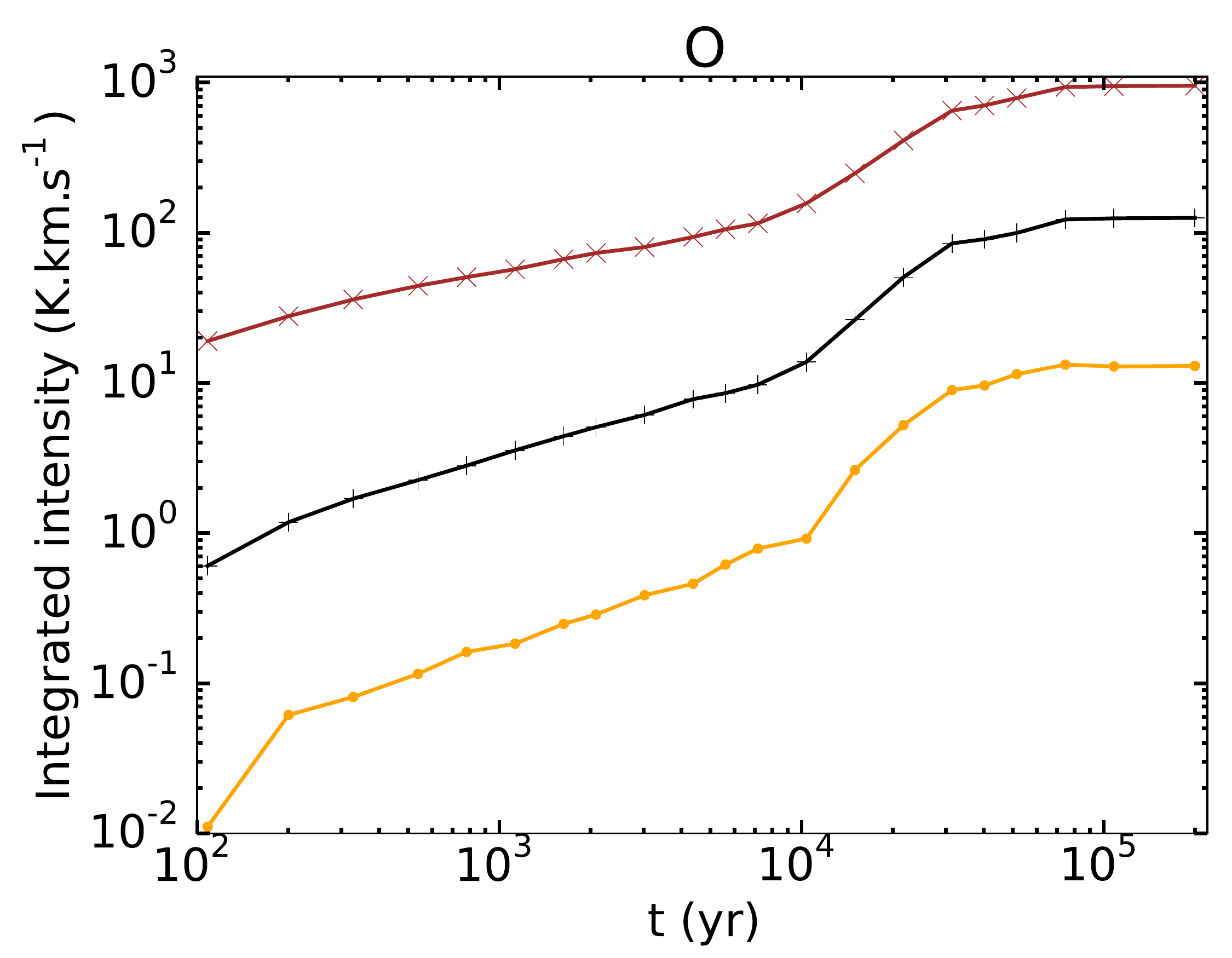}
        \includegraphics[width=0.32\textwidth]{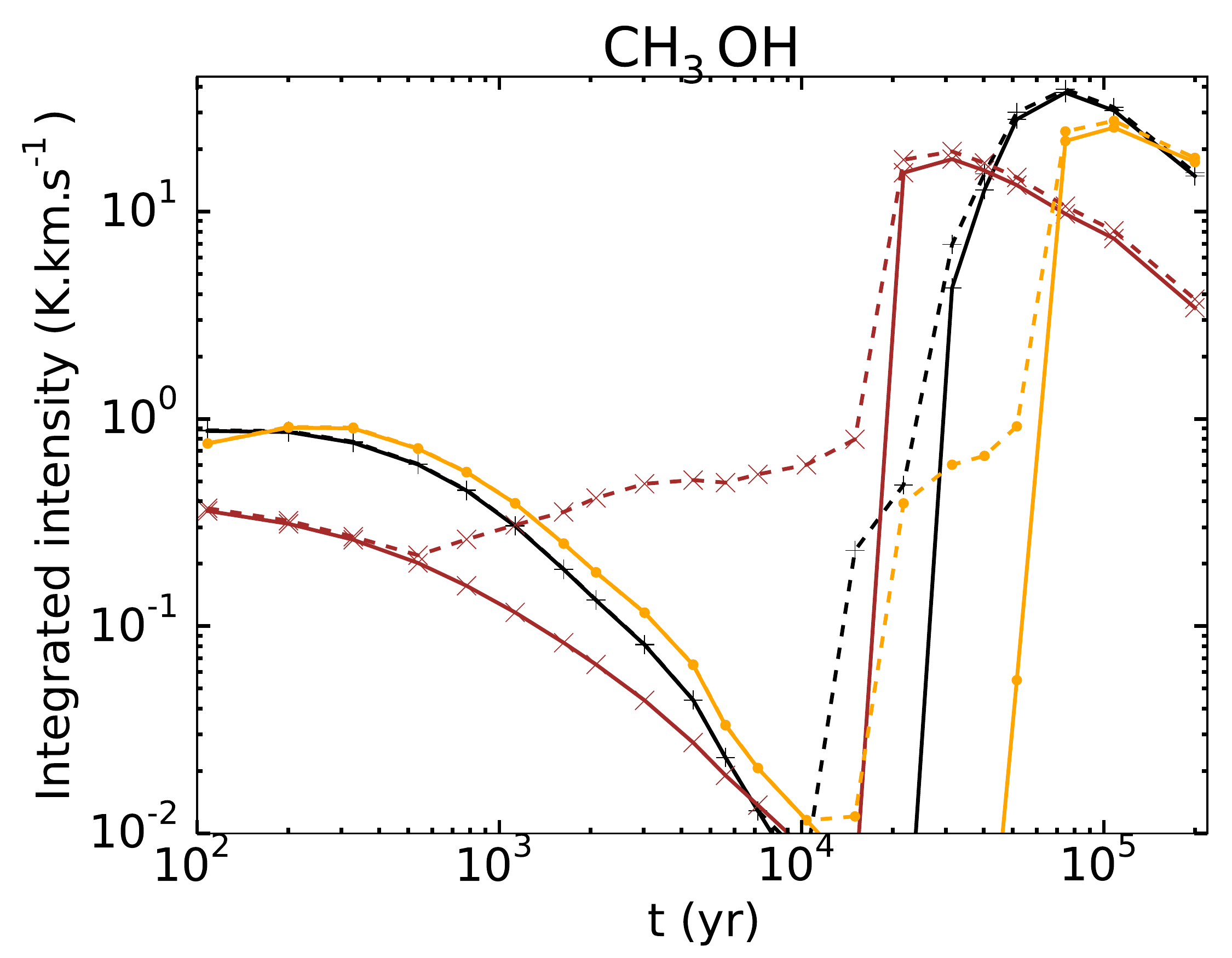} 

        \includegraphics[width=0.32\textwidth]{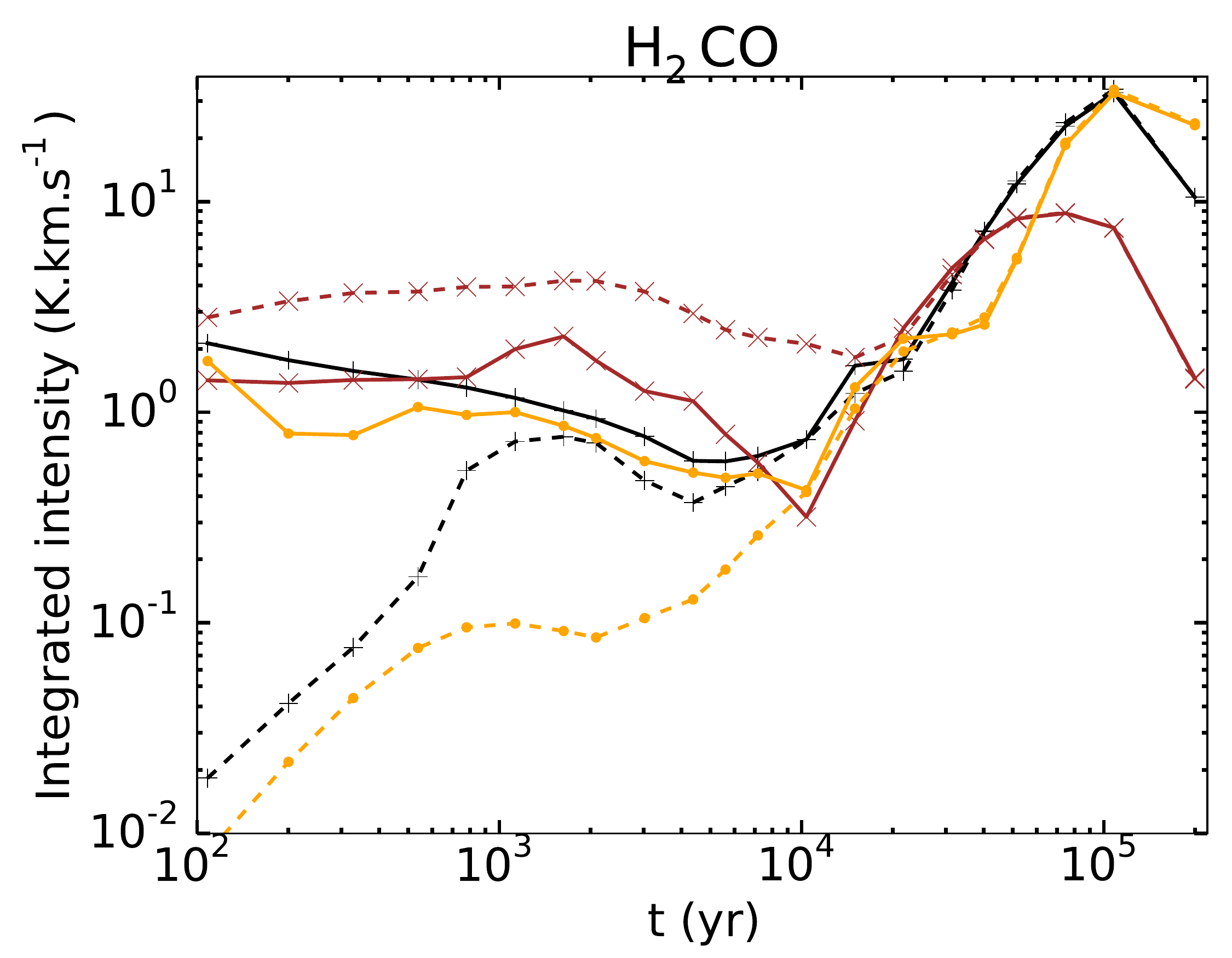}
        \includegraphics[width=0.32\textwidth]{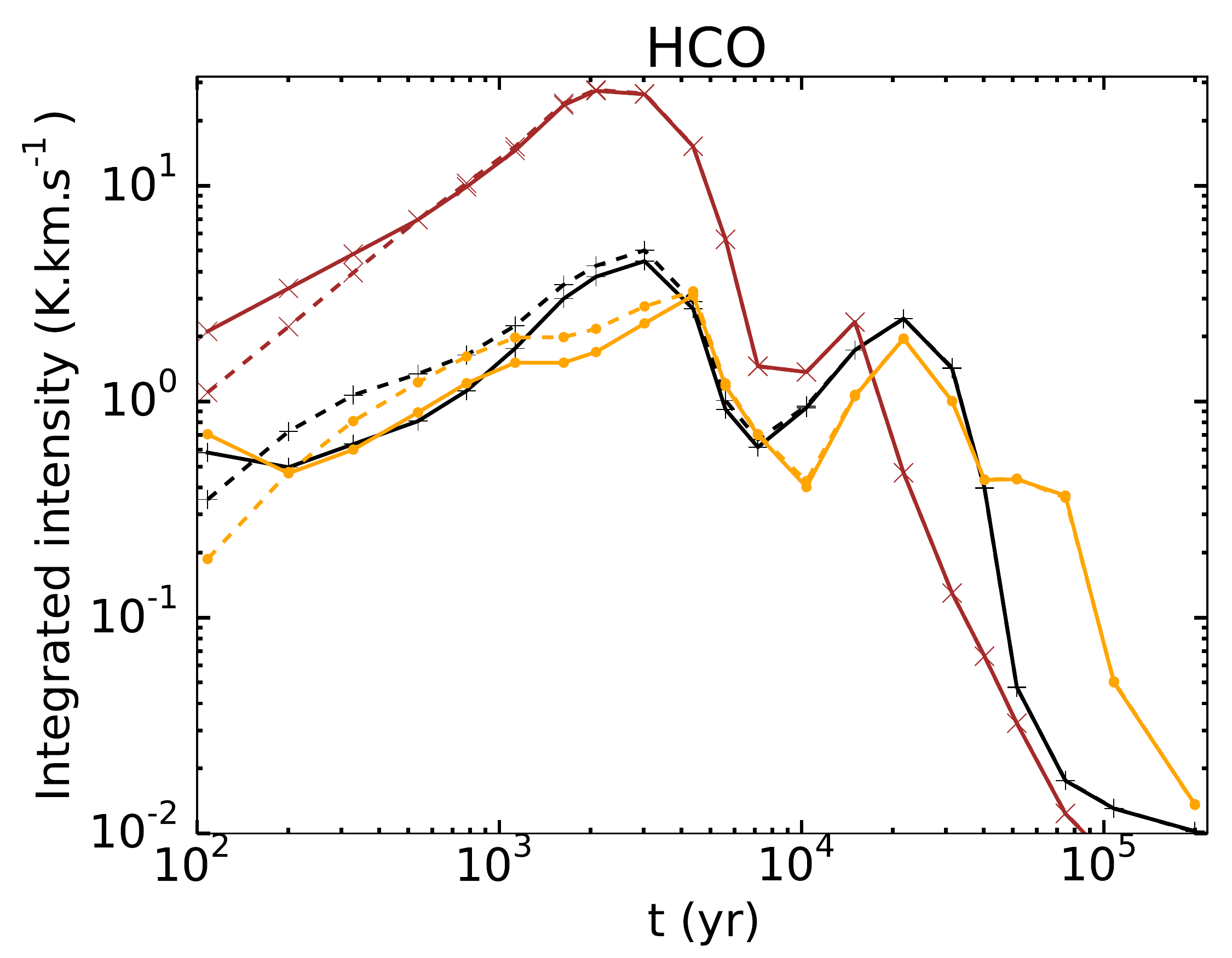}
        \includegraphics[width=0.32\textwidth]{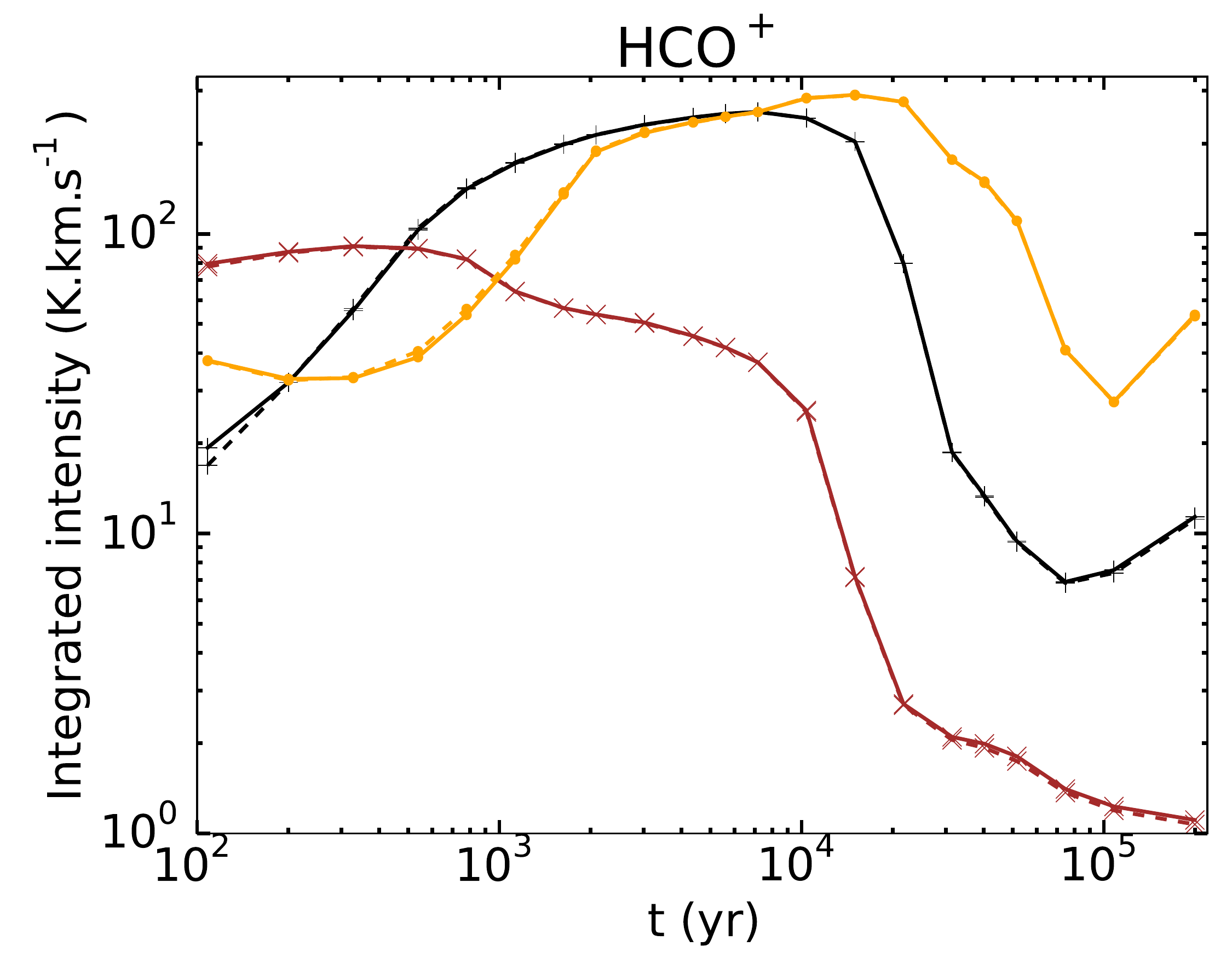} 

        \includegraphics[width=0.32\textwidth]{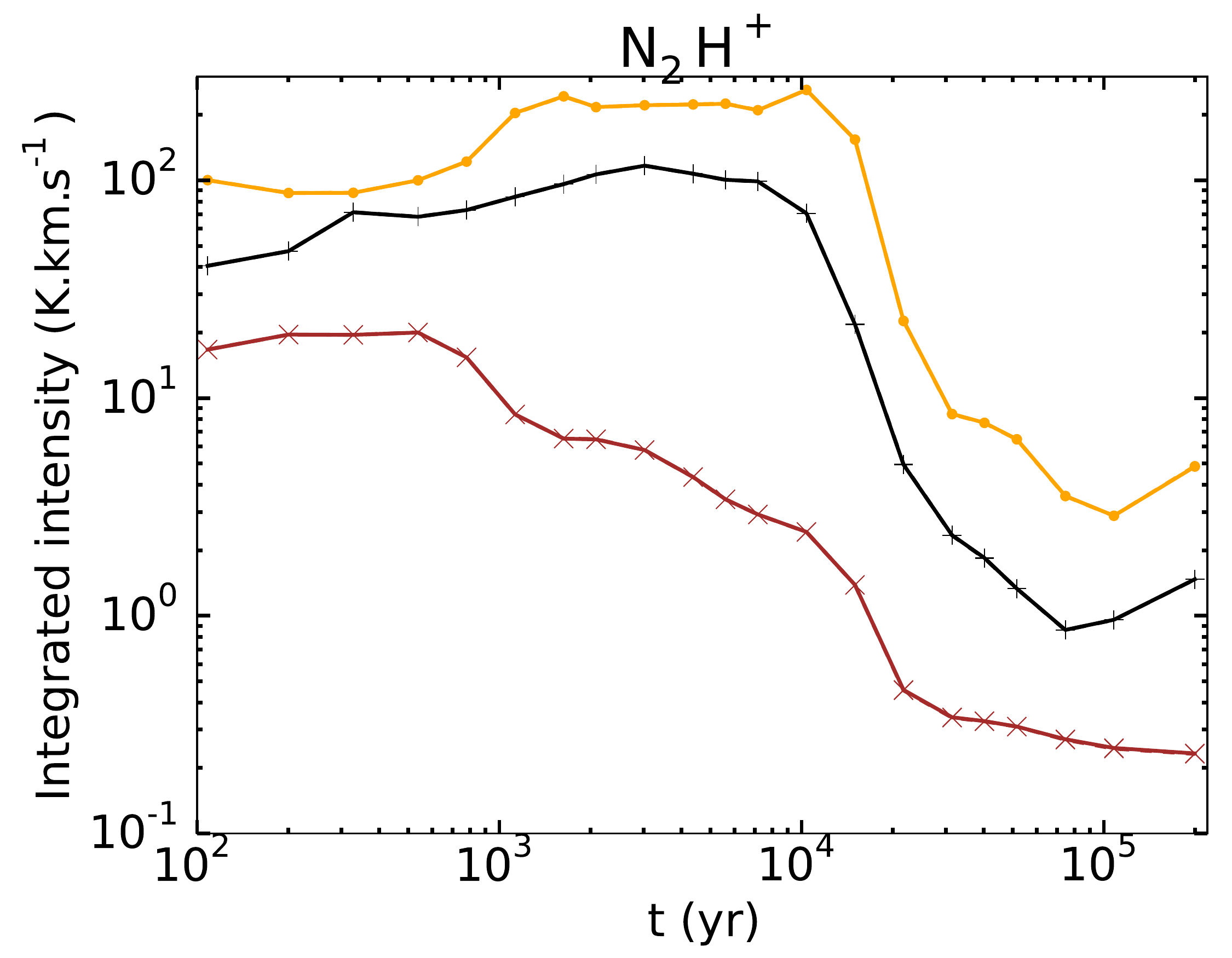}
        \includegraphics[width=0.32\textwidth]{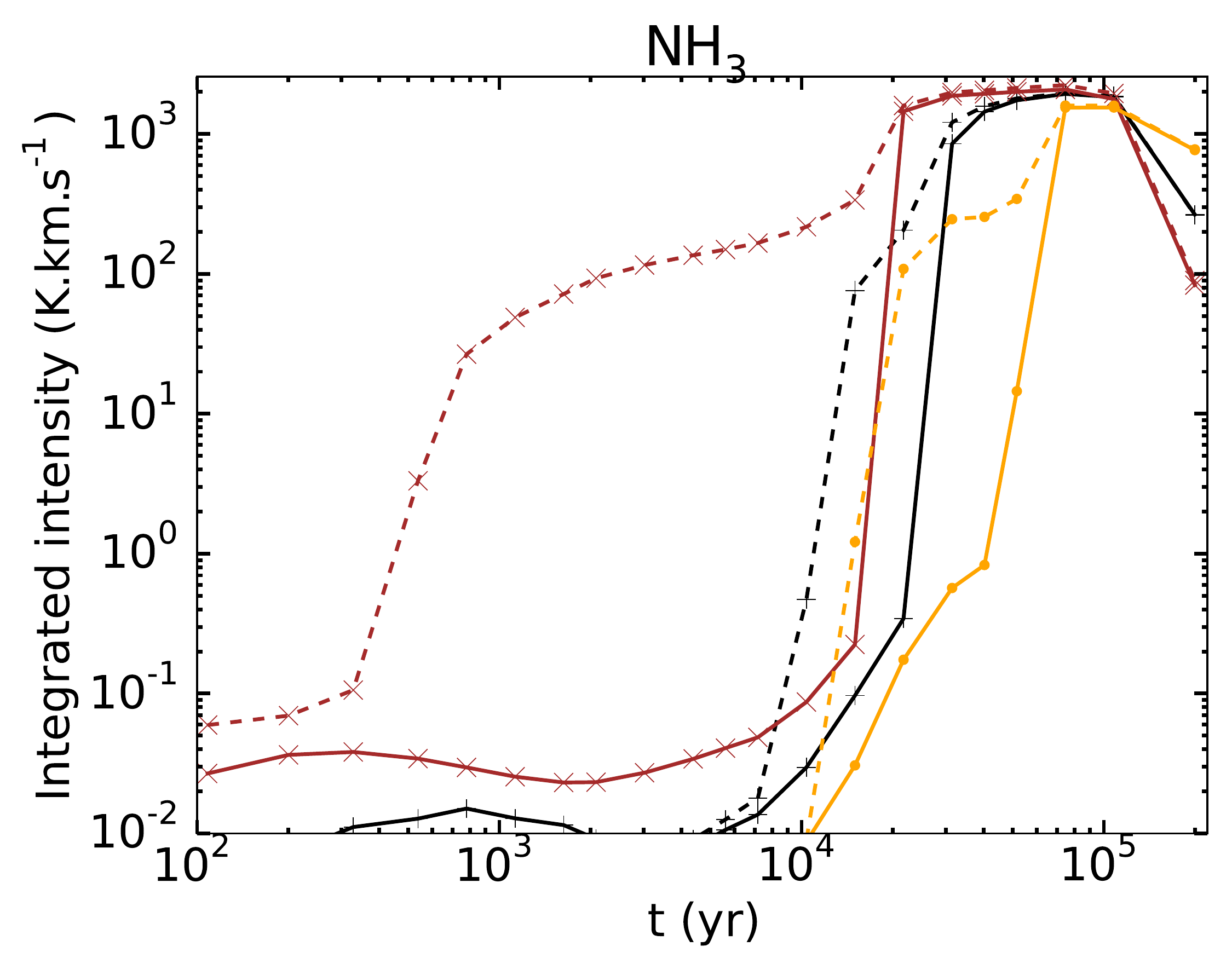}
        \includegraphics[width=0.32\textwidth]{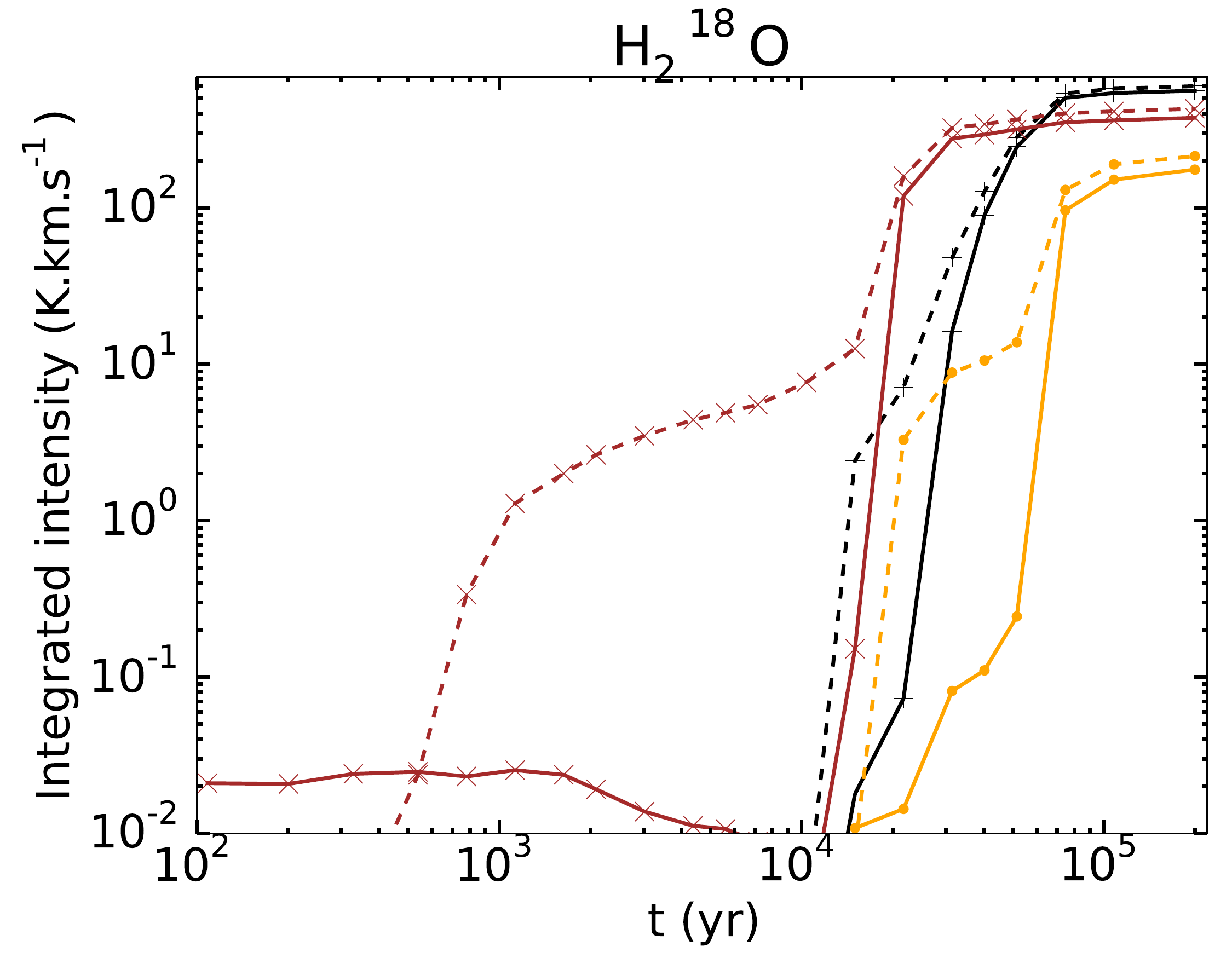} 

        \includegraphics[width=0.32\textwidth]{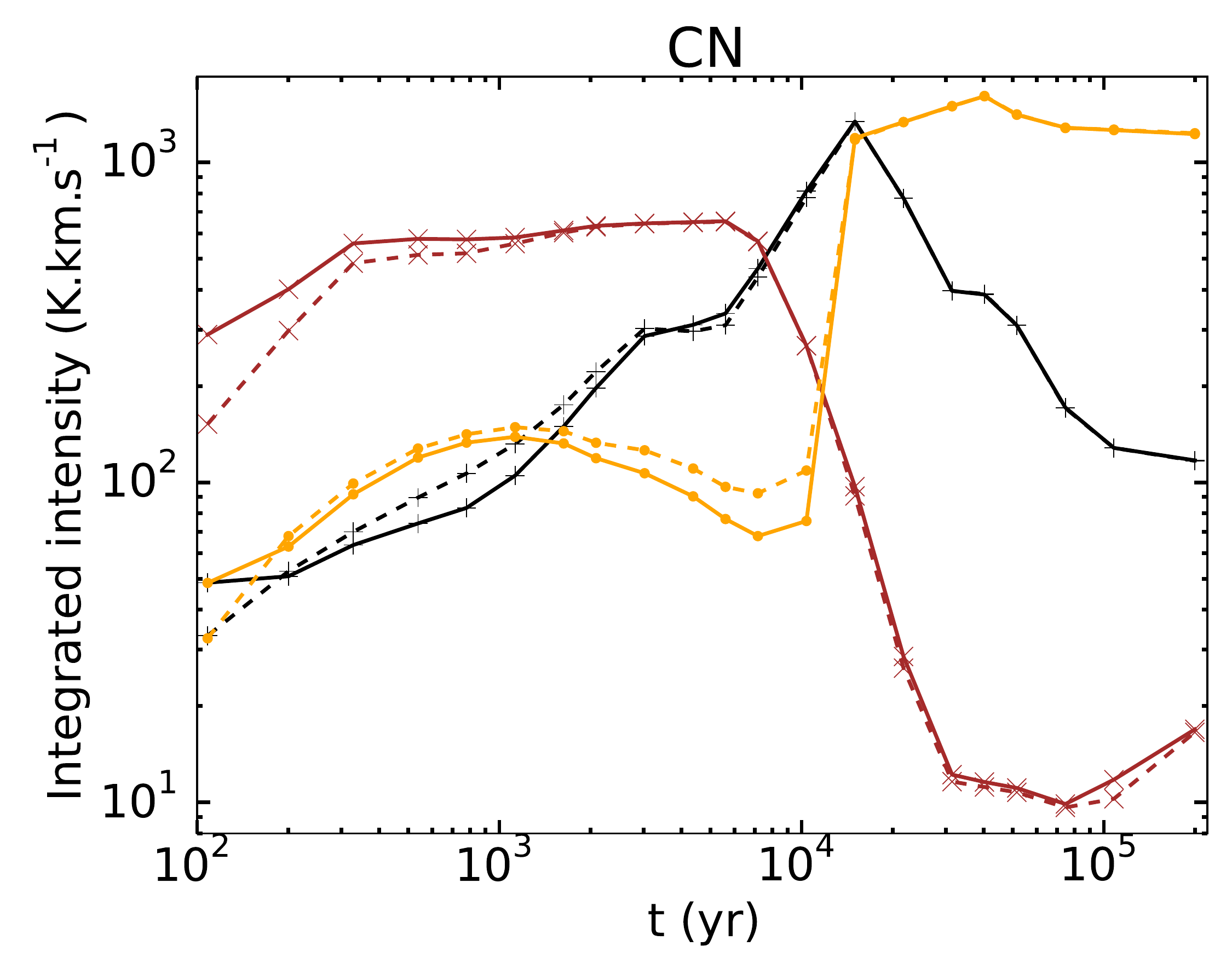} 
        \caption{{\bf Change of the size of the ionized cavity \textit{r0.015}, \textit{r0.05} and \textit{r0.10}:} Time evolution of the integrated intensities of the selected species listed in Table~\ref{tab:selected-molecules}. Model \textit{mHII} is represented with solid line and model \textit{mHHMC} with dashed line. Model \textit{r0.015} is represented in brown, model \textit{r0.05} in black, and model \textit{r0.10} in yellow.}
        \label{apfig:intInt_comp_HIIsize}
\end{figure*}

\begin{figure*}[htbp]
        \includegraphics[width=0.32\textwidth]{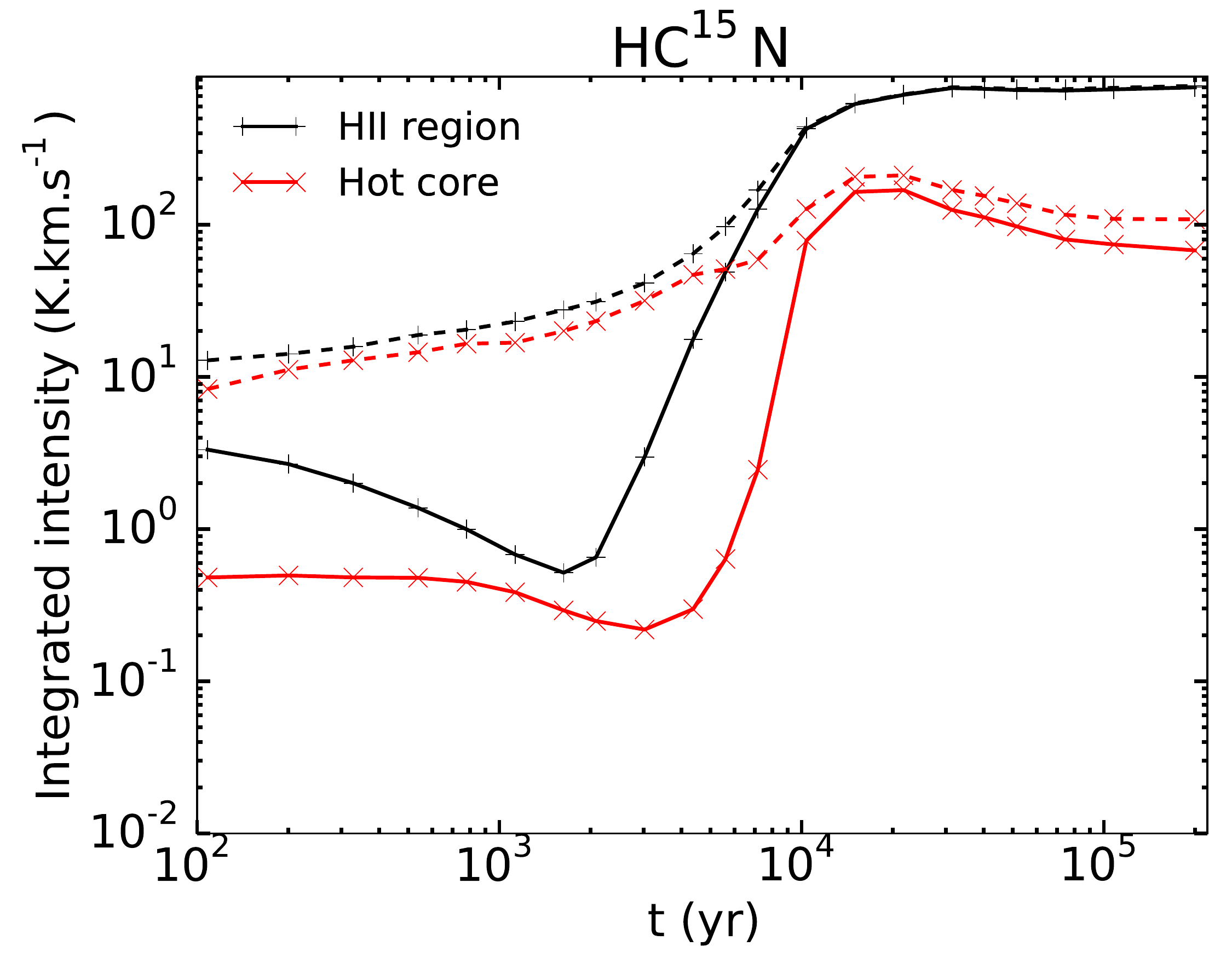}
        \includegraphics[width=0.32\textwidth]{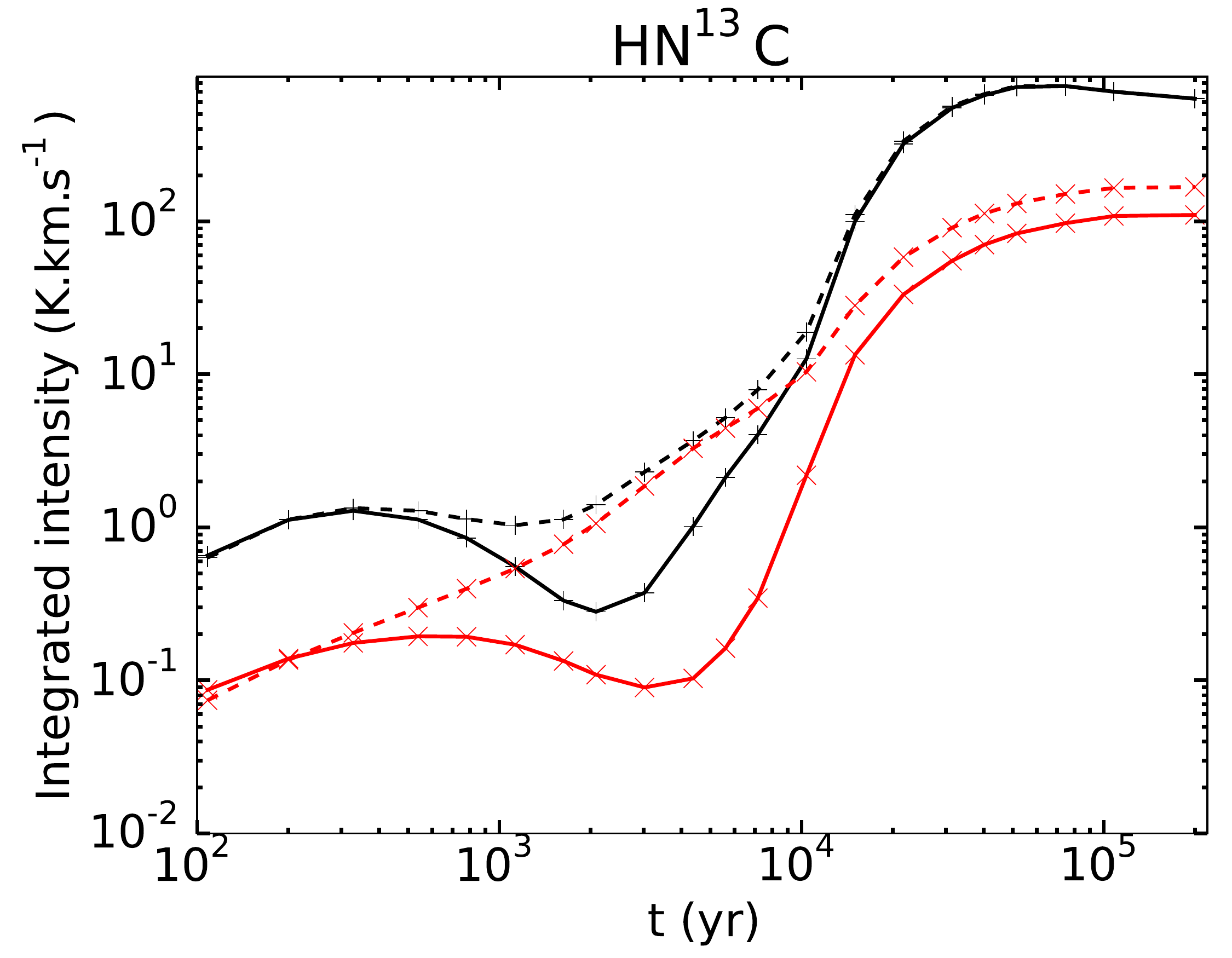}       
        \includegraphics[width=0.32\textwidth]{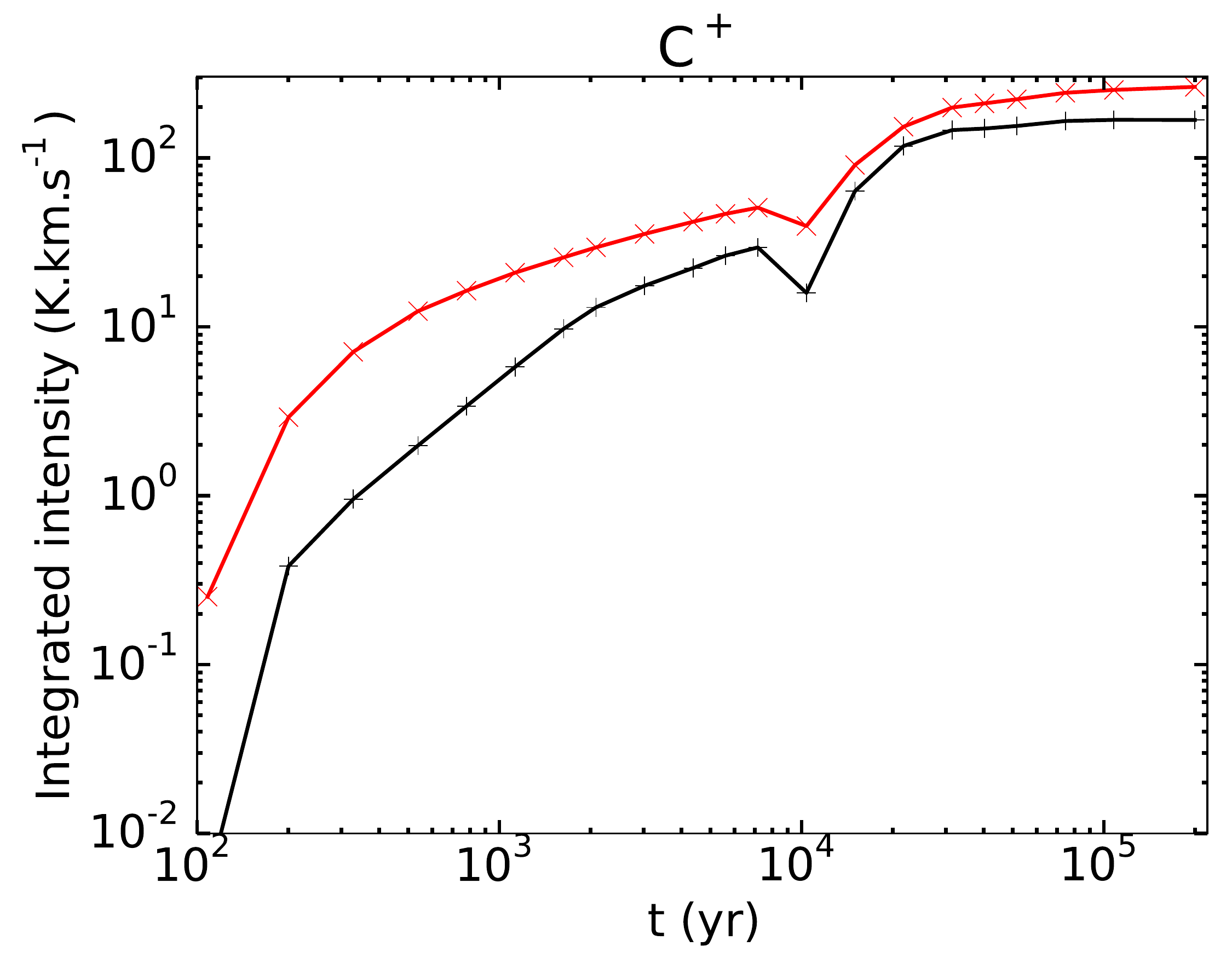} 

        \includegraphics[width=0.32\textwidth]{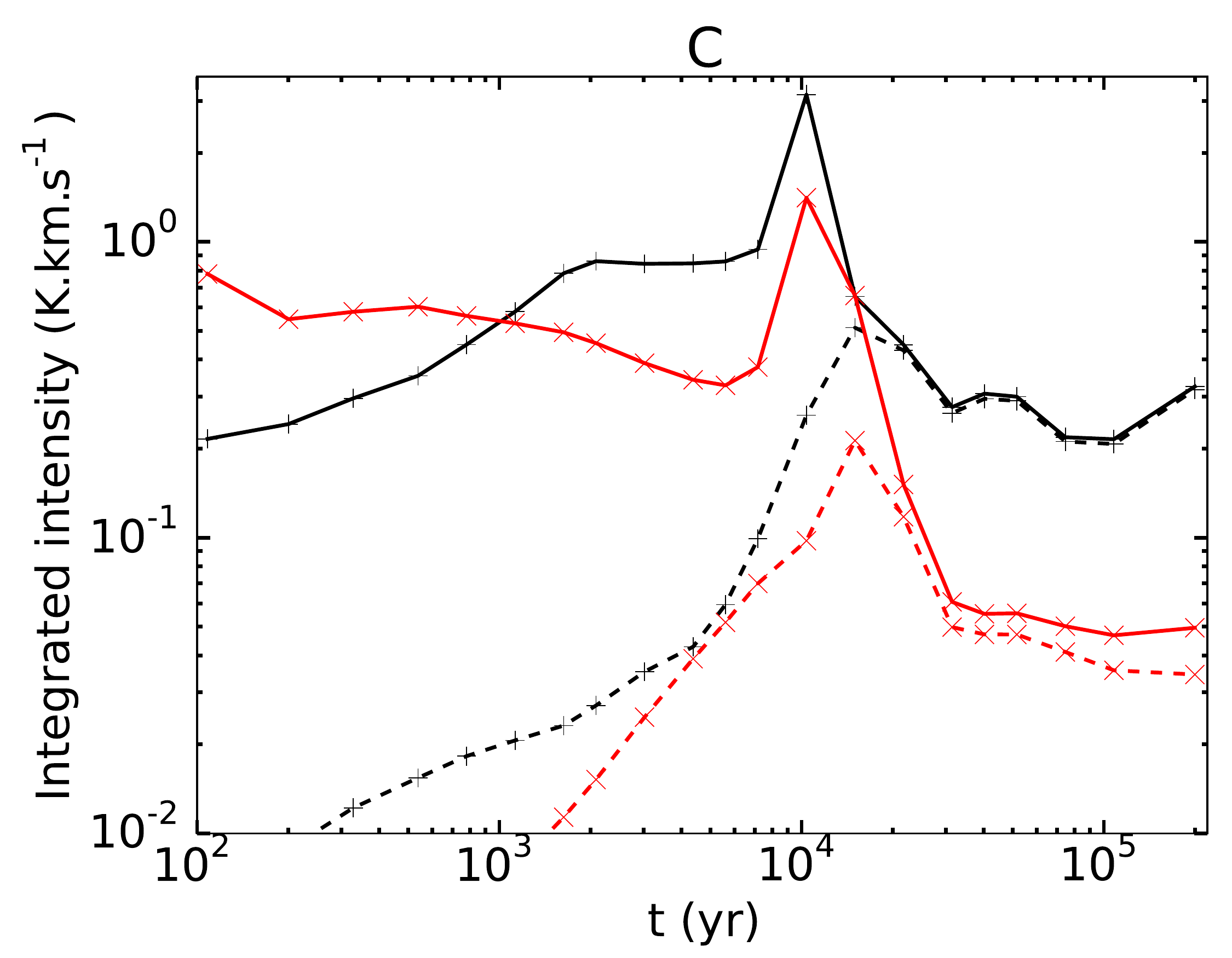}
        \includegraphics[width=0.32\textwidth]{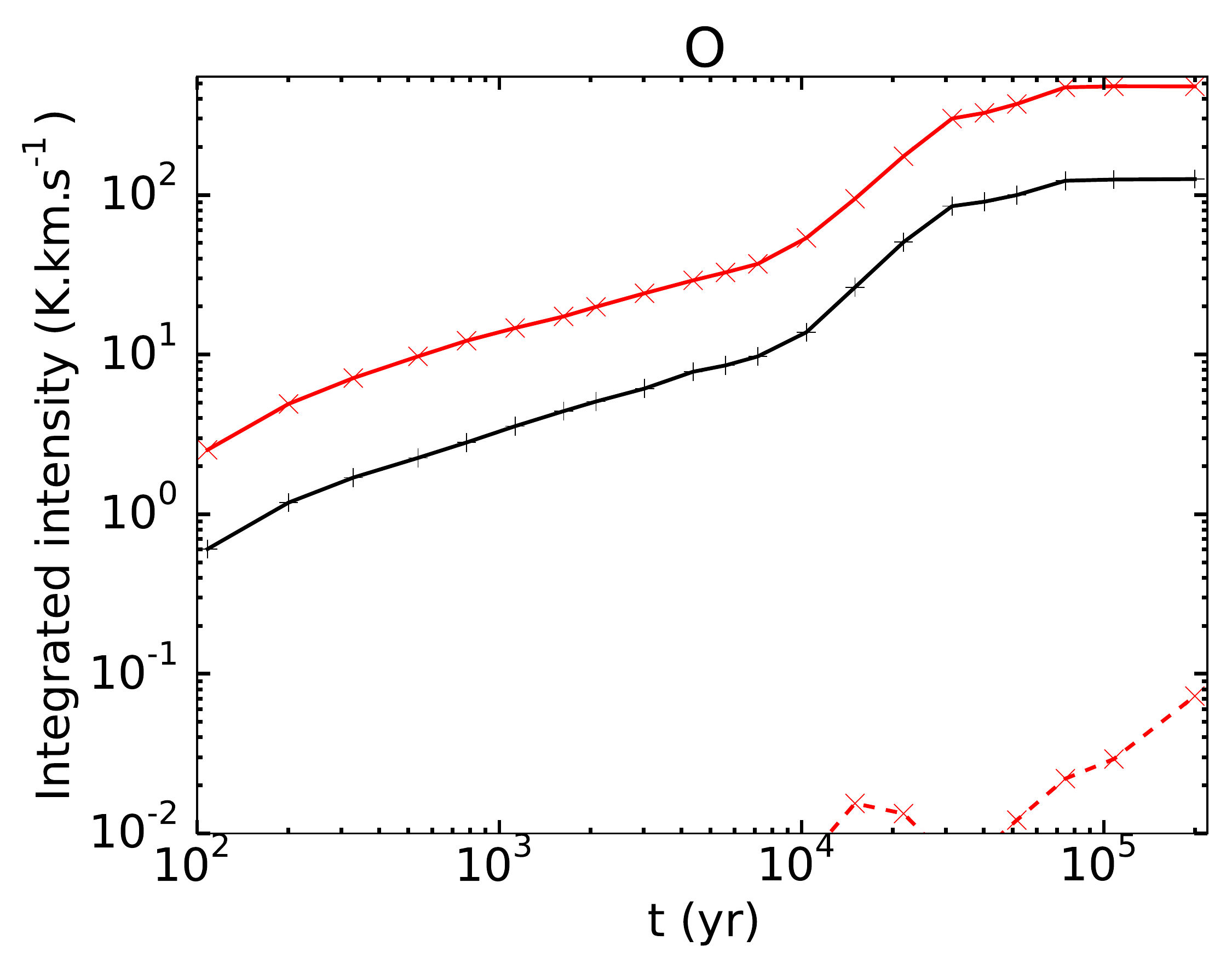}
        \includegraphics[width=0.32\textwidth]{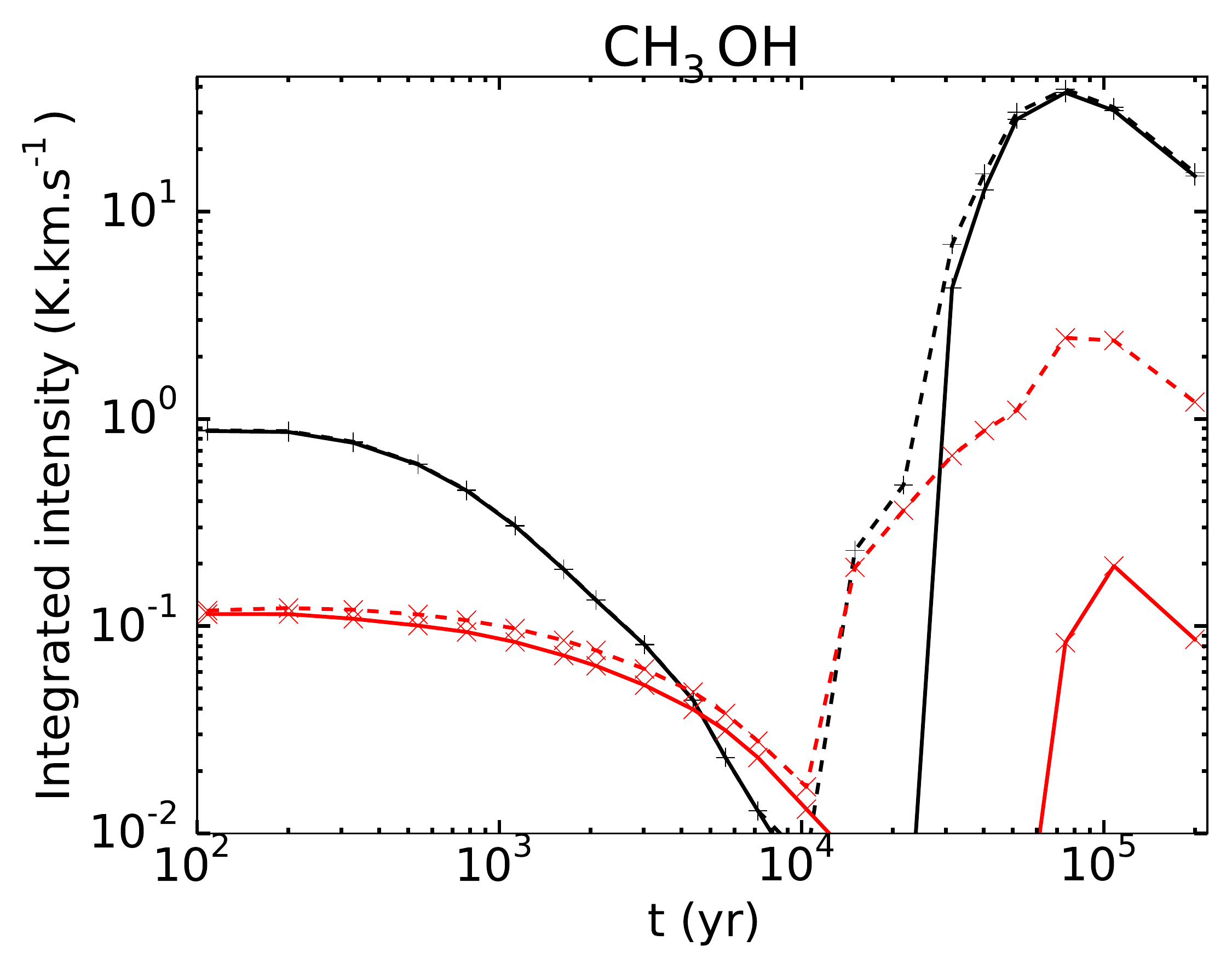} 

        \includegraphics[width=0.32\textwidth]{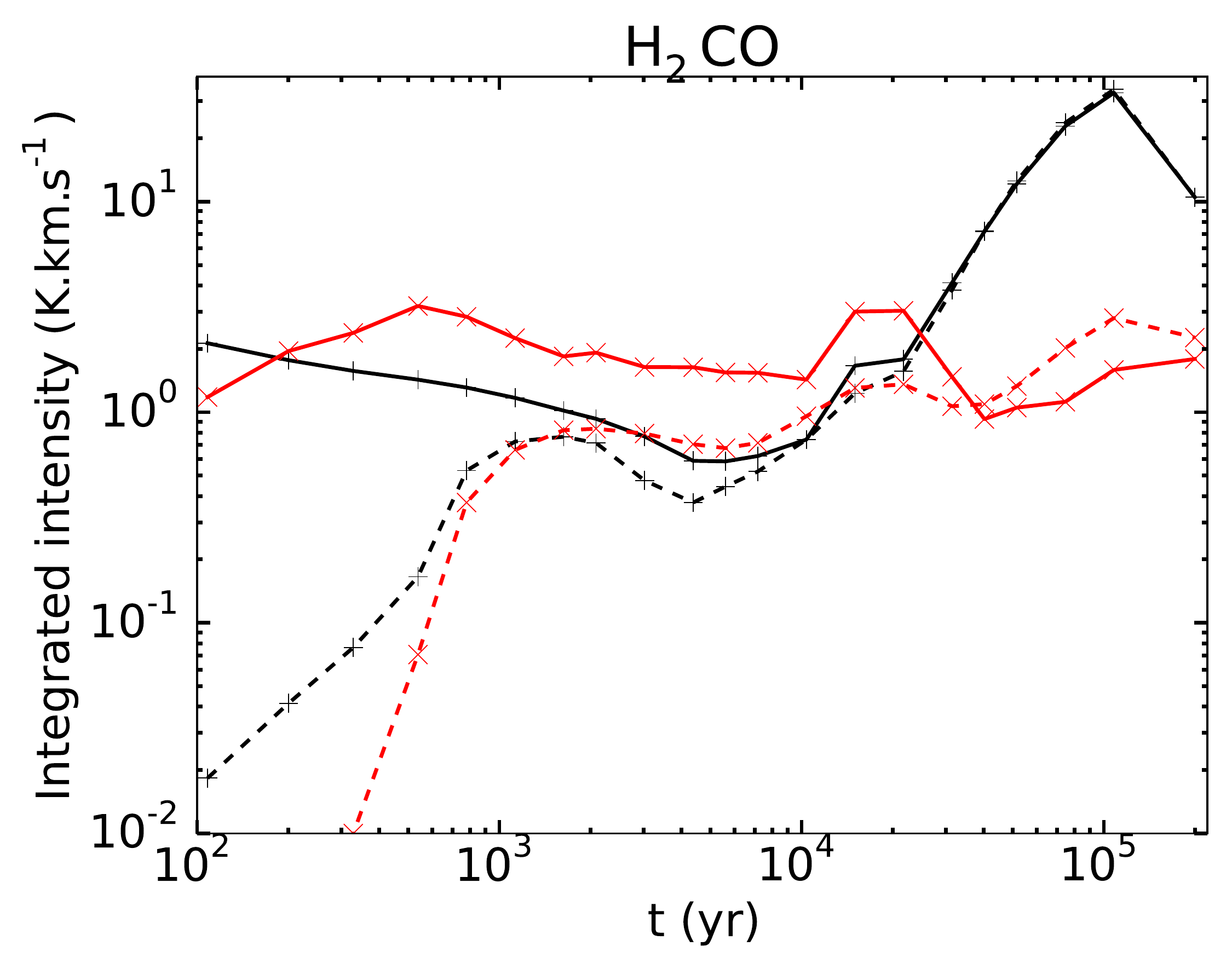}
        \includegraphics[width=0.32\textwidth]{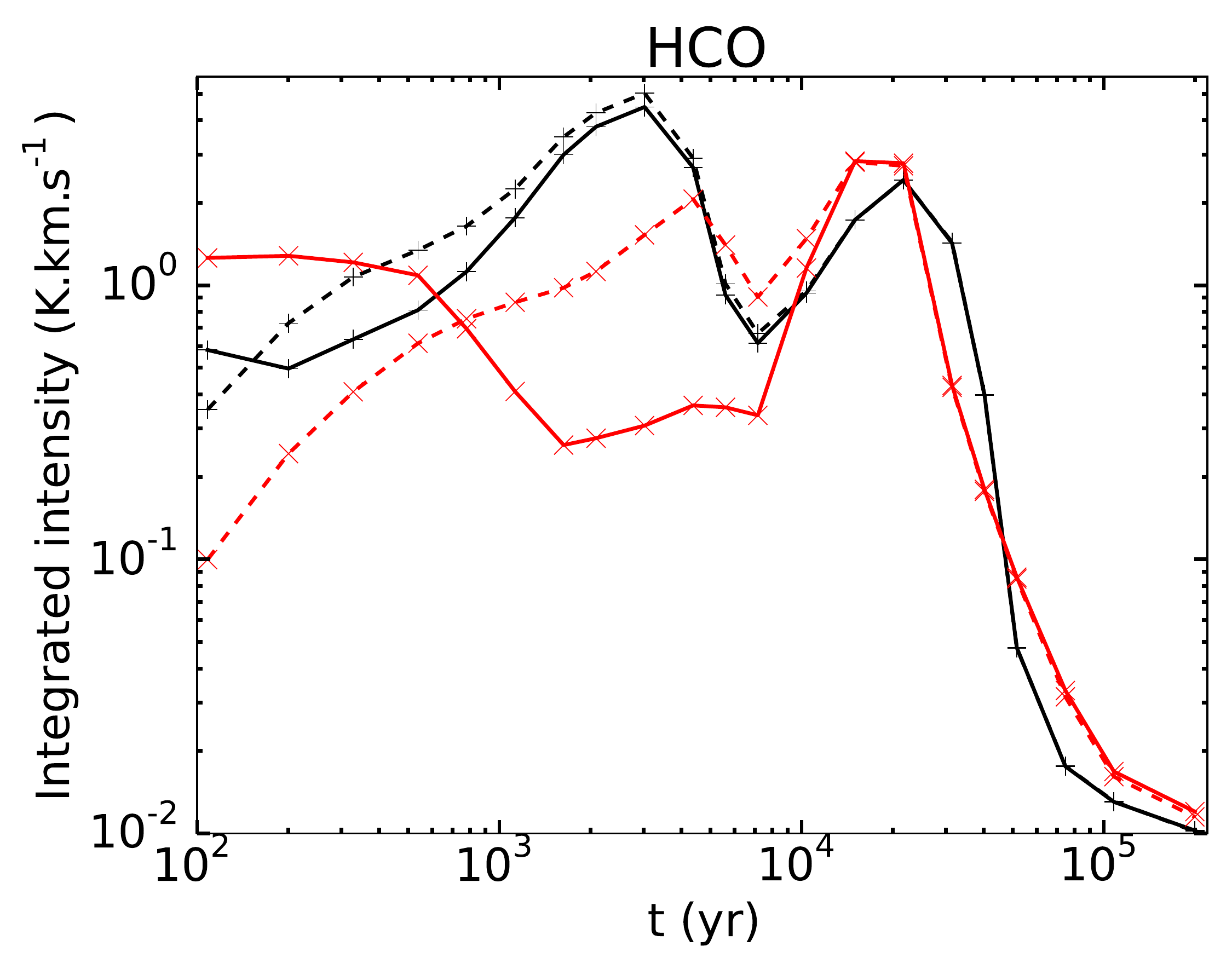}
        \includegraphics[width=0.32\textwidth]{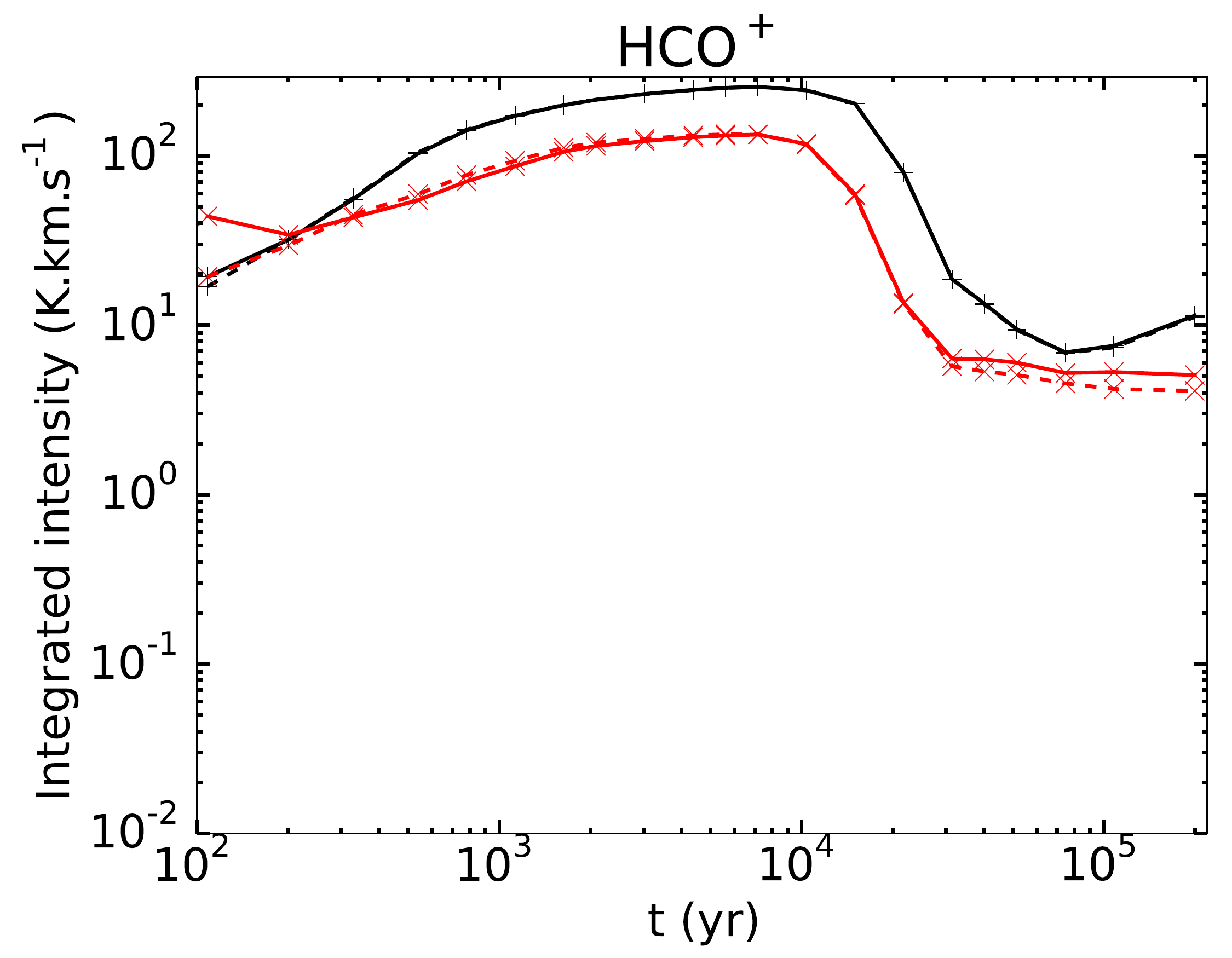} 

        \includegraphics[width=0.32\textwidth]{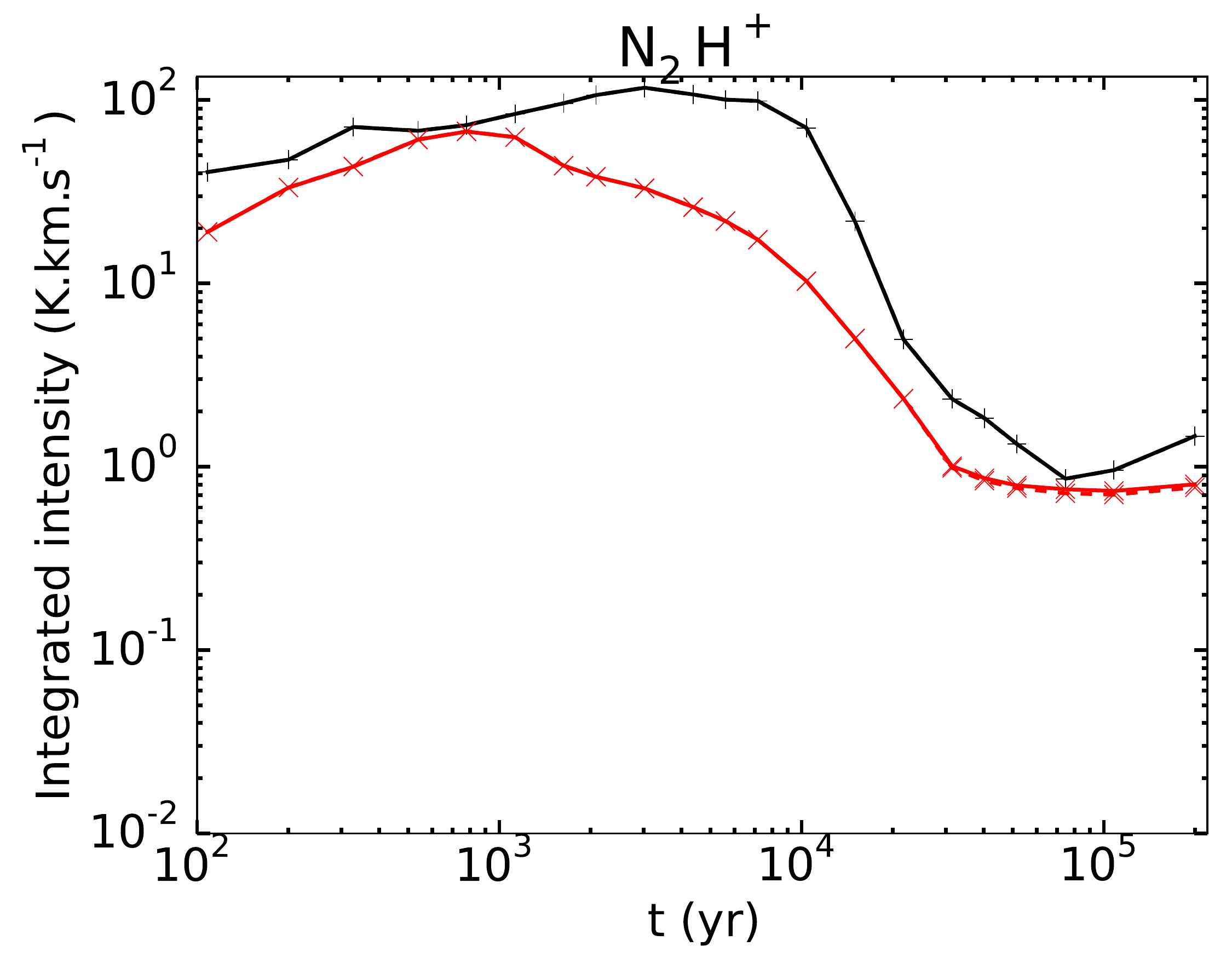}
        \includegraphics[width=0.32\textwidth]{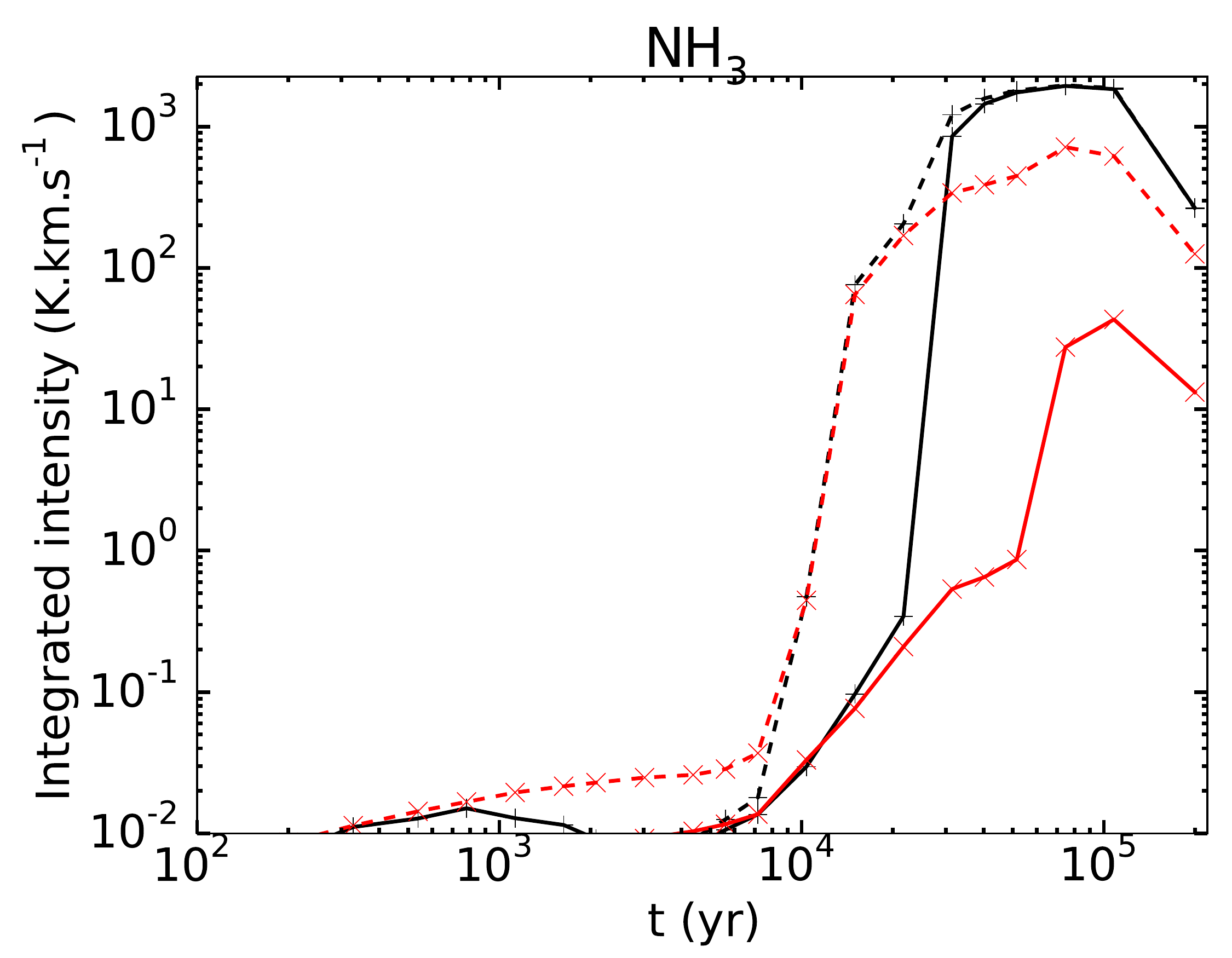}
        \includegraphics[width=0.32\textwidth]{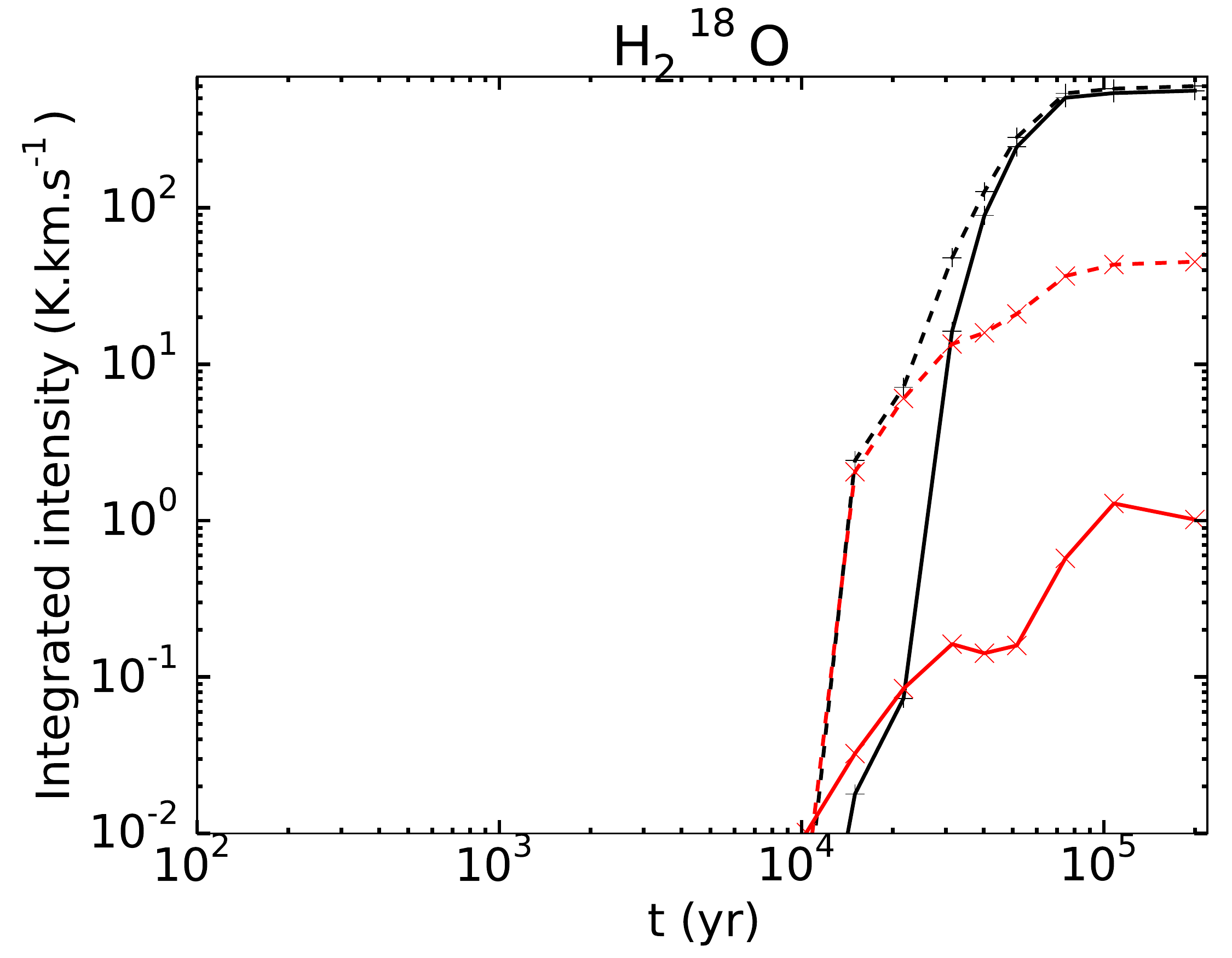} 

        \includegraphics[width=0.32\textwidth]{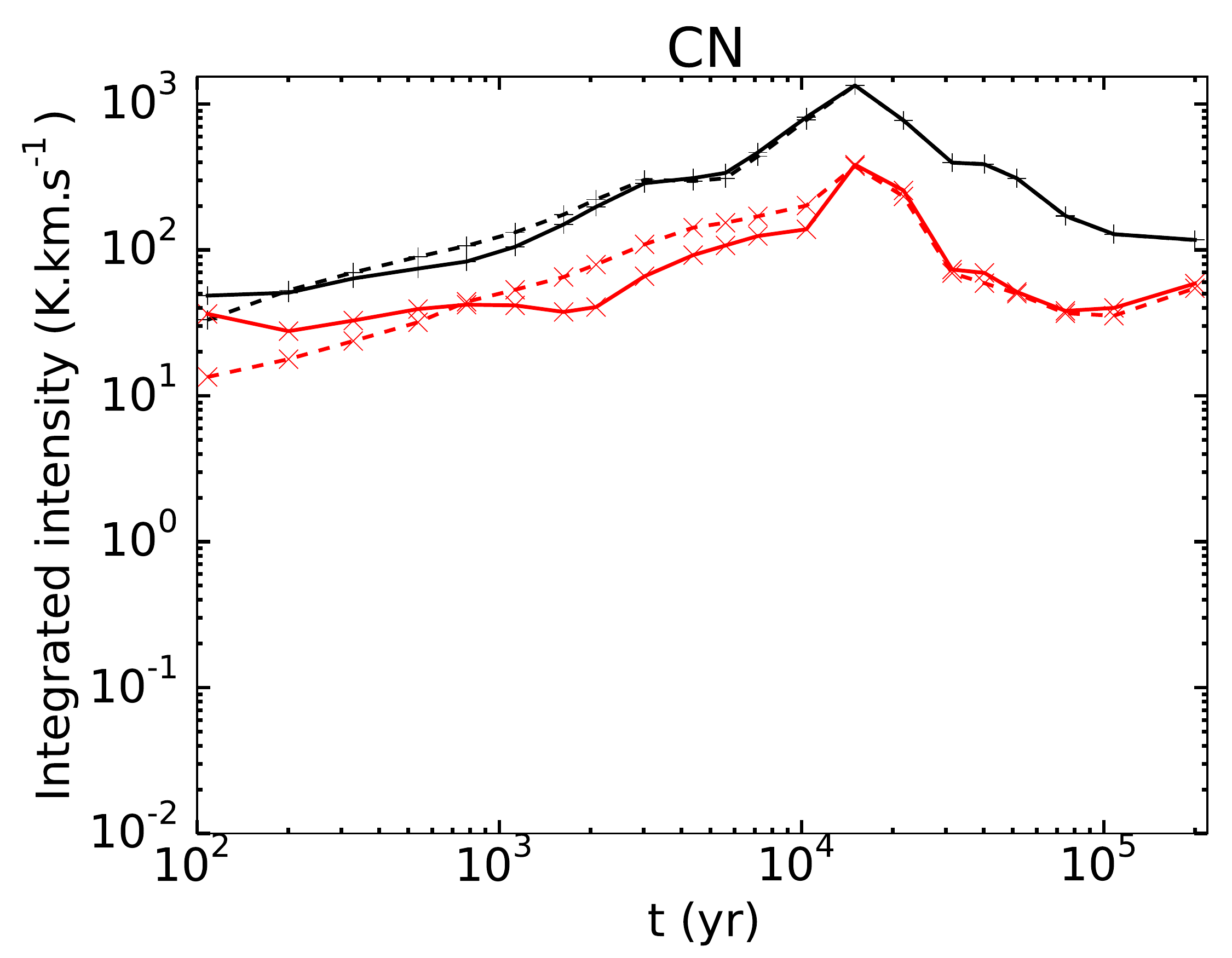} 
        \caption{{\bf Change of density at the ionization front \textit{n7} and \textit{n6}:} Time evolution of the integrated intensities of the selected species listed in Table~\ref{tab:selected-molecules}. Model \textit{mHII} is represented with solid line and model \textit{mHHMC} with dashed line. Model \textit{n7} is represented in black and model \textit{n6} in red.}
        \label{apfig:intInt_comp_HIIdens}
\end{figure*}

\begin{figure*}[htbp]
        \includegraphics[width=0.32\textwidth]{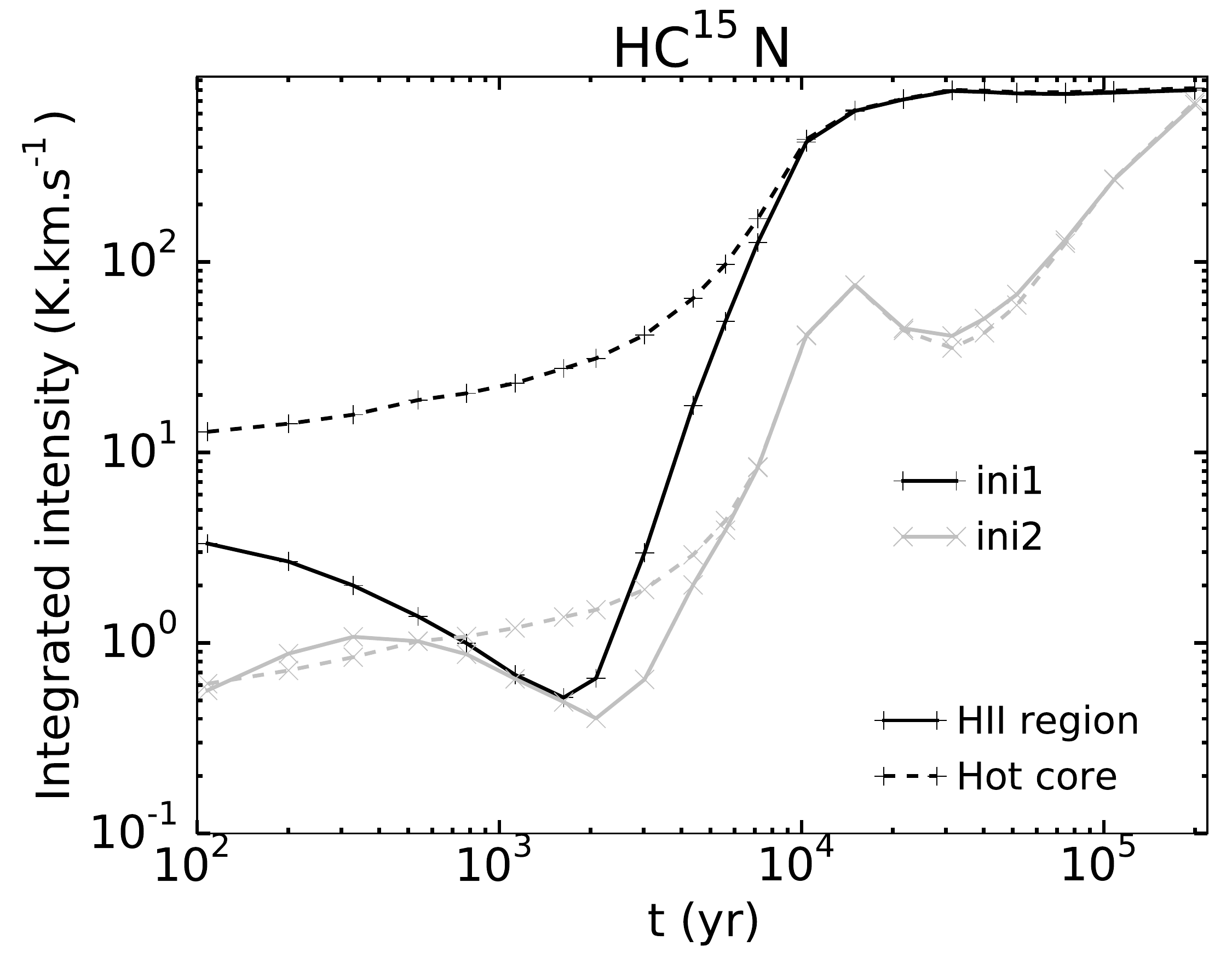}
        \includegraphics[width=0.32\textwidth]{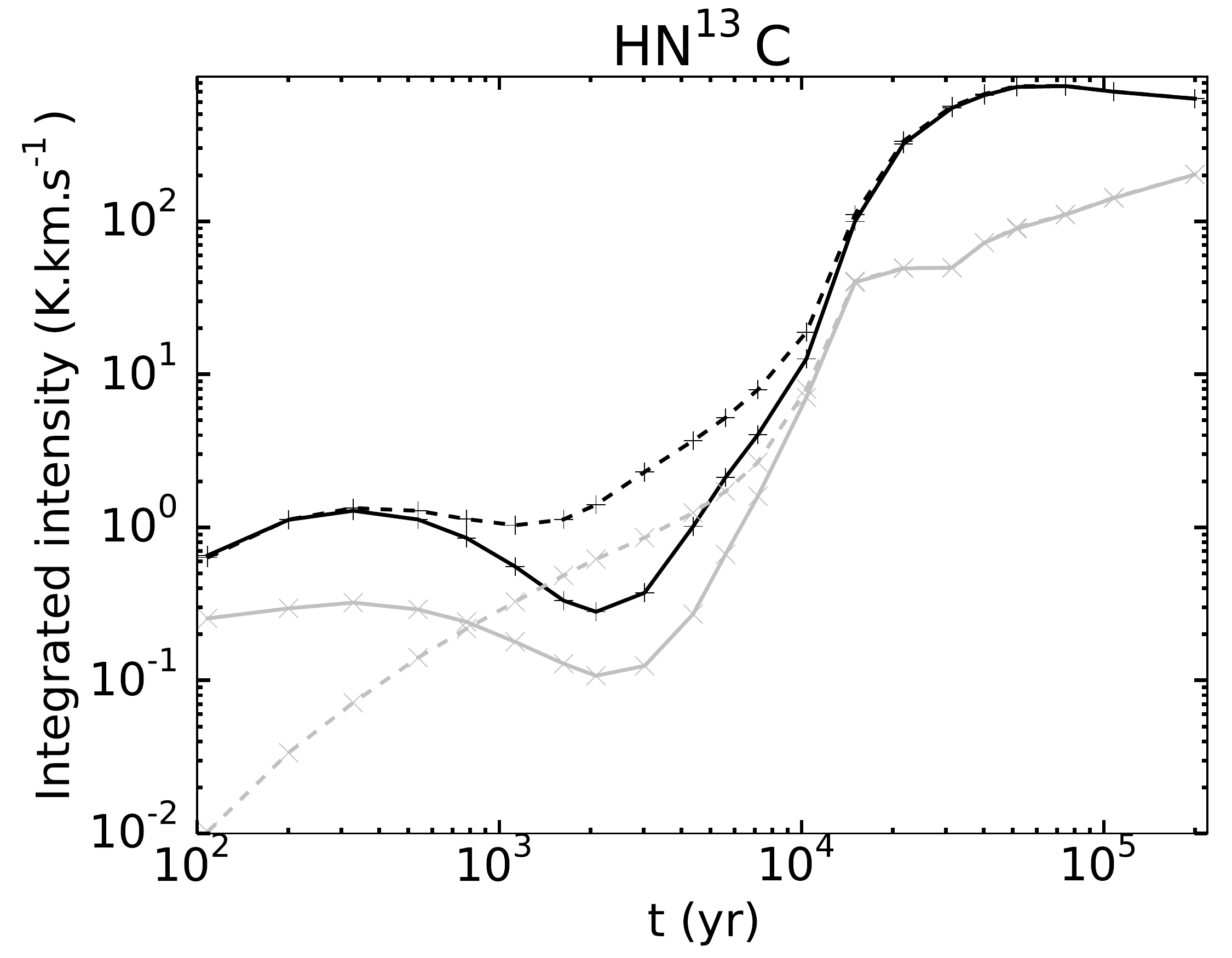}   
        \includegraphics[width=0.32\textwidth]{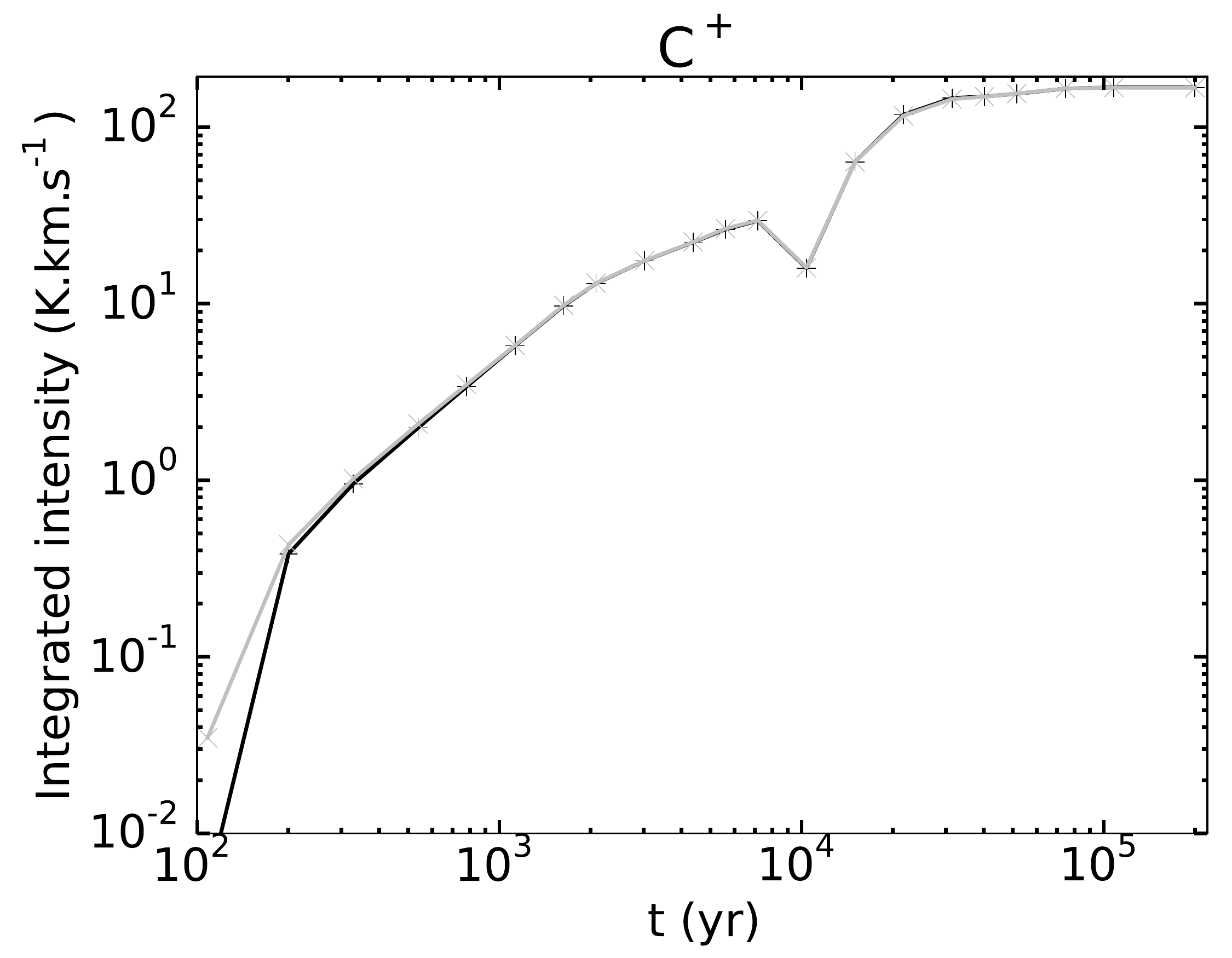} 

        \includegraphics[width=0.32\textwidth]{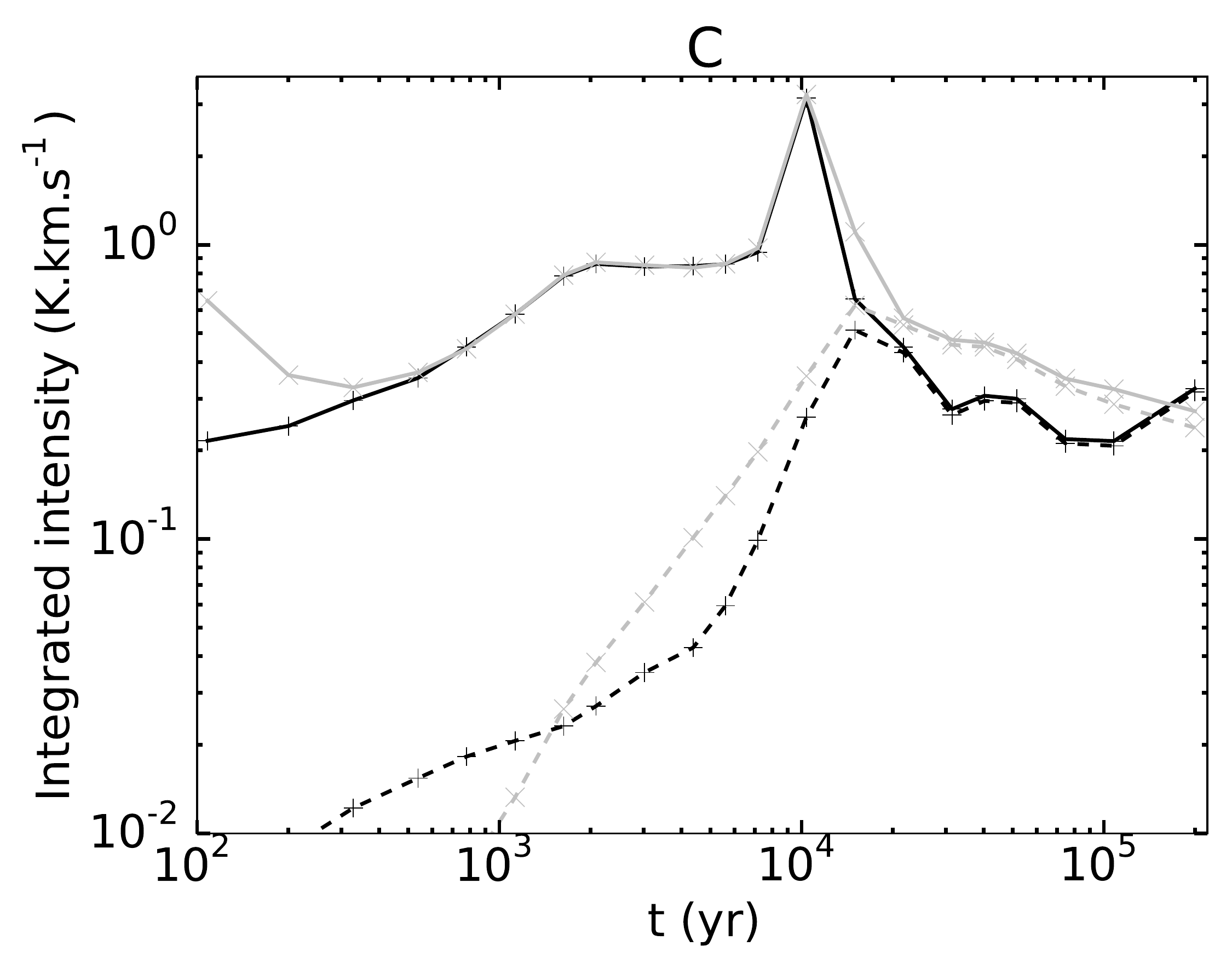}
        \includegraphics[width=0.32\textwidth]{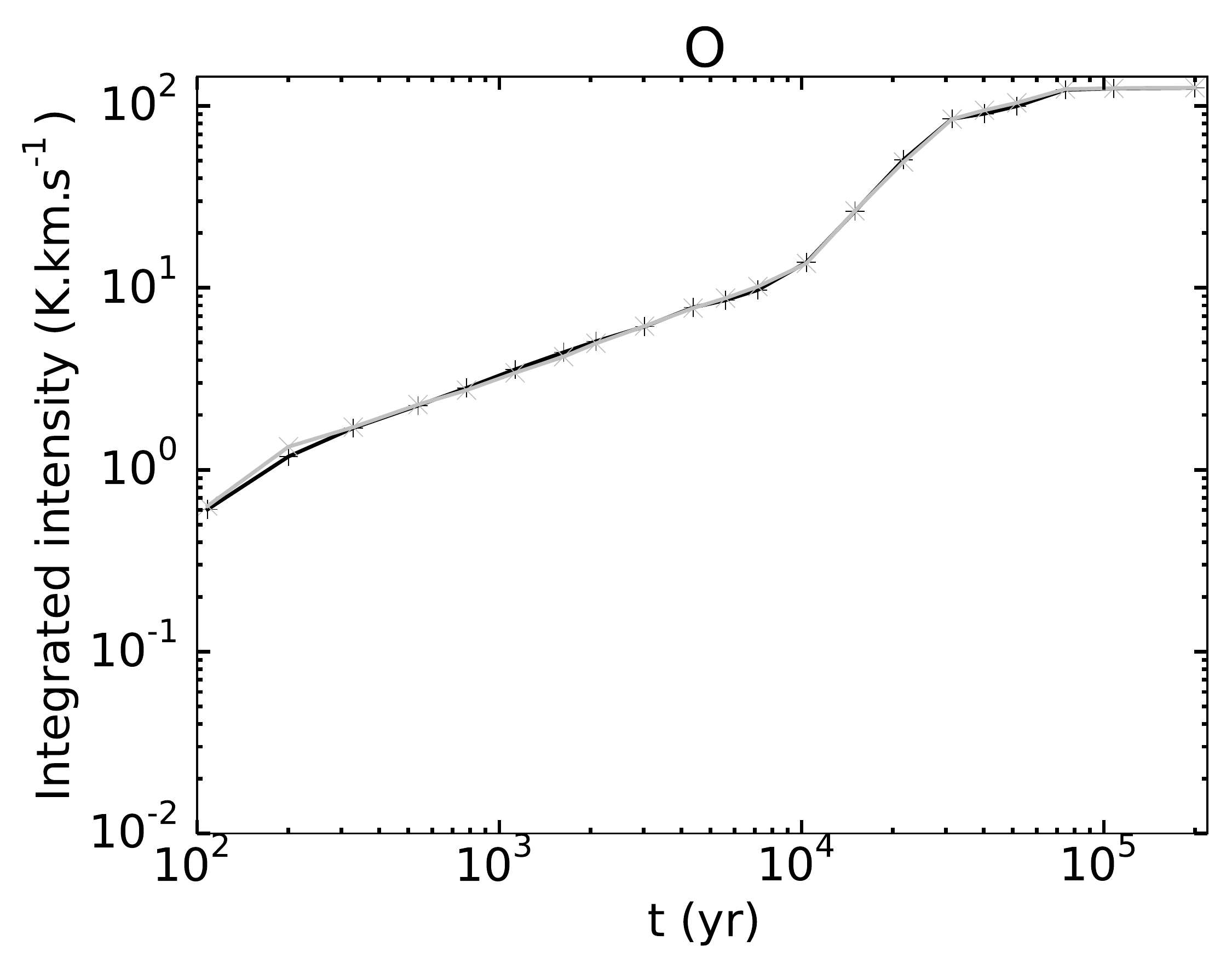}
        \includegraphics[width=0.32\textwidth]{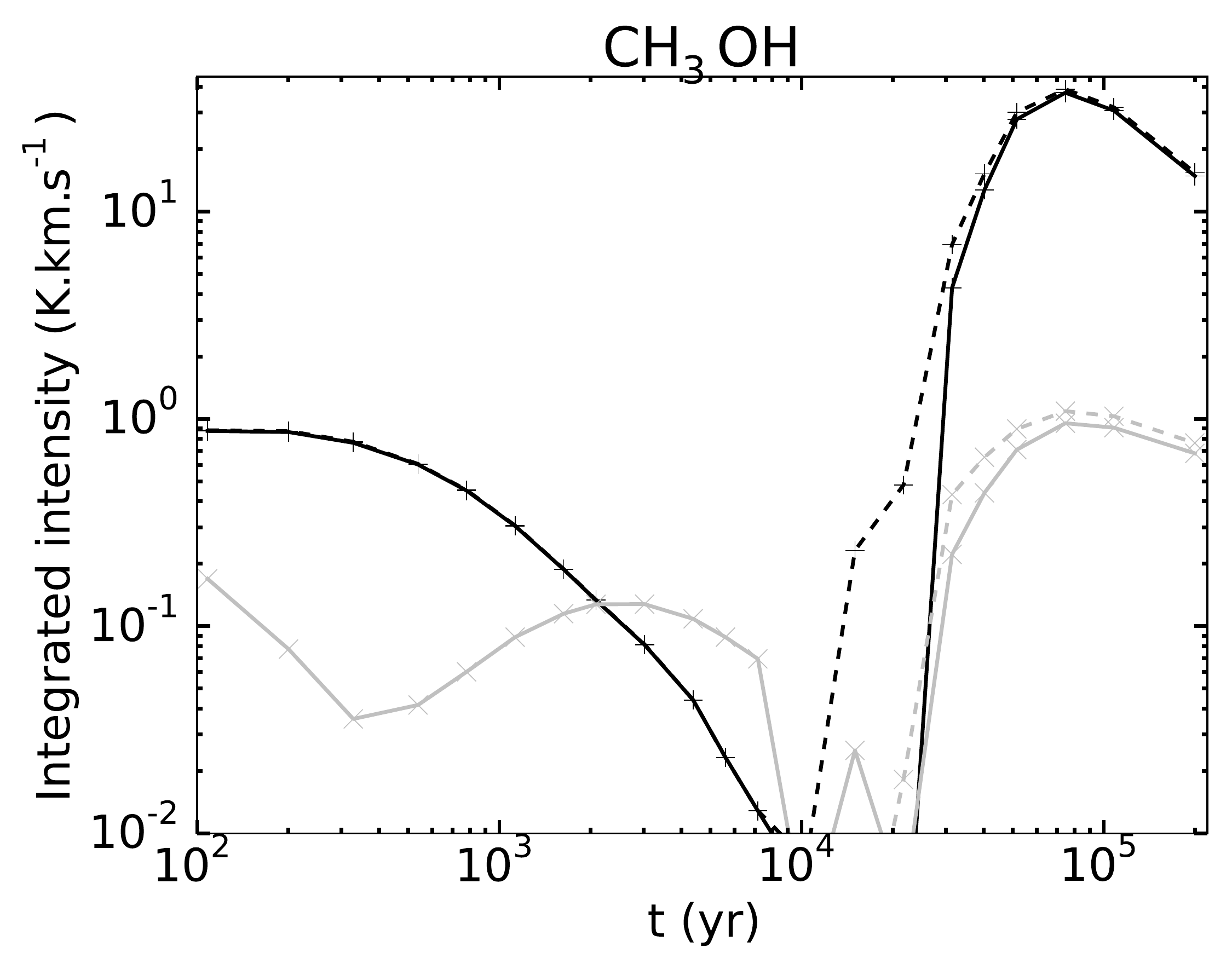} 

        \includegraphics[width=0.32\textwidth]{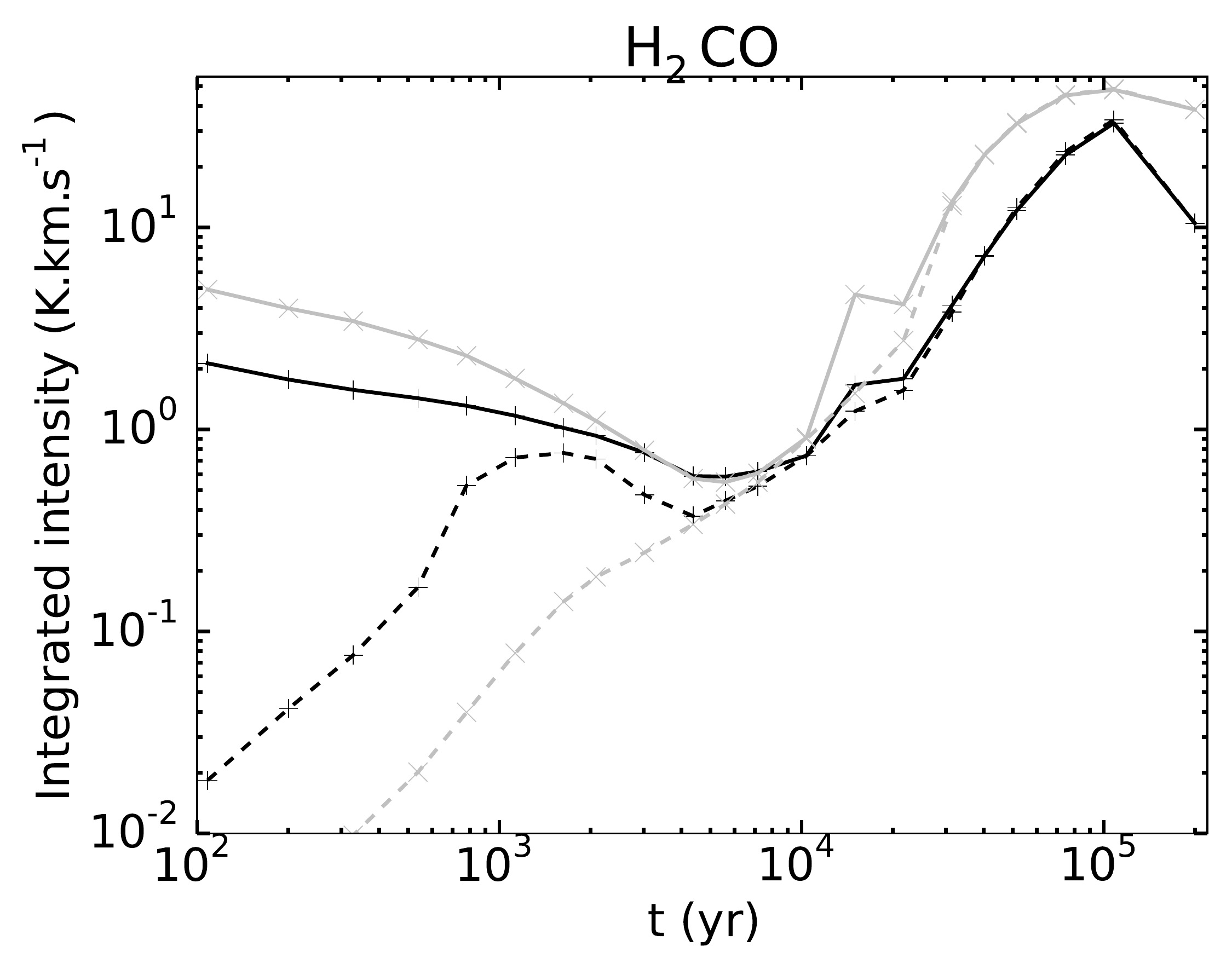}
        \includegraphics[width=0.32\textwidth]{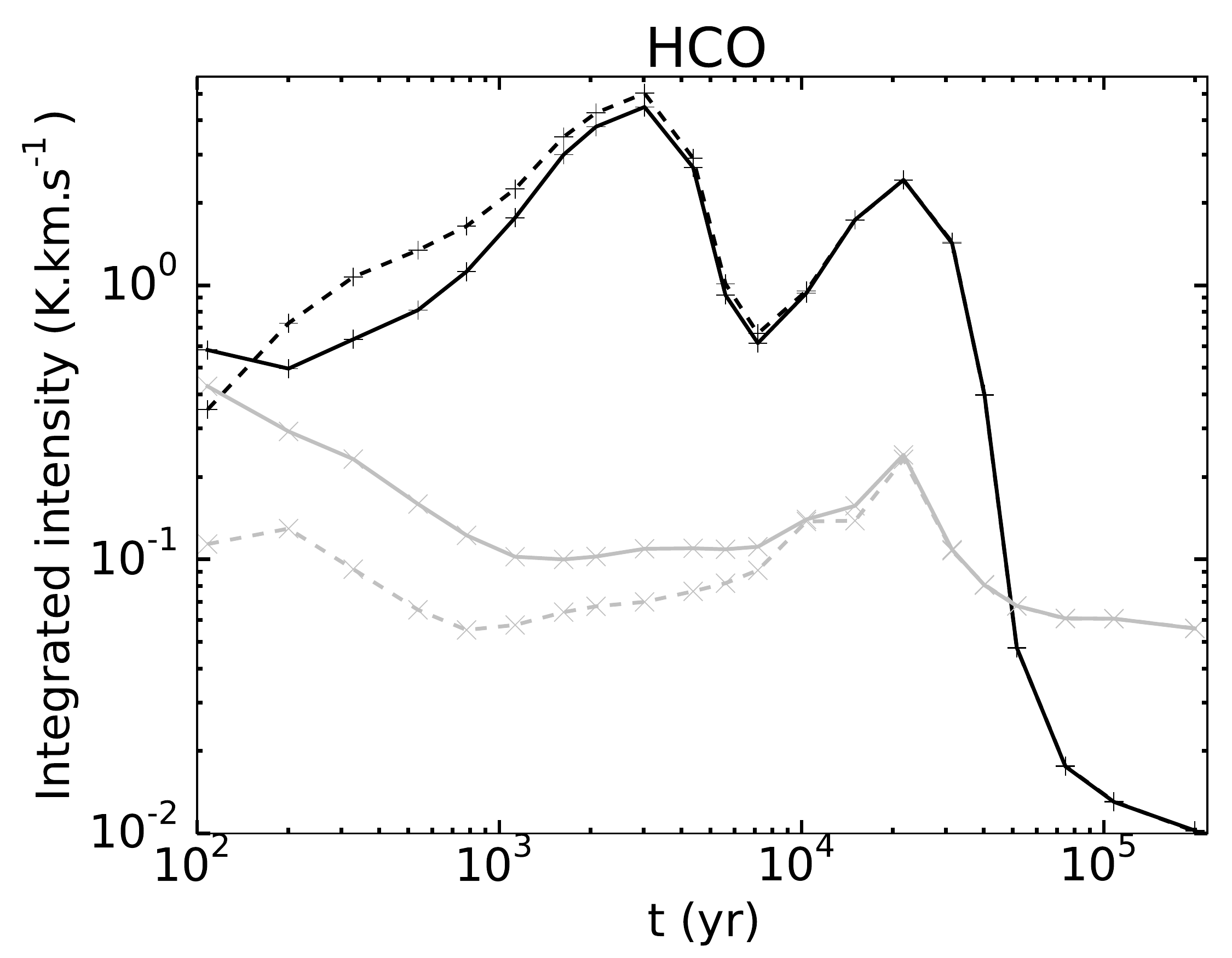}
        \includegraphics[width=0.32\textwidth]{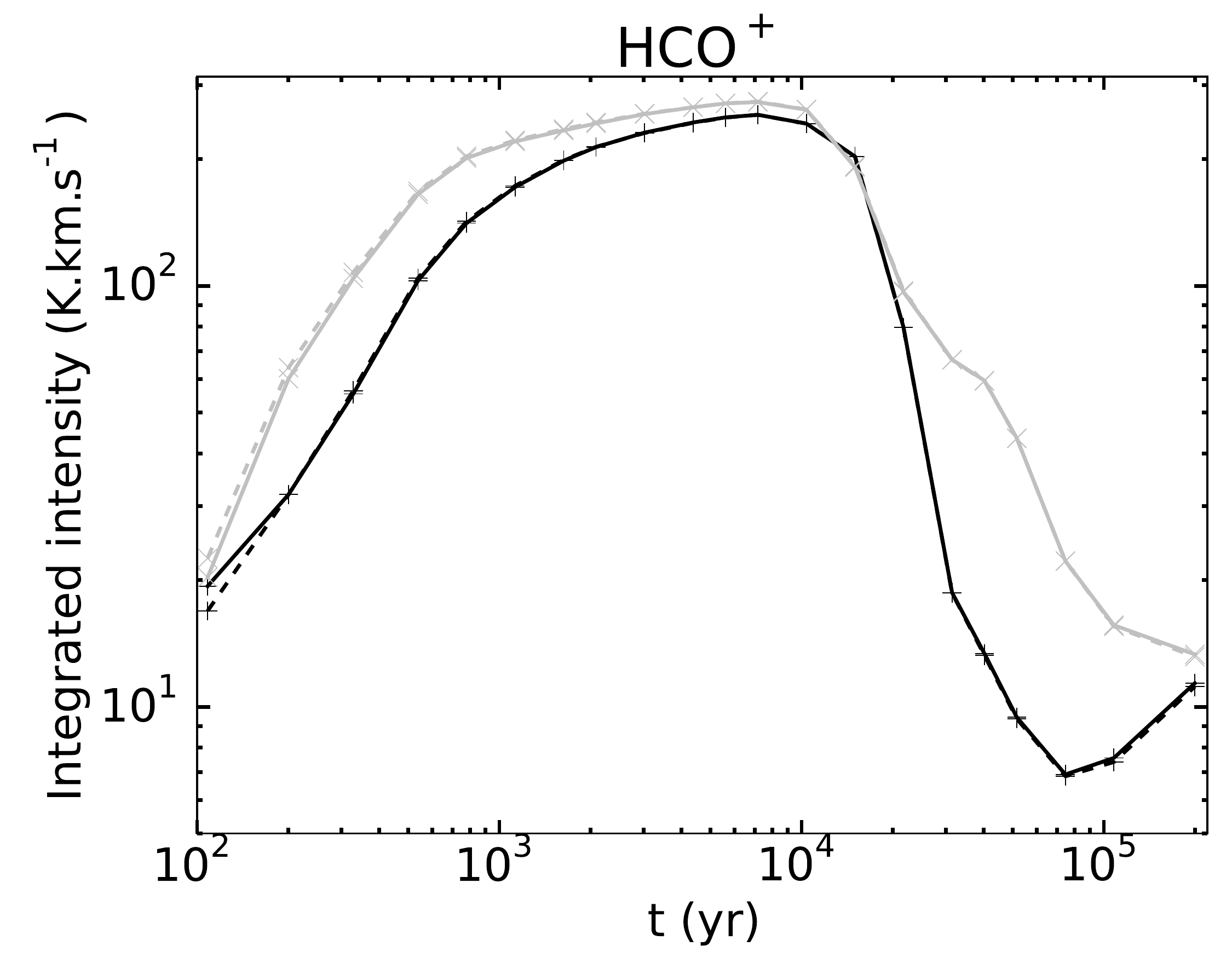} 

        \includegraphics[width=0.32\textwidth]{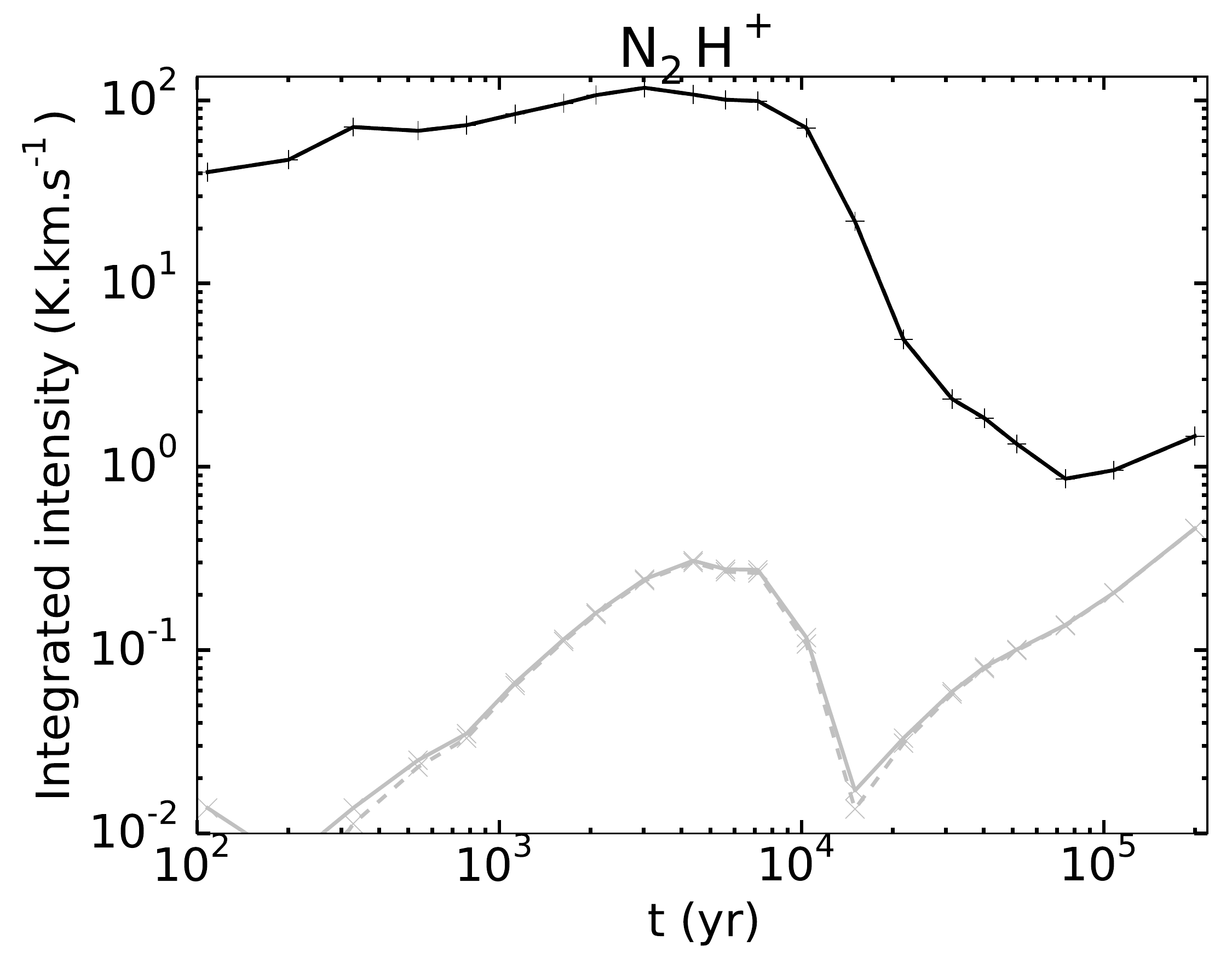}
        \includegraphics[width=0.32\textwidth]{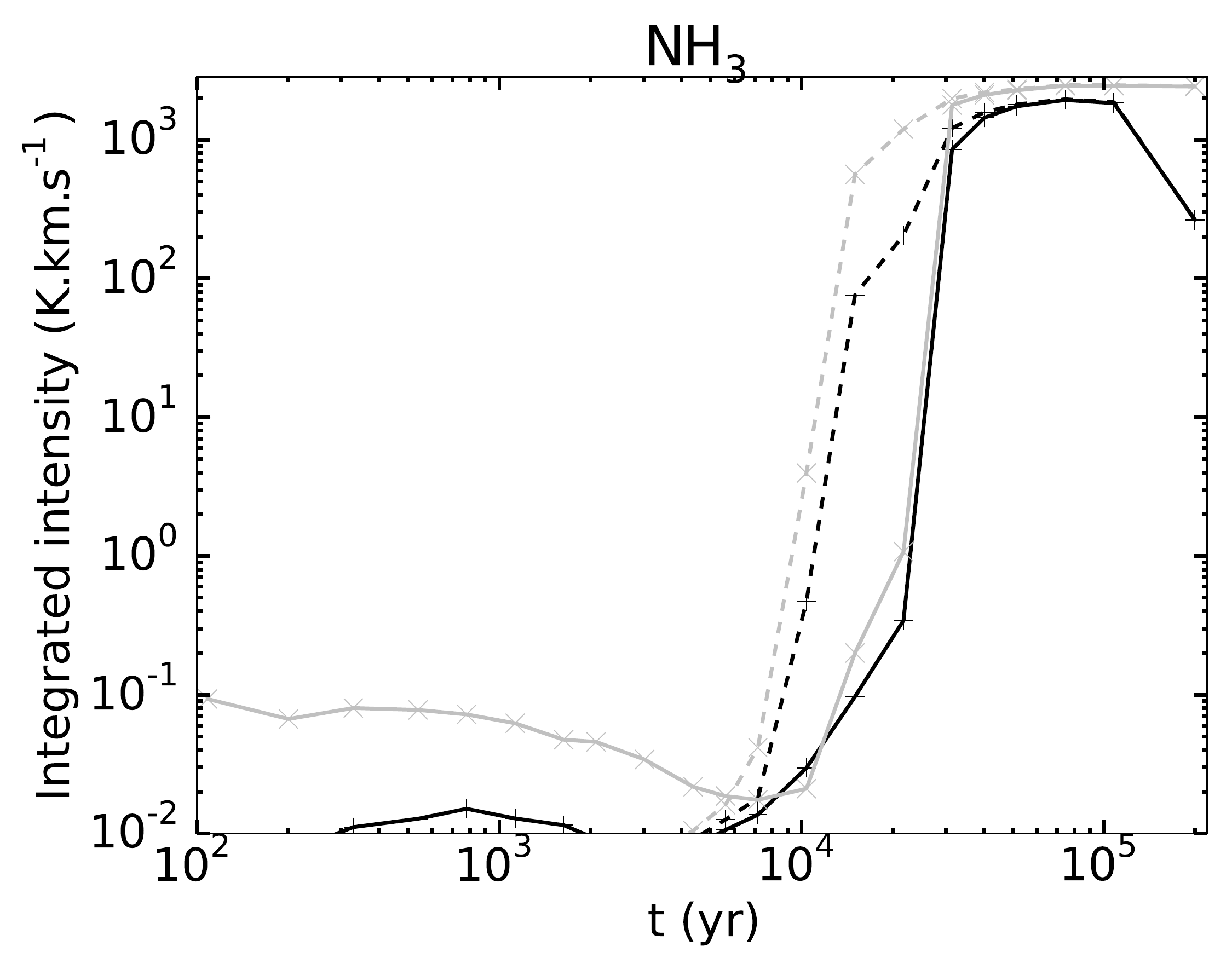}
        \includegraphics[width=0.32\textwidth]{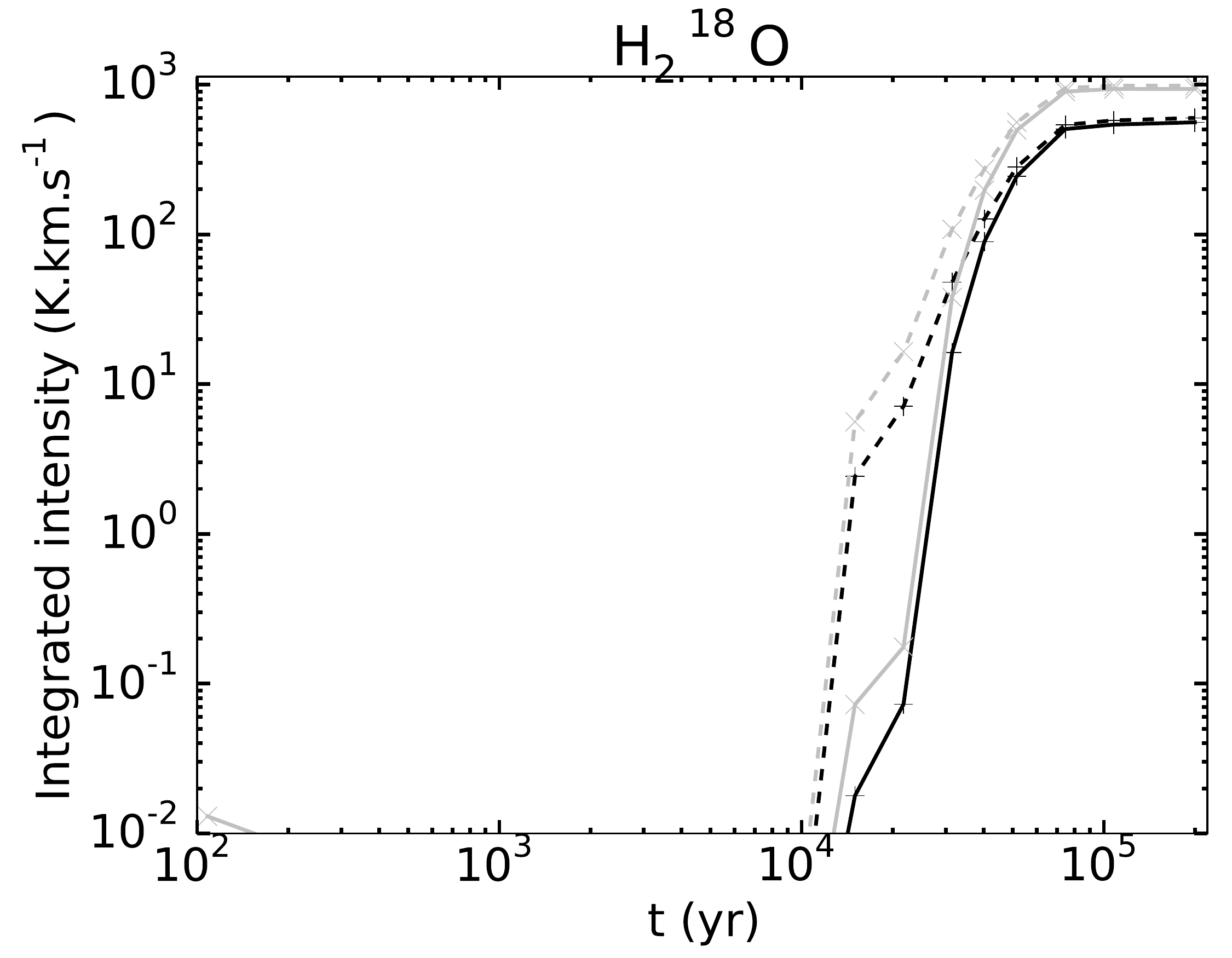} 

        \includegraphics[width=0.32\textwidth]{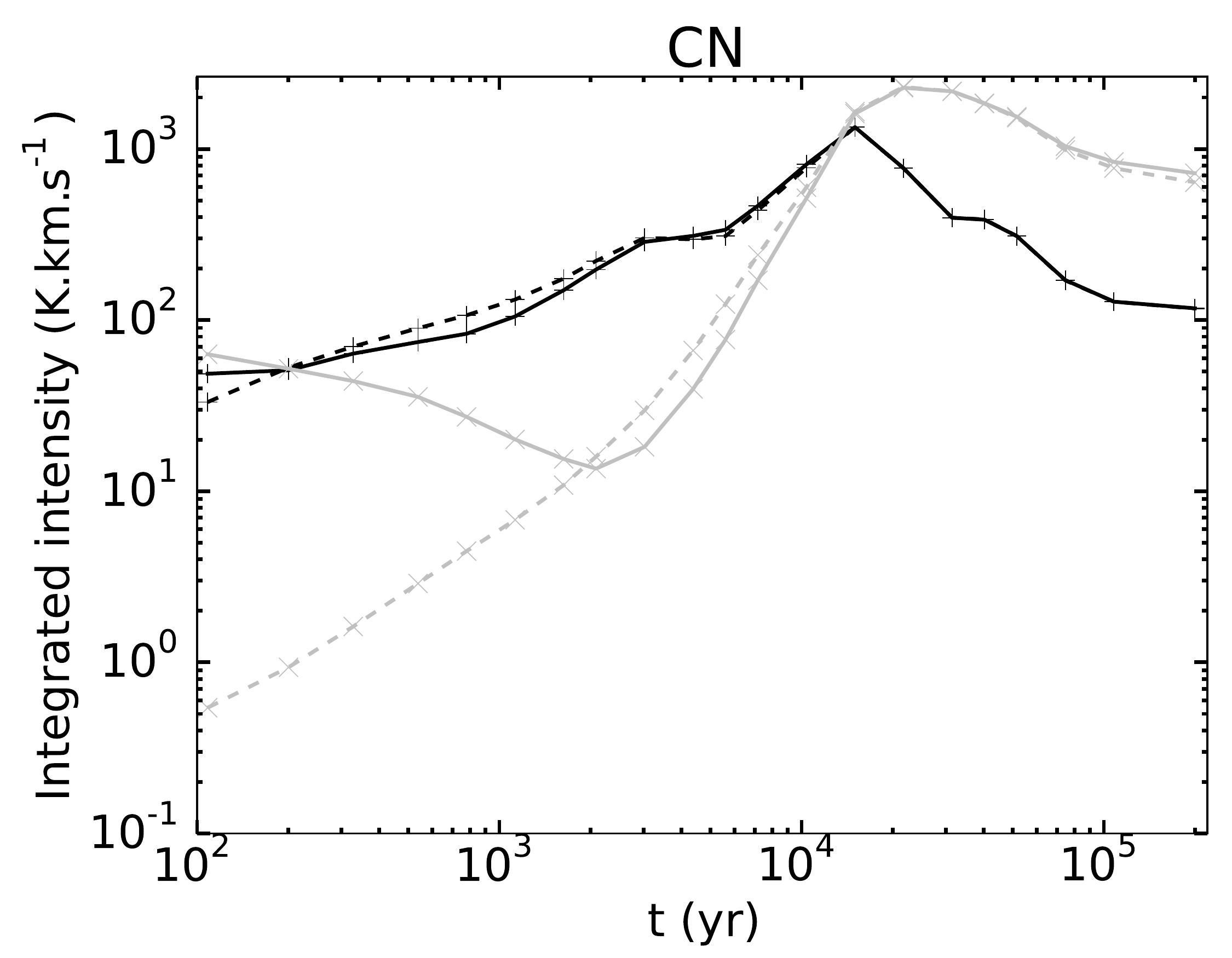} 
        \caption{{\bf Change of the initial abundances \textit{ini1} and \textit{ini2}:} Time evolution of the integrated intensities of the selected species listed in Table~\ref{tab:selected-molecules}. Model \textit{mHII} is represented with solid line and model \textit{mHHMC} with dashed line. Model \textit{ini1} is represented in black and model \textit{ini2} in gray.}
        \label{apfig:intInt_comp_HIIini}
\end{figure*}

\begin{figure*}[htbp]
        \includegraphics[width=0.32\textwidth]{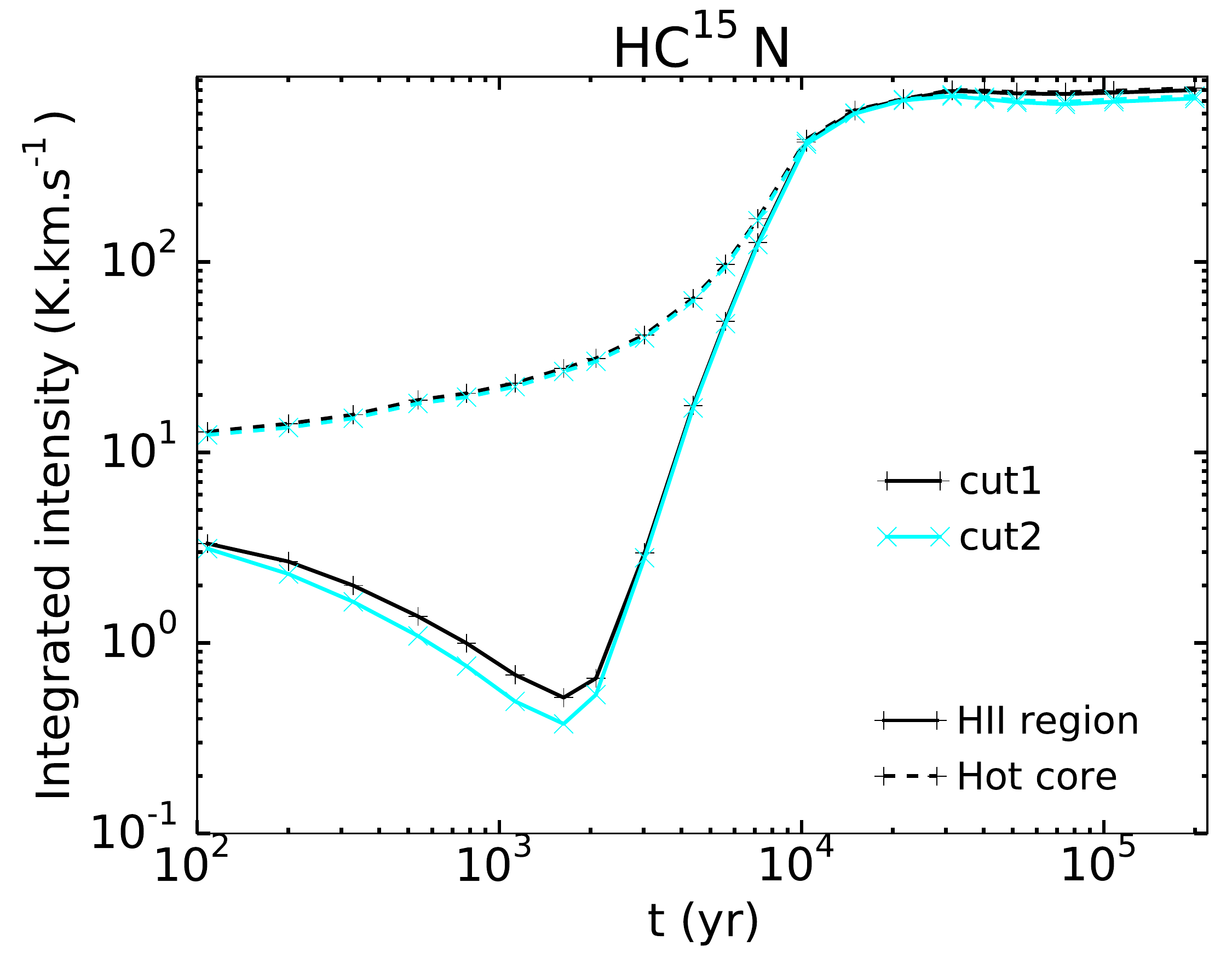}
        \includegraphics[width=0.32\textwidth]{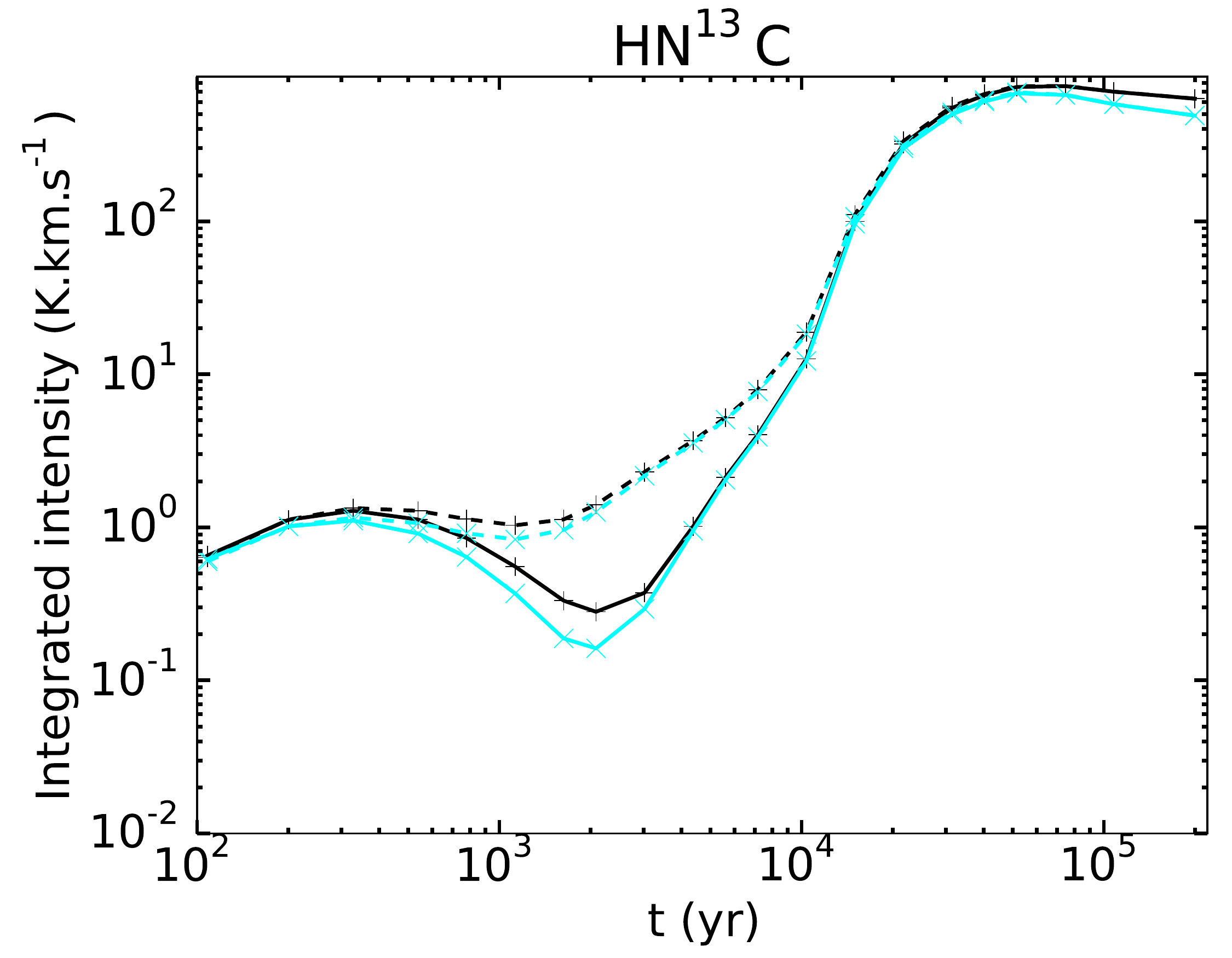}   
        \includegraphics[width=0.32\textwidth]{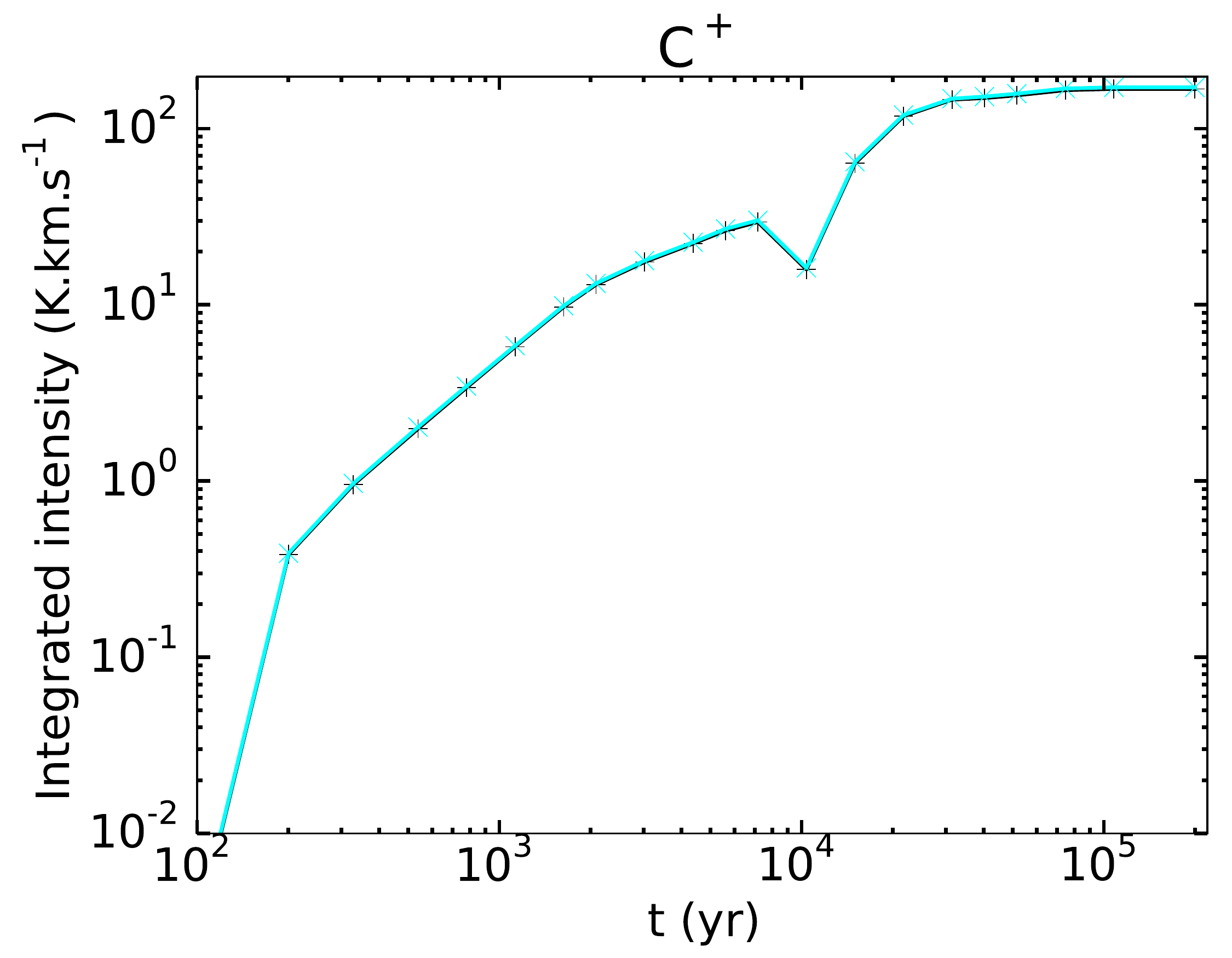} 

        \includegraphics[width=0.32\textwidth]{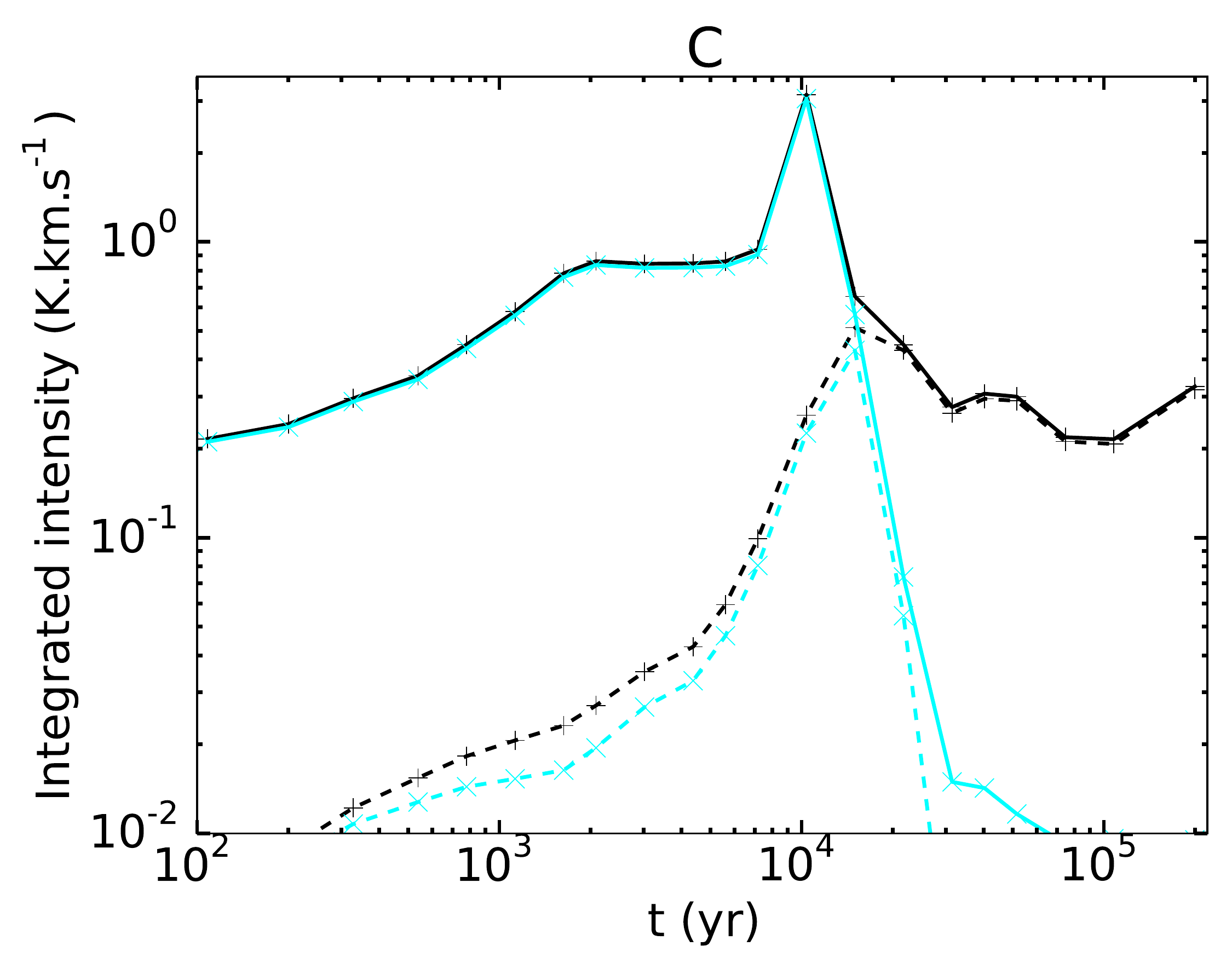}
        \includegraphics[width=0.32\textwidth]{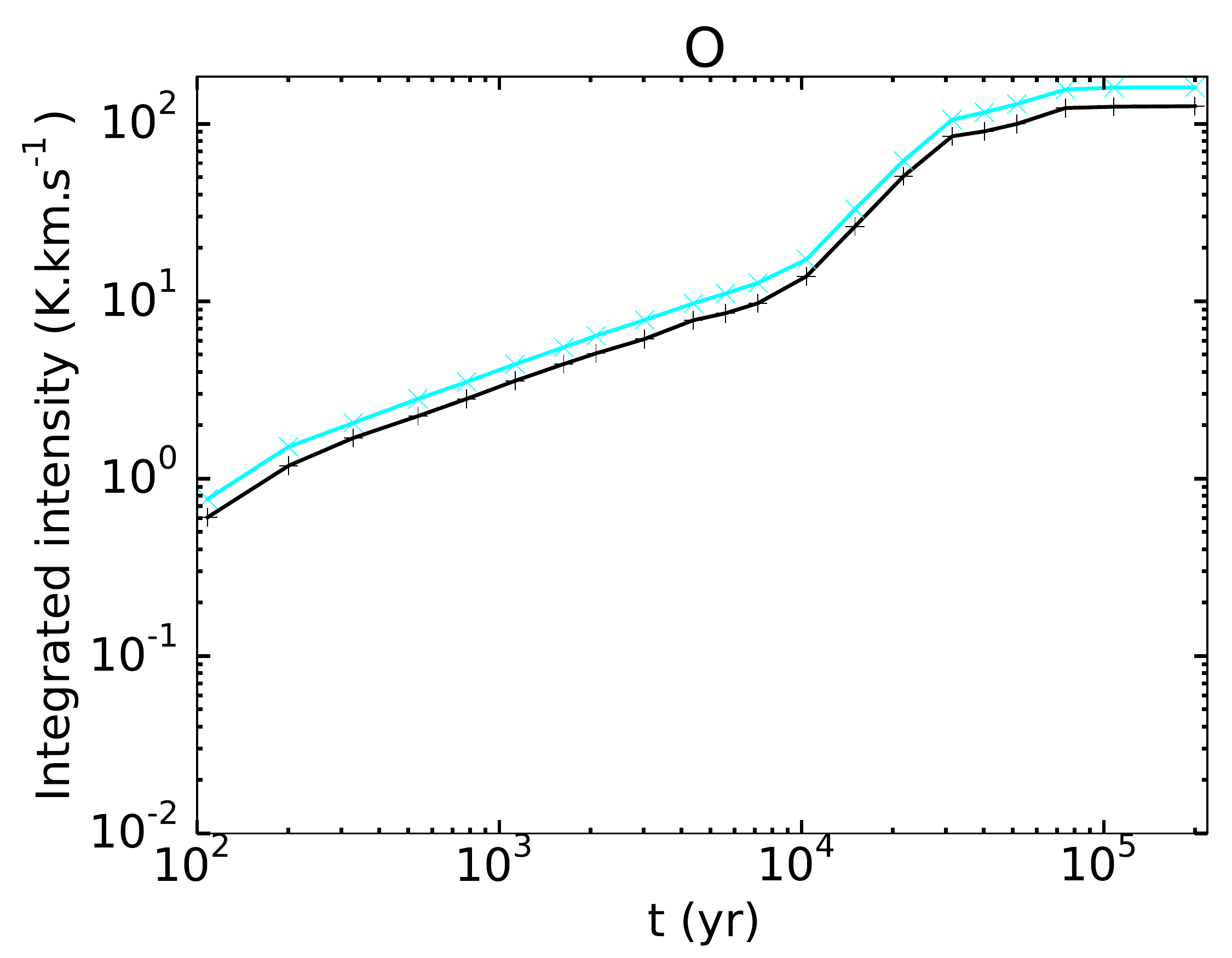}
        \includegraphics[width=0.32\textwidth]{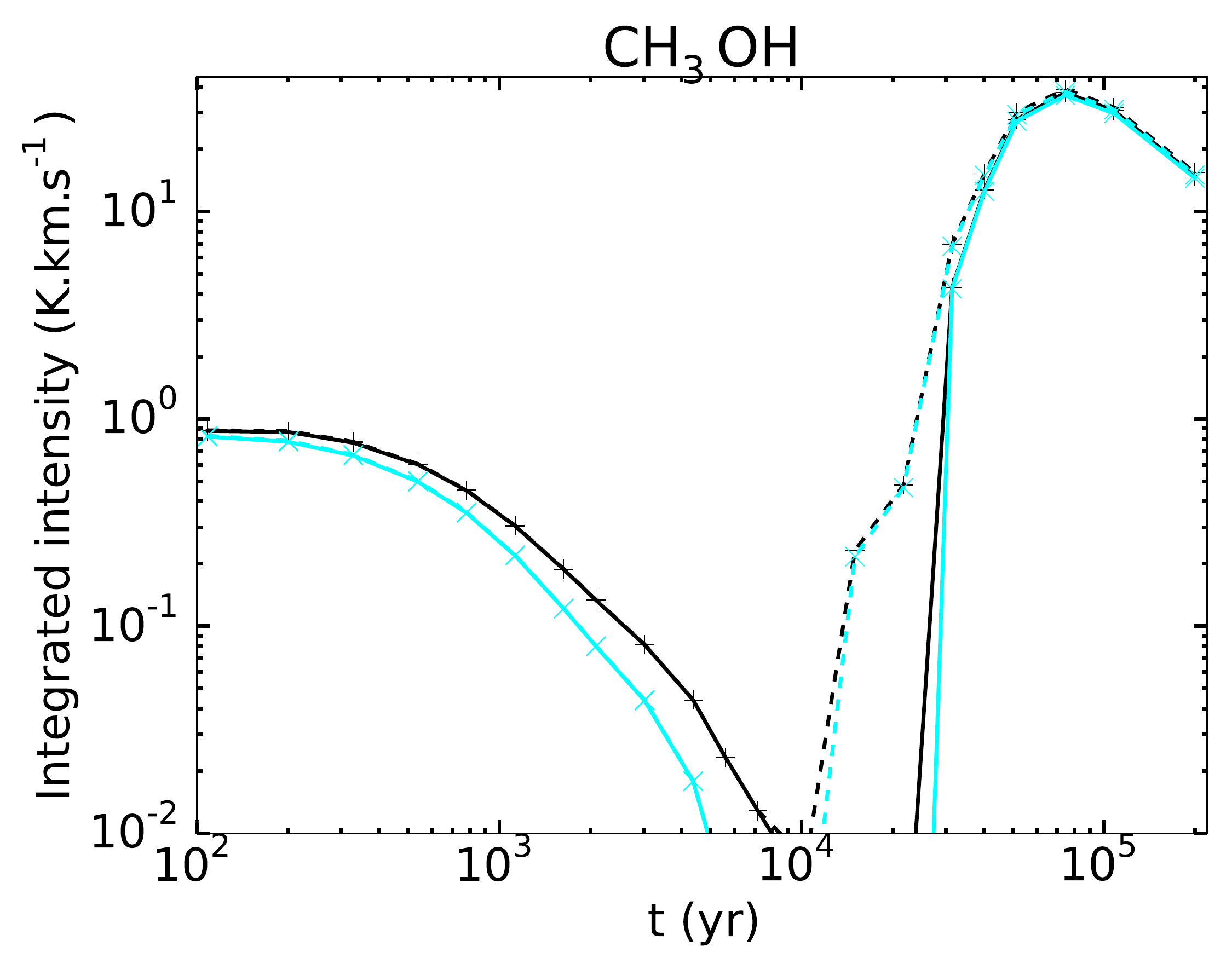} 

        \includegraphics[width=0.32\textwidth]{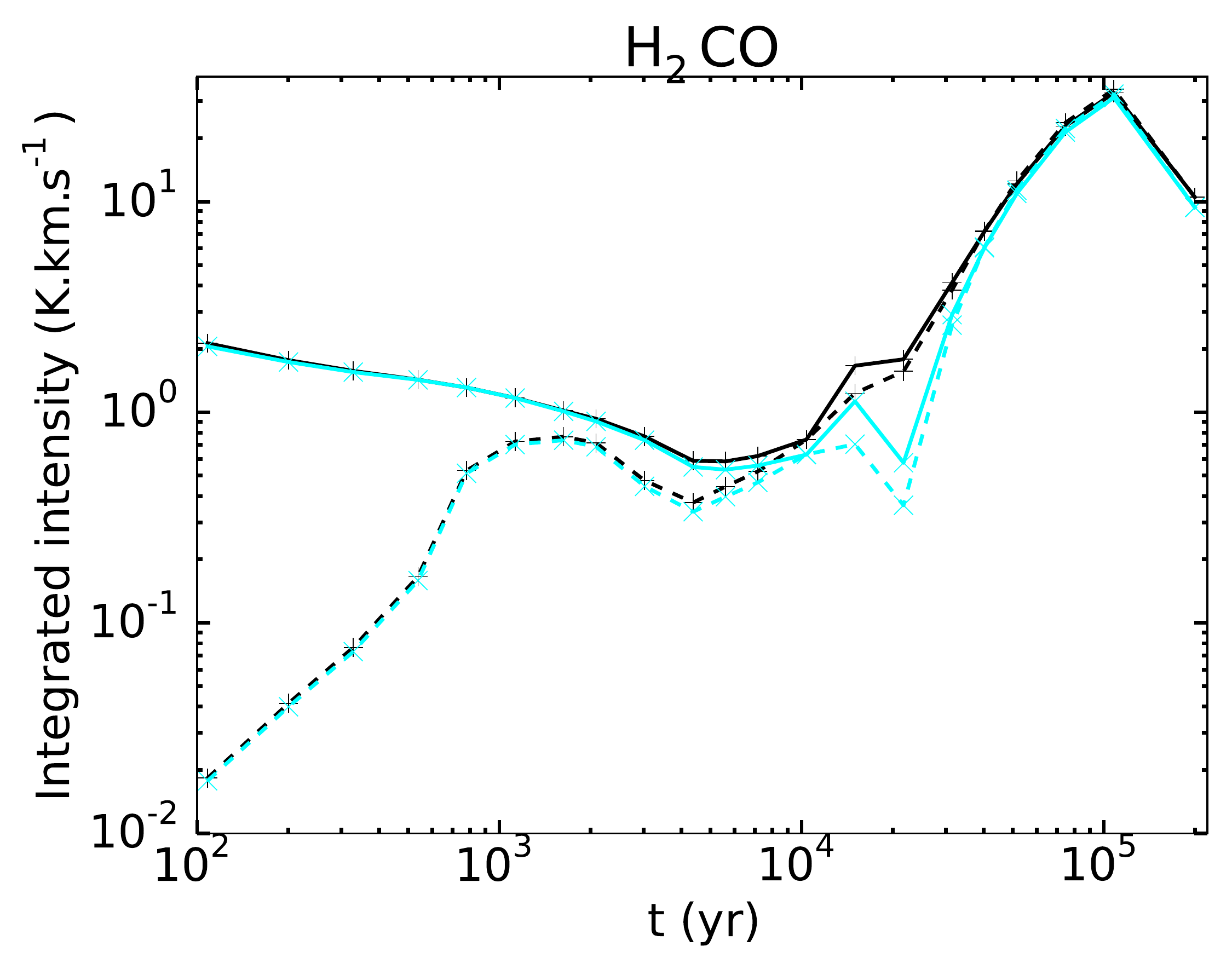}
        \includegraphics[width=0.32\textwidth]{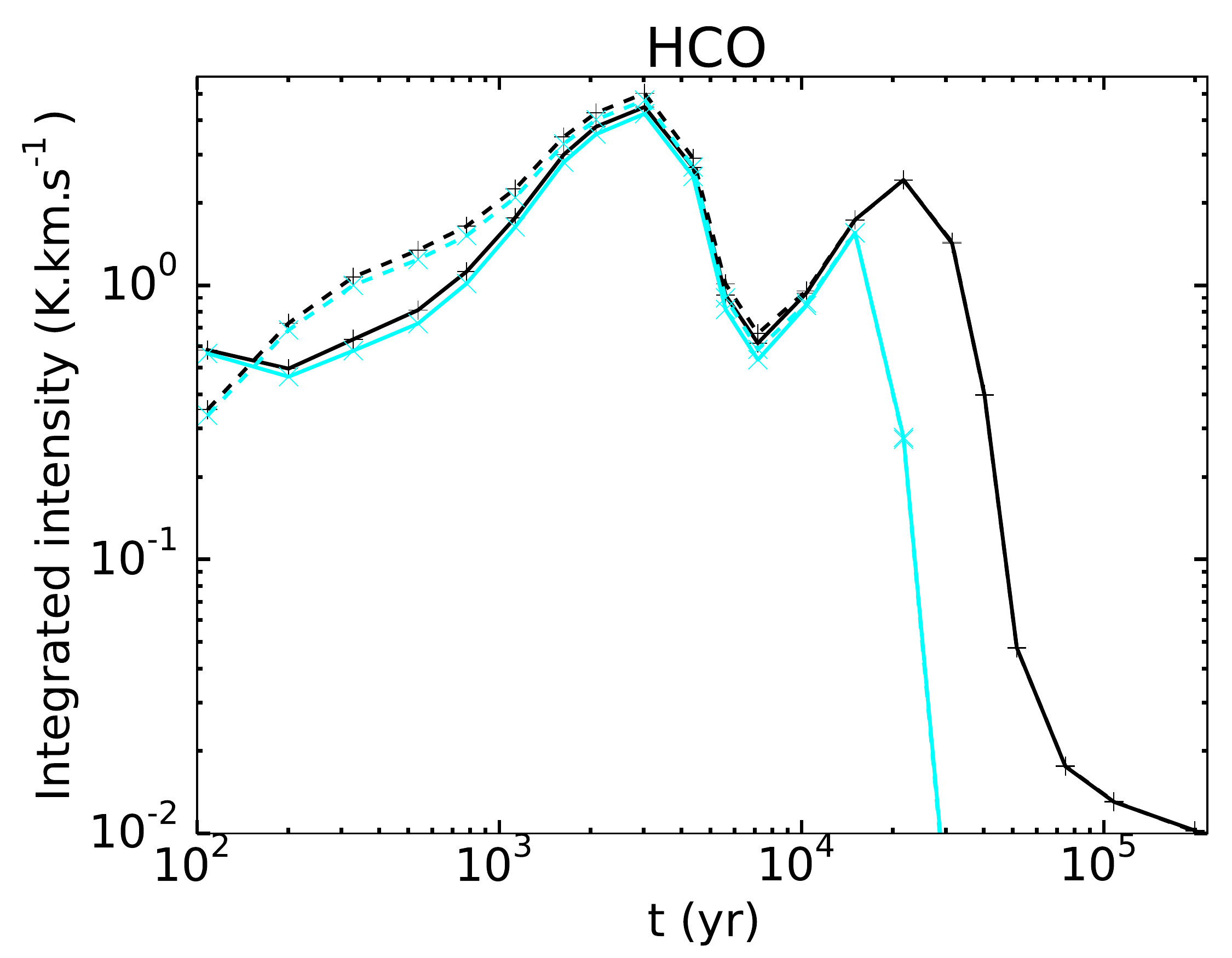}
        \includegraphics[width=0.32\textwidth]{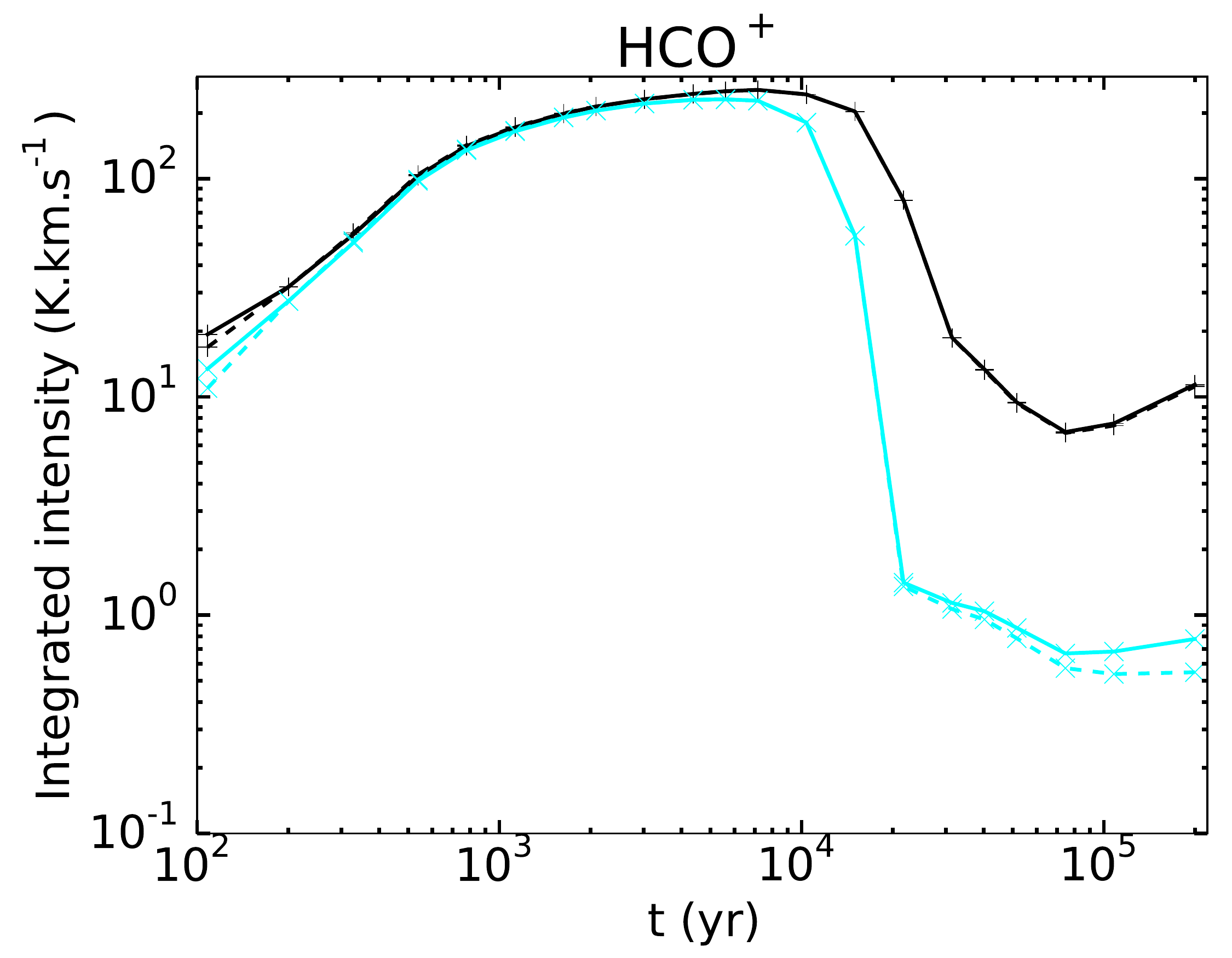} 

        \includegraphics[width=0.32\textwidth]{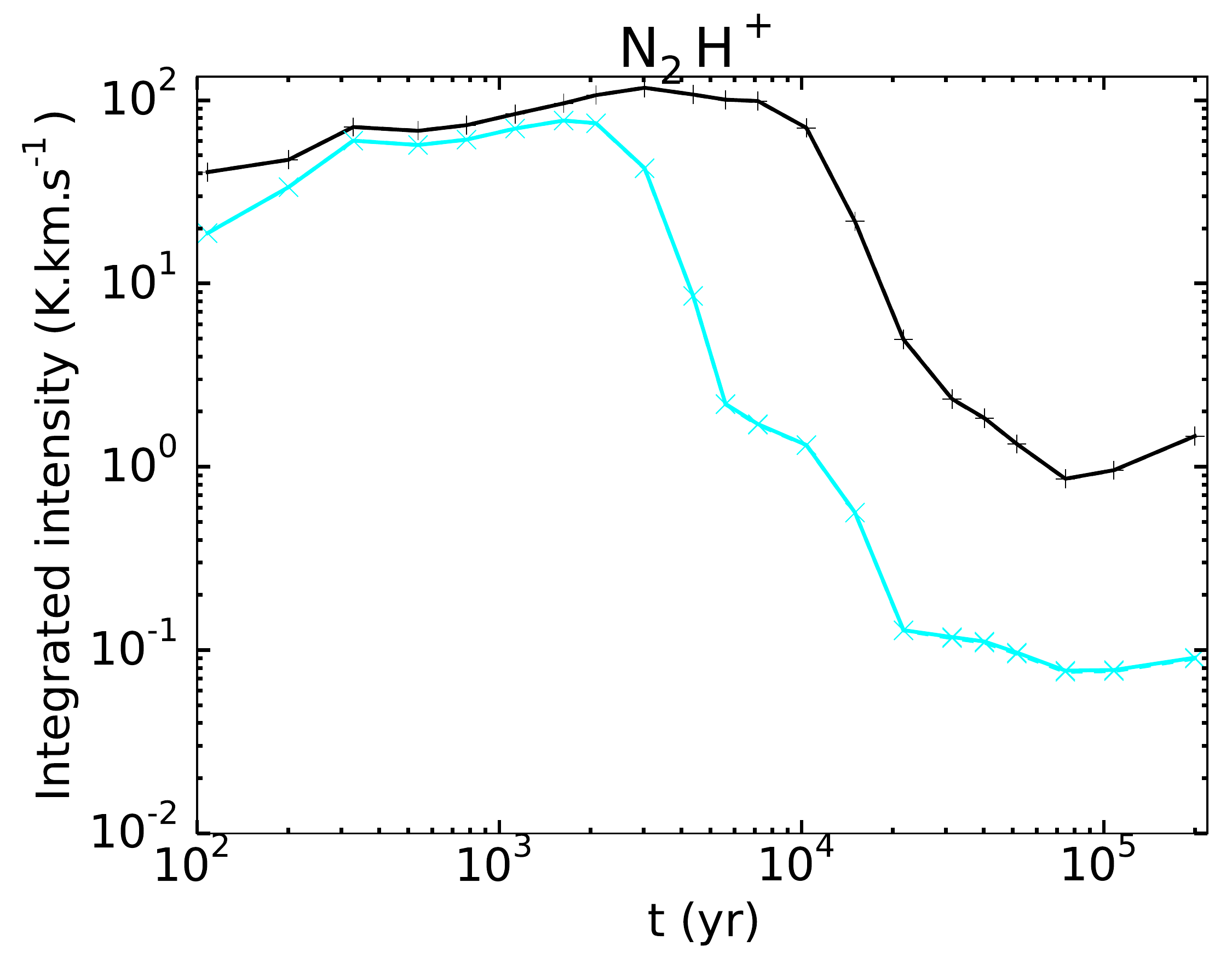}
        \includegraphics[width=0.32\textwidth]{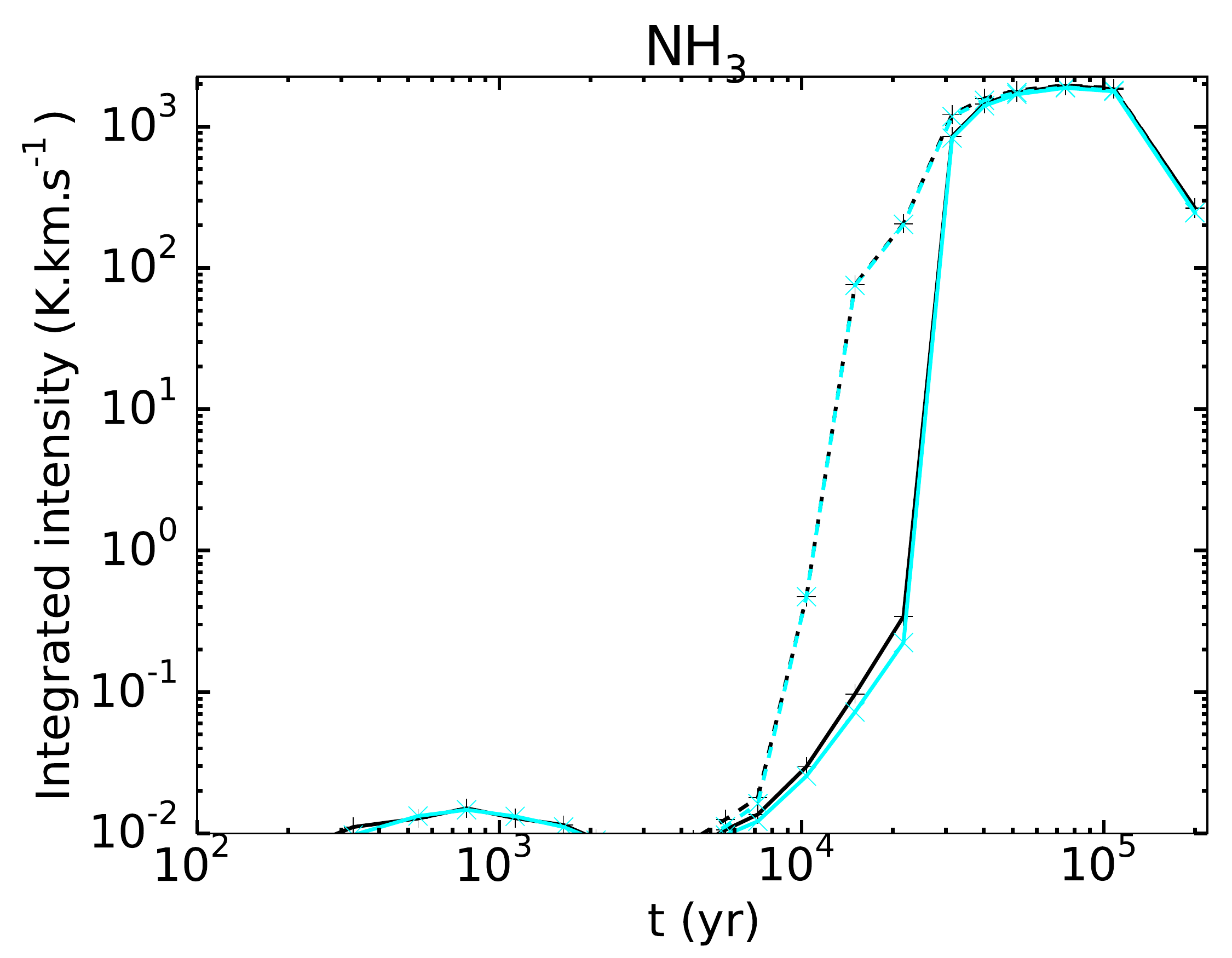}
        \includegraphics[width=0.32\textwidth]{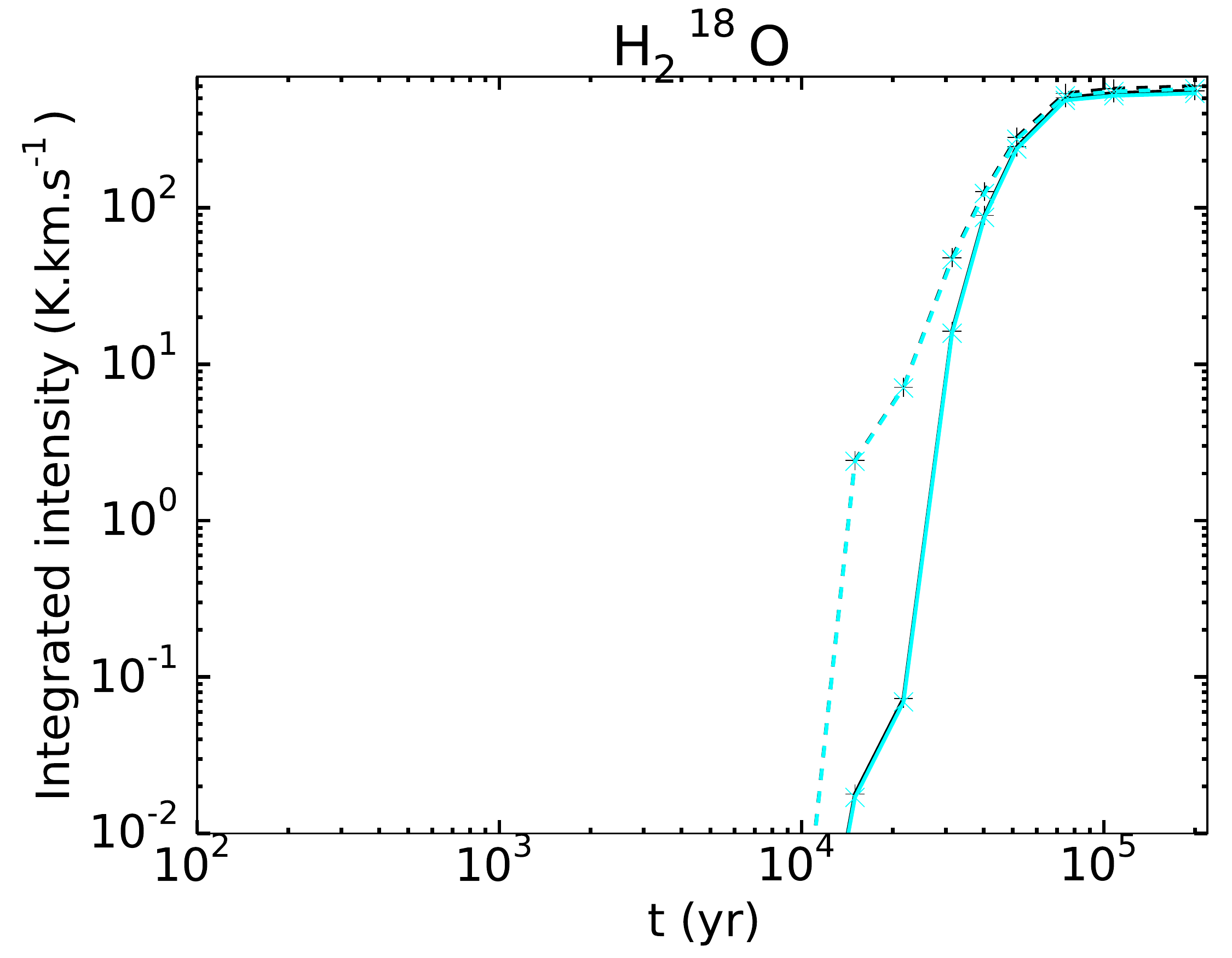} 

        \includegraphics[width=0.32\textwidth]{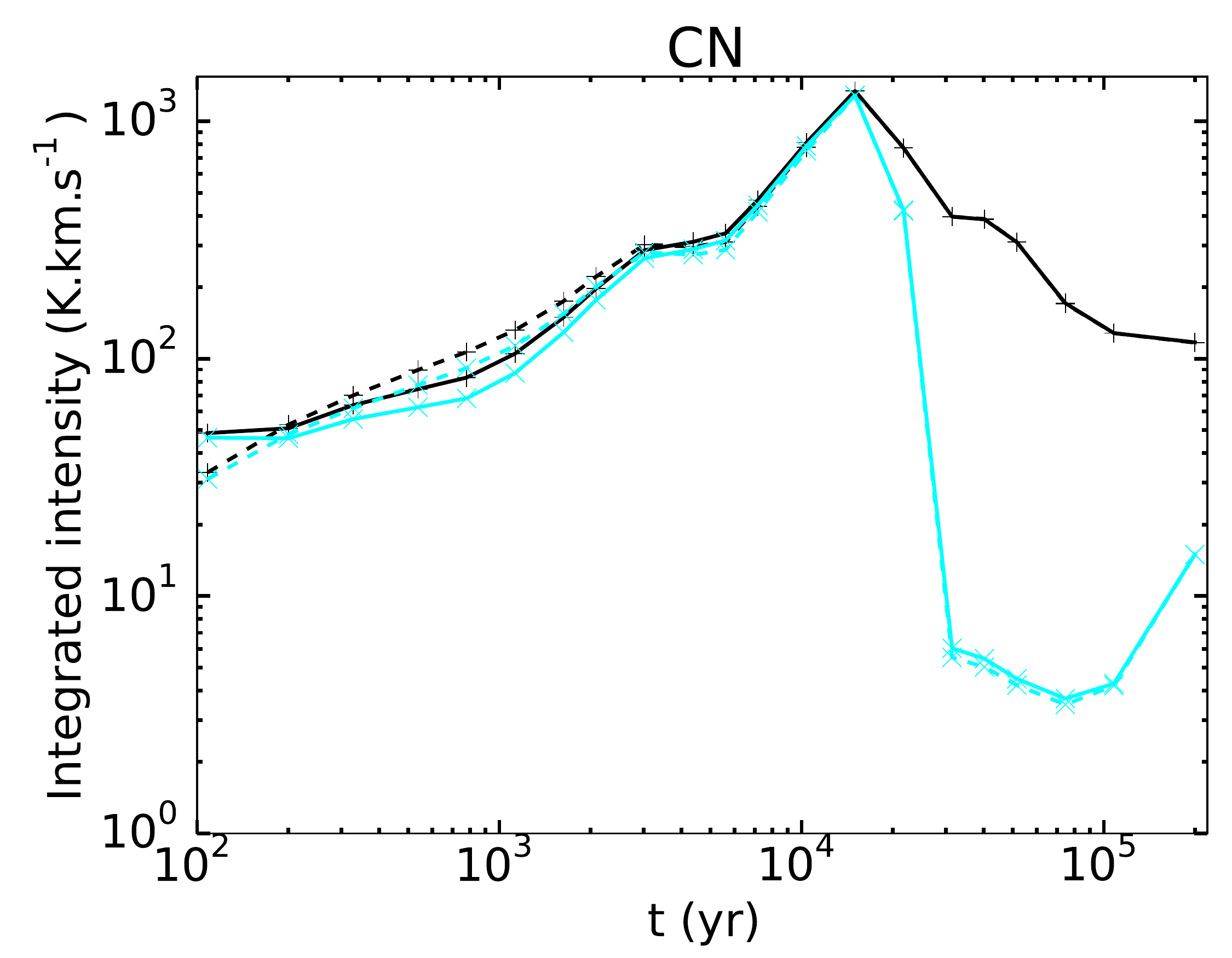} \
        \caption{{\bf Change of the cut-off density of the modeling grid \textit{c1} and \textit{c6}:} Time evolution of the integrated intensities of the selected species listed in Table~\ref{tab:selected-molecules}. Model \textit{mHII} is represented with solid line and model \textit{mHHMC} with dashed line. Model \textit{c1} is represented in black and model \textit{c6} in light blue.}
        \label{apfig:intInt_comp_HIIcut}
\end{figure*}

\end{appendix}

\end{document}